\documentclass[iop,apj,numberedappendix,appendixfloats]{emulateapj} 

\submitted{The Astrophysical Journal, 2012, accepted}

\shorttitle{CO emission from Orion protostars}
\shortauthors{Manoj et al.}

\begin{document} 

\title{ Herschel$^\star$/PACS Spectroscopic Survey of Protostars in Orion: The Origin of Far-Infrared CO Emission}
\thanks{$^\star${\it Herschel} is an ESA space observatory with science instruments provided by European-led Principal Investigator consortia and with important participation from NASA}

\author{P. Manoj\altaffilmark{1}, D. M. Watson\altaffilmark{1}, D. A. Neufeld\altaffilmark{2}, S. T. Megeath\altaffilmark{3}, R. Vavrek\altaffilmark{4}, Vincent Yu\altaffilmark{1},   R. Visser\altaffilmark{5}, E. A. Bergin\altaffilmark{5}, W. J. Fischer\altaffilmark{3}, J. J. Tobin\altaffilmark{6},  A. M. Stutz\altaffilmark{7,8}, B. Ali\altaffilmark{9}, T. L. Wilson\altaffilmark{10}, J. Di Francesco\altaffilmark{11}, M. Osorio\altaffilmark{12}, S. Maret\altaffilmark{13} and C. A. Poteet\altaffilmark{3}}
\affil{$^{1}$Department of Physics and Astronomy, University of Rochester, Rochester, NY 14627, USA; manoj@pas.rochester.edu}
\affil{$^{2}$Department of Physics and Astronomy, Johns Hopkins University, 3400 North Charles Street, Baltimore, MD 21218, USA}
\affil{$^{3}$Department of Physics and Astronomy, University of Toledo, 2801 West Bancroft Street, OH 43606, USA}
\affil{$^{4}$European Space Agency, ESAC/SRE-OAH, 28691 Villanueva de la Ca\~{n}ada P.O. Box 78, Madrid, Spain}
\affil{$^{5}$Department of Astronomy, University of Michigan, 500 Church Street, Ann Arbor, MI 48109, USA}
\affil{$^{6}$Hubble Fellow, National Radio Astronomy Observatory, Charlottesville, VA 22903, USA}
\affil{$^{7}$Max-Planck-Institut f\"{u}r Astronomie, K\"{o}nigstuhl 17, D-69117 Heidelberg, Germany}
\affil{$^{8}$Steward Observatory, University of Arizona, 933 North Cherry Avenue, Tucson, AZ 85721, USA}
\affil{$^{9}$NHSC/IPAC/Caltech, 770 S. Wilson Avenue, Pasadena, CA 91125, USA}
\affil{$^{10}$US Naval Research Laboratory, Code 7210, Washington, DC 20375, USA}
\affil{$^{11}$National Research Council of Canada, Herzberg Institute of Astrophysics, Department of Physics and Astronomy, University of Victoria, Victoria, BC V9E 2E7, Canada}
\affil{$^{12}$Instituto de Astrofisica de Andalucia, CSIC, Camino Bajo de H\'{u}etor 50, E-18008 Granada, Spain}
\affil{$^{13}$Laboratoire d'Astrophysique de Grenoble, Observatoire de Grenoble, Universit\'{e} Joseph Fourier, CNRS, UMR 571 Grenoble, France}

\begin{abstract}

We present far-infrared (57$-$196 $\micron$) spectra of 21 protostars
in the Orion molecular clouds. These were obtained with the
Photodetector Array Camera and Spectrometer (PACS) onboard the {\it
  Herschel} Space observatory, as part of the Herschel Orion Protostar
Survey (HOPS) program. We analyzed the emission lines from rotational
transitions of CO, involving rotational quantum numbers in the range
$J_{up}$ = 14$-$46, using PACS spectra extracted within a projected
distance of $\la$~2000~AU centered on the protostar. The total
luminosity of the CO lines observed with PACS ($L_{\mathrm{CO}}$) is
found to increase with increasing protostellar luminosity
($L_{bol}$). However, no significant correlation is found between
$L_{\mathrm{CO}}$ and evolutionary indicators or envelope properties
of the protostars such as bolometric temperature, $T_{bol}$ or
envelope density. The CO rotational (excitation) temperature implied
by the line ratios increases with increasing rotational quantum number
$J$, and at least 3$-$4 rotational temperature components are required
to fit the observed rotational diagram in the PACS wavelength
range. The rotational temperature components are remarkably invariant
between protostars and show no dependence on $L_{bol}$, $T_{bol}$ or
envelope density, implying that if the emitting gas is in {\it local
  thermodynamic equillibrium}, the CO emission must arise in multiple
temperature components that remain independent of $L_{bol}$ over two
orders of magnitudes. The observed CO emission can also be modeled as
arising from a single temperature gas component or from a medium with
a power-law temperature distribution; both of these require {\it
  sub-thermally} excited molecular gas at low densities
($n\mathrm{(H_2)}$ $\la$ 10$^6 $ cm$^{-3}$) and high temperatures
(T~$\ga$~2000~K). Our results suggest that the contribution from
photodissociation regions (PDRs), produced along the envelope cavity
walls from UV-heating, is unlikely to be the dominant component of the
CO emission observed with PACS. Instead, the `universality' of the
rotational temperatures and the observed correlation between
$L_{\mathrm{CO}}$ and $L_{bol}$ can most easily be explained if the
observed CO emission originates in shock-heated, hot (T~$\ga$~2000~K),
{\it sub-thermally} excited ($n\mathrm{(H_2)}$ $\la$ 10$^6 $
cm$^{-3}$) molecular gas. Post-shock gas at these densities is more
likely to be found within the outflow cavities along the molecular
outflow or along the cavity walls at radii $\ga$ several
100$-$1000~AU.

\end{abstract}

\keywords{circumstellar matter --- ISM: jets and outflows ---
  molecular processes --- stars: formation --- techniques:
  spectroscopic}

\section{Introduction}

The early evolution of protostars is driven by the competition between
infall and outflow \citep{shu87,sp05,mo07,hart09}. Energetic processes
associated with mass accretion and ejection heat up the surrounding
gas to temperatures of several 100~K to several 1000~K
\citep[e.g.,][]{vd09}. The infalling envelope material first lands on
the protostellar disk before accreting onto the central star
\citep{cm81,nh94,hart09}. Ultraviolet radiation produced by accretion
onto a protostar can heat up the circumstellar gas to a few 100~K
\citep{spaans95,vankemp09b}. Protostars drive powerful bipolar jets
and outflows into the surrounding medium, producing shocks, that
compress and heat up the envelope or ambient material to several
1000~K even at relatively large distances ($\ga$ 1000~AU) from the
protostar \citep{bach96,bt99, rb01,hm89,kn96a}. While the lowest
rotational transitions ($J$~$\le$~3) of CO have been widely used to
trace cold (T $\la$ 30~K) molecular gas associated with protostars,
the far-infrared (FIR) CO lines ($J$~$\ga$14) are better diagnostics
of warm and hot molecular gas \citep{watson80, storey81, giannini01,
  nisini02}. These latter lines can in principle be used to
characterize the density, temperature, molecular abundance and spatial
extent of the emitting region \citep{watson85b, drd83, hm89, kn96a}. The
spatial distribution and physical conditions in the emitting gas can
provide vital clues about the heating mechanisms and therefore also on
the various energetic processes associated with the earliest phases of
star formation.

The first detection of the FIR CO lines from an astronomical source
was toward Orion-KL, with the {\it Kuiper Airborne Observatory}
\citep[KAO;][]{watson80}. Because of low sensitivity, the KAO could
only observe the brightest regions around luminous and massive
protostars \citep{storey81, stacey82, stacey83, jaffe87}. These early
observations of very massive protostars suggested that FIR CO emission
arises in gas heated by non-dissociative, magnetohydrodynamic shocks
\citep{watson85b, storey81}. FIR CO lines toward several low-mass
protostars were later observed with the {\it Infrared Space
  Observatory} \citep[ISO;
  e.g.,][]{nisini97,ceccarelli98,nisini99a,benedettini00,saraceno99a,saraceno99b,giannini01,nisini02,vd04}. Modeling
based on the ISO data found that the observed CO emission originates
in molecular gas within a projected distance of $\la$ 1500~AU from the
protostar, with temperatures in the range of 200$-$2000~K and H$_2$
densities in the range of 10$^4$$-$10$^7$ cm$^{-3}$
\citep{ceccarelli98,nisini99a,benedettini00,giannini01,nisini02}. These
studies suggested slow ($v_s$ $\la$ 20~kms$^{-1}$) non-dissociative
shocks as the main mechanism for the heating of the molecular
gas. However, based on the similarity between the observed CO line
ratios of protostars and Herbig~Ae/Be stars \citet{ceccarelli00}
proposed that CO $J_{up}$=14$-$21 lines from protostars probably
originate in the dense Photodissociation Regions (PDRs) created by UV
photons from the central objects. Lack of spatial information (beam
size~$\sim$~80$\arcsec$) and low spectral resolution of the ISO-LWS,
however, prevented any further progress in disentangling the various
components contributing to the FIR CO emission.

With the Photodetector Array Camera and Spectrometer
\citep[PACS;][]{pog10} onboard the {\it Herschel} Space Observatory
\citep{pilbratt10}, it has now been possible to observe the rotational
transitions from the entire CO ladder from $J$=13$-$12 to $J$=49$-$48
in the FIR wavelength range
\citep{vankemp10b,vankemp10a,fich10,herczeg12,goico12} at high spatial
resolution (beam size $\sim$ 10$\arcsec$ at 63 $\micron$) and
unprecedented sensitivity. Modeling of the absolute and relative line
fluxes of more than 30 CO rotational transitions accessible with PACS
can be used to place strong constraints on the density and temperature
of the emitting medium \citep[e.g.][]{neufeld12}. However, the CO
lines observed with PACS are spectrally unresolved
\citep[e.g][]{vankemp10b,vankemp10a,herczeg12} rendering it difficult
to separate out the contribution from various velocity components of
the emitting gas to the total observed line flux. Spectrally resolved
observations of the lowest$-J$~CO lines ($J_{up}~\la~8$) towards
low-mass protostars in the submillimeter wavelengths have shown the
presence of narrow (FWHM~$\la$~2 km~s$^{-1}$) and relatively broad
(FWHM~$\sim$~10$-$30~km~s$^{-1}$) velocity components in the profiles
of these lines. The narrow component was found to dominate the total
line flux for these transitions \citep{vankemp09a,vankemp09b,vd09,
  yildiz12}. For lines with $J_{up} \la 4$, these studies attributed
the narrow component as due to emission from passively heated
protostellar envelope; for CO~$J$~=~6$-$5 and CO~$J$~=~7$-$6
transitions the narrow component was found to be dominated by emission
from UV-photon heated gas.  Early modeling of the unresolved CO lines
observed with Herschel/PACS ($J_{up}~\ga~14$) towards a few protostars
suggested that the emission from UV-heated PDRs dominated the line
flux even in the PACS range up to $J_{up}$ = 20
\citep{vankemp10a}. These studies have shown that two separate
components are required to explain the CO lines observed with
PACS. Emission from PDRs, produced along the envelope cavity walls due
to UV-heating, was found to be the dominant component of the lower$-J$
lines ($J_{up}$ = 14$-$20) accessible to PACS. The higher$-J$ CO lines
($J_{up}$~$\ga$~25) were found to be dominated by the emission from
the small scale shocks along the cavity walls \citep{vankemp10a}. A
more detailed modeling of the CO emission from three protostars with
similar bolometric luminosities by \citet{visser12} essentially
confirmed this picture, except that in these models the PDR component
can dominate emission even for the high$-J$ CO lines, if the density
of the envelope is sufficiently low. Thus, those authors tentatively
suggested an evolutionary trend, in which the CO emission is dominated
by shocks in the youngest source in their sample and by UV-heated gas
in the oldest source with low envelope density \citep{visser12}.
Herschel/PACS spectra of only a few sources have been studied so far,
however. Analysis of a larger sample of protostars with a range in
luminosity, evolutionary status, and envelope properties is required
before the above suggestions can be confirmed. In this paper, we carry
out such a study.

\begin{deluxetable*}{lllccc}
\tablewidth{0pt}
\tablecaption{Protostellar properties \label{sample_tbl}}
\tablehead{
\colhead{HOPS~ID} & \colhead{RA (J2000)} & \colhead{DEC (J2000)} & \colhead{$L_{bol}$} & \colhead{$T_{bol}$} & \colhead{$\rho_{1}$\tablenotemark{a}}\\
\colhead{} & \colhead{} & \colhead{} & \colhead{($L_{\odot}$)} & \colhead{(K)} & \colhead{($\times~10^{-14}$~g~cm$^{-3}$)}
}

\startdata

10&  5 35 9.005 &  -5 58 27.5&    3&   50&       6\\
 11&  5 35 13.409&  -5 57 58.1&    8&   59&      37\\
 30&  5 34 44.062&  -5 41 25.8&    4&   87&       6\\
 32&  5 34 35.450&  -5 39 59.1&    2&   60&       6\\
 56&  5 35 19.466&  -5 15 32.7&   18&   47&     185\\
 60&  5 35 23.328&  -5 12 3.06&   21&   60&       6\\
 68&  5 35 24.305&  -5 8 30.59&    5&   99&      19\\
 84&  5 35 26.570&  -5 3 55.12&   45&   93&       6\\
 85&  5 35 28.183&  -5 3 40.93&   13&  157&       4\\
 87&  5 35 23.472&  -5 1 28.70&   27&   39&     185\\
 91&  5 35 18.914&  -5 0 50.87&    5&   38&     370\\
108&  5 35 27.074&  -5 10 0.37&   57&   41&     185\\
182&  5 36 18.833&  -6 22 10.2&   62&   63&      19\\
203&  5 36 22.838&  -6 46 6.20&   15&   45&      74\\
288&  5 39 55.944&  -7 30 27.9&  103&   57&      37\\
310&  5 42 27.677&  -1 20 1.00&   11&   55&      19\\
329&  5 47 1.606 &  0 17 58.88&    3&   77&       4\\
343&  5 47 59.030&  0 35 32.86&    4&   88&       6\\
368&  5 35 24.725&  -5 10 30.2&   55&  154&       4\\
369&  5 35 26.971&  -5 10 17.1&   20&   42&      90\\
370&  5 35 27.629&  -5 9 33.48&  217&   79&       7\\

\enddata

\tablenotetext{a}{Fiducial envelope density, calculated at 1~AU in the limit of no rotation \citep[see][]{kch93}.}

\end{deluxetable*}

Here we present FIR (57$-$196 $\micron$) spectra of 21 protostars in
the Orion molecular clouds observed with {\it Herschel}/PACS. We
obtained them as part of the {\it Herschel} Orion Protostar Survey
(HOPS), an open time Key program with {\it Herschel}
\citep{fischer10,stanke10,tm12}. Our sample of protostars in Orion is
the largest such sample at a common distance for which FIR spectra
over the entire PACS wavelength range have been obtained and analyzed
so far.  Our analysis begins with a search for possible correlations
between the observed CO emission properties and protostellar
luminosity and evolutionary status. We then model the observed CO
fluxes and line ratios to derive the excitation conditions of the warm
and hot circumstellar gas and investigate the physical origin of CO
emission from protostars.  Our sample is described in
\S~\ref{sample}. Observations and data reduction are described in
\S~\ref{obs}. In \S~\ref{CO_prop}, we present the observed properties
of the CO emission from protostars in Orion and discuss various
correlations found between CO emission and protostellar properties. In
\S~\ref{excitation}, we model the observed CO rotational diagrams to
constrain the physical conditions of the emitting gas and discuss the
results. The physical origin of the FIR CO emission from protostars is
explored in \S~\ref{origin}. Finally, our conclusions are summarized
in \S~\ref{conclude}.

\begin{figure*}
\epsscale{1.1}
\plottwo{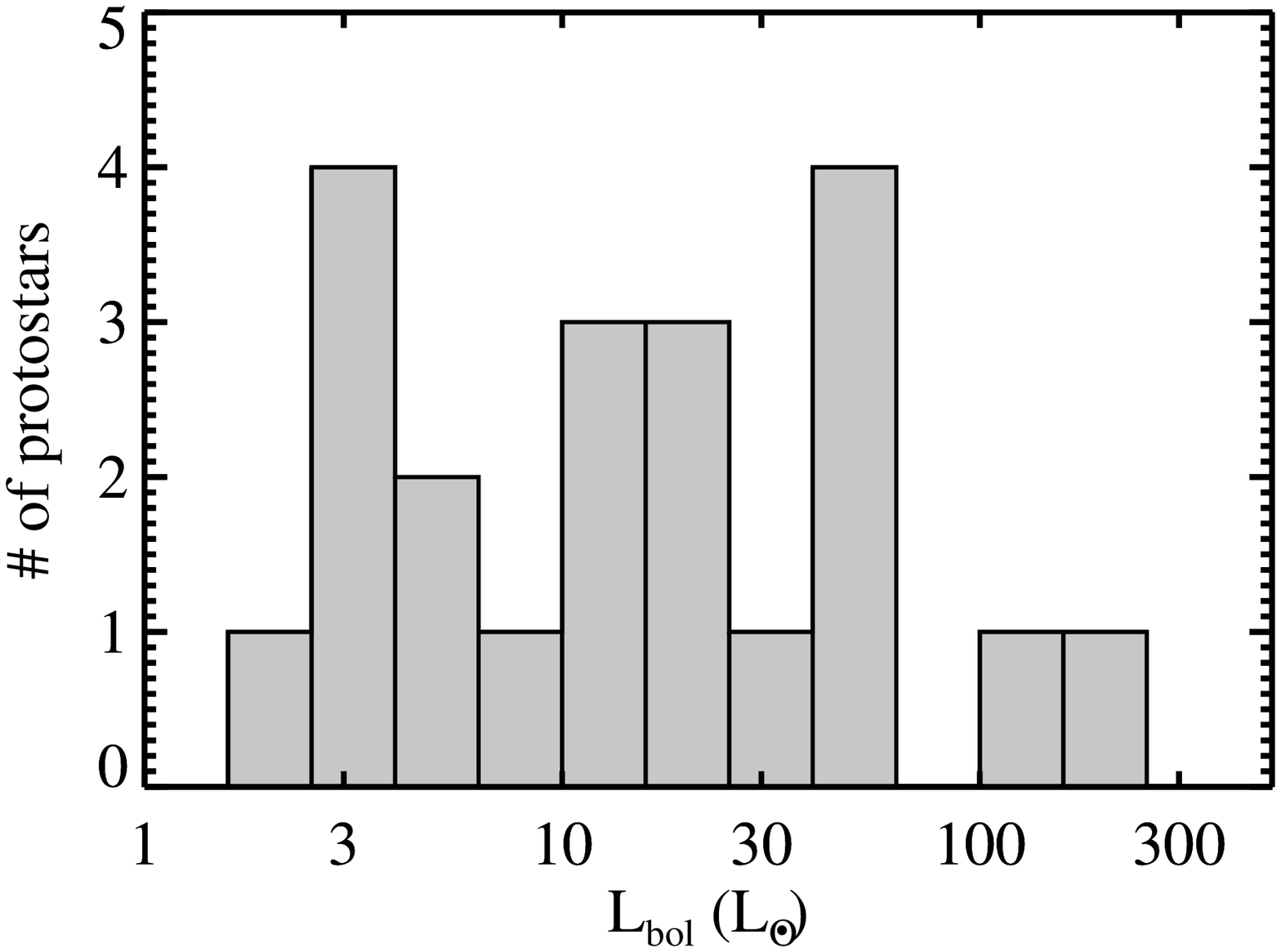}{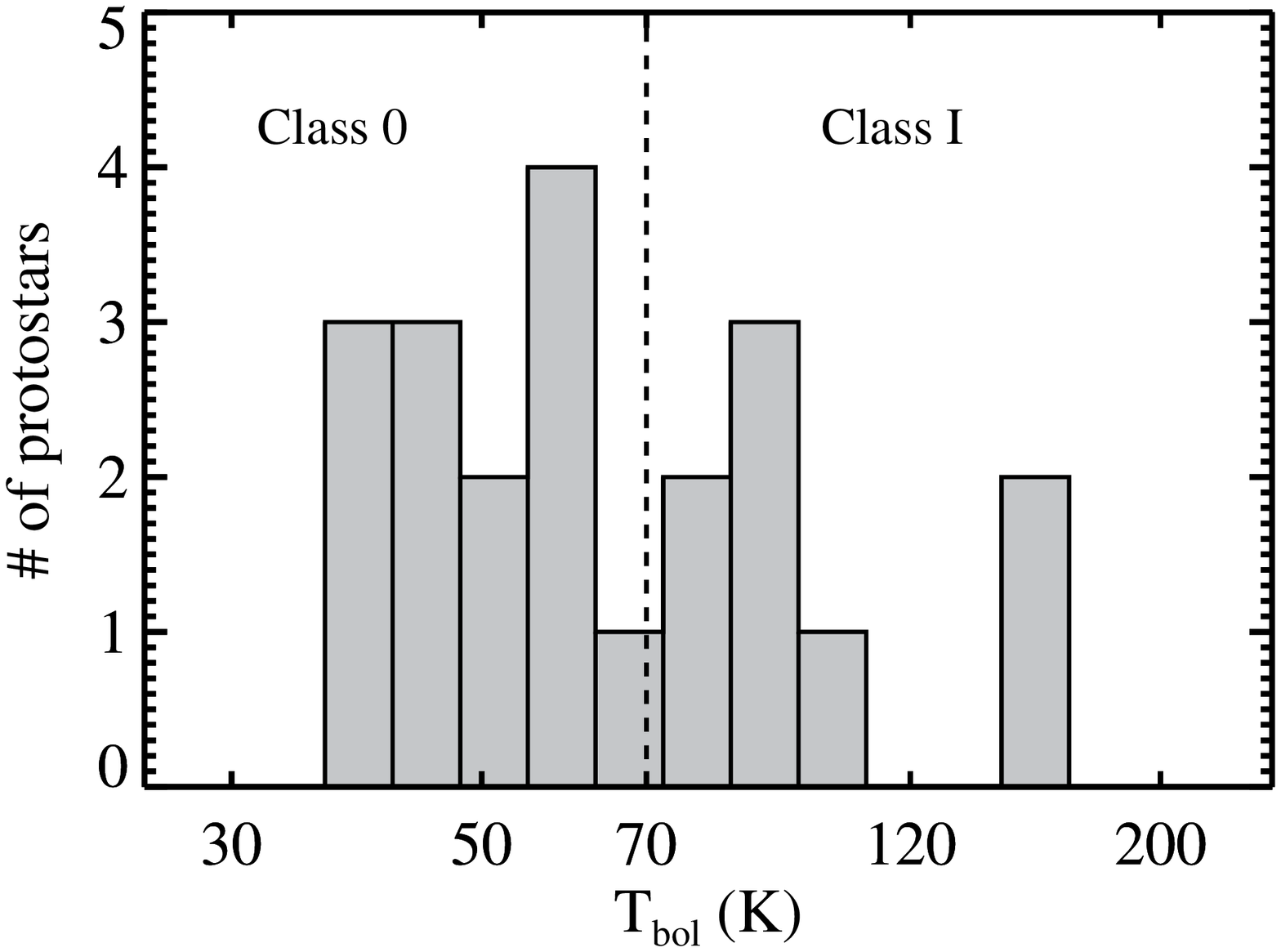}
\caption{Distribution of $L_{bol}$ and $T_{bol}$ of the protostars for which PACS spectra are presented. $T_{bol}$ of 70~K is generally taken as the dividing line between Class~0 and Class~I sources \citep{chen95,evans09} \label{blt_hist}}
\end{figure*}

\begin{figure*}
\epsscale{1.1}
\plottwo{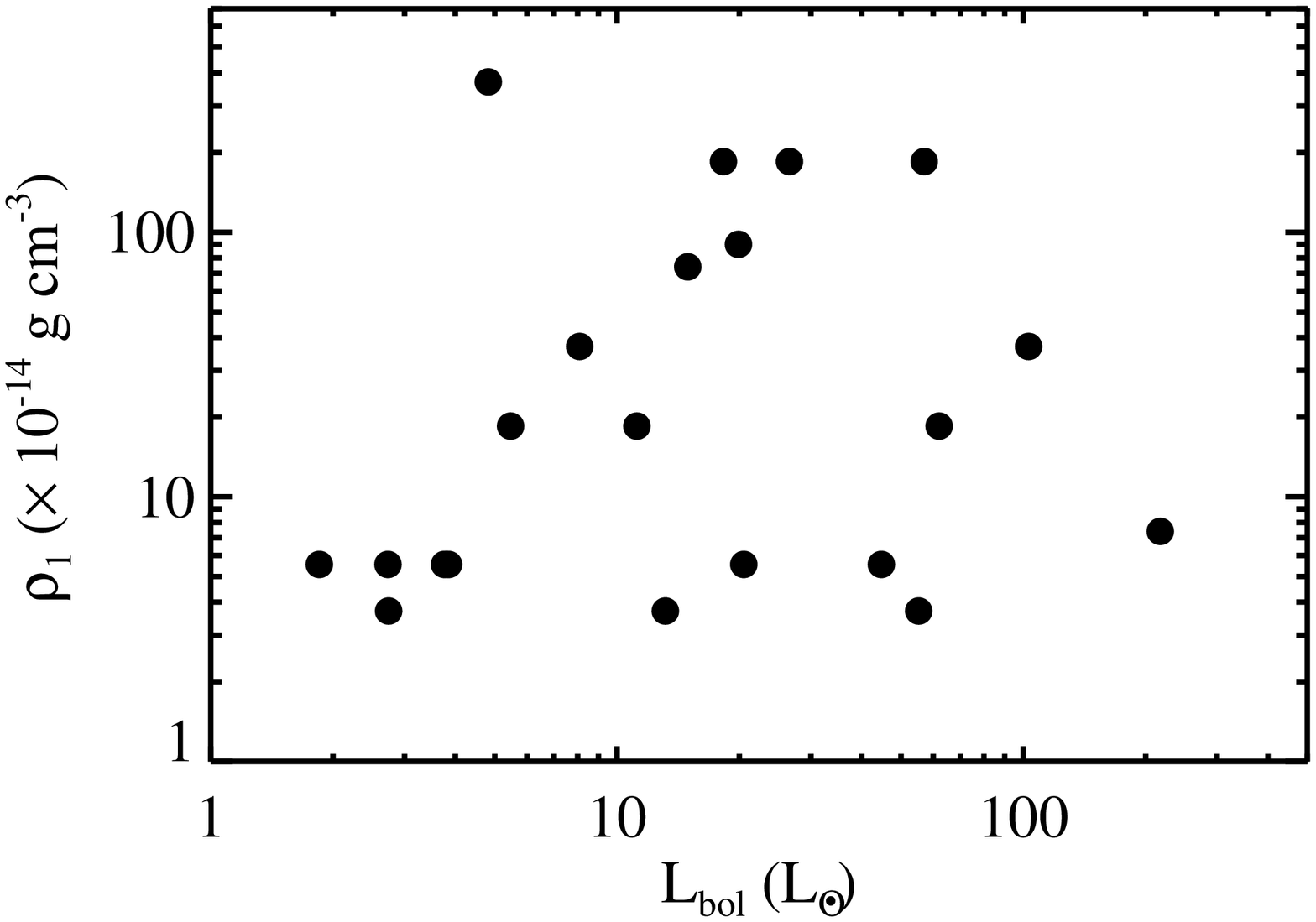}{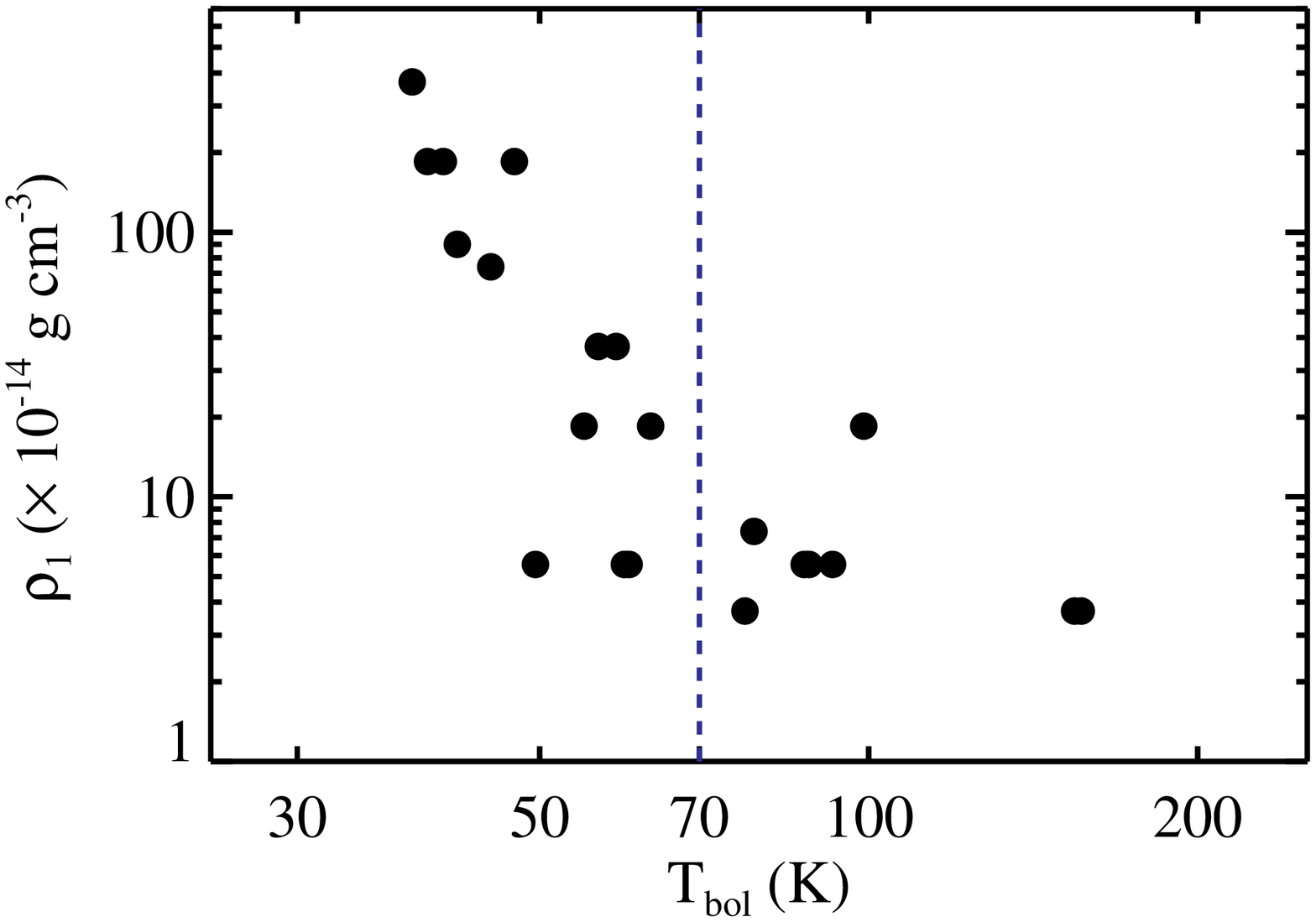}
\caption{$\rho_{1}$ (see text, \S~2), as a function of $L_{bol}$ and
  $T_{bol}$. \label{rho1_lbol_tbol}}
\end{figure*}

\section{HOPS spectroscopic  Sample} \label{sample}

The HOPS sample for the PACS spectroscopic observations was selected
from 300 {\it Spitzer} identified protostars in Orion with detectable
24 micron emission \citep{kry12,tm12}. We first selected objects with
$F_{\nu}$~$\ge$~7~mJy at 24~$\micron$ with Class~0 or Class~I spectral
indices and other strong indications, such as deep silicate and ice
features in the {\it Spitzer} IRS spectra, that they are indeed
protostars. In this list we searched for objects that resemble, at
infrared wavelengths, the nearly face-on protostar NGC~1333~IRAS~4B:
evidence of compact ($\la$~10~$\arcsec$ from the continuum position at
$\ga$~24~$\micron$) scattered light or H$_2$ line emission at 2.3,
3.6, 4.5 or 24~$\micron$ and no evidence for long jets or dust lanes
suggestive of a near-edge-on orientation. Our motive here was for the
sample to contain many objects for which our view to the dense inner
regions would be relatively unextinguished. We initially identified 44
such protostars out of the parent sample of 300, all of which we
observed with {\it Spitzer} IRS in its highest spectral resolution
\citetext{Watson et al. 2012, in preparation}. From these 44 sources,
we chose 31 objects with bolometric luminosity uniformly sampled from
0.5 to 30~$L_{\odot}$ to observe with PACS. An additional 5 objects
with bolometric luminosity in the range of $\sim$20$-$200~$L_{\odot}$
were added to the PACS target list without regard to orientation. In
total, 36 protostars were observed with PACS in the spectroscopy mode
as part of the HOPS program.  Here we present PACS spectra of the
brightest 21 sources in our sample, roughly uniformly sampling
bolometric luminosity in the range of $\sim$1$-$200~$L_{\odot}$. The
FIR spectra of the full HOPS sample will be presented in a later
paper. Due to the difficulty in determining the precise orientation of
the protostars, many of the sources which were suspected to be close
to face-on may actually be observed at a more inclined orientation.  A
future paper will use modeling of the SEDs as well as HST 1.6~$\mu$m
images to better constrain the inclinations for the HOPS protostars.

The observed spectral energy distributions (SEDs) of the protostars in
our sample are presented in Appendix~\ref{appen_a}.  Through several
ancillary observing programs that complement HOPS, we obtained
extensive wavelength coverage for the sources in our sample resulting
in well-sampled SEDs. From these SEDs, we computed the bolometric
luminosity ($L_{bol}$) and bolometric temperature ($T_{bol}$) of each
source. $L_{bol}$ was calculated by integrating the SED over
frequency. Since the SEDs are sampled at a finite number of
frequencies, we interpolated over the intervening frequencies
following the midpoint interpolation method described in
\citet{enoch09}. The flux upper limits were removed and we
interpolated over them before integration. Further, the SEDs were
extrapolated from the non-upper limit flux at the longest observed
wavelength (in most cases 870~$\micron$ or 350~$\micron$) using
$F_{\nu}$~$\propto$~$\nu^2$.  $T_{bol}$, which is defined as the
temperature of a blackbody with the same mean frequency as the source
SED, was computed from the mean frequency of the source SED, following
the method of \citet{ml93} and \citet{enoch09}. Given our wide
wavelength coverage, and, in particular, given that the peak of the
SED is well characterized, our integration method accounts for most of
the luminosity, and the $L_{bol}$ and $T_{bol}$ are well constrained.
The dominant source of uncertainity in $L_{bol}$ and $T_{bol}$ is from
the finite sampling errors, typically $\sim$10$-$15\%. One of the
sources in our sample, HOPS~369, has been modeled as a double source,
i.e., a disk-dominated source plus a colder, envelope-dominated source
by \citet[the source SOF~4 in the paper]{adams12} . The envelope
source dominate the observed SED only at wavelengths
$\ga$~37~$\micron$ \citep[see Figure~2 in][]{adams12}; so we
integrated the SED longward of 37~$\micron$ to obtain $L_{bol}$ and
$T_{bol}$ for HOPS~369. The $L_{bol}$ and $T_{bol}$ computed for the
protostars in our sample are listed in Table~\ref{sample_tbl}.

The distribution of the bolometric luminosities and temperatures of
the protostars in our sample are presented in
Figure~\ref{blt_hist}. Protostars in our sample span a large range in
$L_{bol}$; from $\sim$ 1.9~$L_{\odot}$ to 217~$L_{\odot}$. They
generally are in different evolutionary phases. Protostars are
generally classified in to Class~0 and Class~I sources, two different,
but related, evolutionary phases. Class~0 sources are the youngest
protostars and have most of their mass still in the infalling envelope
\citep{andre93,andremont94}; they are highly obscured objects,
invisible -- in some cases -- even at mid-infrared wavelengths. The
more-evolved Class~I sources have envelopes of mass less than the
protostar itself and are visible in the near-infrared wavelengths
\citep{lada87, wilk89}. $T_{bol}$ of 70~K is generally taken as the
dividing line between Class~0 and Class~I sources
\citep{chen95,evans09}. Based on this criterion, our sample has 13
Class~0 sources and 8 Class~I sources.

\begin{deluxetable*}{lccccll}
\tablewidth{0pt}
\tablecaption{Log of PACS spectroscopy observations \label{log_tbl}}
\tablehead{
\colhead{HOPS~ID} & \colhead{OBSID} & \colhead{OD} & \colhead{Date} & \colhead{Total time} &\colhead{Observing mode} &\colhead{Primary wavelength}\\
\colhead{} & \colhead{} & \colhead{} & \colhead{} & \colhead{(s)} &\colhead{}  &\colhead{ranges observed }\\
\colhead{} & \colhead{} & \colhead{} & \colhead{} & \colhead{(s)} &\colhead{}  &\colhead{($\micron$)}\\
}

\startdata

     10&     1342215693&    665&      10 Mar 2011&   3079&     Pointed/Unchopped&      57-71 \& 102-142          \\
       &     1342215694&    665&      10 Mar 2011&   1930&     Pointed/Unchopped&                 71-98   \\[0.3cm]
     11&     1342204115&    482&      07 Sep 2010&   5156&      Pointed/Chop-Nod&      71-98 \& 102-142          \\
       &     1342204116&    482&      07 Sep 2010&   3369&      Pointed/Chop-Nod&                 57-71   \\[0.3cm]
     30&     1342215697&    665&      10 Mar 2011&   3079&     Pointed/Unchopped&      57-71 \& 102-142          \\
       &     1342215698&    665&      10 Mar 2011&   1930&     Pointed/Unchopped&                 71-98   \\[0.3cm]
     32&     1342192115&    303&      12 Mar 2010&   3129&      Pointed/Chop-Nod&                57-71           \\
       &     1342192116&    303&      12 Mar 2010&   5020&      Pointed/Chop-Nod&      71-98 \& 102-142   \\[0.3cm]
     56&     1342227334&    833&      24 Aug 2011&   3079&     Pointed/Unchopped&      57-71 \& 102-142          \\
       &     1342227335&    833&      24 Aug 2011&   1930&     Pointed/Unchopped&                 71-98   \\[0.3cm]
     60&     1342227611&    839&      30 Aug 2011&   8933&     Mapping/Unchopped&                 57-71          \\
       &     1342227613&    839&      30 Aug 2011&   8933&     Mapping/Unchopped&                 57-71          \\
       &     1342227616&    839&      30 Aug 2011&   6083&     Mapping/Unchopped&      71-98 \& 102-142          \\
       &     1342227618&    839&      30 Aug 2011&   6083&     Mapping/Unchopped&     71-98 \& 102-142    \\[0.3cm]
     68&     1342226202&    823&      14 Aug 2011&   3079&     Pointed/Unchopped&      57-71 \& 102-142          \\
       &     1342226203&    823&      14 Aug 2011&   1930&     Pointed/Unchopped&                71-98    \\[0.3cm]
     84&     1342227338&    833&      25 Aug 2011&   3079&     Pointed/Unchopped&      57-71 \& 102-142          \\
       &     1342227339&    833&      25 Aug 2011&   1930&     Pointed/Unchopped&                 71-98   \\[0.3cm]
     85&     1342215657&    664&       8 Mar 2011&   8933&     Mapping/Unchopped&                 57-71          \\
       &     1342215659&    664&       8 Mar 2011&   8933&     Mapping/Unchopped&                 57-71          \\
       &     1342215662&    664&       8 Mar 2011&   6083&     Mapping/Unchopped&      71-98 \& 102-142          \\
       &     1342215664&    664&       9 Mar 2011&   6083&     Mapping/Unchopped&      71-98 \& 102-142   \\
     87&     1342215650&    664&       8 Mar 2011&   3079&     Pointed/Unchopped&      57-71 \& 102-142          \\
       &     1342215651&    664&       8 Mar 2011&   1930&     Pointed/Unchopped&                 71-98   \\[0.3cm]
     91&     1342215654&    664&       8 Mar 2011&   3079&     Pointed/Unchopped&      57-71 \& 102-142          \\
       &     1342215655&    664&       8 Mar 2011&   1930&     Pointed/Unchopped&                71-98    \\[0.3cm]
    108&     1342239690&   1020&      27 Feb 2012&   3079&     Pointed/Unchopped&      57-71 \& 102-142          \\
       &     1342239693&   1020&      27 Feb 2012&   1930&     Pointed/Unchopped&                 71-98   \\[0.3cm]
    182&     1342226753&    826&      18 Aug 2011&   3079&     Pointed/Unchopped&      57-71 \& 102-142          \\
       &     1342226754&    826&      18 Aug 2011&   1930&     Pointed/Unchopped&                 71-98   \\[0.3cm]
    203&     1342191363&    290&      28 Feb 2010&   3129&      Pointed/Chop-Nod&                57-71           \\
       &     1342191364&    290&      28 Feb 2010&   5020&      Pointed/Chop-Nod&      71-98 \& 102-142   \\[0.3cm]
    288&     1342227754&    838&      29 Aug 2011&   8933&     Mapping/Unchopped&                 57-71          \\
       &     1342227756&    838&      29 Aug 2011&   8933&     Mapping/Unchopped&                 57-71          \\
       &     1342227759&    838&      29 Aug 2011&   6083&     Mapping/Unchopped&      71-98 \& 102-142          \\
       &     1342227761&    838&      29 Aug 2011&   6083&     Mapping/Unchopped&      71-98 \& 102-142   \\
    310&     1342215689&    665&      10 Mar 2011&   3079&     Pointed/Unchopped&      57-71 \& 102-142          \\
       &     1342215689&    665&      10 Mar 2011&   1930&     Pointed/Unchopped&                71-98    \\[0.3cm]
    329&     1342217855&    690&      04 Apr 2011&   3079&     Pointed/Unchopped&      57-71 \& 102-142          \\
       &     1342217856&    690&      04 Apr 2011&   1930&     Pointed/Unchopped&                71-98    \\[0.3cm]
    343&     1342217851&    690&      04 Apr 2011&   3079&     Pointed/Unchopped&      57-71 \& 102-142          \\
       &     1342217852&    690&      04 Apr 2011&   1930&     Pointed/Unchopped&                 71-98   \\[0.3cm]
    368&     1342227340&    833&      25 Aug 2011&   3079&     Pointed/Unchopped&      57-71 \& 102-142          \\
       &     1342227341&    833&      25 Aug 2011&   1930&     Pointed/Unchopped&                 71-98   \\[0.3cm]
    369&     1342227344&    833&      25 Aug 2011&   3079&     Pointed/Unchopped&      57-71 \& 102-142          \\
       &     1342227345&    833&      25 Aug 2011&   1930&     Pointed/Unchopped&                 71-98   \\[0.3cm]
    370&     1342227764&    838&      29 Aug 2011&   8933&     Mapping/Unchopped&                 57-71          \\
       &     1342227766&    838&      30 Aug 2011&   8933&     Mapping/Unchopped&                 57-71          \\
       &     1342227769&    838&      30 Aug 2011&   6083&     Mapping/Unchopped&      71-98 \& 102-142          \\
       &     1342227771&    838&      30 Aug 2011&   6083&     Mapping/Unchopped&      71-98 \& 102-142          \\

\enddata
\end{deluxetable*}

To follow the evolution of the physical conditions in protostellar
envelopes, we estimated the envelope densities of the protostars in
our sample by modeling their SEDs. We assume the density follows the
rotating collapse solution of \citet{tsc84}.  While the density varies
throughout the envelope, we adopt the approach of \citet{kch93} and
use a fiducial density $\rho_1$, calculated at 1 AU in the limit of no
rotation, to track the envelope density.  We computed $\rho_{1}$ by
fitting the observed SEDs of the protostars with a large grid of
models created with the Monte Carlo radiative transfer code of
\citet{whitney03}. The detailed description of the grid and the
fitting procedure can be found in \citet{tobin08}, \citet{ali10} and
\citet{fischer10,fischer12}.  For six sources in our sample (HOPS~60,
68, 108, 368, 369, 370), SED modeling results have already been
published \citep{adams12,poteet11}; for these objects, values of
$\rho_{1}$ obtained from our modeling agreed with those published
within a factor of 2$-$3. For the composite SED of HOPS~369, we
adopted the $\rho_{1}$ value derived for the envelope source by
\citet{adams12}. The envelope $\rho_{1}$ values for our sample are
also listed in Table~\ref{sample_tbl} and are plotted as a function of
$L_{bol}$ and $T_{bol}$ in Figure~\ref{rho1_lbol_tbol}. As can be
readily seen, $\rho_{1}$ appears to be uncorrelated with
$L_{bol}$. The {\it Spearman} rank coefficient, $r_s$, for the
correlation is 0.27 and the probability that that this correlation
could be produced by two random variables of the same sample size,
$P(r_s)$, is 23\%. On the other hand, $\rho_{1}$ is tightly correlated
with $T_{bol}$; the {\it Spearman} correlation coefficient, $r_s =
-$~0.82 and $P(r_s)$~$\ll$~0.001\%, indicating that the correlation is
statistically significant. The envelope density generally decreases
with increasing $T_{bol}$ suggesting that $T_{bol}$ is a good
indicator of the evolutionary stage of the protostars in our sample
\citep[also see][]{chen95,evans09}.

\begin{figure}
\epsscale{1.1}
\plotone{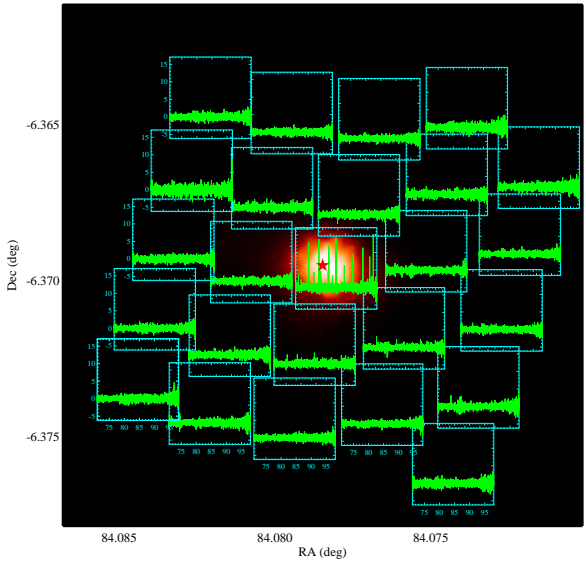}
\caption{PACS spectral map of HOPS~182 in the B2B (72$-$98~$\micron$)
  spectral band. Within the square boxes, which represents the
  5$\times$5 spaxel array, the observed spectra at that location is
  shown. The PACS spectroscopic field is overlaid on the PACS
  70$\micron$ image of HOPS~182 obtained as part of the HOPS
  program. The star symbol represents the MIPS 24$\micron$ coordinates
  of HOPS~182, which was used for telescope pointing. \label{pacs_ifu}}
\end{figure}

\begin{figure}
\epsscale{1.1}
\plotone{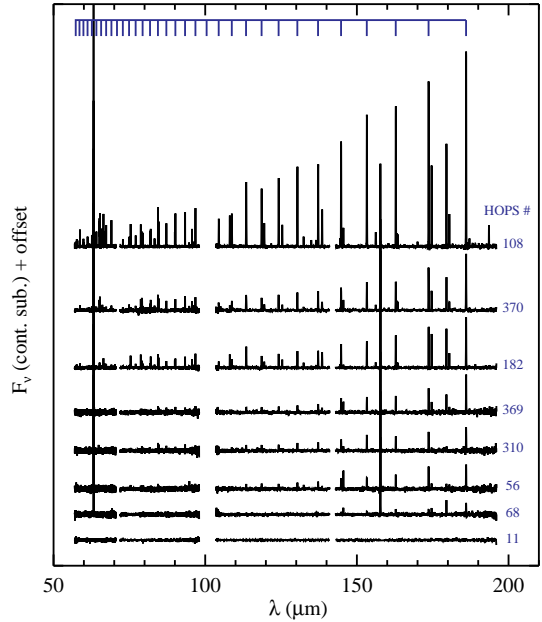}
\caption{Continuum subtracted PACS spectra of 8 protostars which are
  representative of the entire sample. The central wavelengths of the
  observed CO lines ($J_{up}$=14$-$46) are indicated at the top. The
  spectra have been offset for better viewing. The continuum has been
  subtracted by applying a median filter to the spectra.  An arbitrary
  offset has been added to the spectra to separate them in the
  plot. \label{pacs_spectra}}
\end{figure}

\begin{figure*}
\epsscale{1.0}
\plottwo{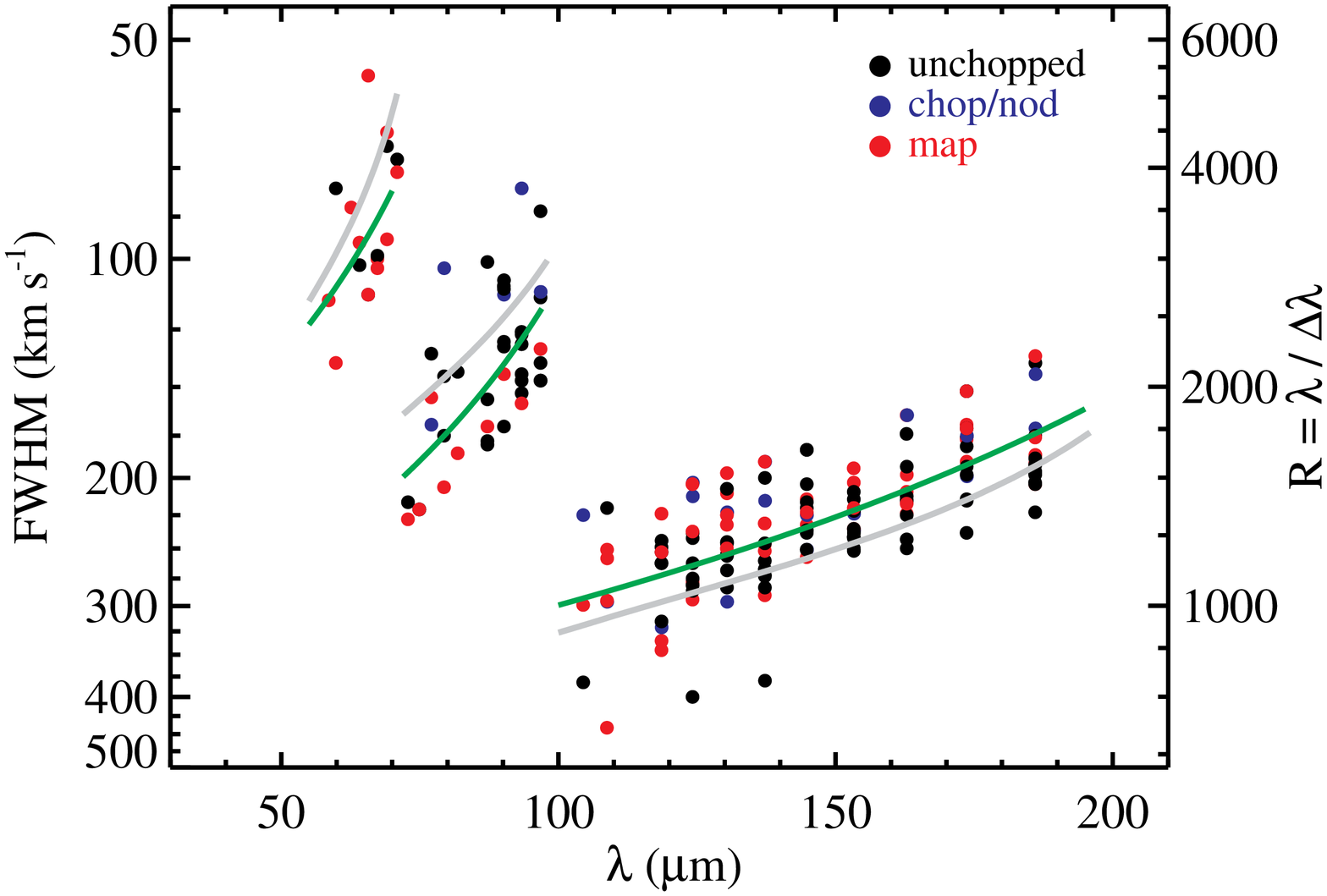}{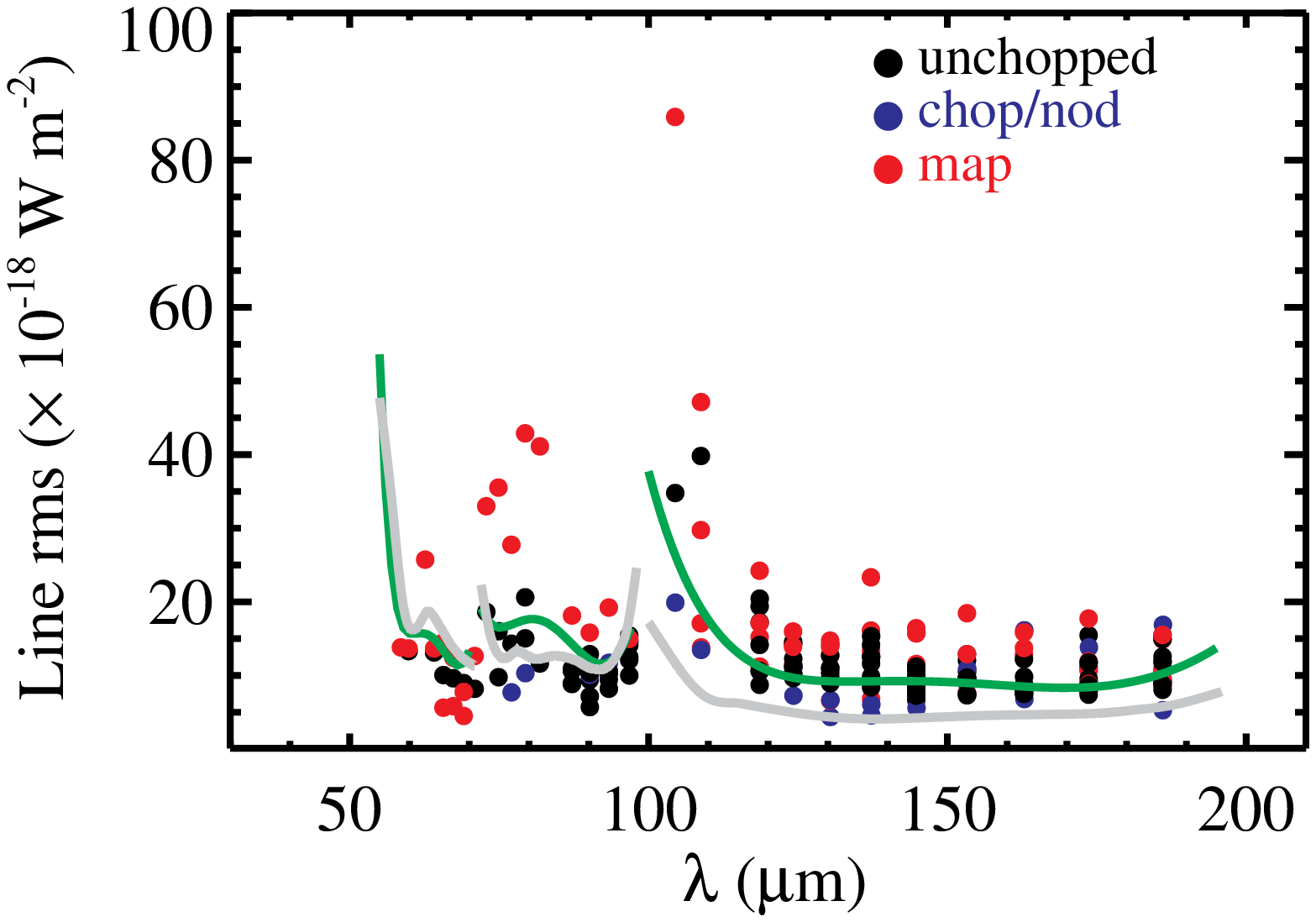}
\caption{Measured CO line widths {\it(left)} and 1-$\sigma$ line
  sensitivity of our observations {\it(right)}. Lines detected at
  3-$\sigma$ level or higher are shown. Pointed/unchopped (black solid
  circles), pointed/chop-nod (blue circles), and mapping/unchopped (red
  circles) observations are identified. The gray solid lines
  correspond to HSpot predictions for our observational set up. The
  green solid lines are fits to the observed points in different
  spectral bands. \label{fwhm_rms}}
\end{figure*}

\begin{figure}
\epsscale{1.1}
\plotone{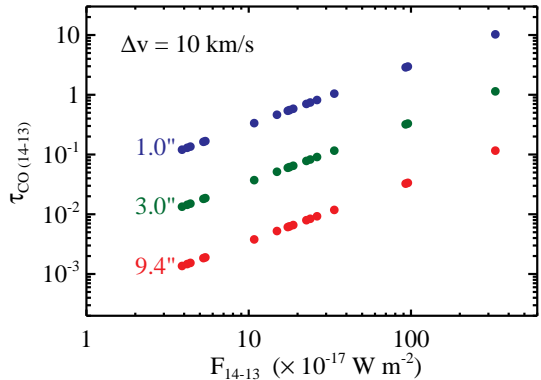}
\caption{ Optical depth, $\tau$, of the CO~$J=14-13$ line as a
  function of observed line flux for three angular sizes (for $d =
  420$ pc to Orion) of the emitting region. The optical depths shown
  are for a CO line width ($FWHM$) of 10 kms$^{-1}$. \label{tau}}
\end{figure}

\begin{figure*}
\centering
\resizebox{0.45\textwidth}{!}{\includegraphics{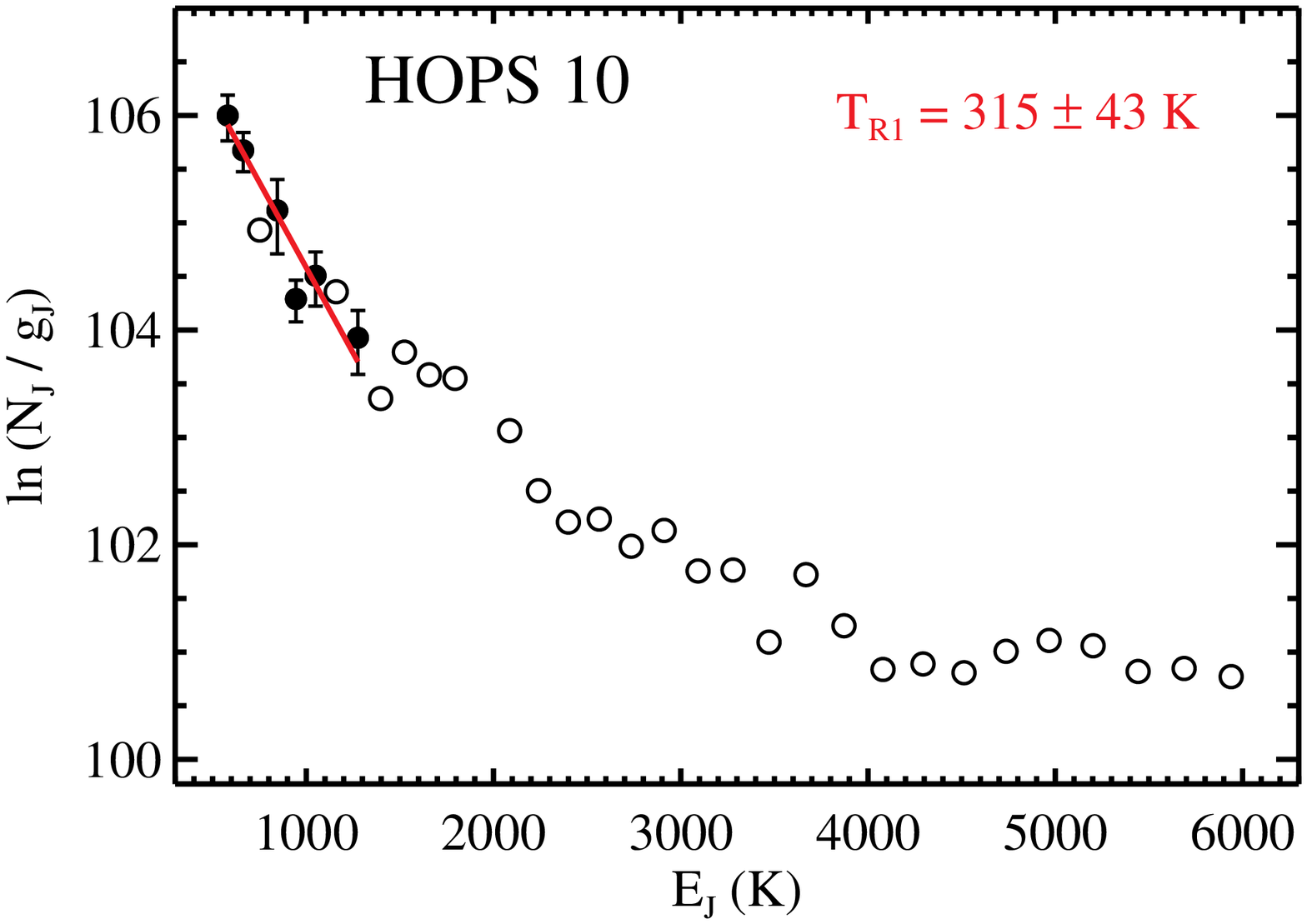}}
\resizebox{0.45\textwidth}{!}{\includegraphics{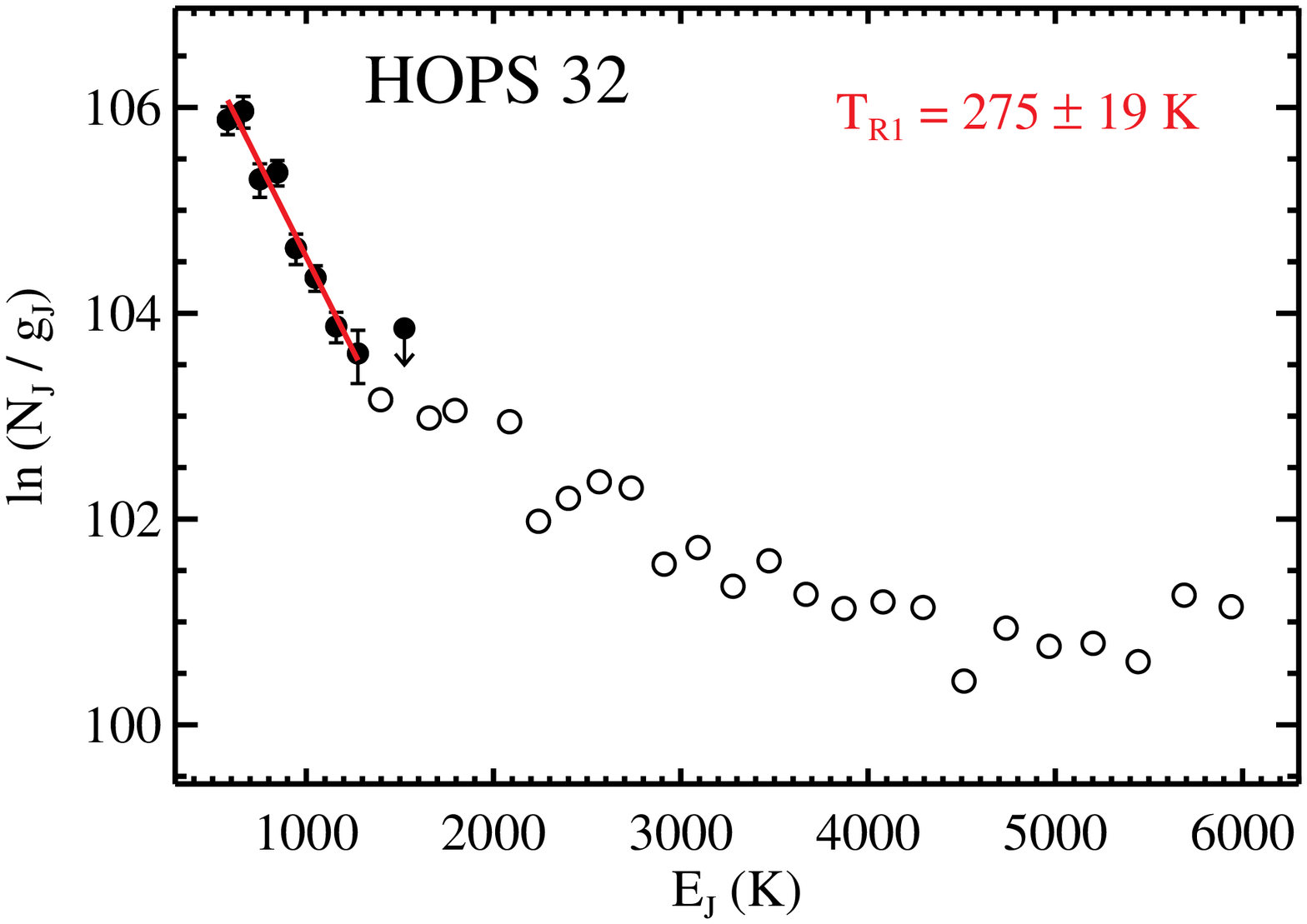}}
\resizebox{0.45\textwidth}{!}{\includegraphics{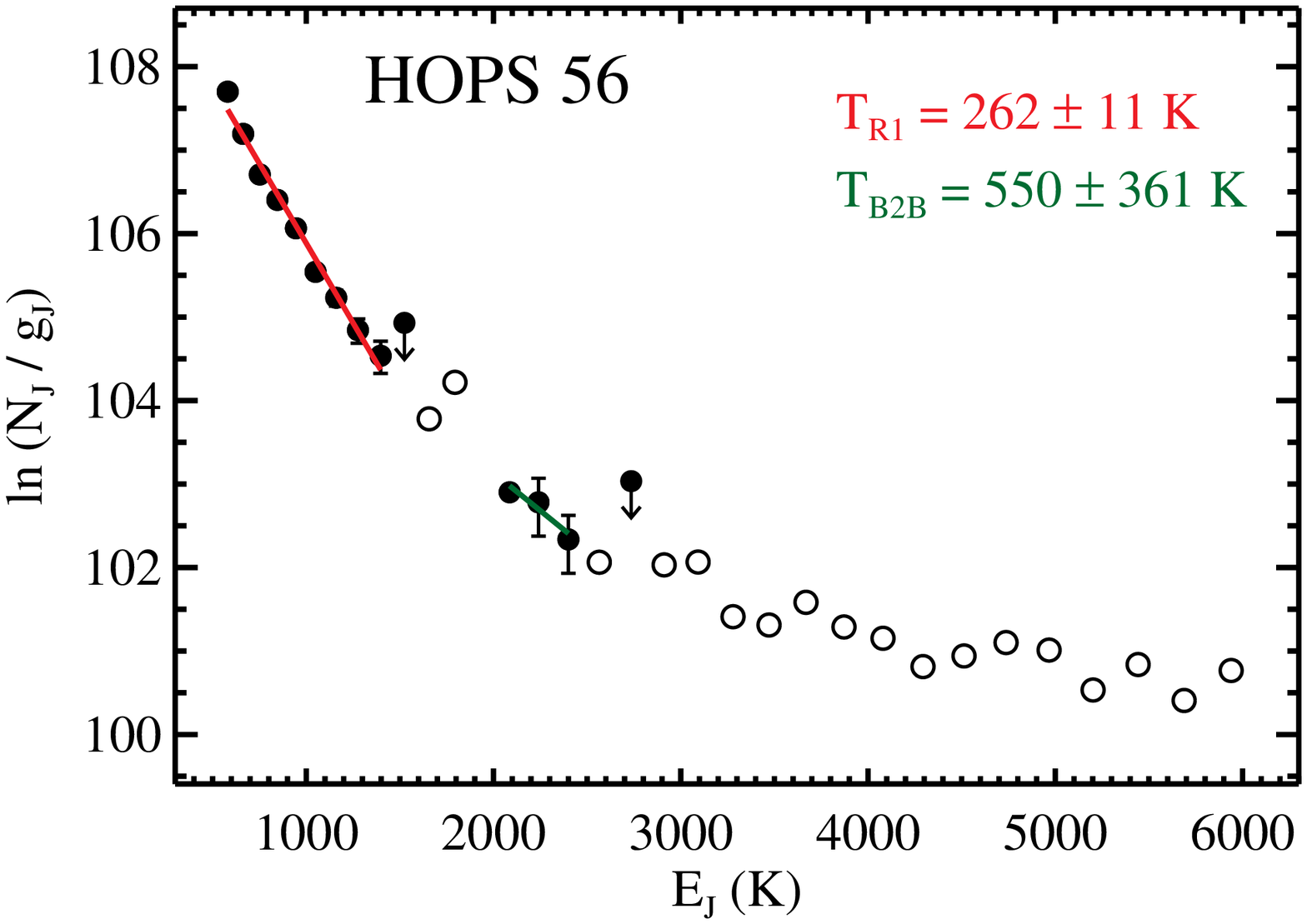}}
\resizebox{0.45\textwidth}{!}{\includegraphics{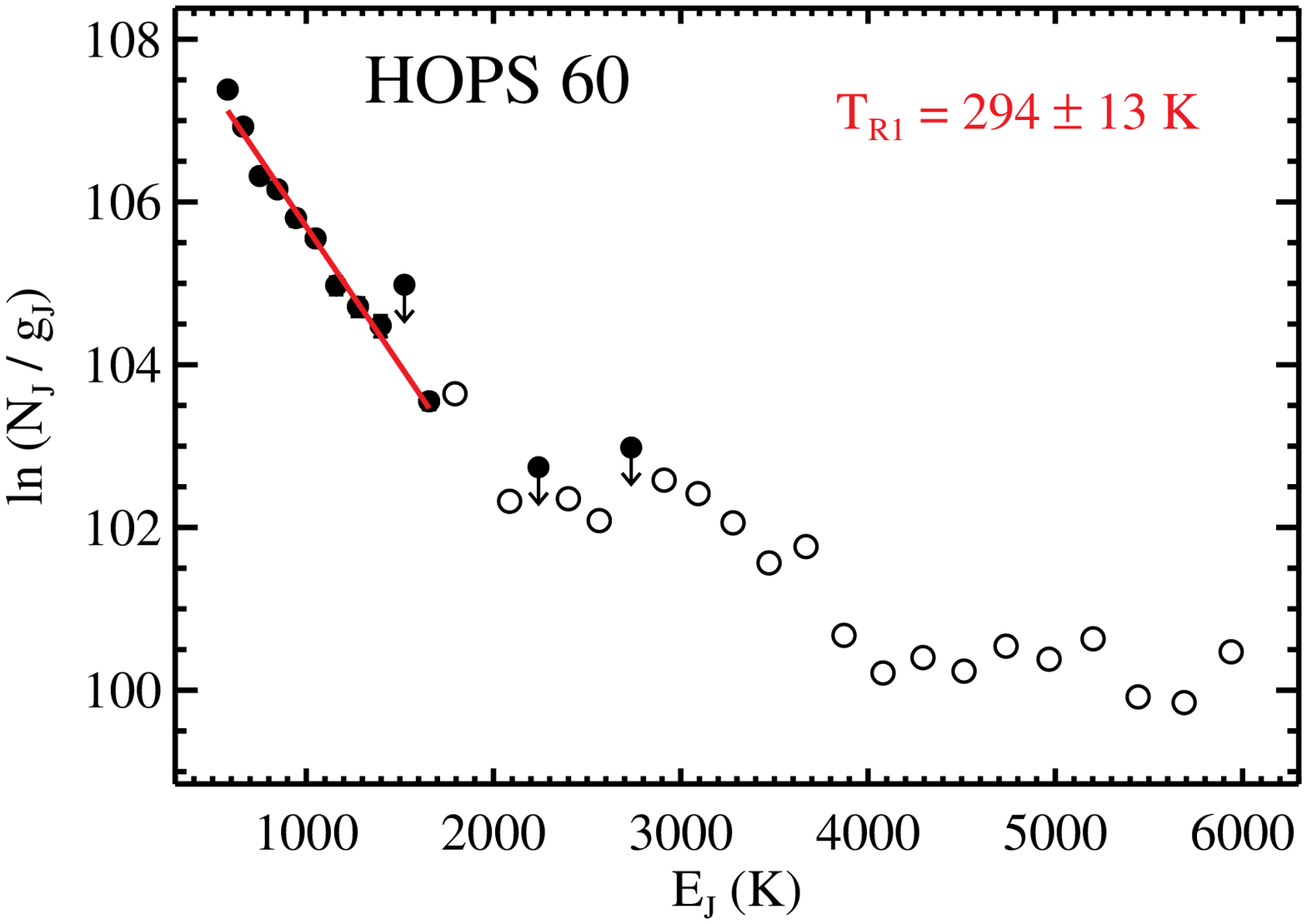}}
\resizebox{0.45\textwidth}{!}{\includegraphics{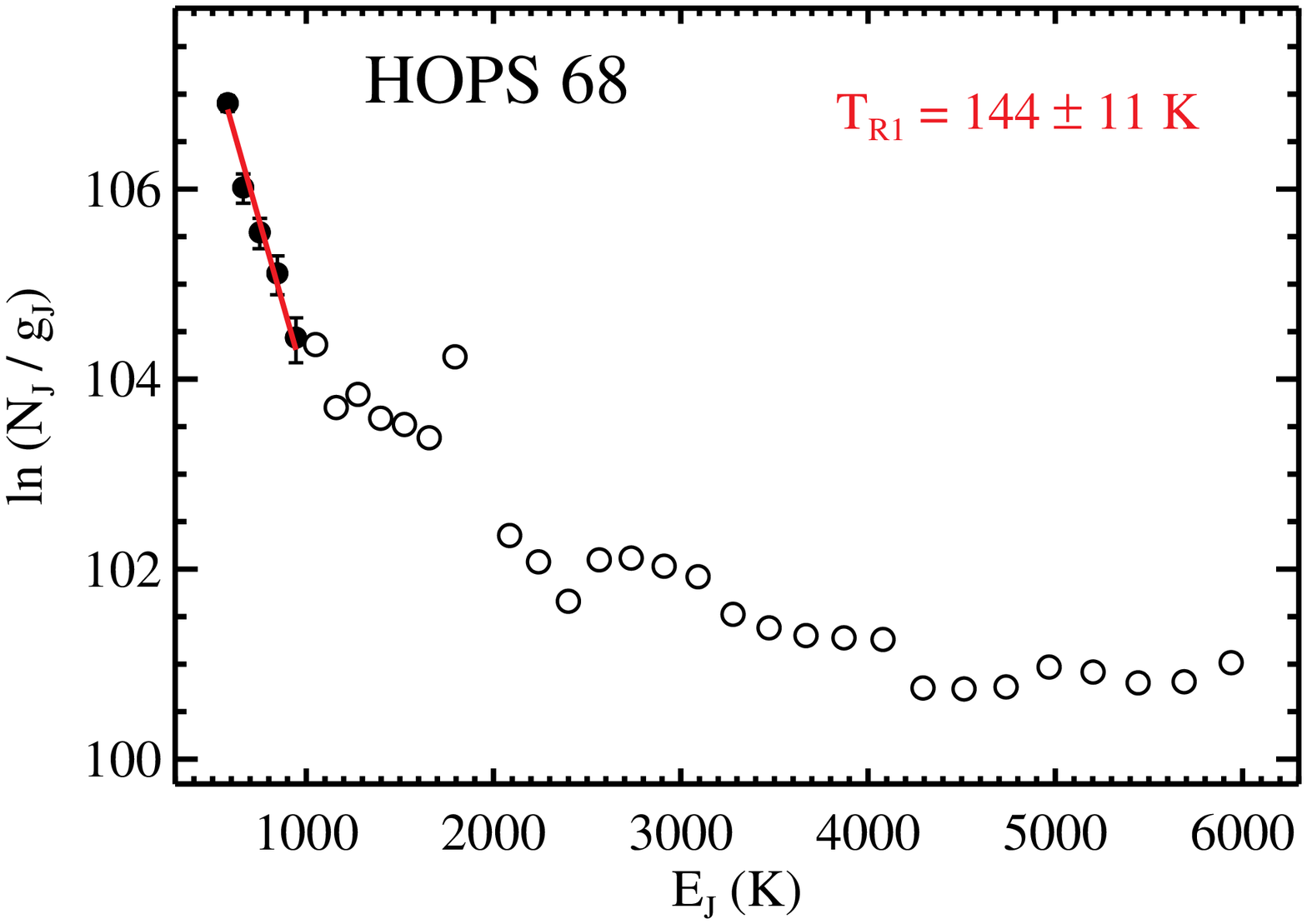}}
\resizebox{0.45\textwidth}{!}{\includegraphics{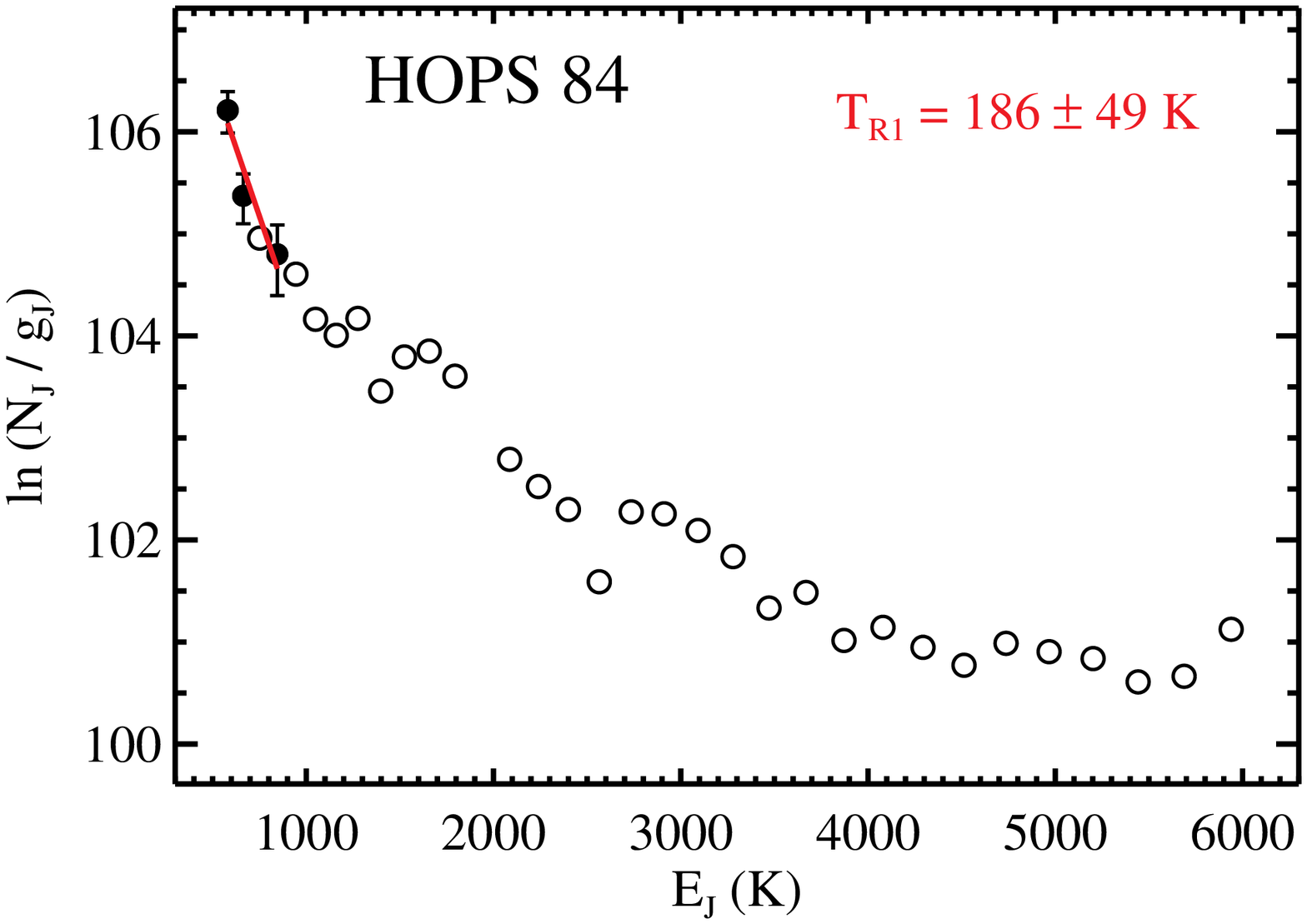}}
\caption{CO rotational excitation diagrams for the 17 protostars
  towards which CO lines are detected. Linear fits to the CO lines in
  the R1 (550~K~$\le~E_J/k~\le$~1800~K; $J_{up}$~=~14$-$25) band (red
  solid line), the B2B (2000~K~$\le~E_J/k~\le$~3700~K;
  $J_{up}$~=~27$-$36) band (green solid line), and the B3A
  (3700~K~$\le~E_J/k~\le$~6000~K; $J_{up}$~=~37$-$46) band (blue solid
  line) and the average rotational temperatures derived,
  $T_{\mathrm{R1}}$, $T_{\mathrm{B2B}}$, and $T_{\mathrm{B3A}}$ are
  shown.  Downward arrows indicate upper limits to the fluxes of CO
  lines which are blended with a nearby line. Open circles correspond
  to 3$\sigma$ upper limits for the non-detections. Sources in which
  no CO lines are detected are not shown. \label{rot_diagram}}
\end{figure*}

\begin{figure*}
\centering
\addtocounter{figure}{-1}
\resizebox{0.45\textwidth}{!}{\includegraphics{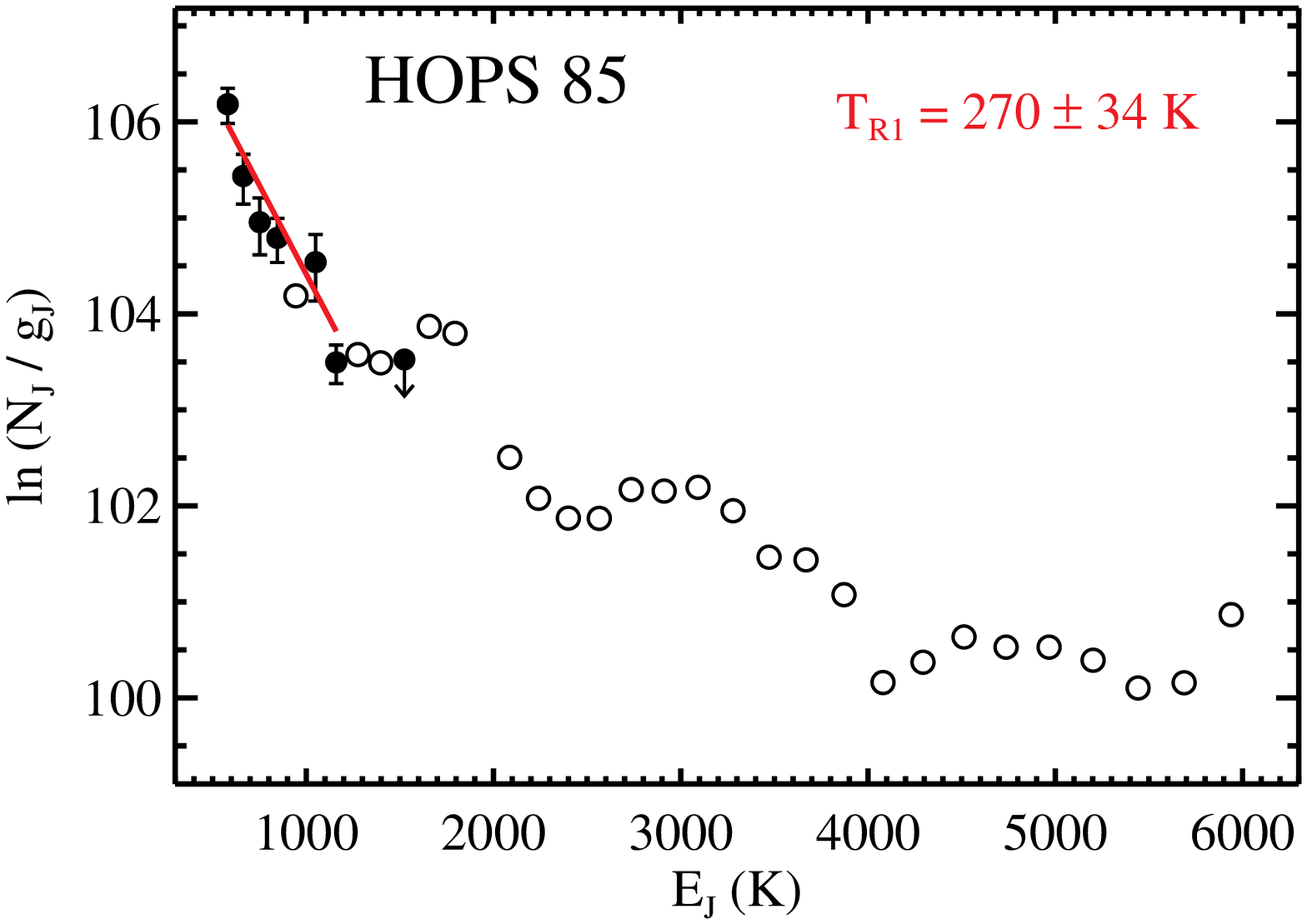}}
\resizebox{0.45\textwidth}{!}{\includegraphics{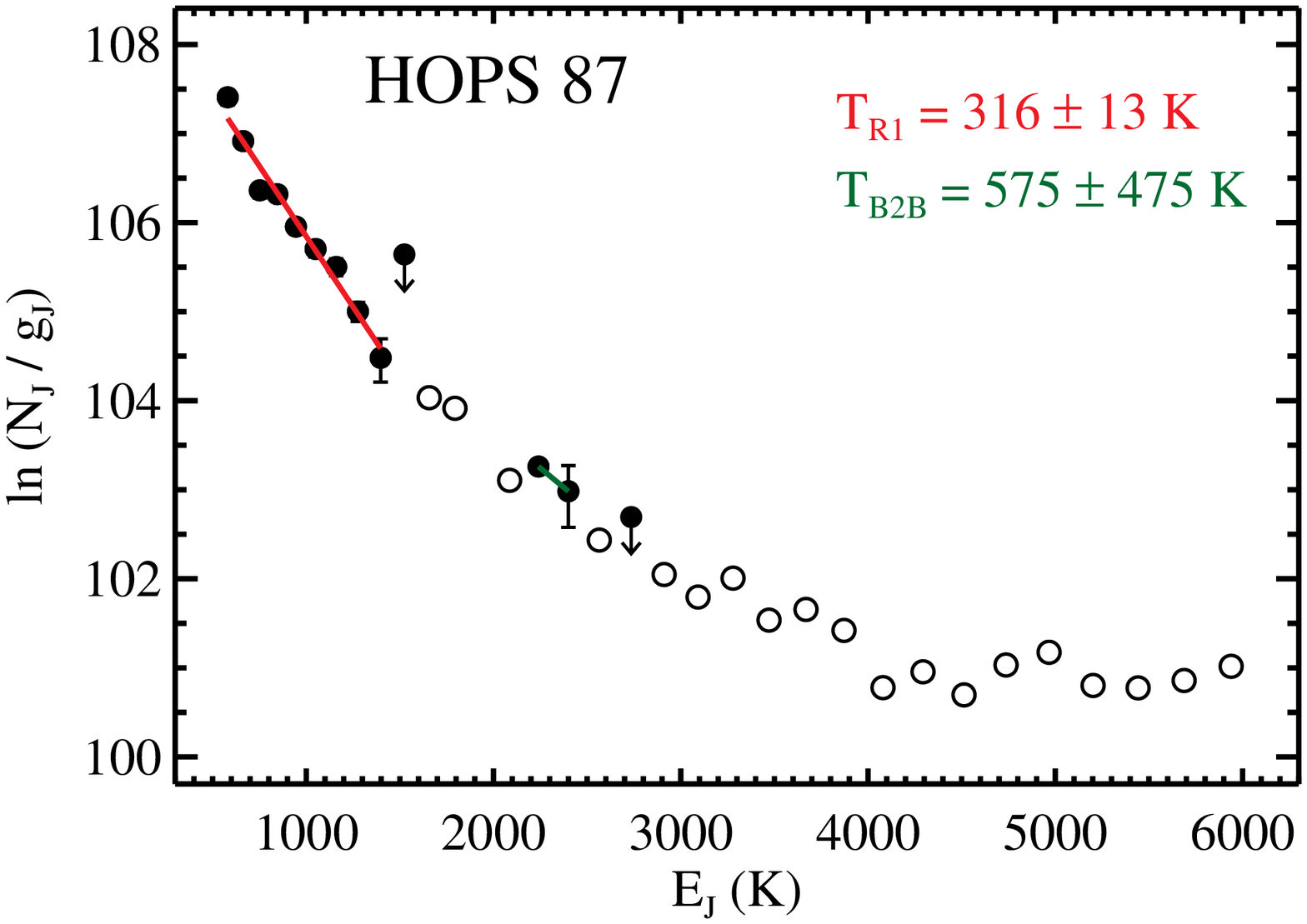}}
\resizebox{0.45\textwidth}{!}{\includegraphics{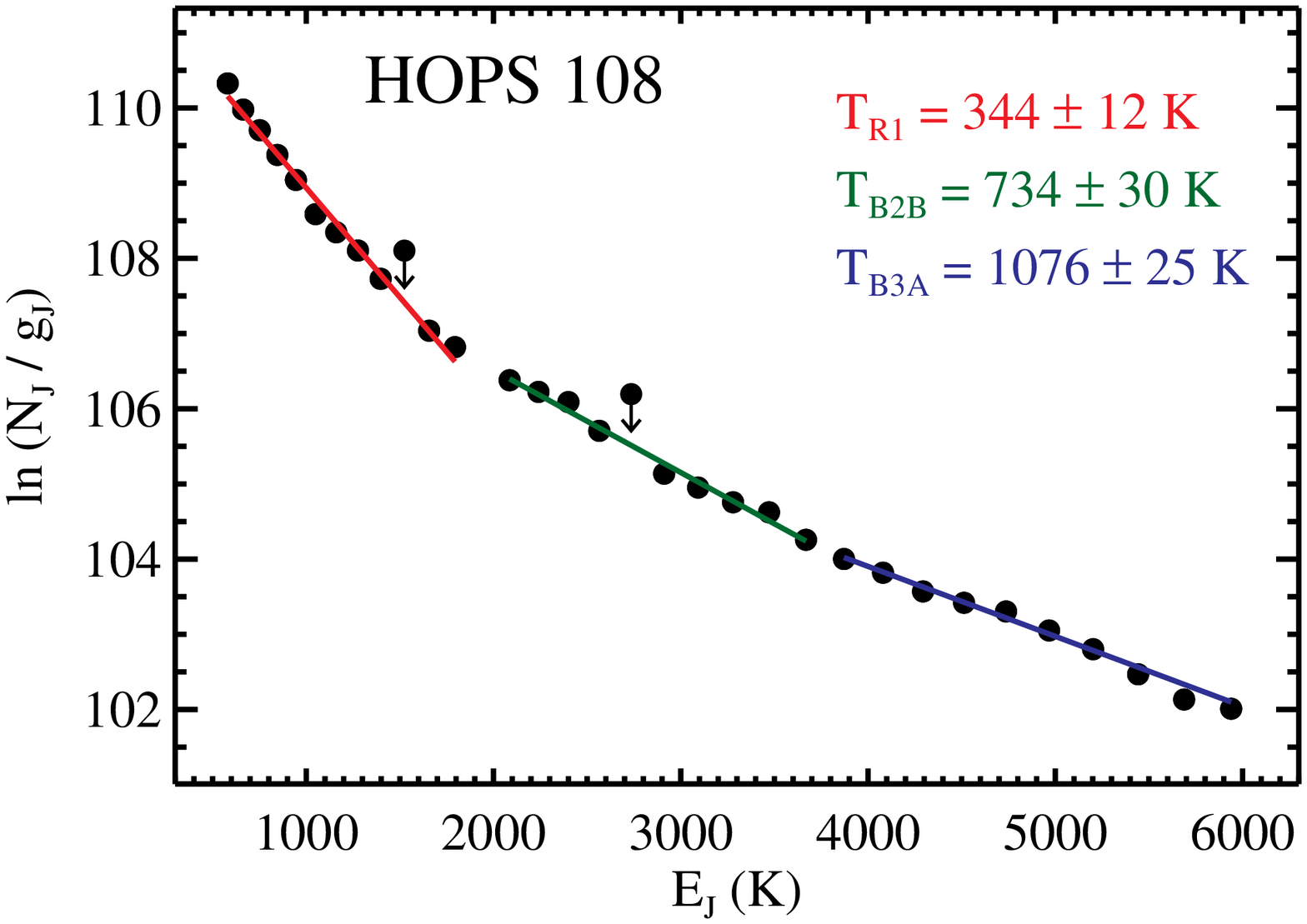}}
\resizebox{0.45\textwidth}{!}{\includegraphics{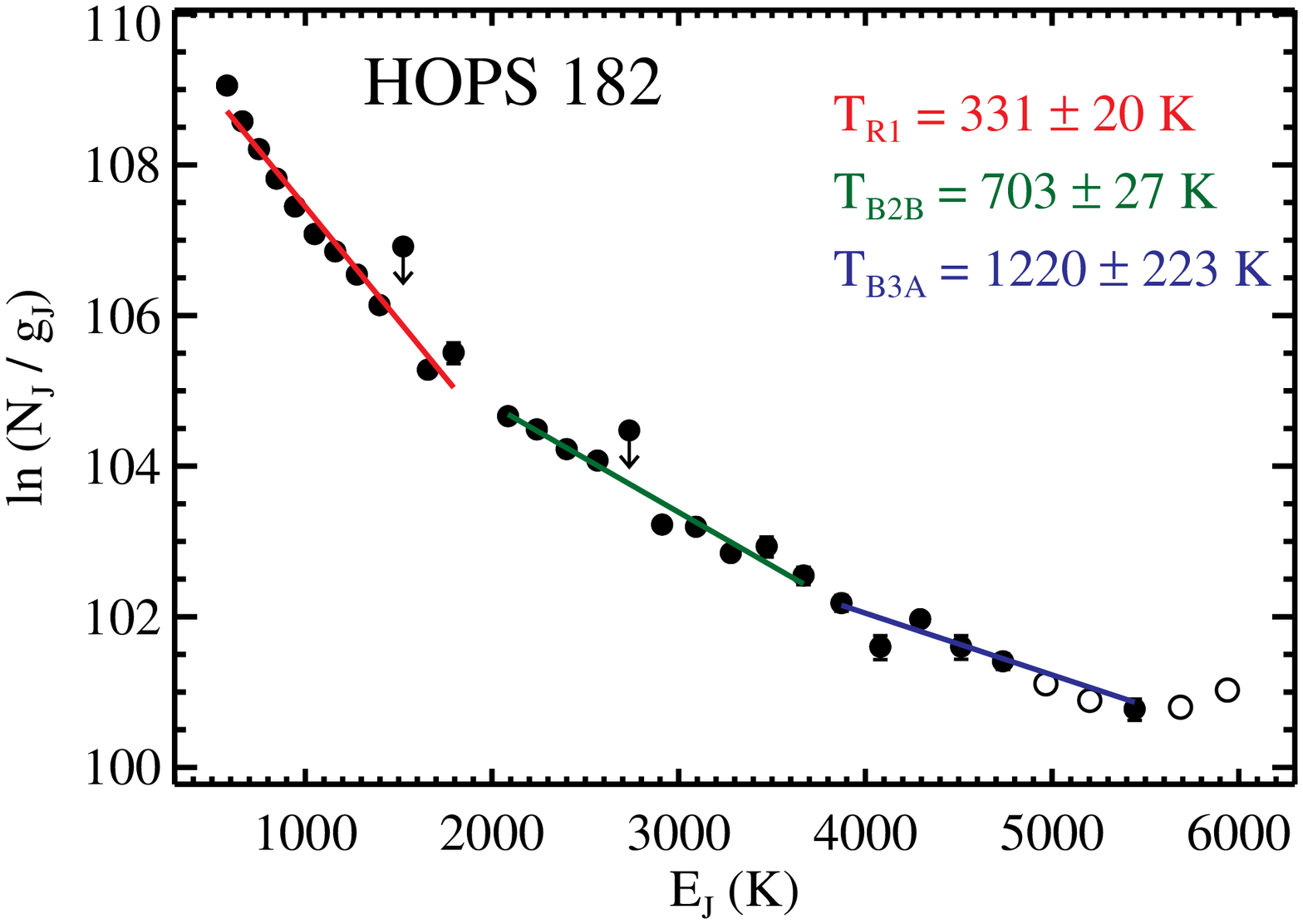}}
\resizebox{0.45\textwidth}{!}{\includegraphics{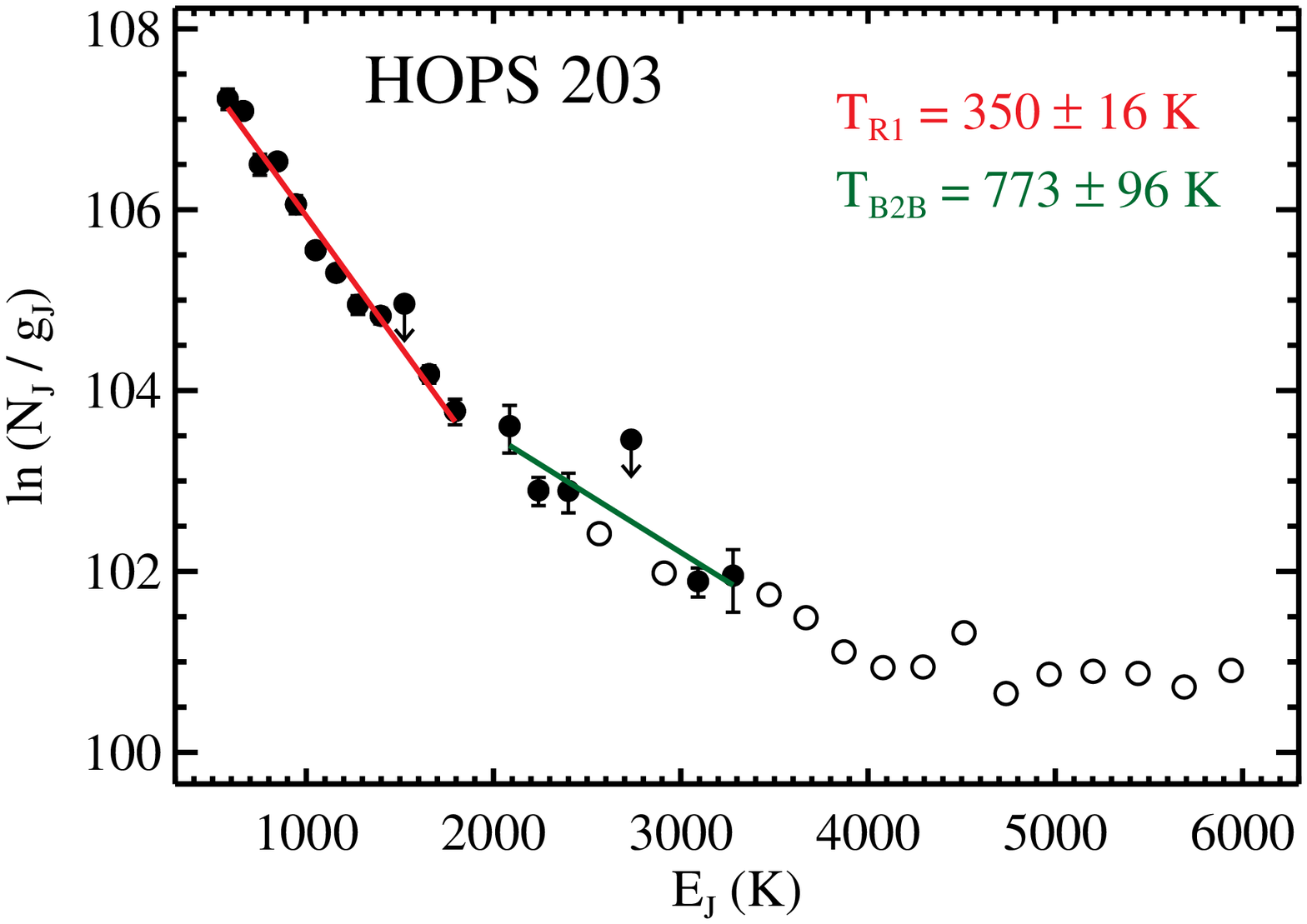}}
\resizebox{0.45\textwidth}{!}{\includegraphics{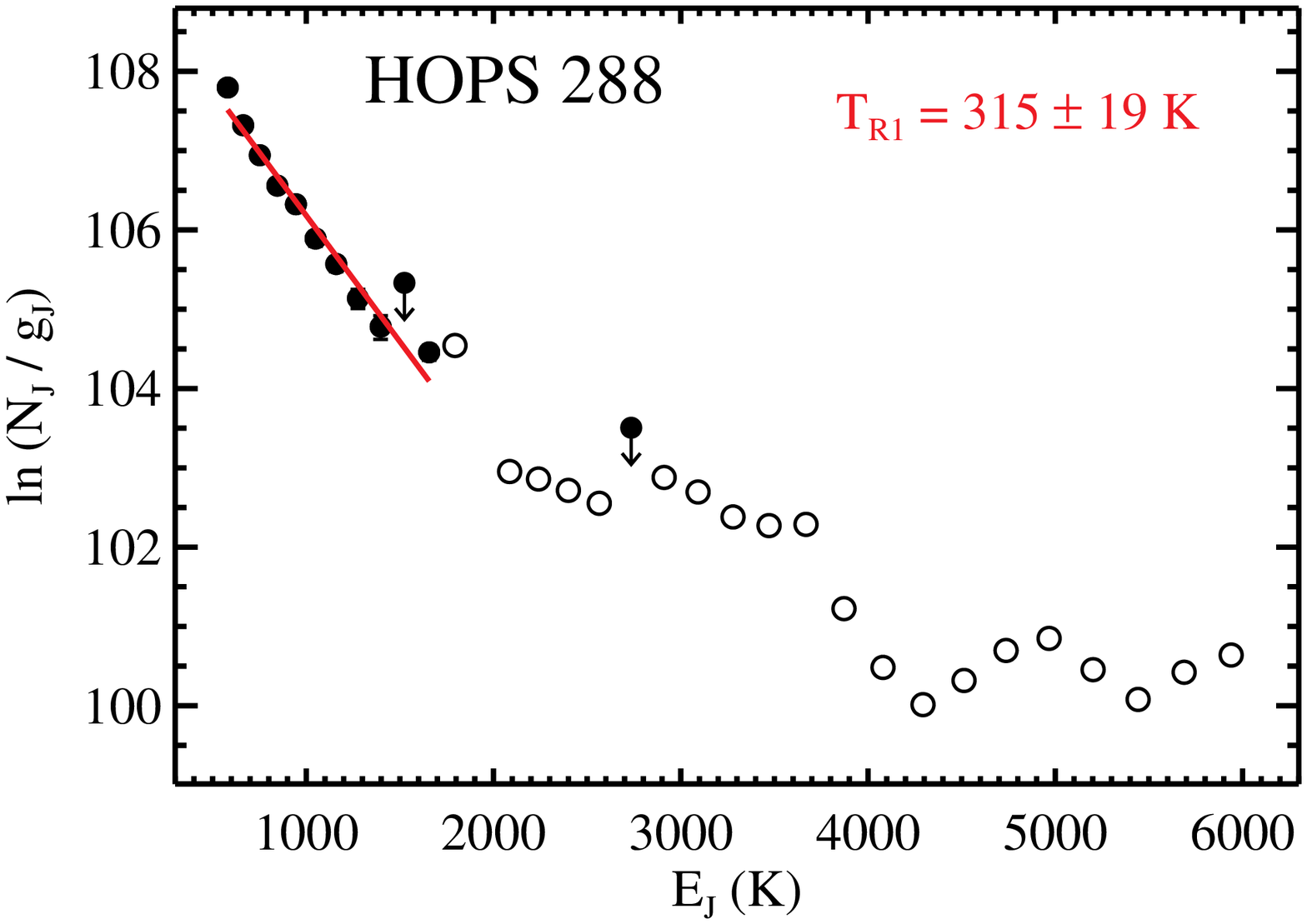}}
\resizebox{0.45\textwidth}{!}{\includegraphics{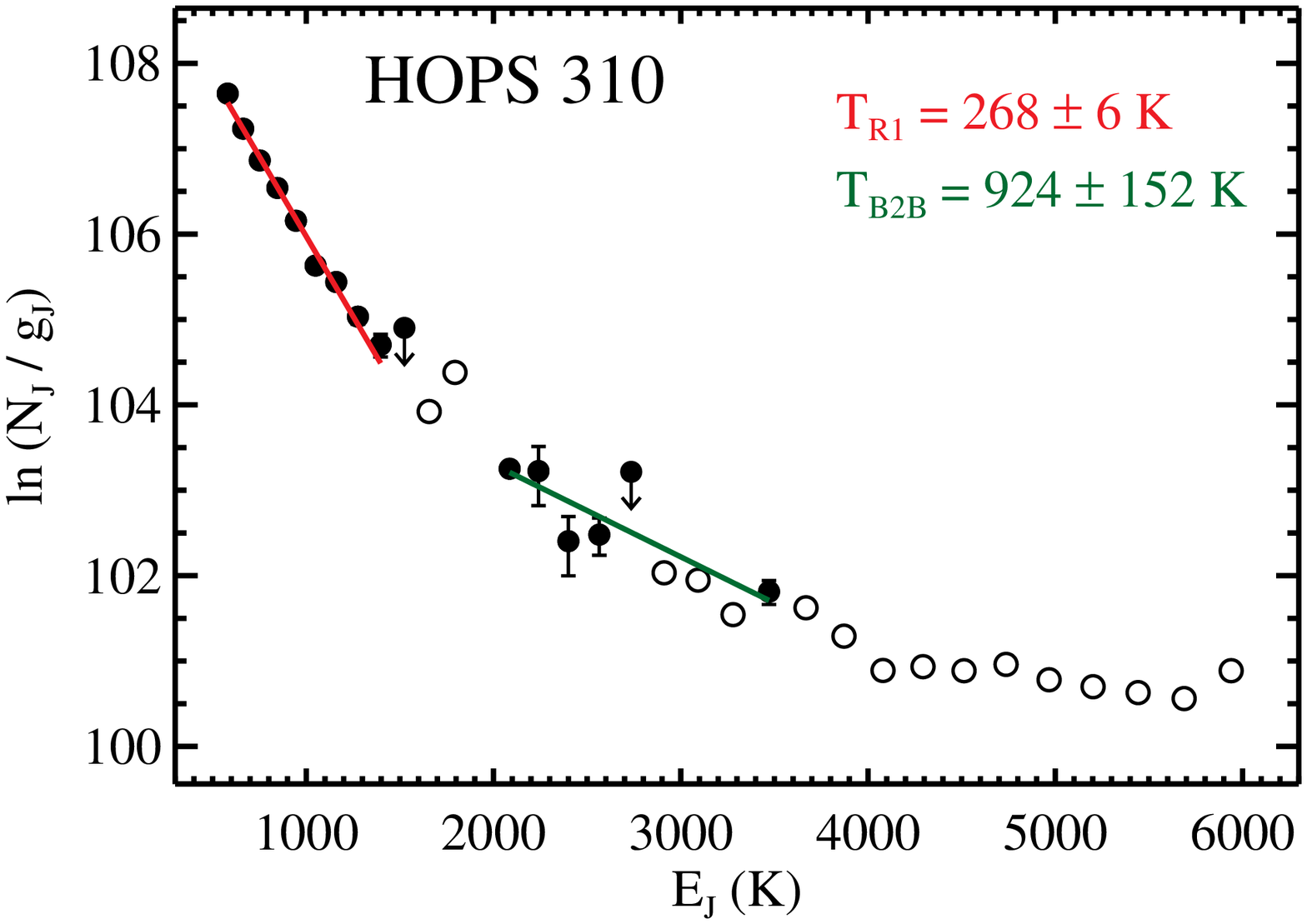}}
\resizebox{0.45\textwidth}{!}{\includegraphics{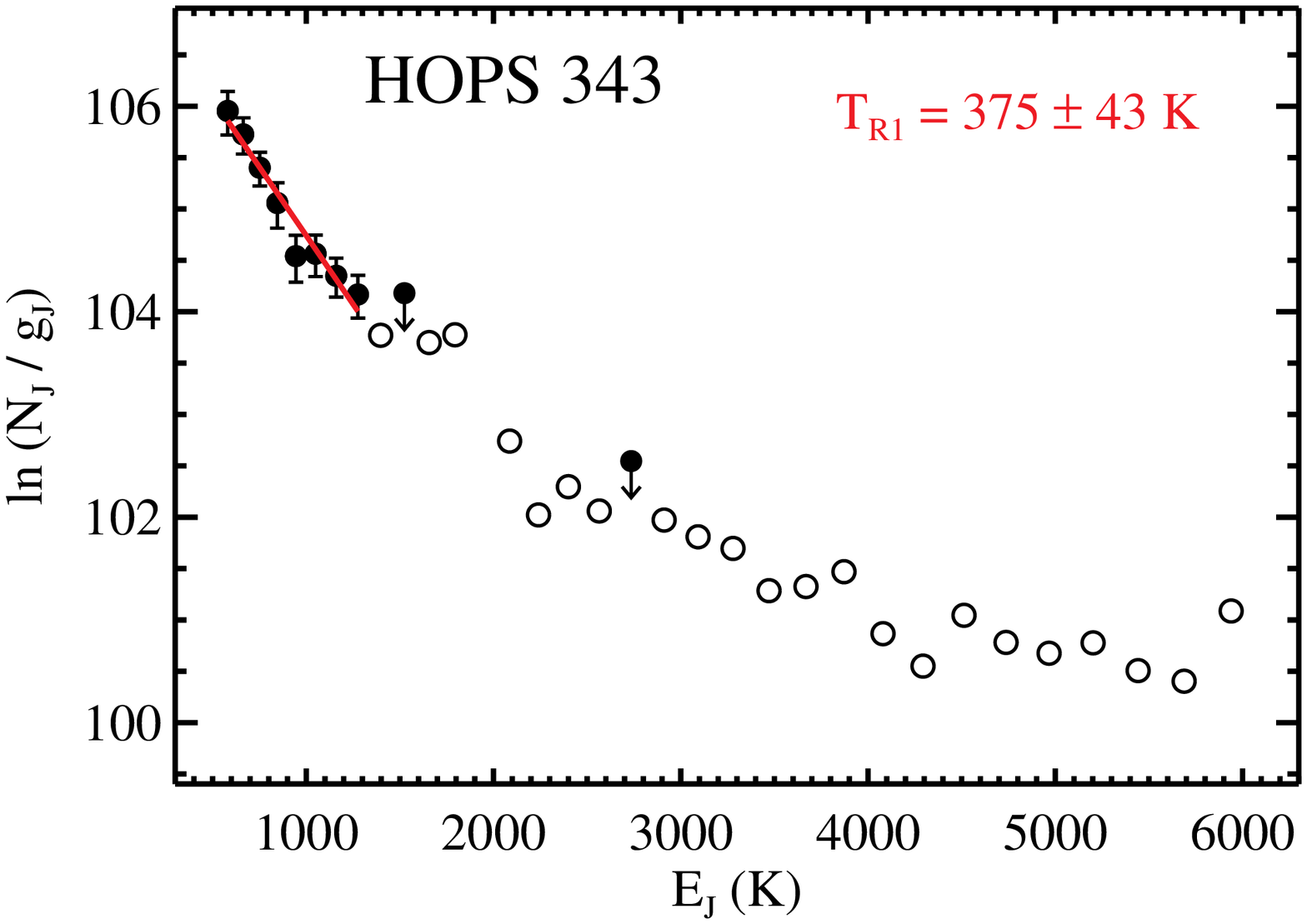}}
\caption{continued ... }
\end{figure*}

\begin{figure*}
\centering
\addtocounter{figure}{-1}
\resizebox{0.45\textwidth}{!}{\includegraphics{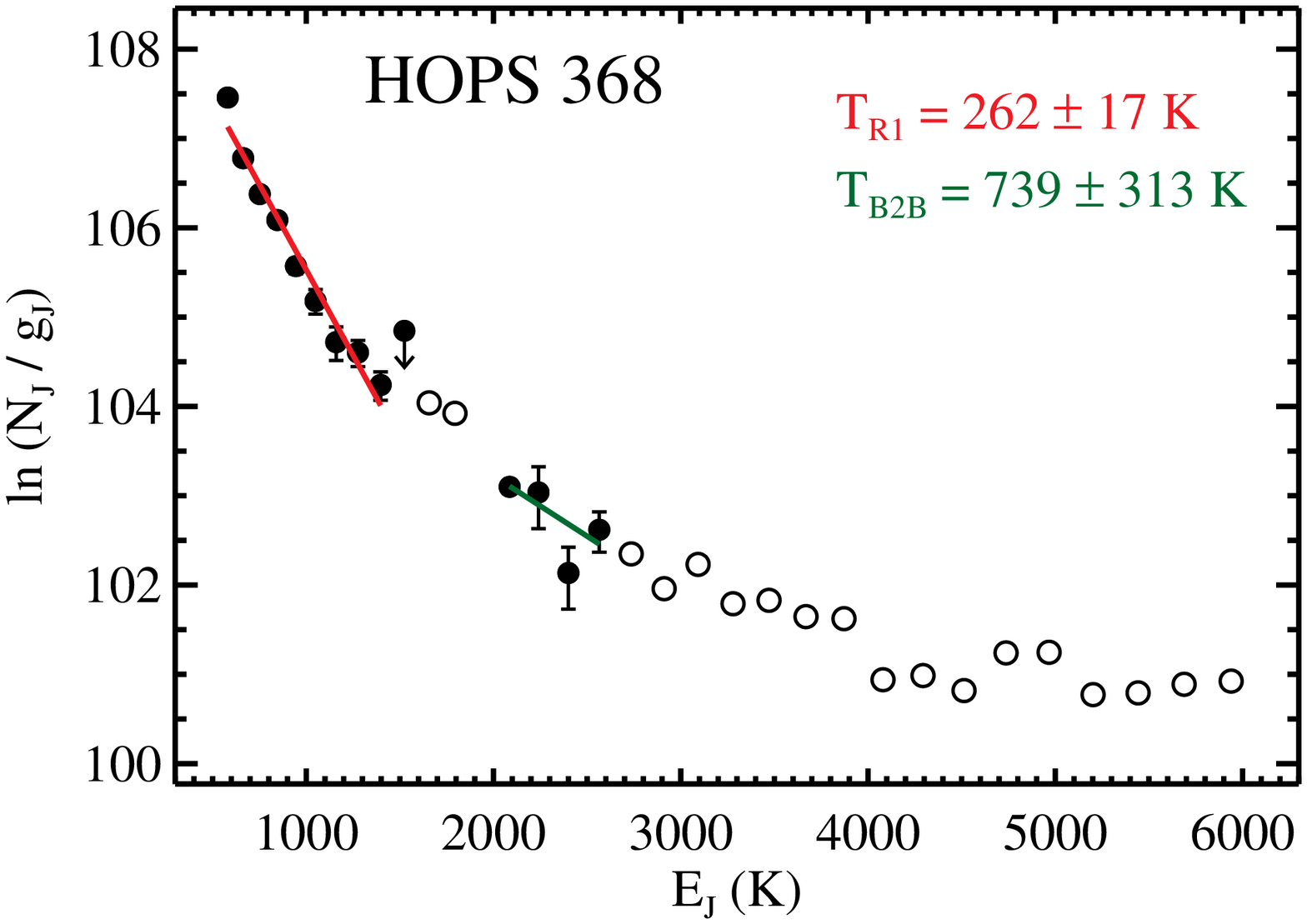}}
\resizebox{0.45\textwidth}{!}{\includegraphics{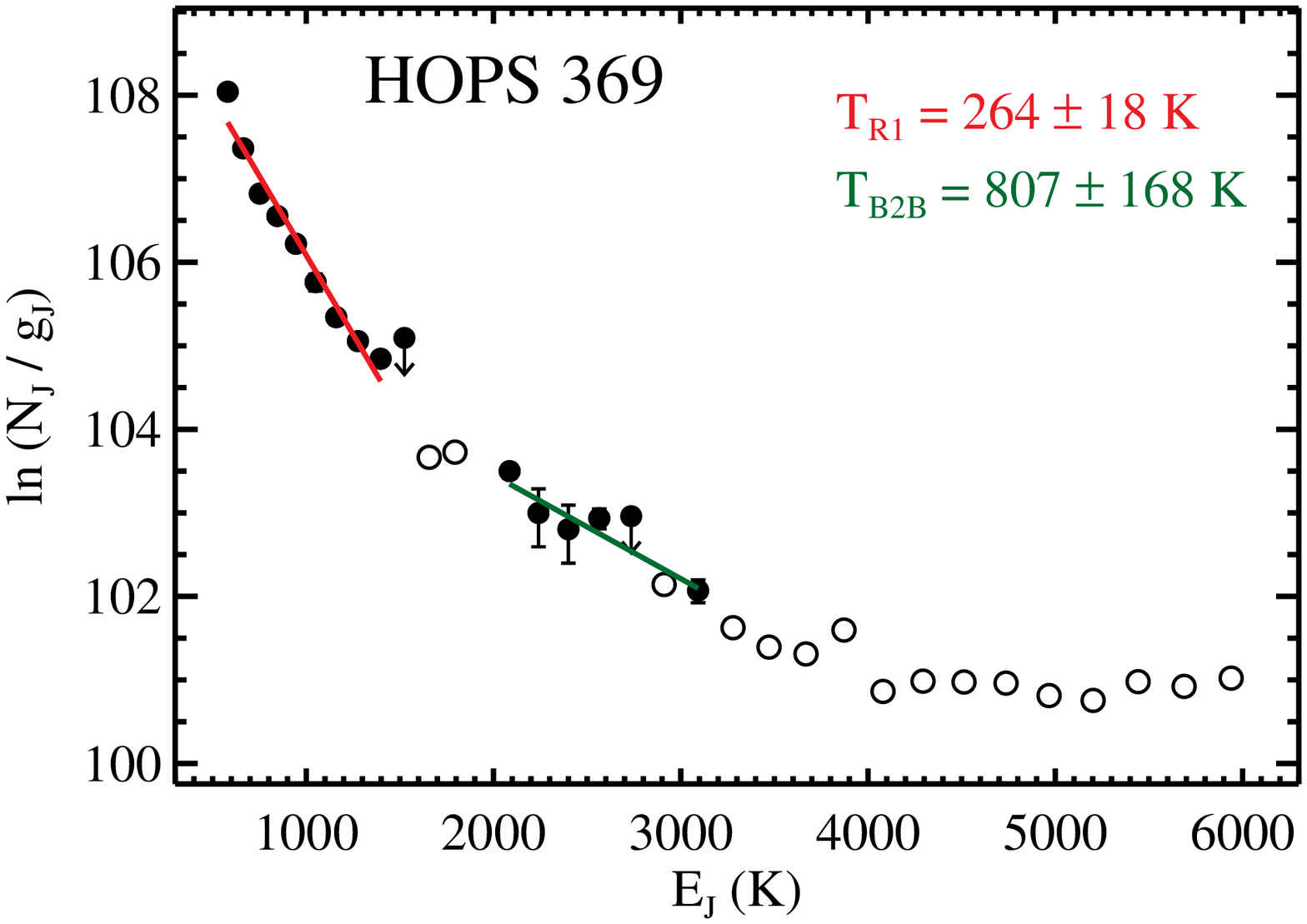}}
\resizebox{0.45\textwidth}{!}{\includegraphics{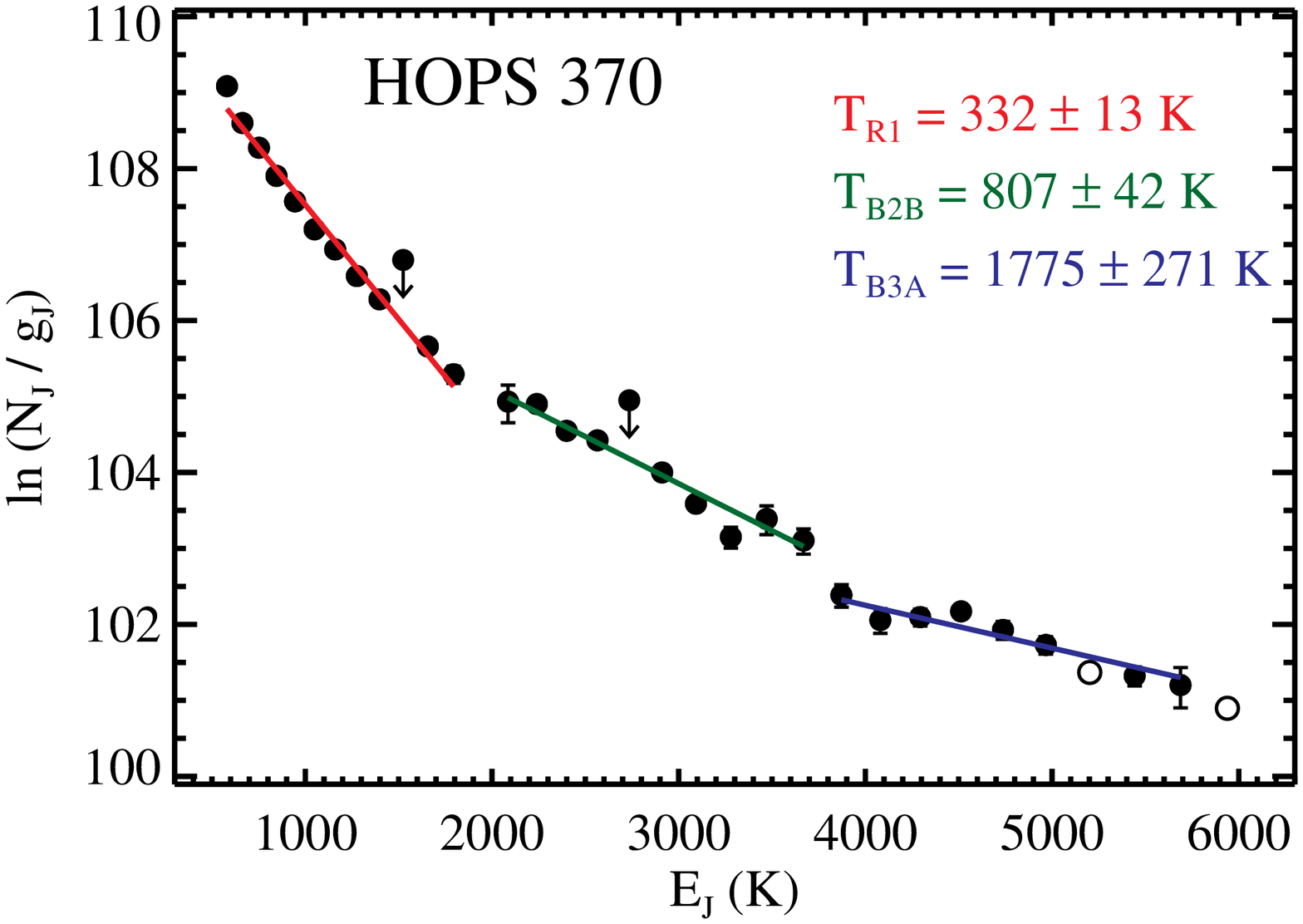}}
\caption{continued ... }
\end{figure*}

\section{Observations and Data Reduction} \label{obs}

The log of {\it Herschel} PACS observations is provided in
Table~\ref{log_tbl}. All spectra were obtained in the range scan mode
to achieve a total wavelength coverage of 57~$\micron$ to
196~$\micron$. The wavelength range of 57$-$71~$\micron$ was observed
in the 3$^{rd}$ grating order (B3A), the range 71$-$98~$\micron$ in
the 2$^{nd}$ order (B2B), and the range 102$-$196~$\micron$ in the
1$^{st}$ order (R1). Nyquist sampling, with grating step size
of 6.25 spectral pixels was used to cover the broad wavelength ranges
in each spectral band. Seventeen of the sources were observed in the
pointed mode and four were observed in the full PACS spatial
resolution mapping mode. The pointed observations of the bright
sources ($F_{70}~\ga$~10~Jy) were carried out in the unchopped mode
and those of the fainter sources ($F_{70}~\la$~10~Jy) in the chop/nod
mode. The mapping observations were done in the unchopped mode. For
the chop/nod observations, a maximum chopper throw of 6$\arcmin$ was
used. For unchopped observations, off-positions free of cloud emission
within 2$\degr$ from the target were used to subtract the instrument
background. These emission-free off-positions were determined from the
CO $J = 3-2$ maps of the region \citetext{Di Francesco et al. 2012, in
  preparation}.

All spectra were reduced using the Herschel Interactive Processing
Environment (HIPE) version 8.0 \citep{ott10}. We used the pipeline
scripts to obtain the final spectral maps over an area of 47$\arcsec$
$\times$ 47$\arcsec$ centered on each source.  An example of the PACS
spectral map is shown in Figure~\ref{pacs_ifu}, where the 5$\times$5
spatial pixel (``spaxel'') array and the spectrum corresponding to
each spaxel is displayed. The telescope pointing in all our
observations was good: we mapped the continuum in each of the PACS
spectral bands and found that the continuum peak was well centered on
the central spaxel within the pointing uncertainty of
$\sim$~2$\arcsec$. The 1-D spectrum used in the analysis was extracted
from the central spaxel (9.4$\arcsec$ $\times$ 9.4$\arcsec$) where the
continuum emission from the protostar peaked. For the mapping
observations, we extracted the 1-D spectrum within an 9.4$\arcsec$
$\times$ 9.4$\arcsec$ aperture centered on the continuum peak. The
$FWHM$ of the spectrometer beam is a function of wavelength and ranges
from $\sim$~10$\arcsec$ at 75~$\micron$ to $\sim$~12$\arcsec$ at
160~$\micron$ (see PACS Observer's Manual). We applied a wavelength
dependent PSF-loss correction to the extracted spectrum to account for
the different fractions of the PSF structure seen by the central
spaxel at different wavelengths.

Representative FIR spectra (57$-$196~$\micron$) of a few sources in
our sample are shown in Figure~\ref{pacs_spectra}. At the distance of
Orion ($\sim$~420~pc), the extracted 1-D spectra corresponds to the
emission from within a radius of $\sim$ 2000~AU from the
protostars. Our PACS spectroscopic observations were designed to cover
the rotational transitions in the ground vibrational state of CO
ranging from $J = 14-13$ to $J = 46-45$. Several of these lines are
seen in emission in the final extracted spectrum of most objects in
our sample, indicating the presence of warm CO gas associated with
these protostars. In four objects, HOPS~11, 30, 91, and 329, we do not
detect any CO lines above the 3$\sigma$ rms level in the PACS
wavelength range. In the remaining 17 objects, several CO lines are
detected, ranging from 3 lines in HOPS~84 up to 30 lines in HOPS~370
and HOPS~108 (see Figure~\ref{pacs_spectra}). We measured the fluxes
of the observed CO lines in the spectra of the protostars by fitting
them with Gaussian profiles. The measured CO line widths and the line
sensitivities achieved from our observations are shown in
Figure~\ref{fwhm_rms}.

\begin{deluxetable*}{lcccccc}
\tablewidth{0pt}
\tablecaption{CO Rotational temperatures and luminosity \label{rot_tbl}}
\tablehead{
\colhead{} & \colhead{$T_{\mathrm{LR1}}$} & \colhead{$T_{\mathrm{SR1}}$} & \colhead{$T_{\mathrm{R1}}$} & \colhead{$T_{\mathrm{B2B}}$} & \colhead{$T_{\mathrm{B3A}}$} & \colhead{$L_{\mathrm{CO}}$}\\
\colhead{HOPS~ID} & \colhead{(K)} & \colhead{(K)} & \colhead{(K)}  & \colhead{(K)} & \colhead{(K)} & \colhead{($\times$~10$^{-3}$ $L_{\odot}$)}\\
}
\startdata

 10&     238 $\pm$ 44 &    391 $\pm$ 280 &     315 $\pm$ 43 &          \nodata &          \nodata &            1.3\\
 11&          \nodata &          \nodata &          \nodata &          \nodata &          \nodata &      $\le$~0.7\\
 30&          \nodata &          \nodata &          \nodata &          \nodata &          \nodata &      $\le$~0.8\\
 32&     307 $\pm$ 46 &     282 $\pm$ 91 &     275 $\pm$ 19 &          \nodata &          \nodata &            1.8\\
 56&     218 $\pm$ 10 &     338 $\pm$ 58 &     262 $\pm$ 11 &    550 $\pm$ 361 &          \nodata &            8.2\\
 60&     221 $\pm$ 11 &     303 $\pm$ 27 &     294 $\pm$ 13 &          \nodata &          \nodata &            6.2\\
 68&     144 $\pm$ 11 &          \nodata &     144 $\pm$ 11 &          \nodata &          \nodata &            1.6\\
 84&     186 $\pm$ 49 &          \nodata &     186 $\pm$ 49 &          \nodata &          \nodata &            0.7\\
 85&     186 $\pm$ 37 &          \nodata &     270 $\pm$ 34 &          \nodata &          \nodata &            1.2\\
 87&     270 $\pm$ 19 &     307 $\pm$ 49 &     316 $\pm$ 13 &    575 $\pm$ 475 &          \nodata &            7.6\\
 91&          \nodata &          \nodata &          \nodata &          \nodata &          \nodata &      $\le$~0.8\\
108&      290 $\pm$ 3 &      410 $\pm$ 4 &     344 $\pm$ 12 &     734 $\pm$ 30 &    1076 $\pm$ 25 &          238.0\\
182&      230 $\pm$ 3 &     413 $\pm$ 15 &     331 $\pm$ 20 &     703 $\pm$ 27 &   1220 $\pm$ 223 &           48.4\\
203&     322 $\pm$ 34 &     443 $\pm$ 35 &     350 $\pm$ 16 &     773 $\pm$ 96 &          \nodata &            9.7\\
288&     234 $\pm$ 11 &     362 $\pm$ 45 &     315 $\pm$ 19 &          \nodata &          \nodata &           10.0\\
310&     252 $\pm$ 11 &     375 $\pm$ 49 &      268 $\pm$ 6 &    924 $\pm$ 152 &          \nodata &            9.8\\
329&          \nodata &          \nodata &          \nodata &          \nodata &          \nodata &      $\le$~0.7\\
343&     260 $\pm$ 51 &    570 $\pm$ 417 &     375 $\pm$ 43 &          \nodata &          \nodata &            2.0\\
368&      193 $\pm$ 8 &     386 $\pm$ 86 &     262 $\pm$ 17 &    739 $\pm$ 313 &          \nodata &            6.4\\
369&      191 $\pm$ 6 &     398 $\pm$ 52 &     264 $\pm$ 18 &    807 $\pm$ 168 &          \nodata &           11.0\\
370&      239 $\pm$ 4 &     374 $\pm$ 17 &     332 $\pm$ 13 &     807 $\pm$ 42 &   1775 $\pm$ 271 &           58.0\\
\enddata 
\tablecomments{$T_{\mathrm{LR1}}$ is computed for the CO
  lines in the range $J_{up}$=14$-$18, $T_{\mathrm{SR1}}$ for
  $J_{up}$=19$-$25, $T_{\mathrm{R1}}$ for $J_{up}$=14$-$25,
  $T_{\mathrm{B2B}}$ for $J_{up}$=27$-$36 and $T_{\mathrm{B3A}}$ for
  $J_{up}$=37$-$46.}
\end{deluxetable*}

\section{Observed properties of FIR CO emission from protostars} \label{CO_prop}

\subsection{CO excitation diagrams \& rotational temperatures } \label{trot}
Since several CO lines are detected for many objects in the PACS
wavelength range, it is useful to construct a rotational excitation
diagram, which shows the relative populations of CO rotational states
derived from the observed line fluxes. In this diagram, the natural
logarithm of the detected number of CO molecules in a rotational state
divided by the degeneracy of that state, ln$(N_J/g_J)$, is plotted as
a function of the energy, $E_J$ \citep[see, e.g.,][]{gl99}. If the CO
emission is optically thin then the number of CO molecules in the
$J^{th}$ rotational state is given by,
\[N_J = (4~\pi~d^2~F_{J,J-1}) / (h~\nu_{J,J-1}~A_{J,J-1})\] 
where $F_{J,J-1}$, $\nu_{J,J-1}$ and $A_{J,J-1}$ are the line flux,
frequency and Einstein A-coefficient, respectively, corresponding to
the transition $J \rightarrow J-1$ and $d$ is the distance to the
source. The FIR CO lines observed with PACS are likely to be optically
thin if the size of the emitting region is sufficiently large
($\ga$1$\arcsec$). Figure~\ref{tau} shows the optical depths of the
CO~$J=14-13$ line observed toward the protostars in our sample for
three assumed angular sizes of emission. To compute the optical depth,
we assumed an excitation temperature of 300~K and a line width
($FWHM$) of 10 kms$^{-1}$.  The observed CO~$J=14-13$ transition,
which is the lowest excitation line observed and the most likely one
to be optically thick, has $\tau$ $\gg$ 1 only if the size of the
emitting region is $\ll$ 1$\arcsec$ ($\ll$ 420~AU at the distance of
Orion).  The extent of the observed CO emission is likely to be much
larger ($\sim$ 1000~AU) than this \citep{visser12}. Moreover,
  \citet{yildiz12} have shown that even for the CO~$J$~=~6$-$5
  line the optical depths are low ($\tau~<$~2) for broad velocity
  component ($FWHM$~$\ga$~10$-$15~km~s$^{-1}$). The higher excitation
  ($J_{up}~\ge~$14) CO lines observed with PACS are likely to be
  broader than this, which would make the line optical depths even
  lower. Thus, the optically thin assumption appears valid and the
above equation can be used to compute $N_J$ and to construct the
rotational diagrams.  The CO rotational diagrams for the 17 protostars
with CO detections are shown in Figure~\ref{rot_diagram}.

\begin{figure}
\centering
\epsscale{1.1}
\plotone{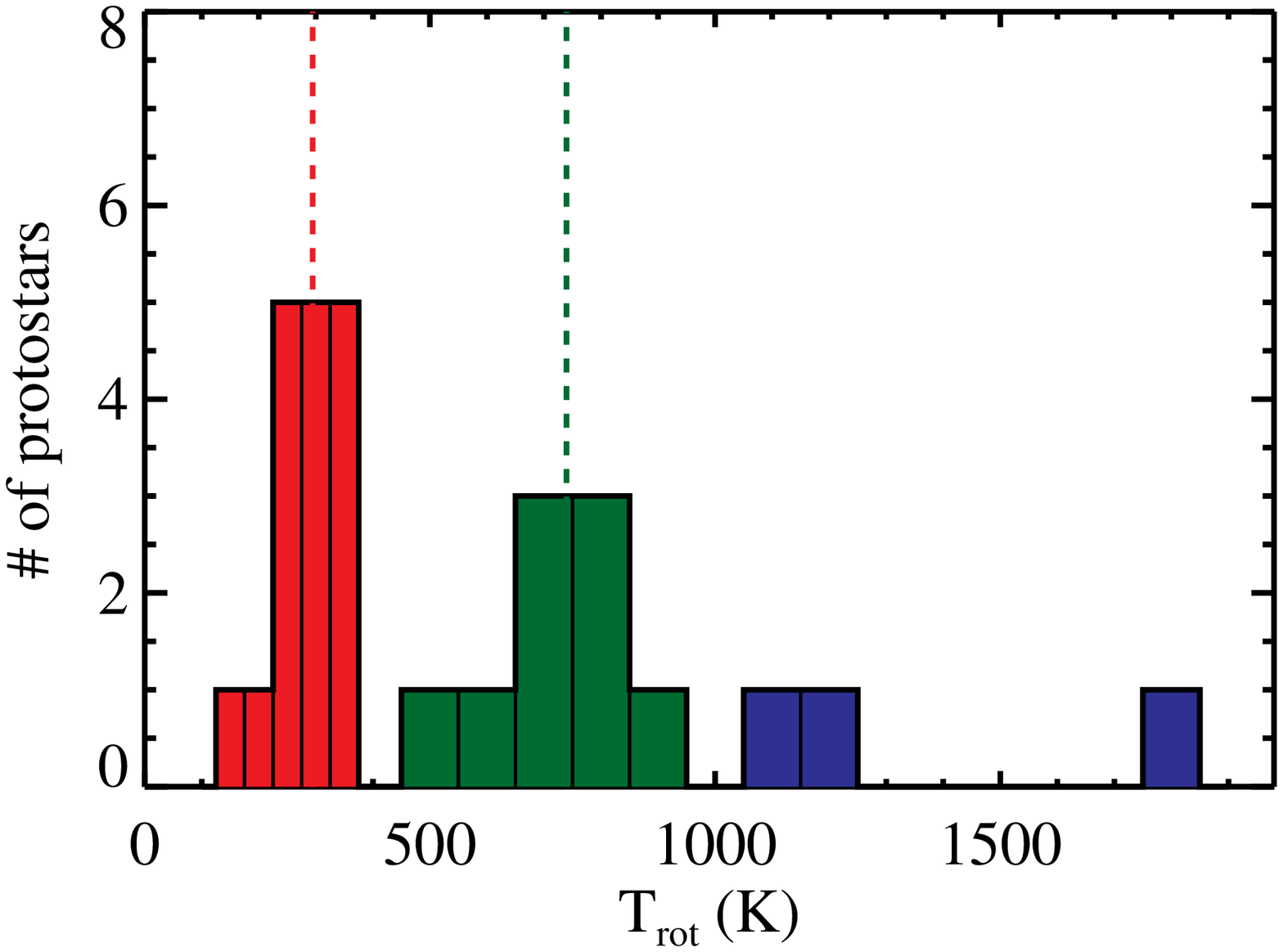}
\caption{ Distribution of the average CO rotational temperatures, 
  $T_{\mathrm{R1}}$~(red), $T_{\mathrm{B2B}}$~(green) and
  $T_{\mathrm{B3A}}$~(blue). Dashed lines indicate median values of
  $T_{\mathrm{R1}}$~(red) and $T_{\mathrm{B2B}}$~(green). \label{trb_hist}}
\end{figure}

The observed rotational diagrams of the protostars typically show a
positive curvature, which is more apparent in cases where more than 10
lines are detected. The rotational diagrams of a few other protostars
whose {\it Herschel}/PACS spectra have been published
\citep{vankemp10a,fich10, herczeg12} also show similar positive
curvature as noted by \citet{neufeld12}. This means that the
rotational temperature defined as
$T_{rot}=-(k$~dln[$N_J/g_J$]/d$E_J)^{-1}$ increases monotonically with
$E_J$. An average rotational temperature can be computed for a
relatively small range in $E_J$ by fitting a straight line to the
points in the observed rotational diagram.  We begin our analysis by
computing the average slopes of the observed rotational diagrams for
the PACS spectral bands R1 (102$-$190 $\micron$;
550~K~$\le~E_J/k~\le$~1800~K; $J_{up}$~=~14$-$25), B2B (72$-$98
$\micron$; 2000~K~$\le~E_J/k~\le$~3700~K; $J_{up}$~=~27$-$36), and B3A
(57$-$71 $\micron$; 3700~K~$\le~E_J/k~\le$~6000~K;
$J_{up}$~=~37$-$46). The best linear fit for the average slope was
obtained by minimizing $\chi^2$; from these slopes, we obtained the
average rotational temperatures, $T_{\mathrm{R1}}$, $T_{\mathrm{B2B}}$
and $T_{\mathrm{B3A}}$ which are shown in Figure~\ref{rot_diagram} and
are listed in Table~\ref{rot_tbl}. This method of computing $T_{rot}$
for CO lines observed within the same spectral band has the advantage
of minimizing possible systematic uncertainties introduced by any
mismatch in the relative flux calibration between different spectral
bands.

\begin{figure}
\centering
\epsscale{1.1}
\plotone{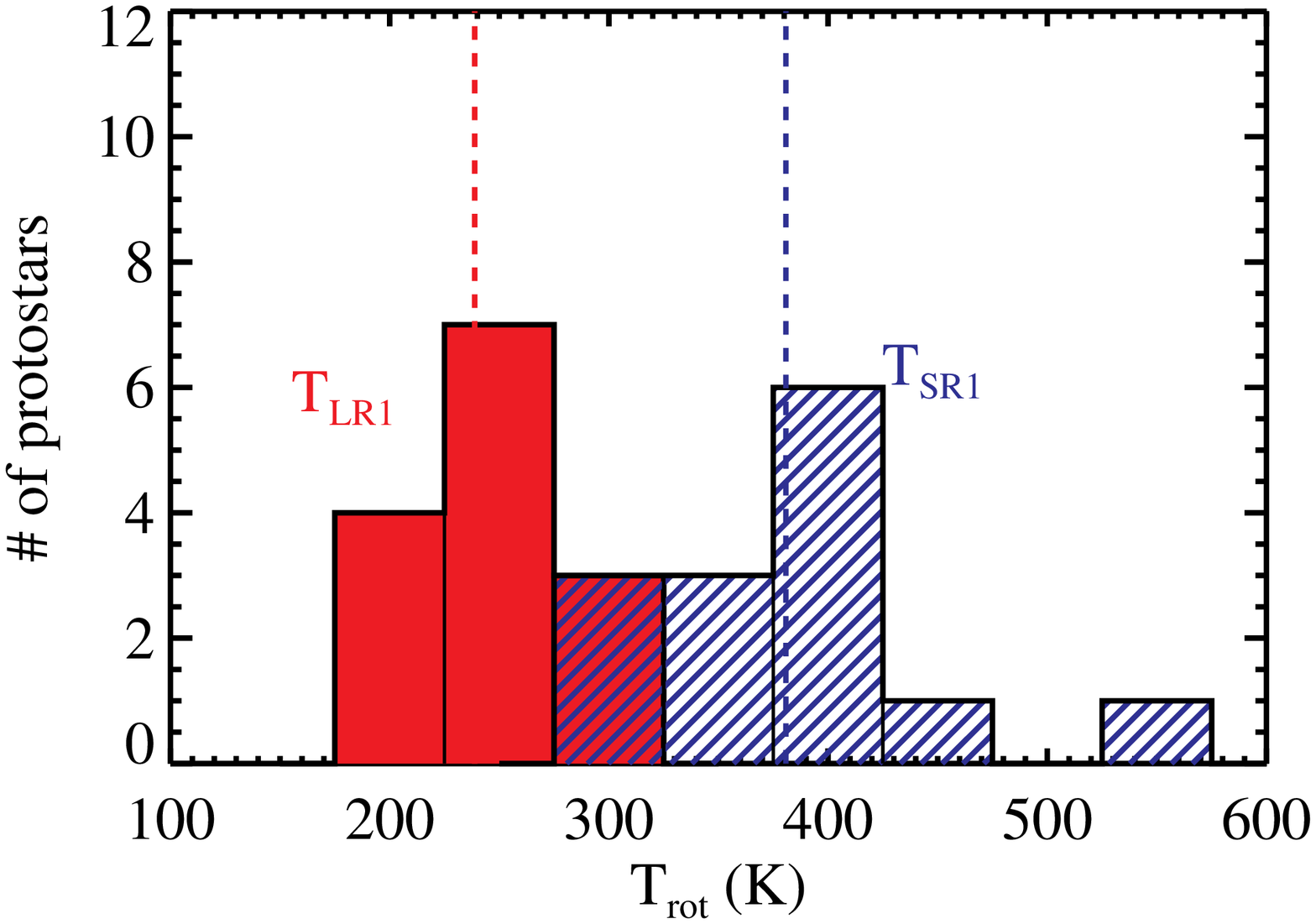}
\caption{Distribution of rotational temperatures $T_{\mathrm{LR1}}$ (red solid bars) and $T_{\mathrm{SR1}}$ (blue line bars). The dashed lines represent the median values of $T_{\mathrm{LR1}}$ (red) and $T_{\mathrm{SR1}}$ (blue). \label{tlsr1}}
\end{figure}

\begin{figure*}
\centering
\epsscale{0.96}
\plottwo{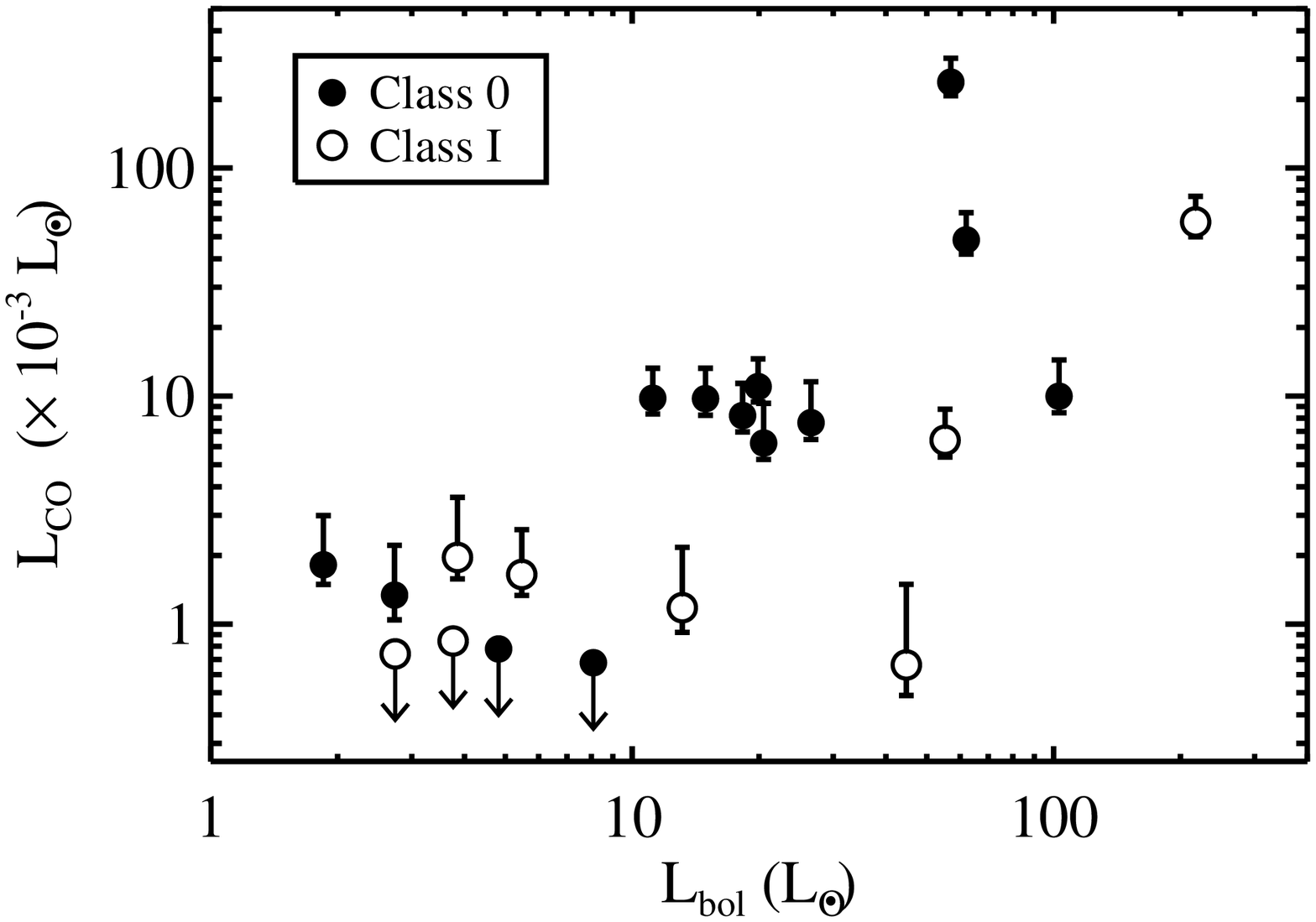}{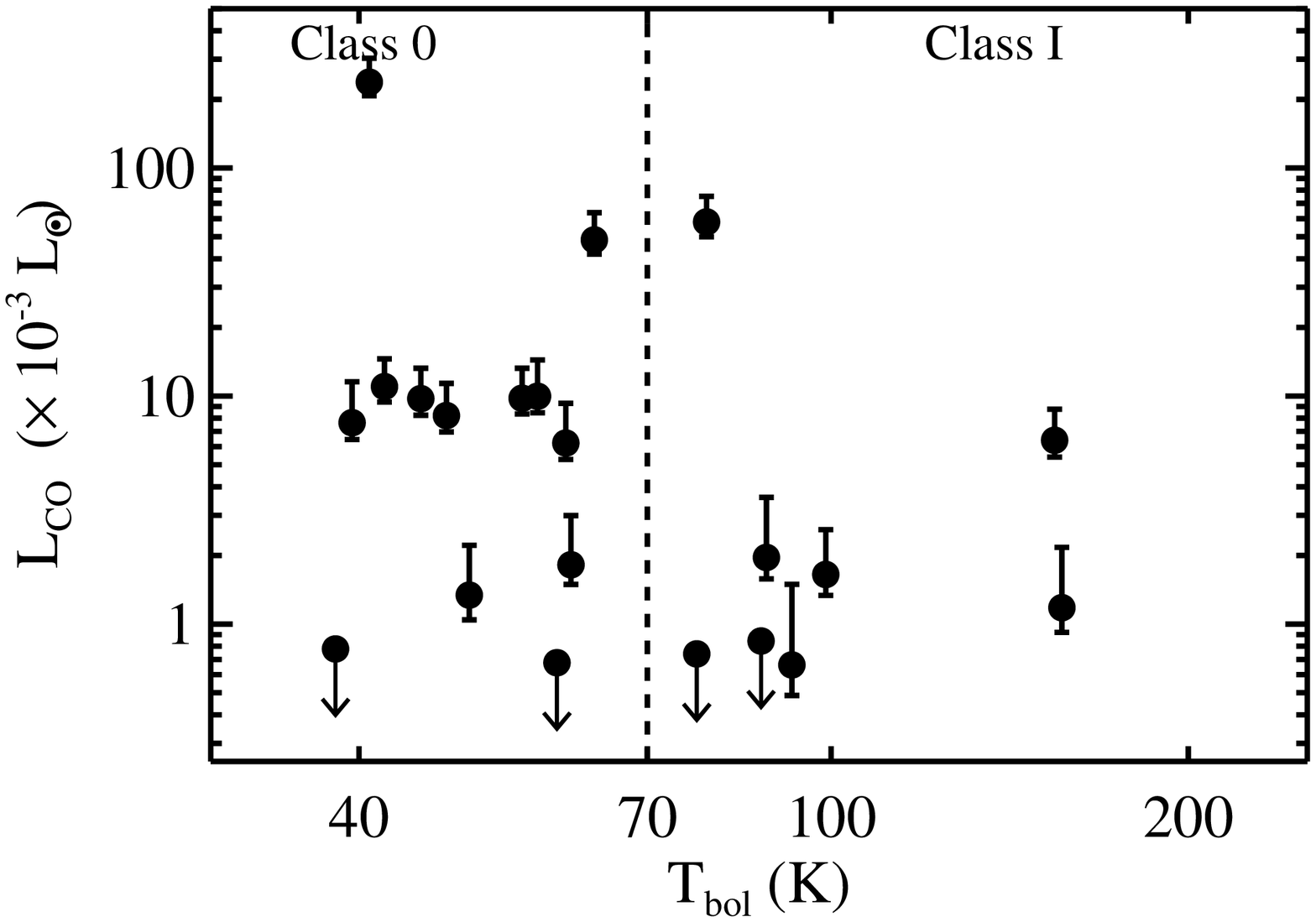}
\epsscale{0.48}
\plotone{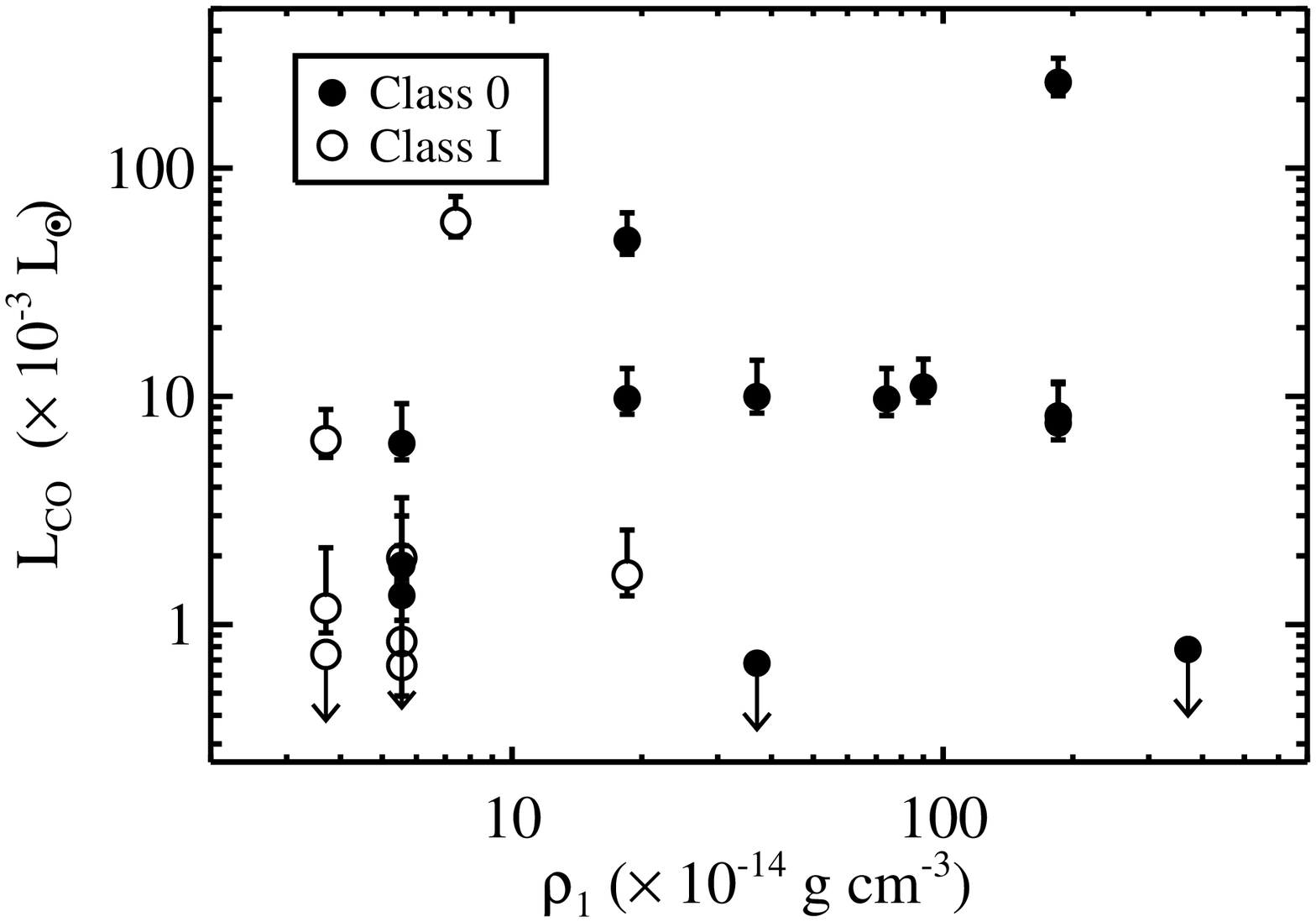}
\caption{Total luminosity of the CO lines detected within the PACS
  wavelength range with S/N~$\ge~3\sigma$ as a function of $L_{bol}$,
  $T_{bol}$ and $\rho_{1}$.  The lower error bars are measurement
  uncertainties. The upper error bars indicate the upper limits to
  $L_{\mathrm{CO}}$ obtained by adding the upper limits to the
  non-detected CO lines.  Downward arrows indicate upper limits to
  $L_{\mathrm{CO}}$ for sources where no CO lines are detected in the
  PACS wavelength range. \label{CO_lum}}
\end{figure*}

\begin{figure*}
\centering
\epsscale{0.96}
\plottwo{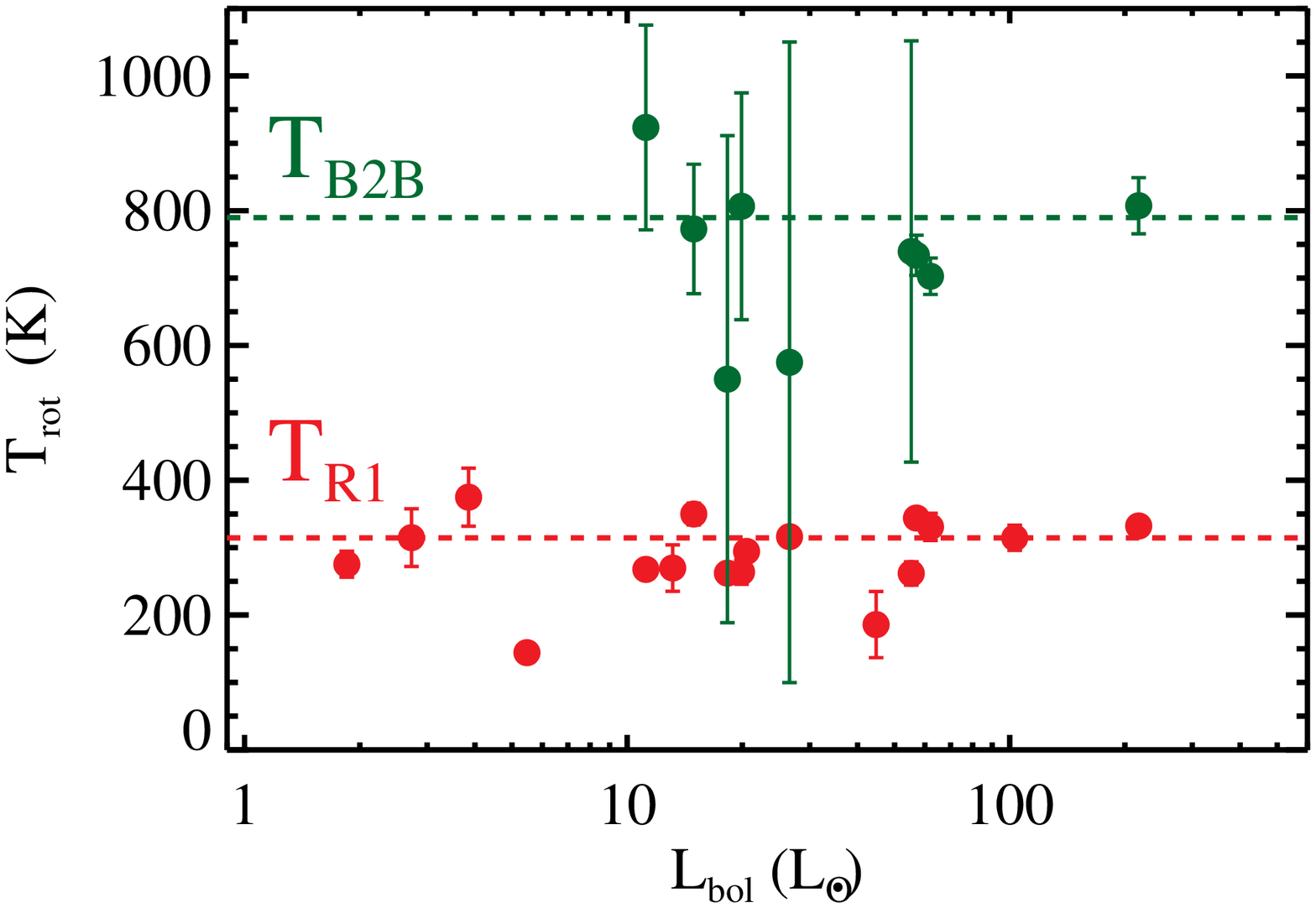}{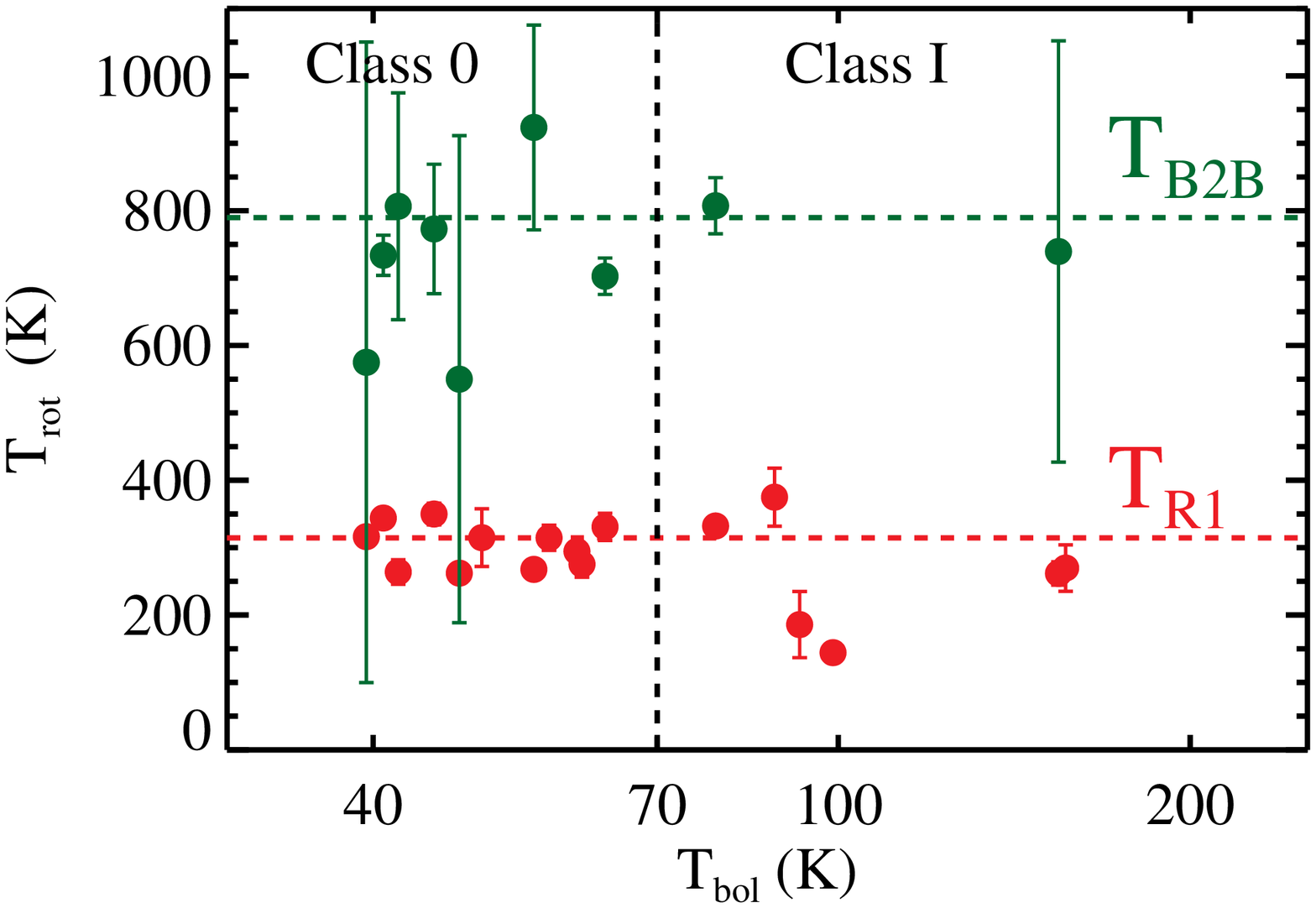}
\epsscale{0.48}
\plotone{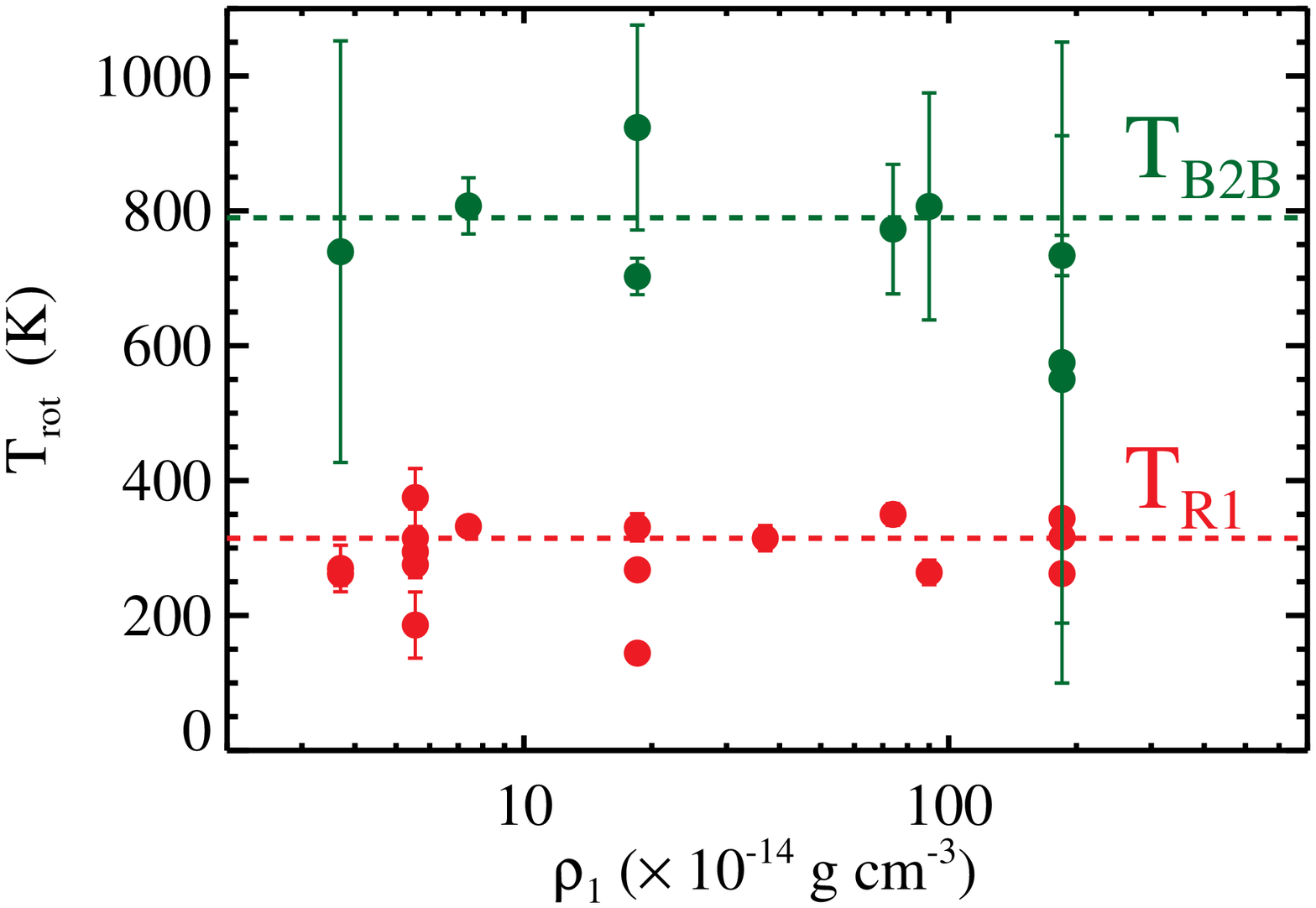}
\caption{CO rotational temperatures $T_{\mathrm{R1}}$ (red solid
  circles) and $T_{\mathrm{B2B}}$ (green solid circles) plotted
  against $L_{bol}$, $T_{bol}$,  and $\rho_{1}$. The dashed
  lines represent the median values of $T_{\mathrm{R1}}$ (red) and
  $T_{\mathrm{B2B}}$ (green).  \label{trot_fig}}
\end{figure*}

Figure~\ref{trb_hist} shows the distribution of the rotational
temperatures $T_{\mathrm{R1}}$, $T_{\mathrm{B2B}}$, and
$T_{\mathrm{B3A}}$. $T_{\mathrm{R1}}$ has a narrow distribution
ranging from 144~K to 375~K with a median value of 294~K. The
uncertainties in $T_{\mathrm{R1}}$ are relatively small with an
average value of 21~K.  $T_{\mathrm{B2B}}$ ranges from 550~K to 924~K
and has a median value of 739~K; it is less well determined and has
relatively large uncertainties (185~K). $T_{\mathrm{B3A}}$ could be
determined only for three sources and ranges from 1076~K to 1775~K
(see Table~\ref{rot_tbl} and Figure~\ref{trb_hist}). Since, as pointed
out above, $T_{rot}$ increases monotonically with $E_J$ (or $J$), the
estimated average rotational temperatures generally increase with the
number of lines detected in a spectral band. The spread in rotational
temperatures listed in Table~\ref{rot_tbl} is partly due to this
behavior. For instance,for eleven sources for which more than 9 CO
lines are detected in the R1 band, the corresponding $T_{\mathrm{R1}}$
values show a much smaller range (262~K to 350~K with a median value
of 315~K). Similarly, for six sources in which 5 or more lines are
detected in the B2B band, the corresponding $T_{\mathrm{B2B}}$ values
ranges from 703~K to 924~K. Thus, the observed CO rotational
temperatures, in particular $T_{\mathrm{R1}}$, is very similar for the
protostars in our sample.

\begin{figure}
\centering
\epsscale{1.1}
\plotone{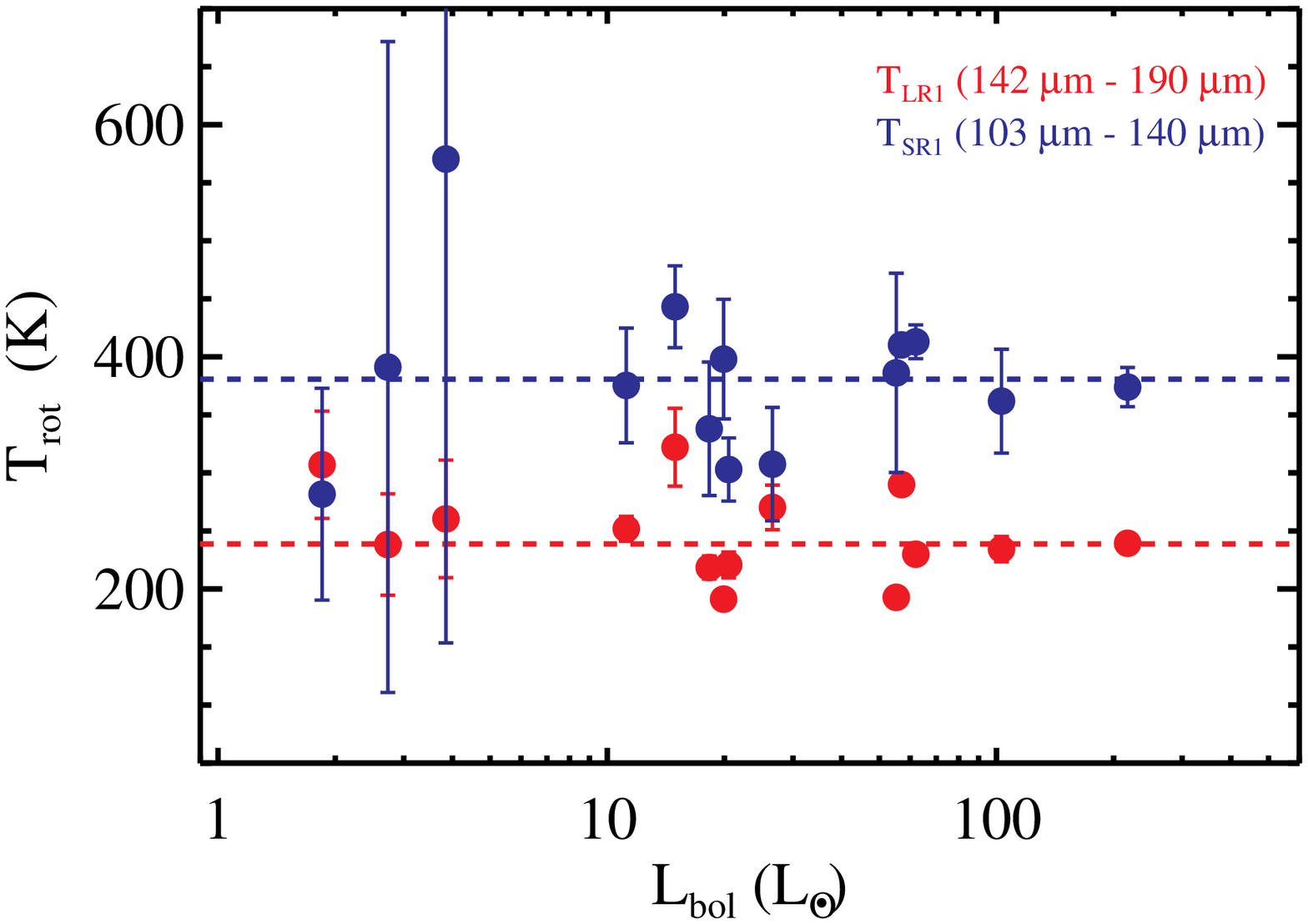}
\caption{$T_{\mathrm{LR1}}$ (red) and $T_{\mathrm{SR1}}$ (blue) as  functions of $L_{bol}$. \label{tlsr1_lbol}}
\end{figure}

We also computed the average rotational temperatures separately for
the PACS spectral bands Long R1 (142$-$190 $\micron$;
550~K~$\le~E_J/k~\le$~1000~K; $J_{up}$~=~14$-$18) and Short R1
(103$-$140 $\micron$; 1000~K~$\le~E_J/k~\le$~2000~K;
$J_{up}$~=~19$-$25) which we will call $T_{\mathrm{LR1}}$ and
$T_{\mathrm{SR1}}$. The rotational temperatures are computed only if
there are more than two lines detected in a given spectral band. The
distribution of $T_{\mathrm{LR1}}$ and $T_{\mathrm{SR1}}$ for sources
where both these temperatures could be determined are presented in
Figure~\ref{tlsr1}.  Although there is some overlap between the
derived values of $T_{\mathrm{LR1}}$ and $T_{\mathrm{SR1}}$, for most
objects there is a significant difference between the two. The median
value of $T_{\mathrm{LR1}}$ is 239~K and that for $T_{\mathrm{SR1}}$
is 386~K; this difference is more than that can be accounted for by
the uncertainties in these quantities. This difference suggests that
even in the R1 spectral band there is a significant curvature in the
rotational diagram for most objects indicating that even the mid$-J$
CO lines ($J_{up}$ = 14 to $J_{up}$ = 25) observed in the PACS R1
spectral band cannot be explained by a single temperature component in
local thermodynamic equilibrium (LTE).

\subsection{Correlation with protostellar properties}  \label{corr}

The CO emission lines detected in the PACS wavelength range have their
origin in the warm gas in the immediate vicinity (i.e., at projected
distance $\la$~2000~AU at 420~pc) of the protostars. Since the energy
that goes into the heating of the circumstellar gas has to be derived
ultimately from the central protostar, it is instructive to search for
possible correlations between the observed characteristics of the CO
emission and protostellar properties.

We first computed the total observed CO luminosity, $L_{\mathrm{CO}}$,
by adding up the fluxes of all the lines detected with S/N $\ge$
3$\sigma$.  $L_{\mathrm{CO}}$ is plotted as a function of $L_{bol}$,
$T_{bol}$, and the envelope density at 1~AU, $\rho_{1}$, for all the
protostars in our sample in Figure~\ref{CO_lum}. For sources where no
CO lines were detected within the PACS wavelength range, an upper
limit to $L_{\mathrm{CO}}$ was computed by adding up the 3$\sigma$
detection limit at the location of each CO line.  To carry out a
correlation analysis accomodating the upper limits, we used the
Astronomical Survival Analysis (ASURV) package \citep{lav92} that
implements the methods presented in \citet{feig85} and
\citet{isobe86}. We used the generalized {\it Kendall's} tau and {\it
  Spearman's} rank order tests to compute the correlation
probabilities between $L_{\mathrm{CO}}$ and the protostellar
properties.  These tests show that $L_{\mathrm{CO}}$ is tightly
correlated with $L_{bol}$. The associated {\it Spearman} rank
correlation coefficient, $r_s$~=~0.71 and the probability that
$L_{\mathrm{CO}}$ and $L_{bol}$ are uncorrelated is 0.2\%; the
generalized {\it Kendall's} tau test gives this probability to be
0.3\%. On the other hand, no statistically significant correlation is
found between $L_{\mathrm{CO}}$ and the protostellar evolutionary
indicators $T_{bol}$ ({\it Spearman} prob. = 17\% and {\it Kendall's}
tau prob. = 18\%) or $\rho_{1}$ ({\it Spearman} prob. = 8\% and {\it
  Kendall's} tau prob. = 9\%).

CO rotational temperatures $T_{\mathrm{R1}}$ and $T_{\mathrm{B2B}}$
(see Section~\ref{trot}) are plotted as a function of $L_{bol}$,
$T_{bol}$ and $\rho_{1}$ for the protostars in our sample in
Figure~\ref{trot_fig}. $T_{\mathrm{R1}}$ is uncorrelated with
$L_{bol}$ and remains constant over more than two orders of magnitude
in $L_{bol}$. $T_{\mathrm{B2B}}$ could be determined only for
protostars with $L_{bol}\ga10L_{\odot}$ as high$-J$ CO lines
($E_J\ge2000~K$) are detected only for these
objects. $T_{\mathrm{B2B}}$ also appears to be uncorrelated with
$L_{bol}$ and within the uncertainties it remains roughly constant
over a smaller range in $L_{bol}$. Further, $T_{\mathrm{R1}}$ and
$T_{\mathrm{B2B}}$ are found to be uncorrelated with the $T_{bol}$ and
$\rho_{1}$ values of the protostars, as shown in
Figure~\ref{trot_fig}. Additionally, Figure~\ref{tlsr1_lbol} shows
that $T_{\mathrm{LR1}}$ and $T_{\mathrm{SR1}}$ also are uncorrelated
with protostellar luminosity; they remain more or less constant over a
large range in $L_{bol}$. This behavior demonstrates that irrespective
of the range in $E_J$ (or $J$) over which the average rotational
temperatures are computed, they remain independent of $L_{bol}$.

In summary, the observed CO line fluxes and luminosities scale with
$L_{bol}$. The relative line fluxes (as quantified by the rotational
temperatures), however, appear independent of $L_{bol}$. Both the CO
line luminosities and the CO rotational temperatures show no
significant correlation with evolutionary indicators or envelope
properties such as $T_{bol}$, and $\rho_{1}$.

\begin{figure*}
\centering
\epsscale{0.8}
\plotone{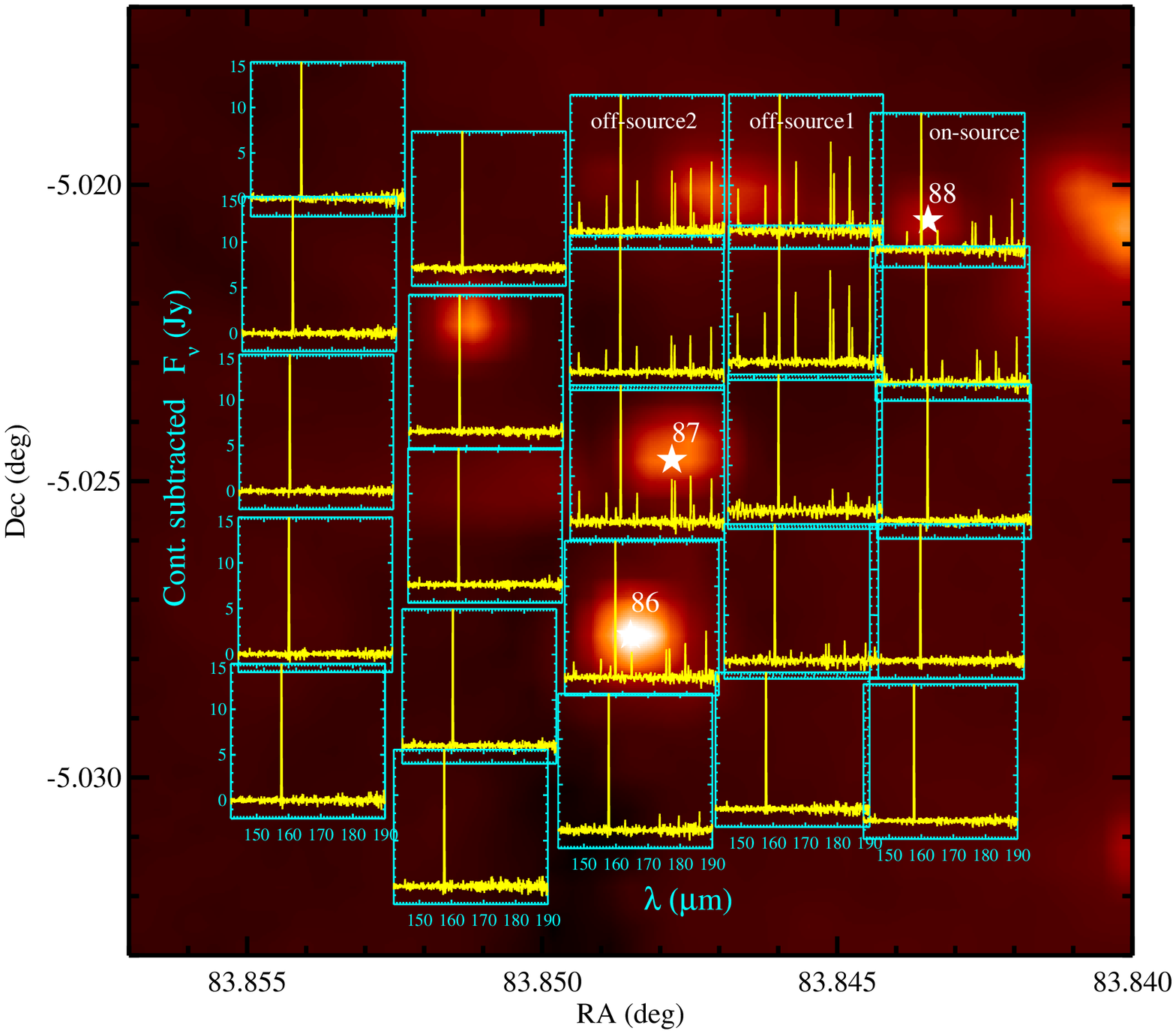}
\caption{ PACS field of the pointed observations centered on HOPS~87
  overlaid on {\it Spitzer}/IRAC 4.5~$\micron$ map. The protostar
  HOPS~88 \citep[MMS~5;][]{chini97} is seen to the north-west within
  the spaxel labeled {\it on-source}. HOPS~88 is driving an outflow in
  the east-west direction. Bright H$_2$ emission from outflow lobes is
  seen on either side of HOPS~88 in the IRAC image.  The redshifted
  lobe of the outflow falls in the two spaxels to the east, which are
  labeled as {\it off-source1} and {\it off-source2}. A continuum
  subtracted spectrum in the spectral band Long R1 (142$-$190
  $\micron$) is displayed in each spaxel. \label{hops88_fov}}
\end{figure*}

\begin{figure*}
\centering
\epsscale{0.45}
\plotone{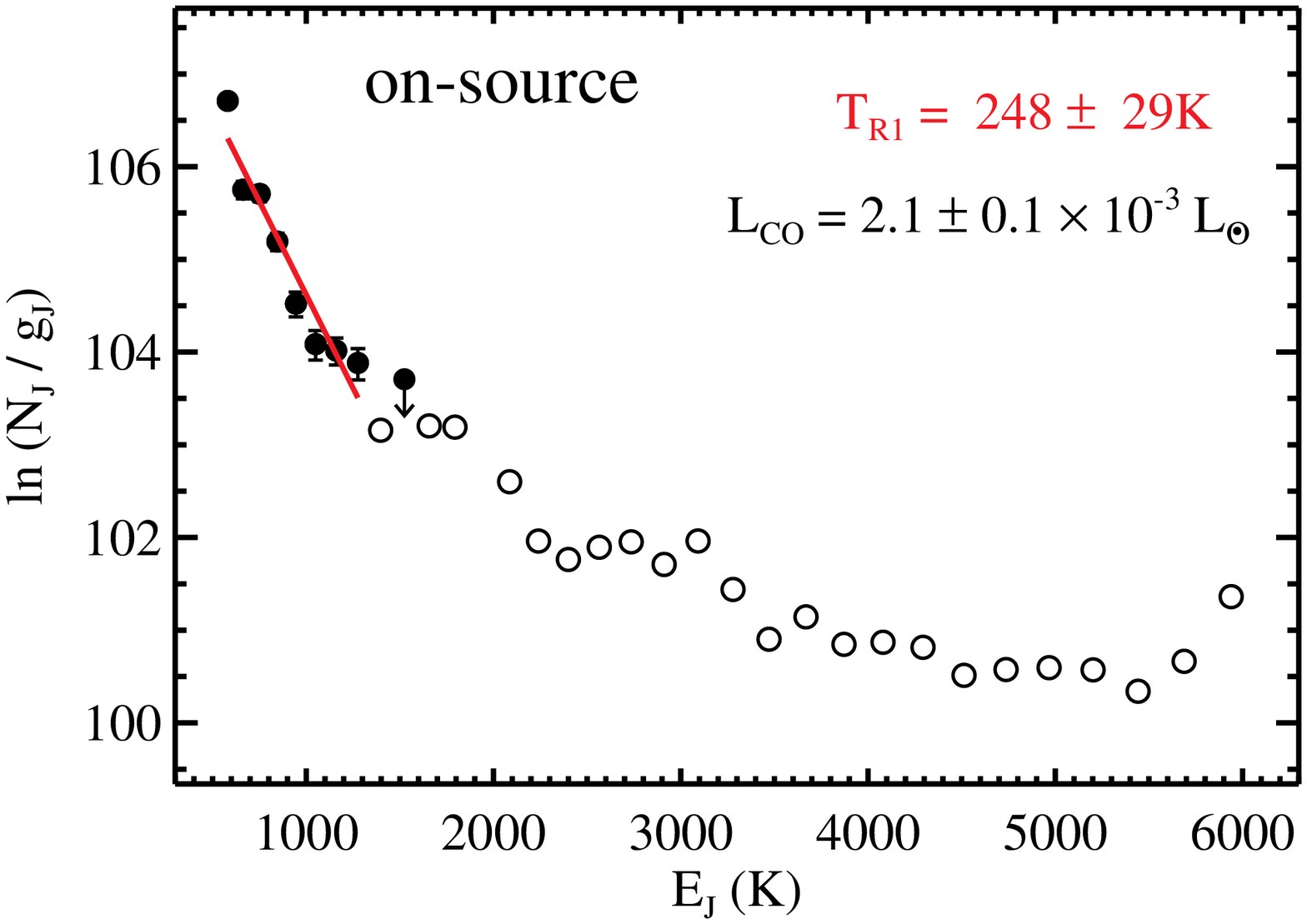}
\epsscale{0.9}
\plottwo{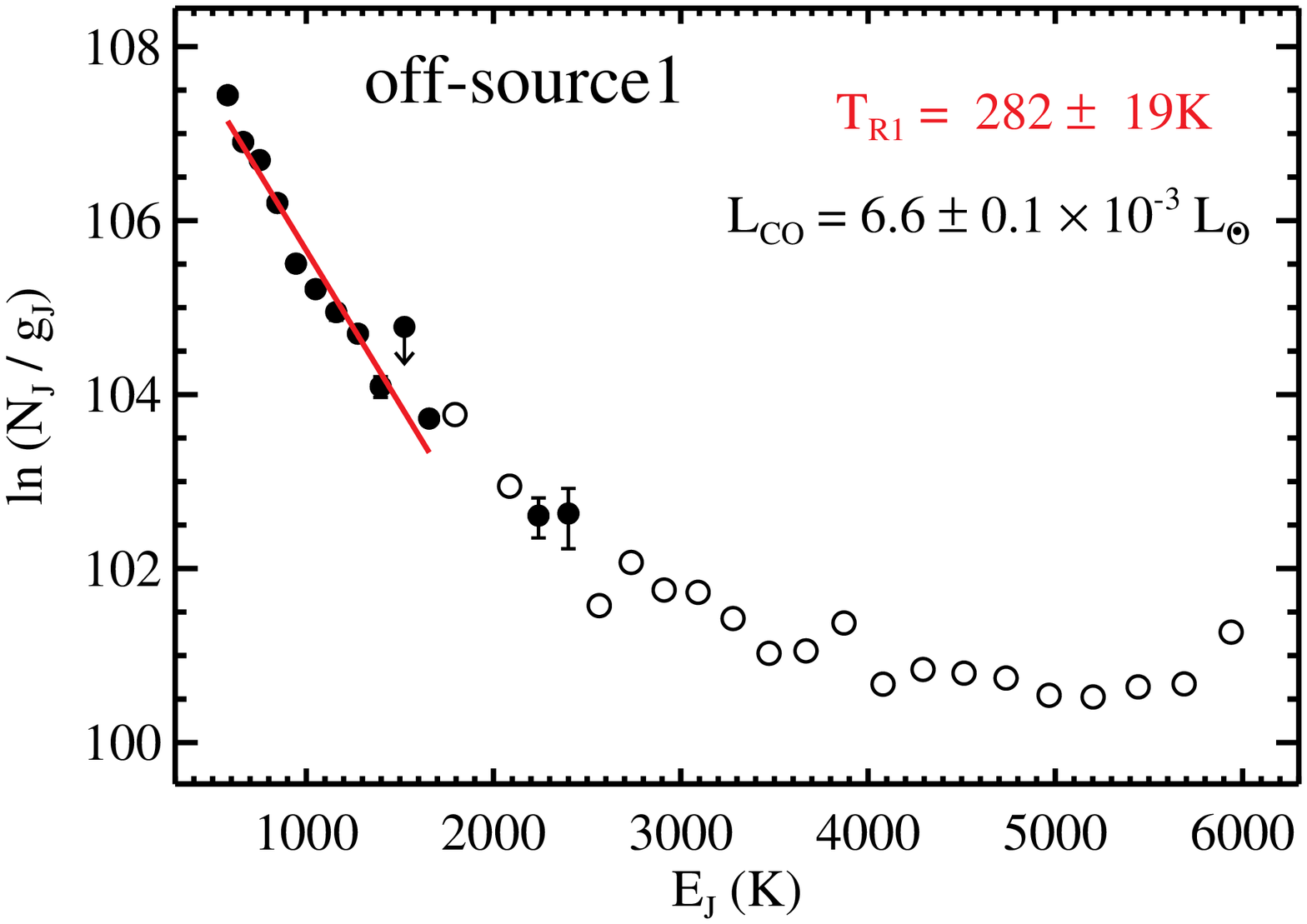}{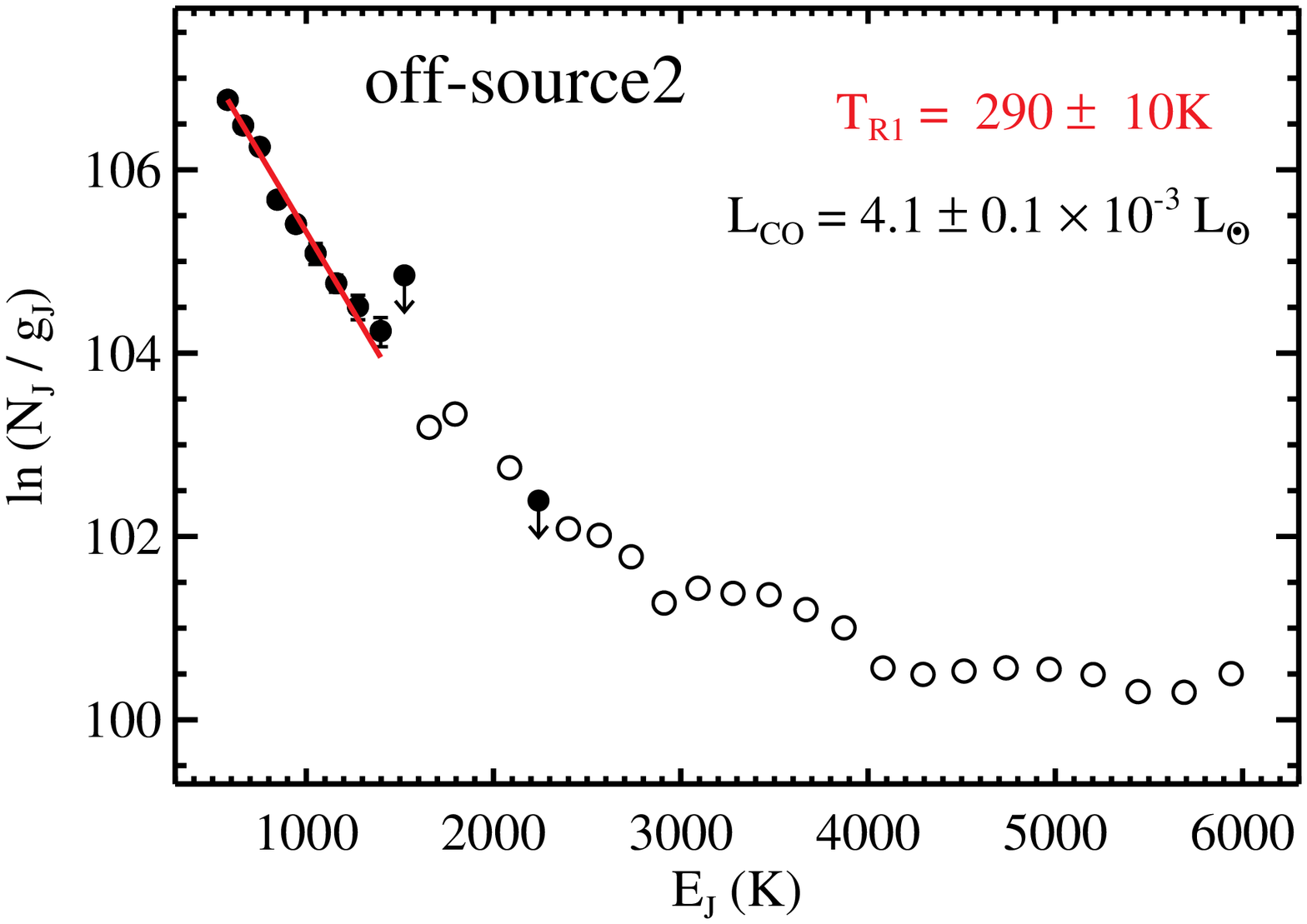}
\caption{CO rotational diagrams for the {\it on-source} spaxel and the
  {\it off-source} outflow spaxels associated with HOPS~88 as shown in
  Figure~\ref{hops88_fov}. \label{hops88_rot}}
\end{figure*}

\subsection{CO emission from an outflow lobe} \label{lobe}

In the PACS field observed toward one of the sources in our sample,
HOPS~87, we detected CO emission from the extended lobe of a molecular
outflow away from the exciting source. The PACS spectral map centered
on HOPS~87 is shown in Figure~\ref{hops88_fov}. Another protostar,
HOPS~88 \citep[a.k.a. MMS~5;][]{chini97}, seen toward the
north-western edge of this field, is driving a bipolar outflow along
the east-west direction \citep{williams03, takahashi08}. The red lobe
of this outflow falls on the two spaxels east of the on-source spaxel
where HOPS~88 is located. Figure~\ref{hops88_fov} shows that the line
emission is brighter in the outflow spaxels labeled ``off-source
1~\&~2'' than in the on-source spaxel. The CO rotational diagrams for
the on-source spaxel and both off-source spaxels corresponding to the
redshifted outflow lobe are shown in Figure~\ref{hops88_rot}. The
rotational temperature, $T_{\mathrm{R1}}$, measured for the on-source
spaxel and for the outflow spaxels are nearly the same, within the
uncertainties. The luminosities of the CO lines detected with PACS at
the outflow lobe positions far away from the protostar are higher than
that from the spaxel where HOPS~88 is located. For example,
$L_{\mathrm{CO}}$ from the off-source~2 position, which is $\sim$
19$\arcsec$ (projected distance $\sim$ 8000~AU) from the protostar, is
about a factor of 2 higher than that for the on-source position (see
Figure~\ref{hops88_rot}). Additionally, the rotational temperature,
$T_{\mathrm{R1}}$, and the CO luminosity, $L_{\mathrm{CO}}$, observed
for the emission from the off-source positions are very similar to
those found for the on-source positions of all the other protostars in
our sample (see Figure~\ref{trb_hist}~\&~\ref{CO_lum}).

\section{Excitation condition of the CO emitting gas} \label{excitation}

Our most striking result is the independence of CO line ratios (or
equivalently rotational temperatures) with respect to protostellar
luminosity, and in a relatively large sample of protostars. The
interpretation of this result and other apparent correlations, or lack
thereof, between the observed properties of the CO emission and the
protostellar properties depends on the physical conditions
(temperature and density) in the emitting molecular gas. The CO
emitting gas will be in LTE if the total gas (molecular hydrogen)
density, $n\mathrm{(H_2)}$, is greater than $n_{cr}$, the critical
density of the CO transitions at a given temperature.  The critical
density of the lowest$-J$ transition ($J=14-13$) that we detect is
2$-$3 $\times$ 10$^6$ cm$^{-3}$ and that of the highest$-J$ transition
($J=46-45$) is 5$-$7 $\times$ 10$^7$ cm$^{-3}$ for temperatures in the
range of 300$-$3000~K \citep{neufeld12,yang10}. Therefore, at
densities $n\mathrm{(H_2)} \ga 10^8$ cm$^{-3}$, the CO rotational
states will be {\it thermalized} and the observed rotational
temperature is the physical temperature of the molecular gas. For gas
densities $\la~10^6$ cm$^{-3}$, the CO excitation is {\it
  sub-thermal} \footnote{In this paper, as in previous papers in the
  literature, the excitation is said to be {\it sub-thermal} if the
  level populations of the upper states are smaller than the values
  that would be obtained in LTE; conversely, the excitation is
  described as {\it thermal} if the relative level populations are
  given by Boltzmann factors appropriate to the gas temperature.  In
  either case, the process of collisional excitation involves
  colliding molecules that are assumed to have a thermal
  (i.e. Maxwell-Boltzmann) distribution of translational energies.}
and the rotational temperature can be significantly different from the
physical temperature of the gas.

\subsection{Modeling of CO emission from protostars} \label{model}

Recently, \citet{neufeld12} has demonstrated that the observed
curvature in the CO rotational diagrams obtained with {\it
  Herschel}/PACS can be described surprisingly accurately by optically
thin FIR CO emission originating in an isothermal medium or a medium
with a power-law distribution of temperatures, both at uniform
density. In the following we use these sets of models to explore the
likely physical conditions of the CO emitting gas in individual
sources.

\subsubsection{Isothermal medium}

We first consider the simple case of an optically thin medium of
uniform temperature and density.  We computed the synthetic CO
rotational diagrams for such a medium for a large range of values of
$T$, $n\mathrm{(H_2)}$, and a column density parameter $\tilde{N}$(CO)
\citep[for details see][]{neufeld12}.  We then compared the model
rotational diagrams with the observed ones to find the best fit
solutions by minimizing the reduced-$\chi^2$.

\begin{figure*}
\epsscale{1.0}
\centering
\plottwo{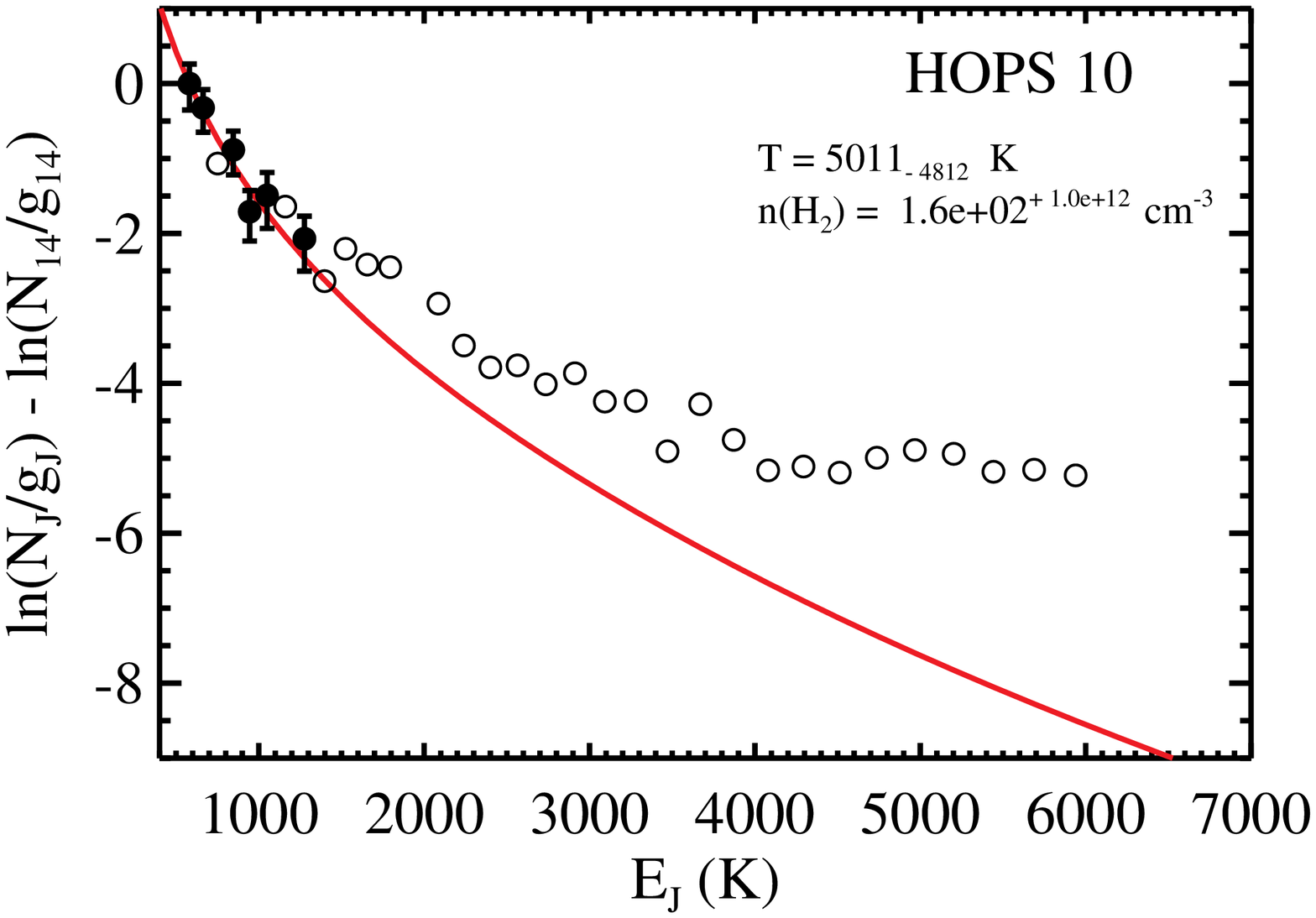}{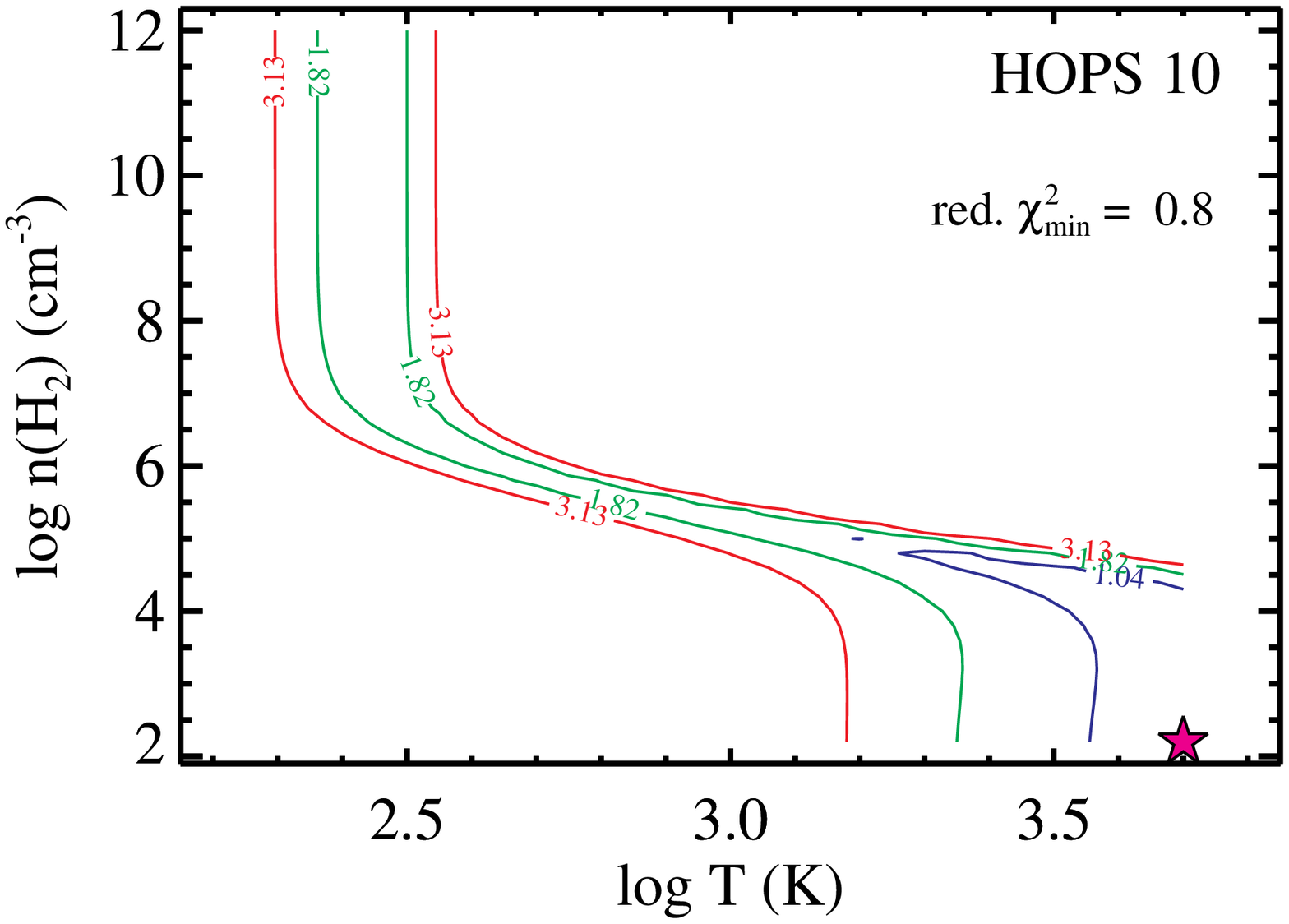}
\plottwo{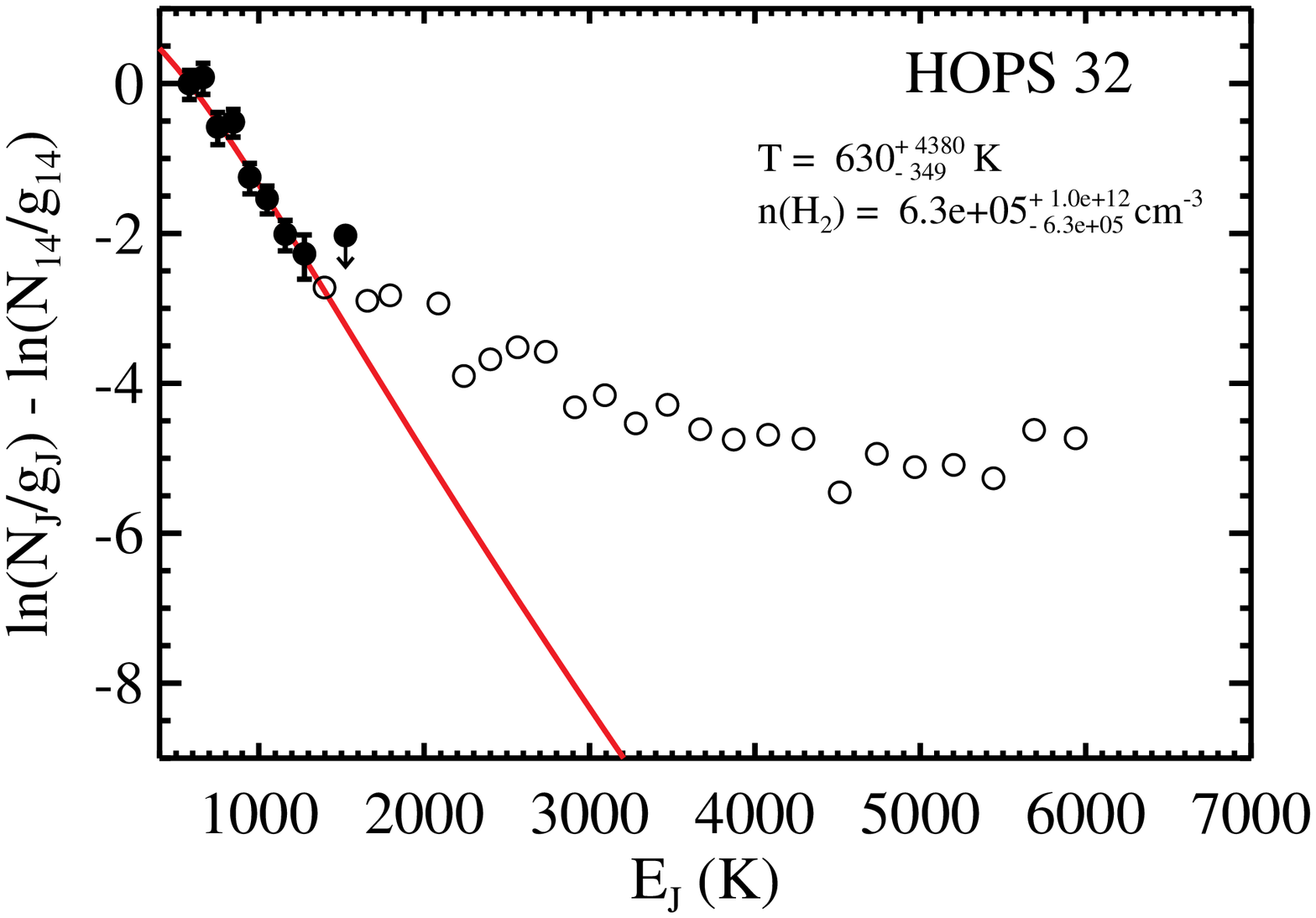}{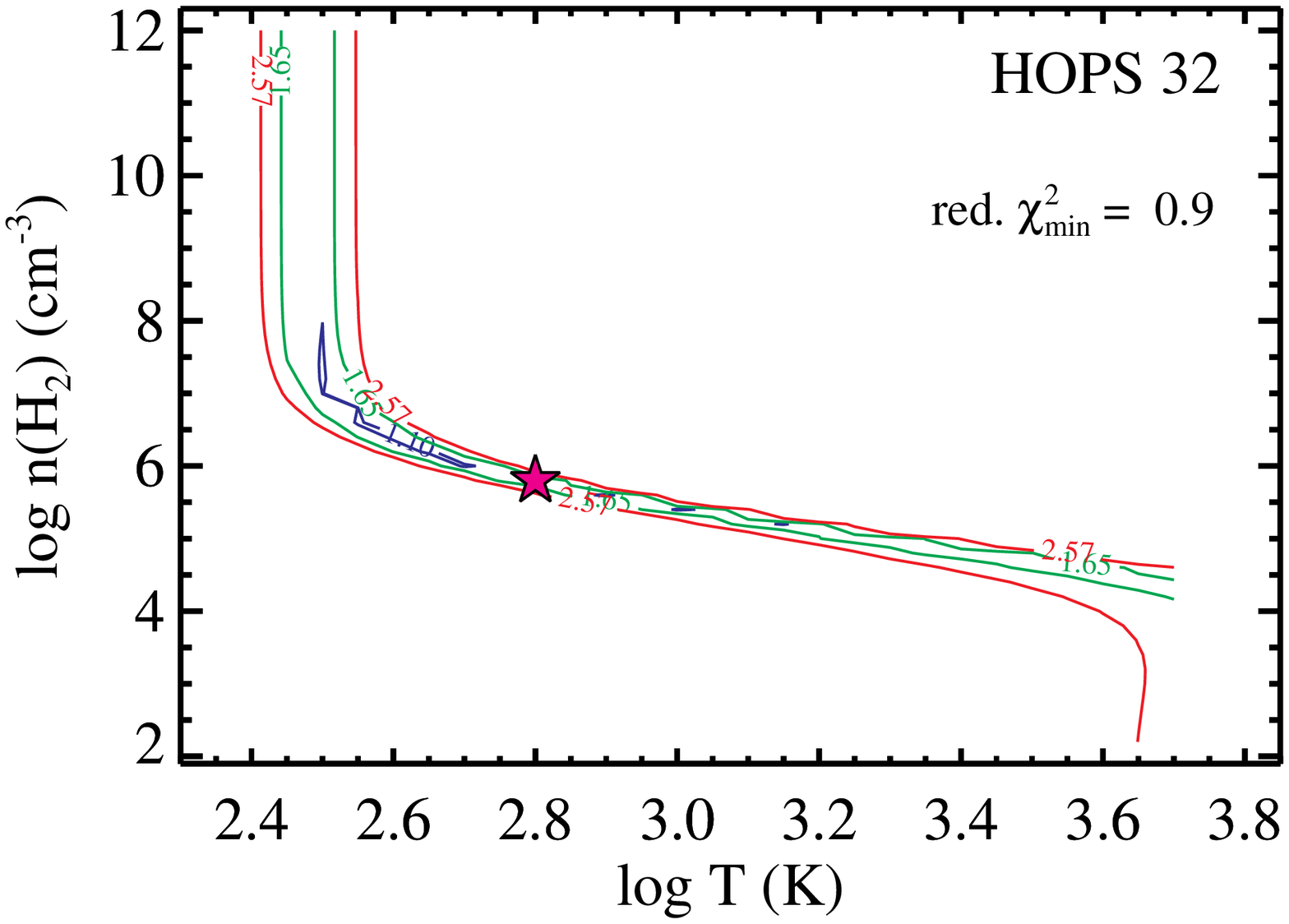}
\plottwo{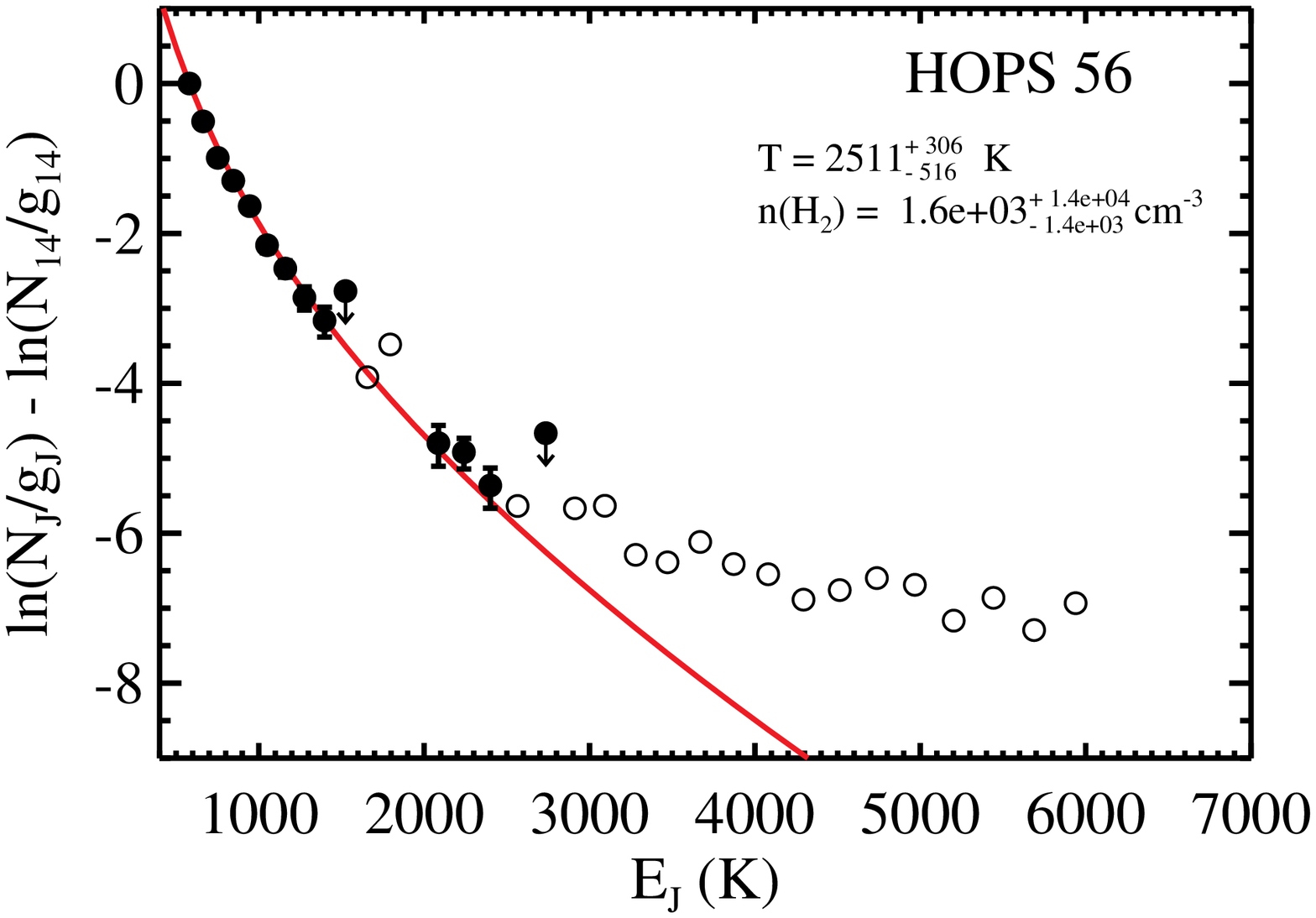}{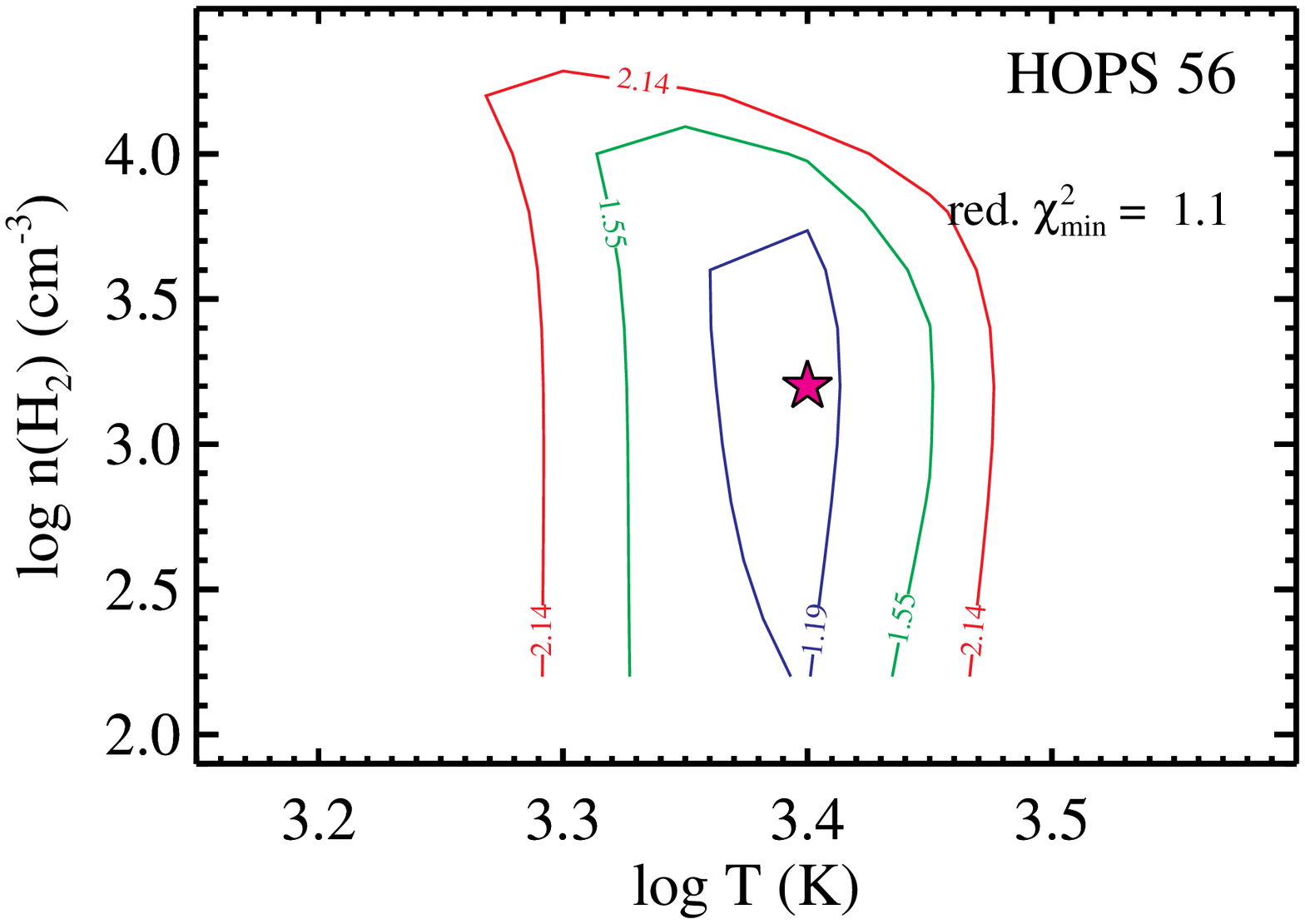}
\plottwo{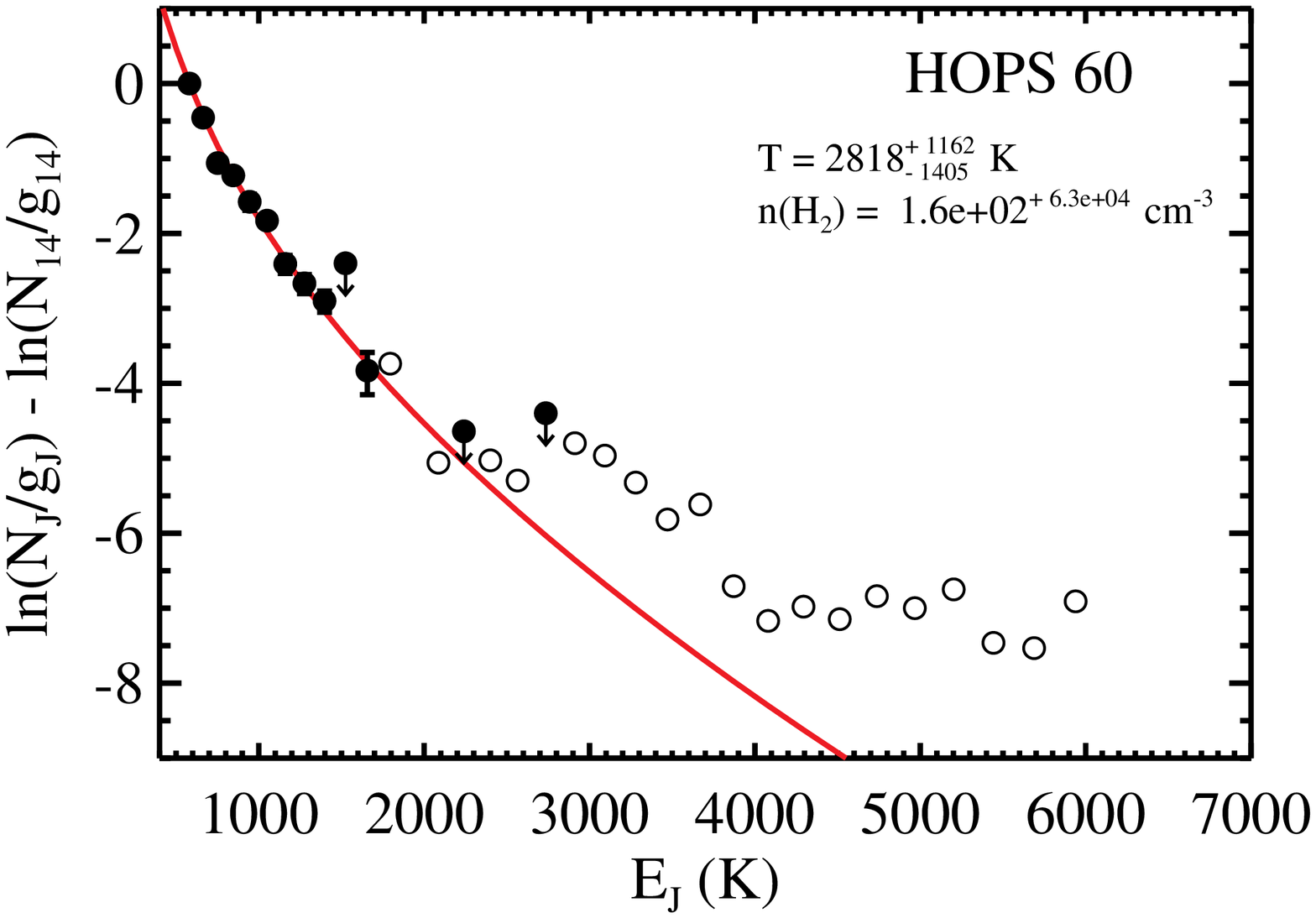}{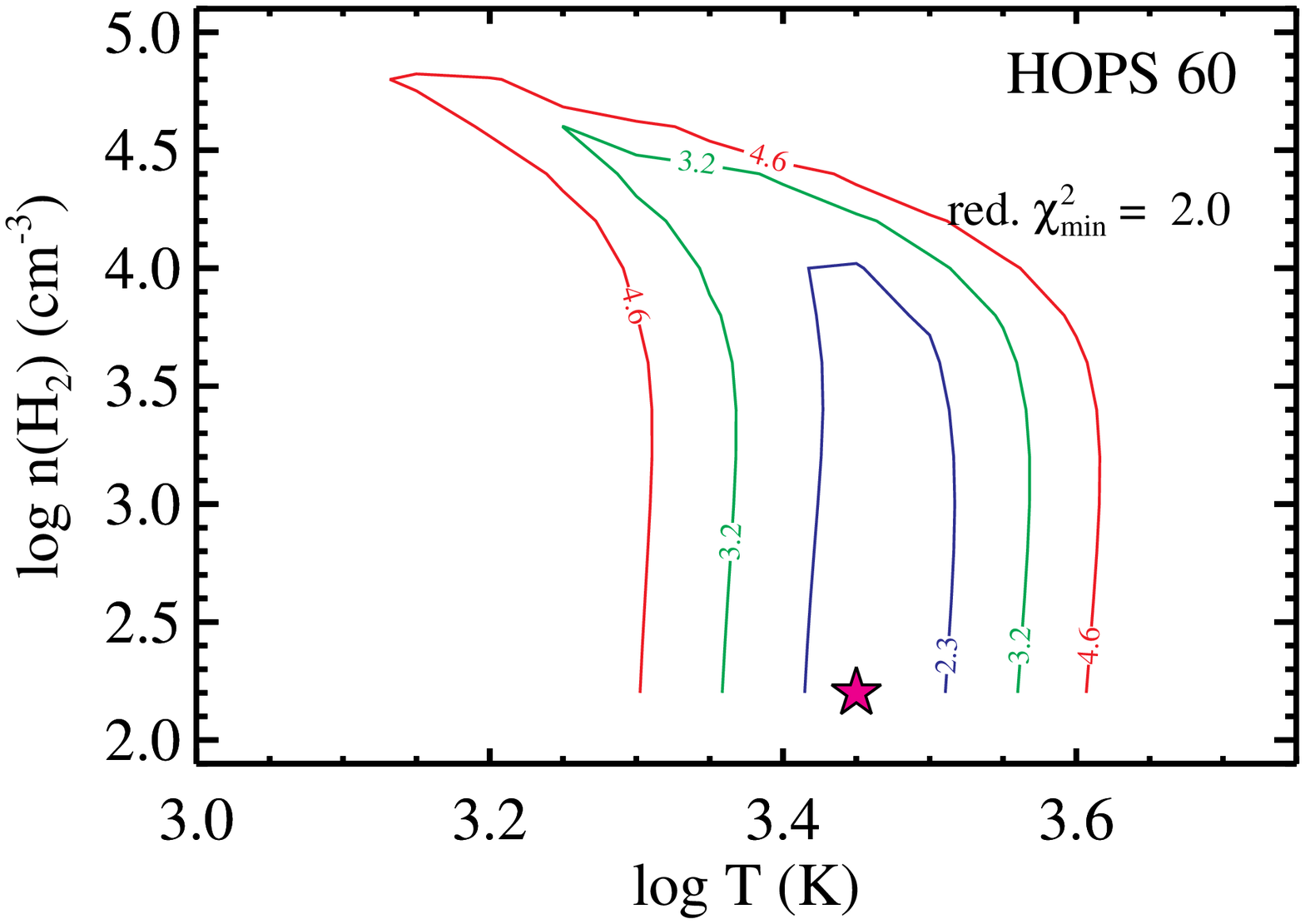}
\caption{ {\it (Left)}~Rotational diagrams of the observed CO emission
  with the best fits for the isothermal medium model overlaid (solid
  red line). Downward arrows indicate upper limits to the fluxes of the CO
  lines which are blended with a nearby line. Open circles correspond
  to 3$\sigma$ upper limits for the non-detections. {\it (Right)}
  Reduced-$\chi^2$ contours for $T$ and $n\mathrm{(H_2)}$. The star
  symbol marks the minimum value of the reduced-$\chi^2$. The contours
  corresponding to 68.3\% (blue), 95.4\% (green) and 99.7\% (red)
  confidence levels are shown.  \label{model_iso_rot}}
\end{figure*}

\begin{figure*}
\centering
\addtocounter{figure}{-1}
\plottwo{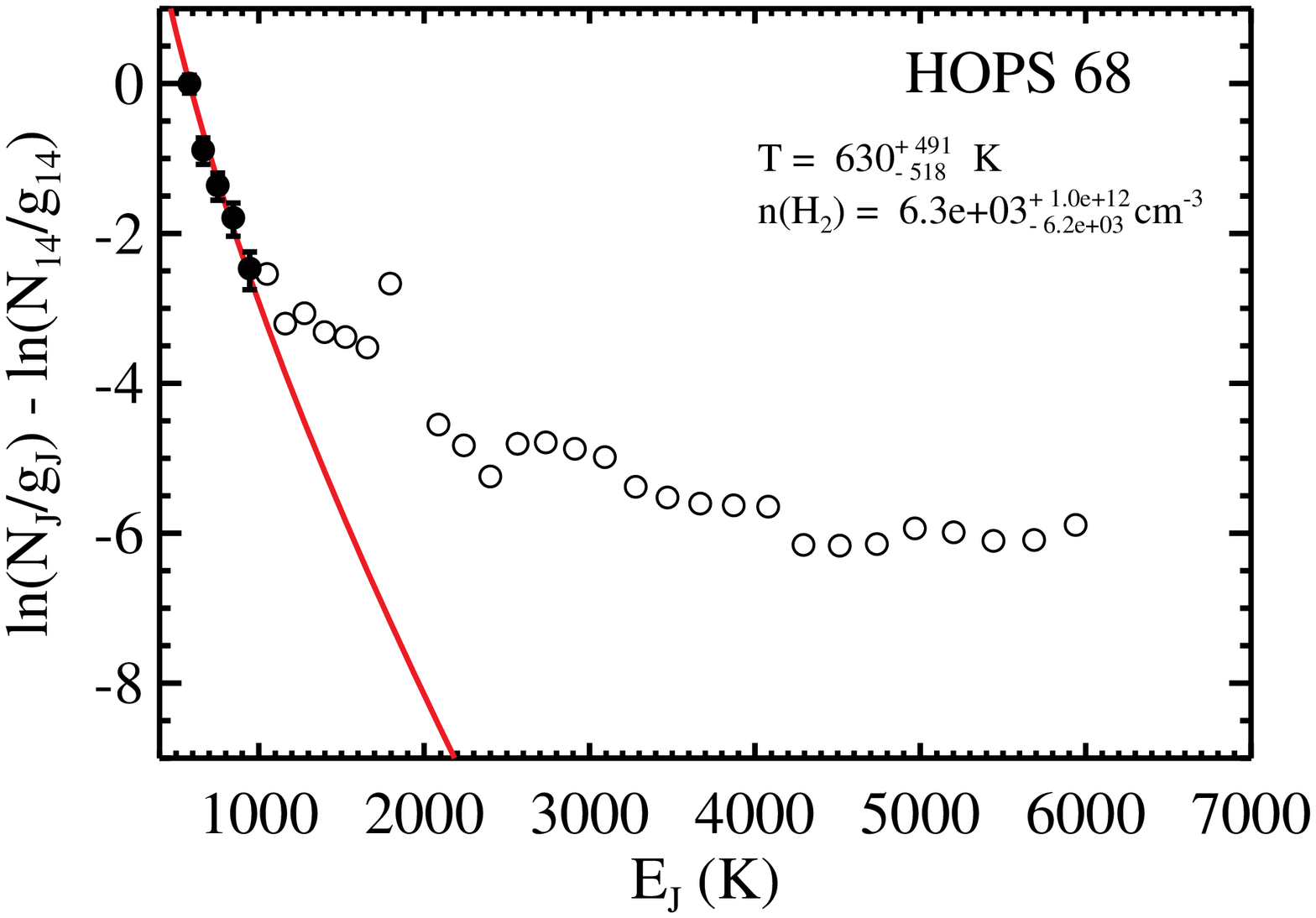}{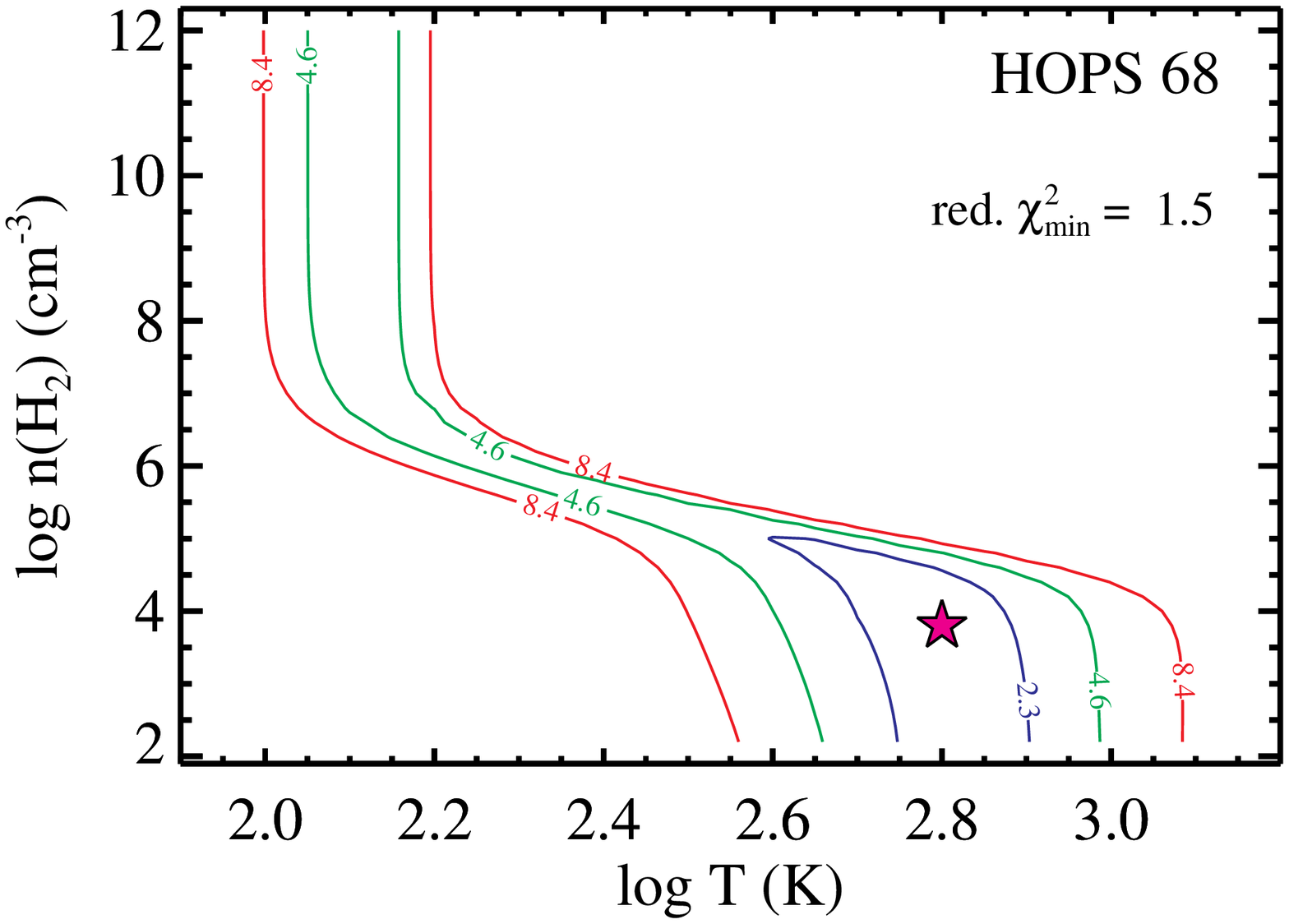}
\plottwo{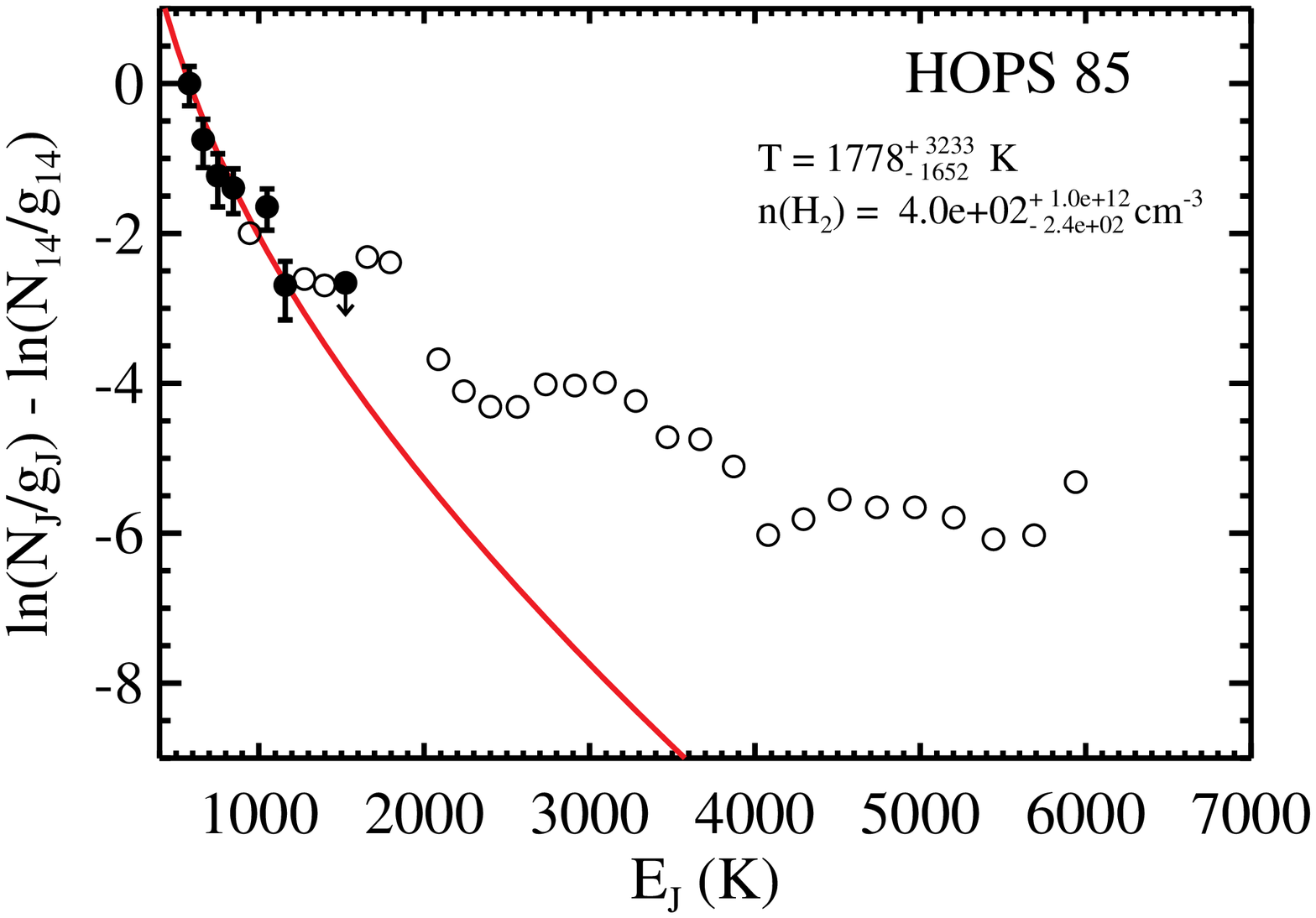}{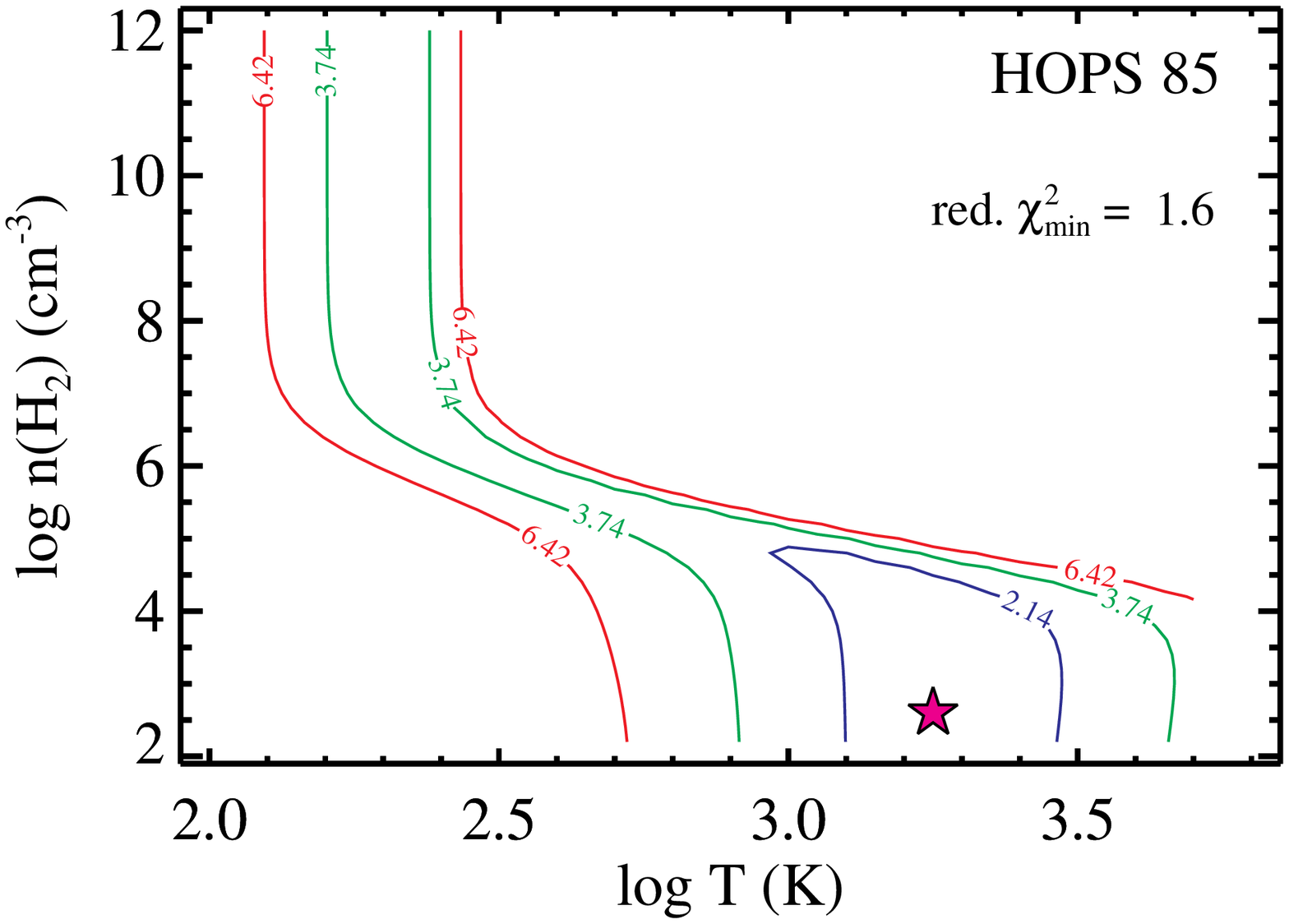}
\plottwo{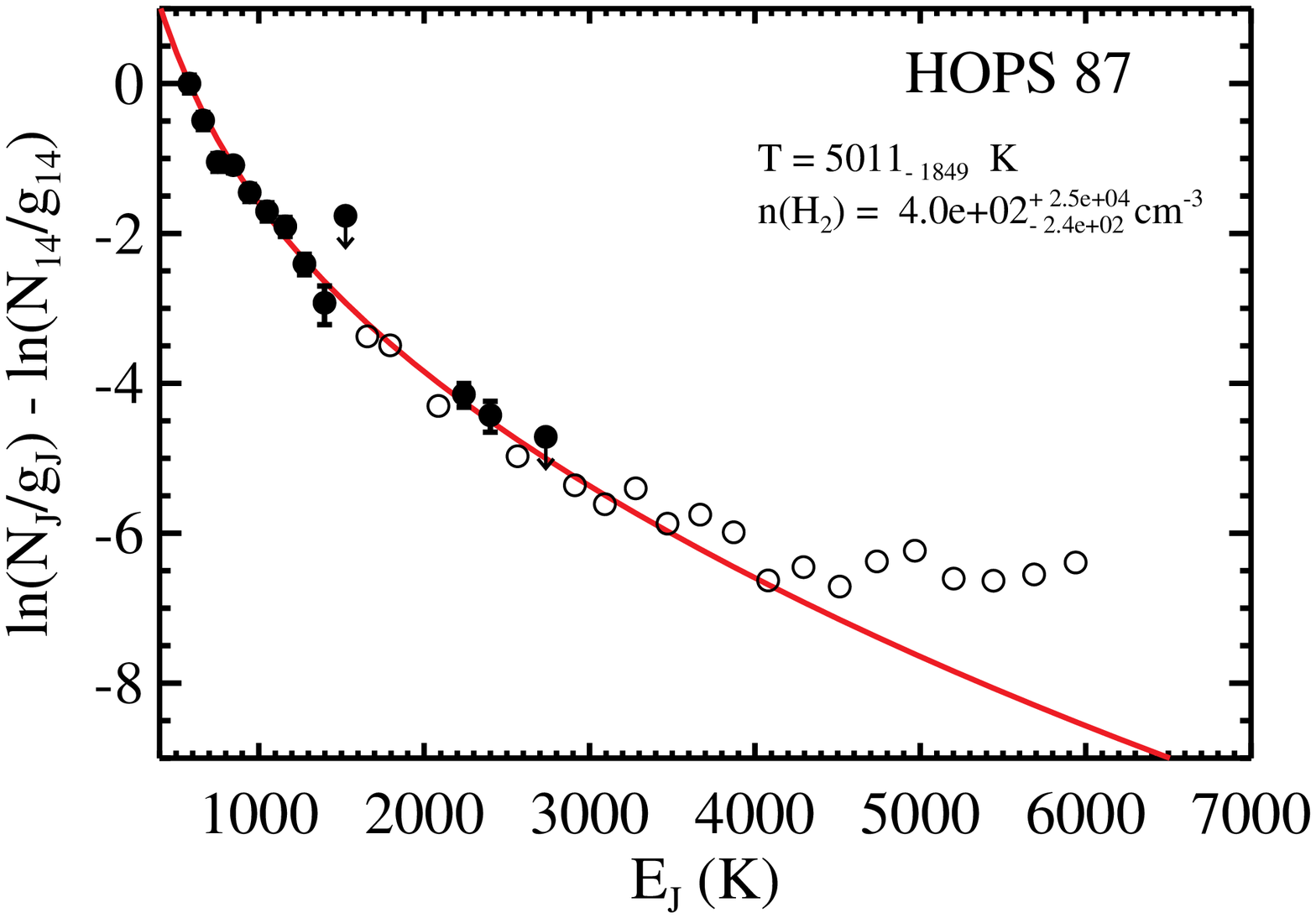}{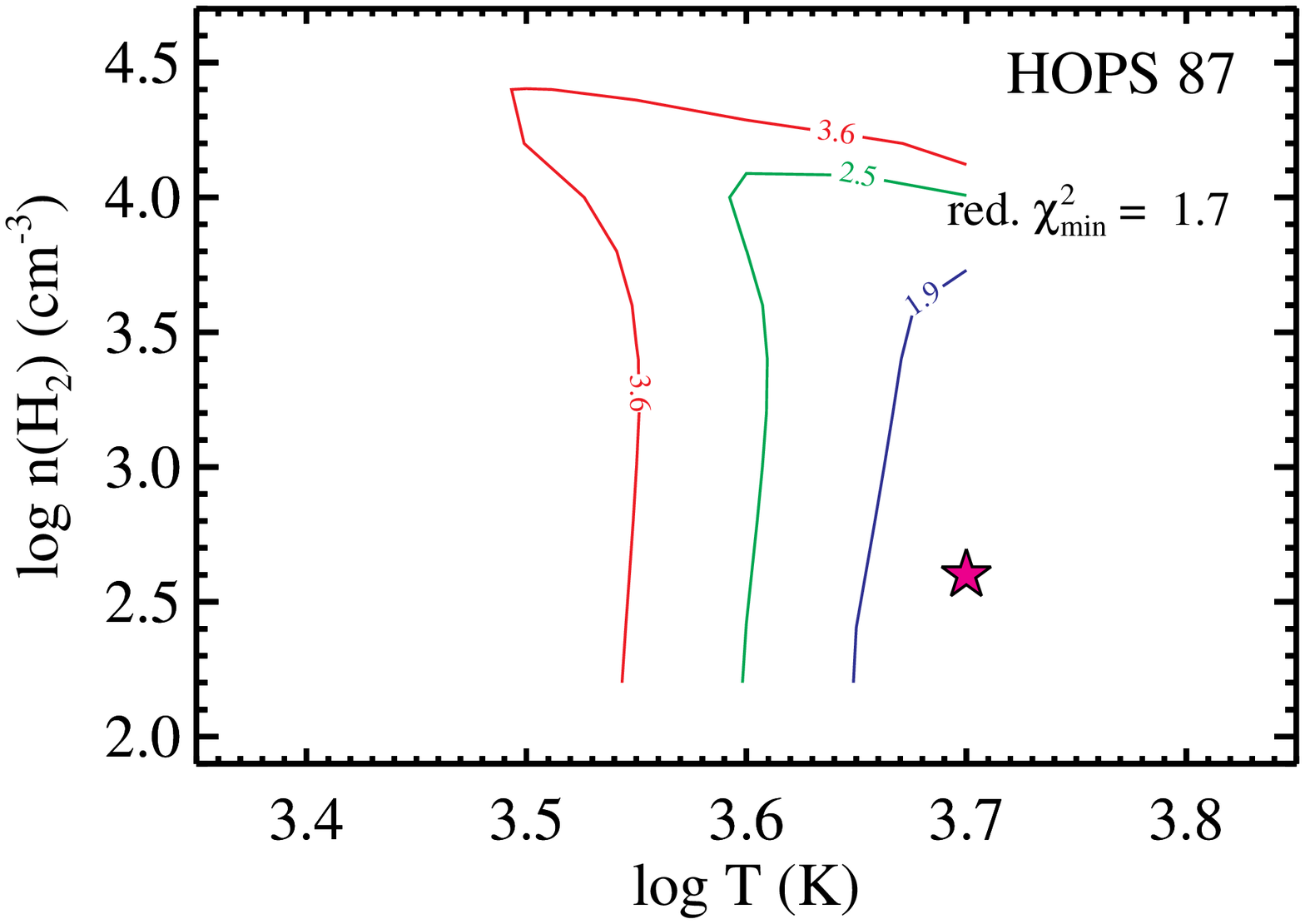}
\plottwo{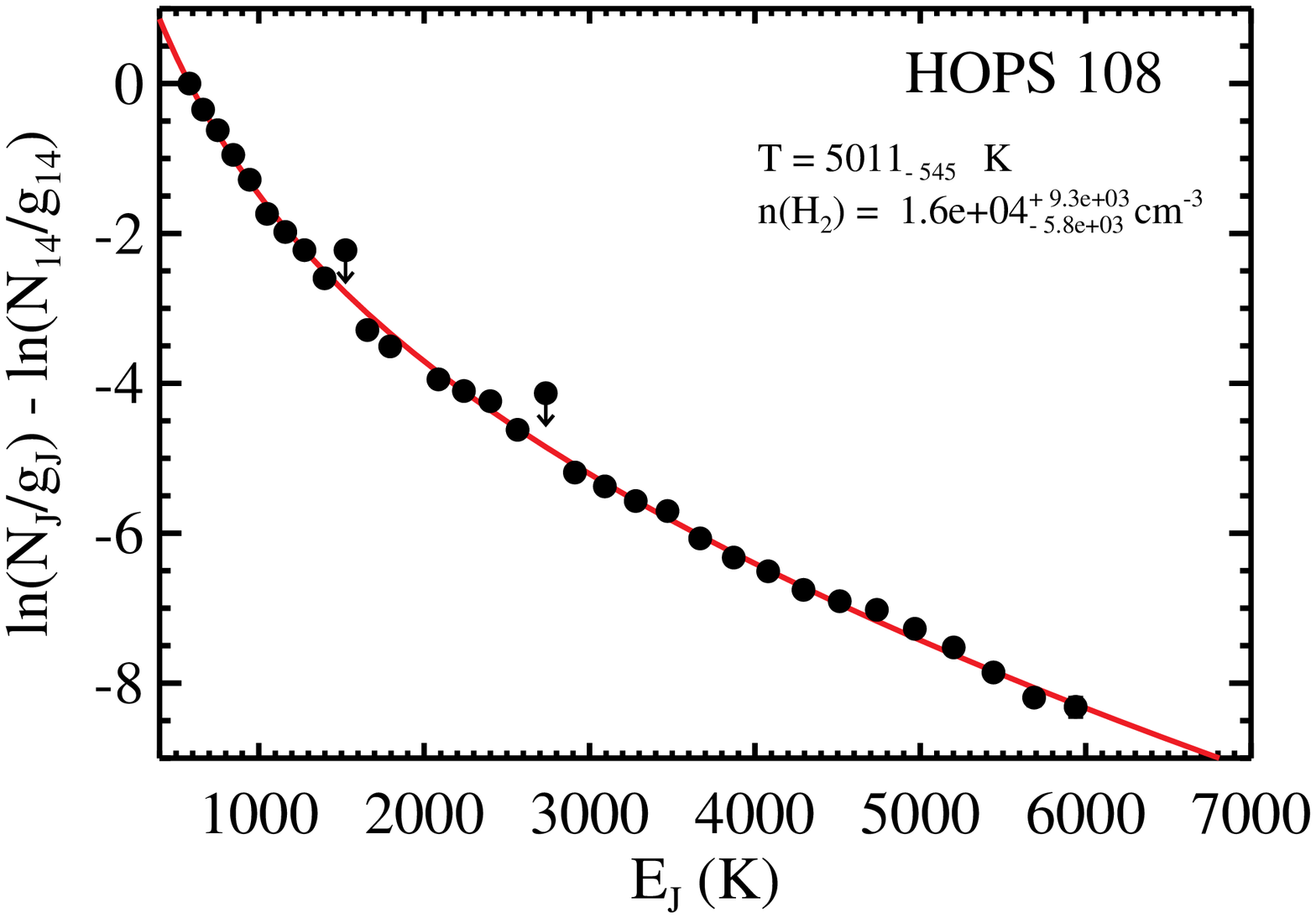}{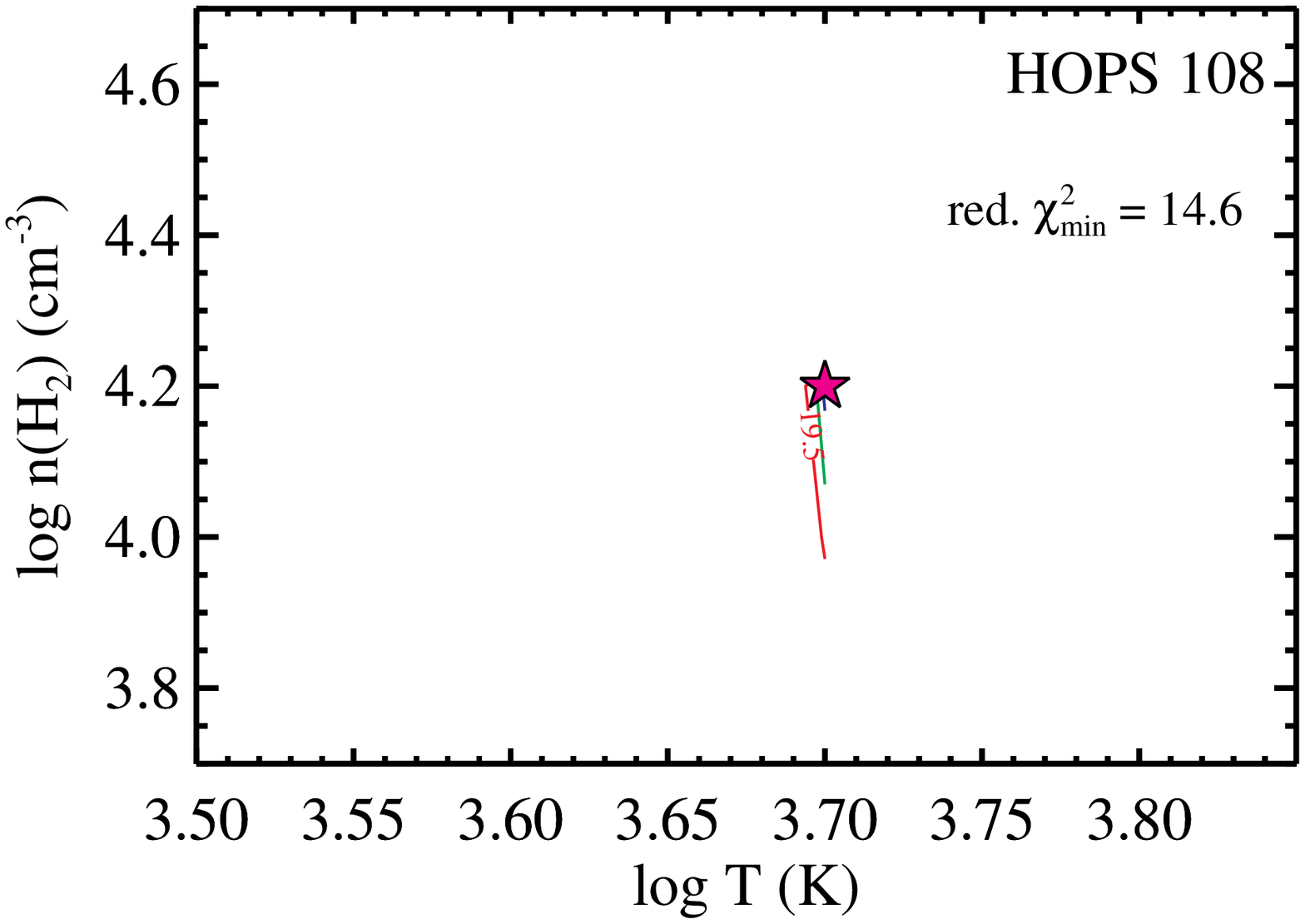}
\caption{ continued ....}
\end{figure*}

\begin{figure*}
\centering
\addtocounter{figure}{-1}
\plottwo{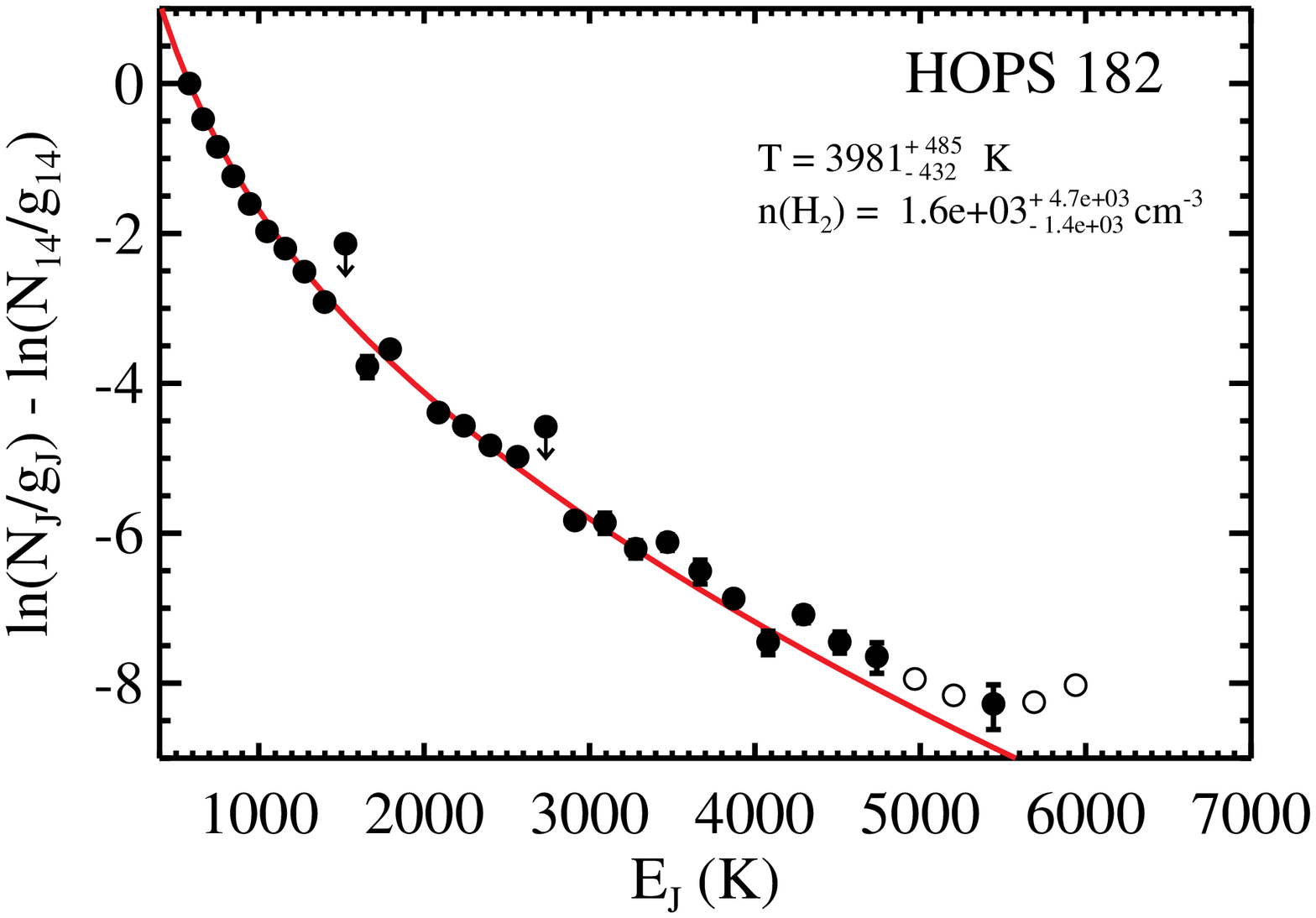}{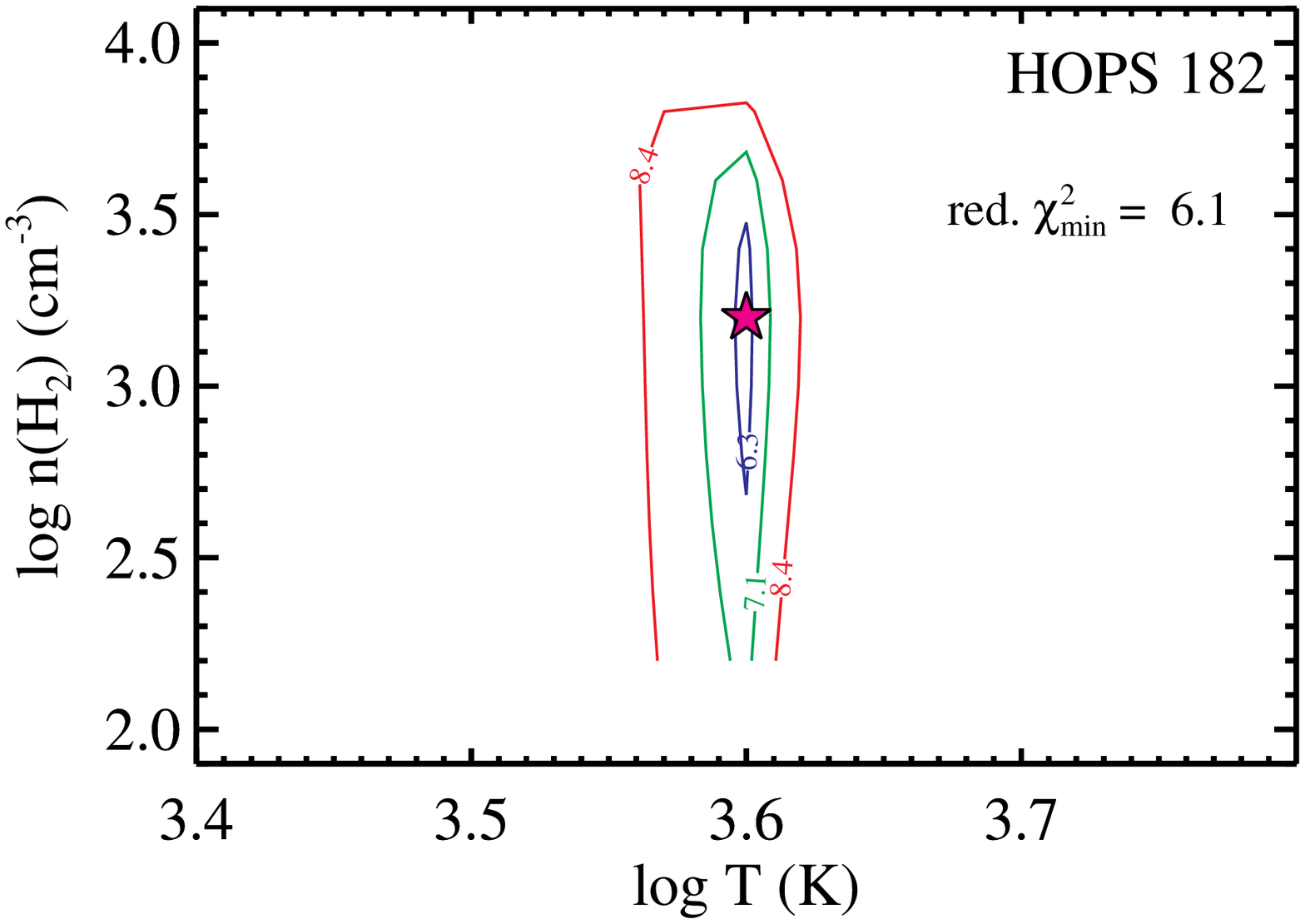}
\plottwo{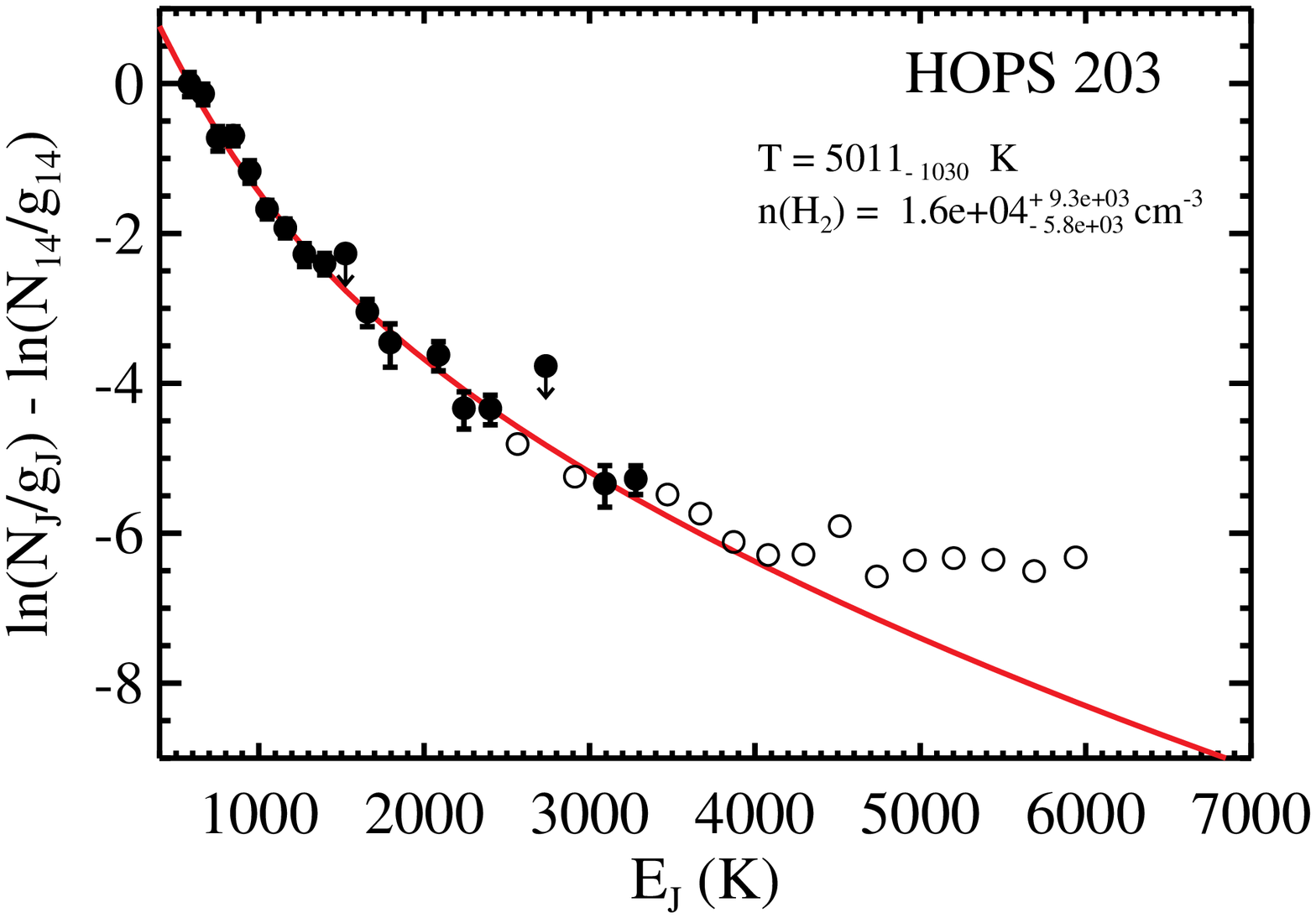}{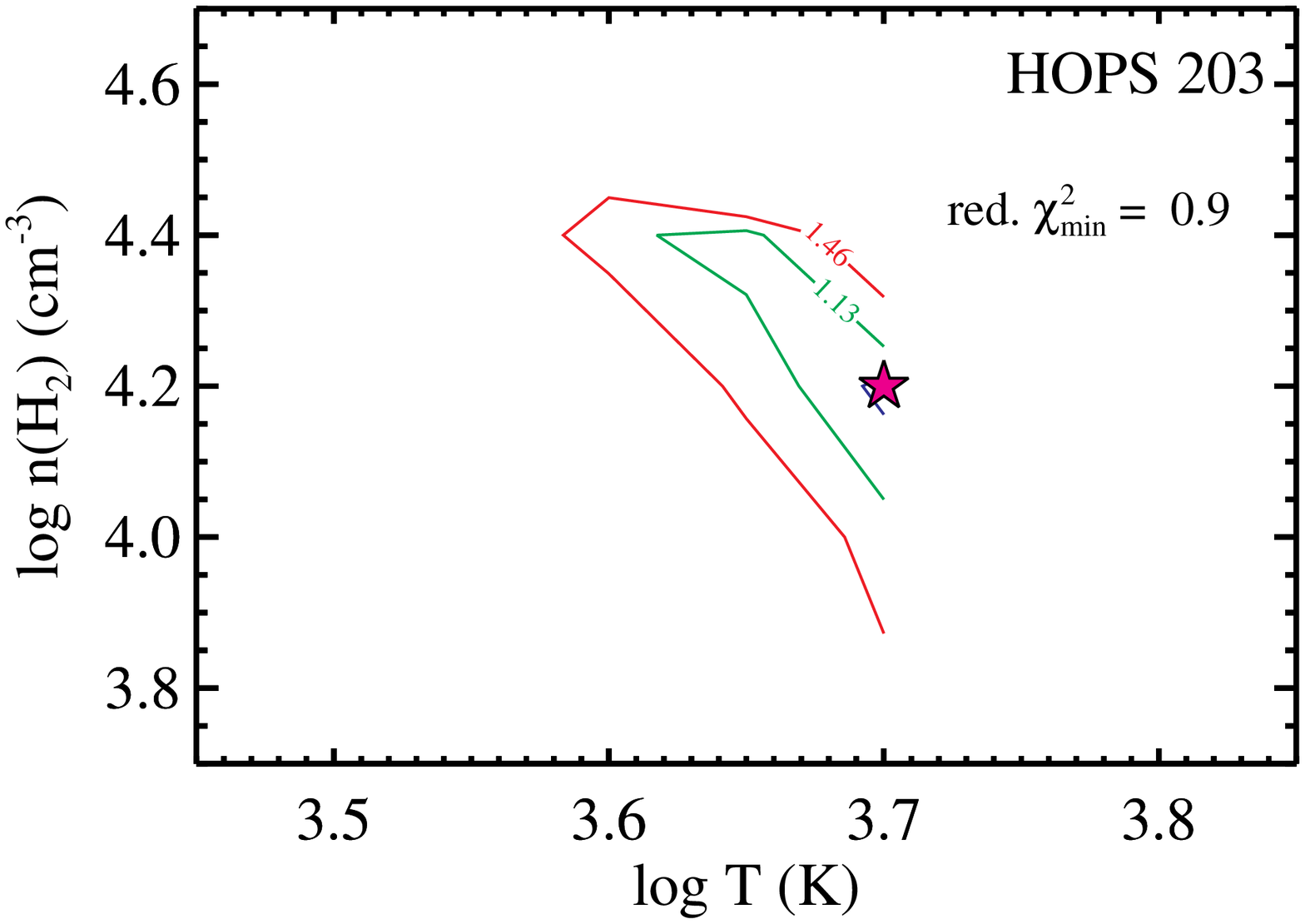}
\plottwo{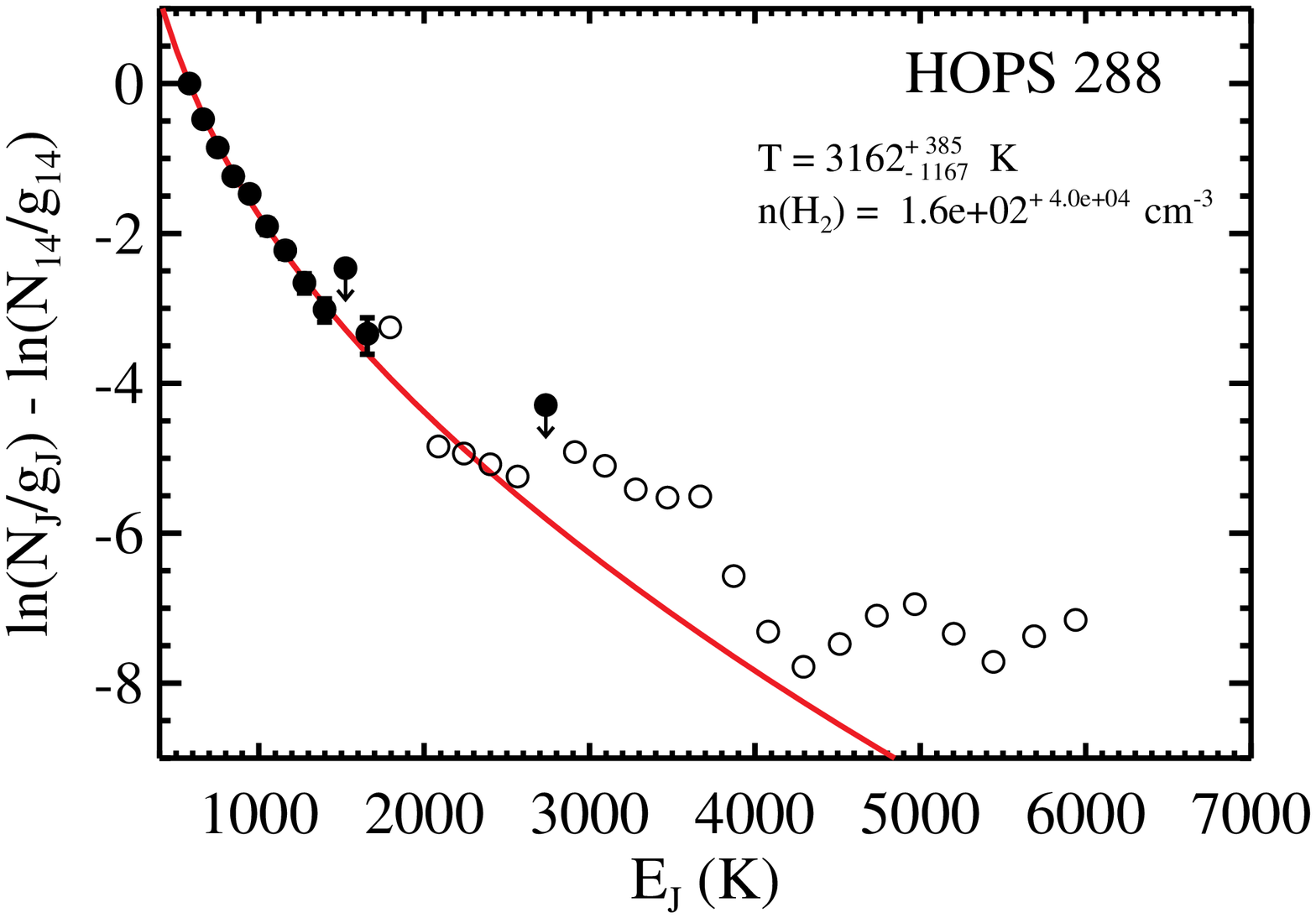}{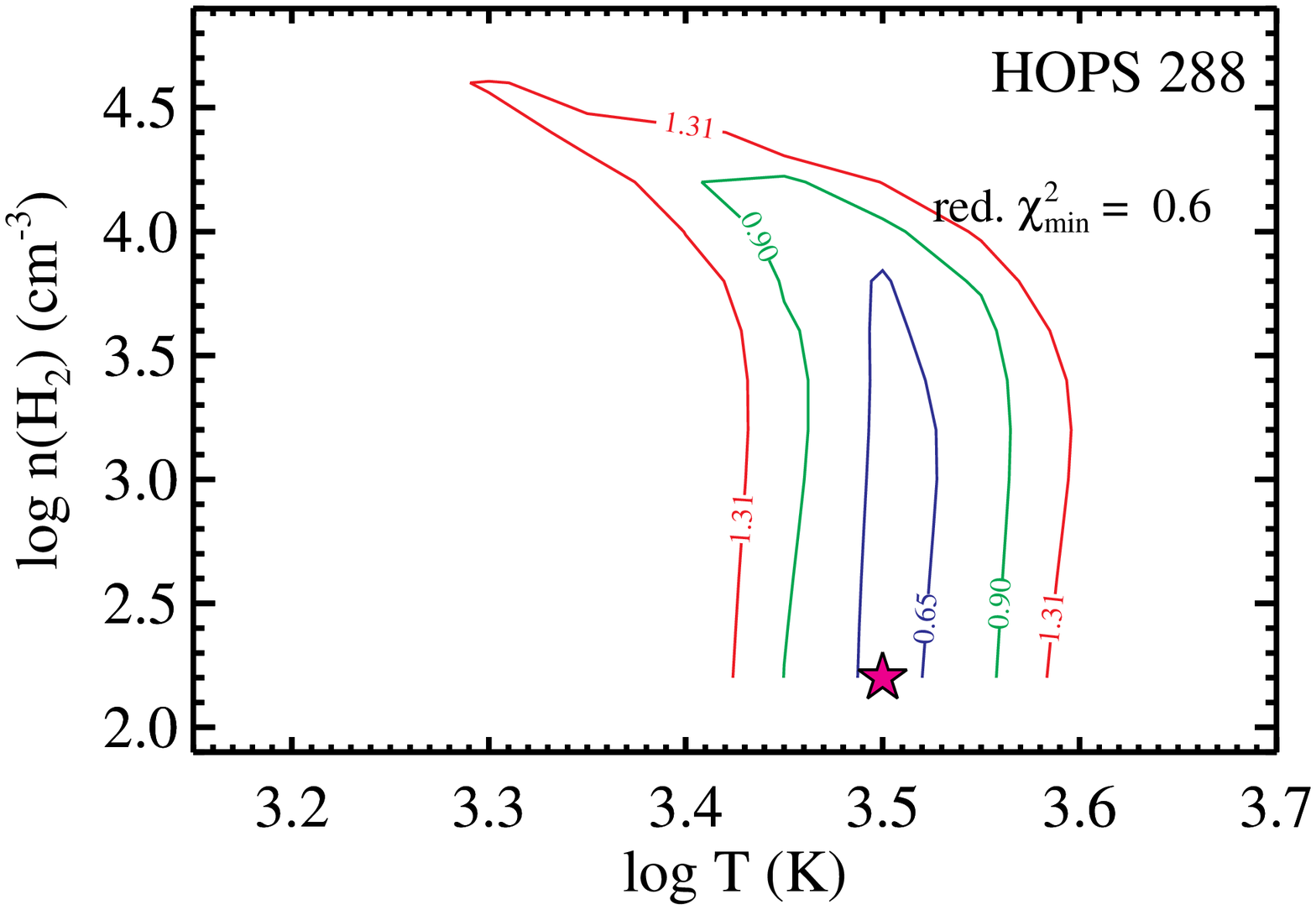}
\plottwo{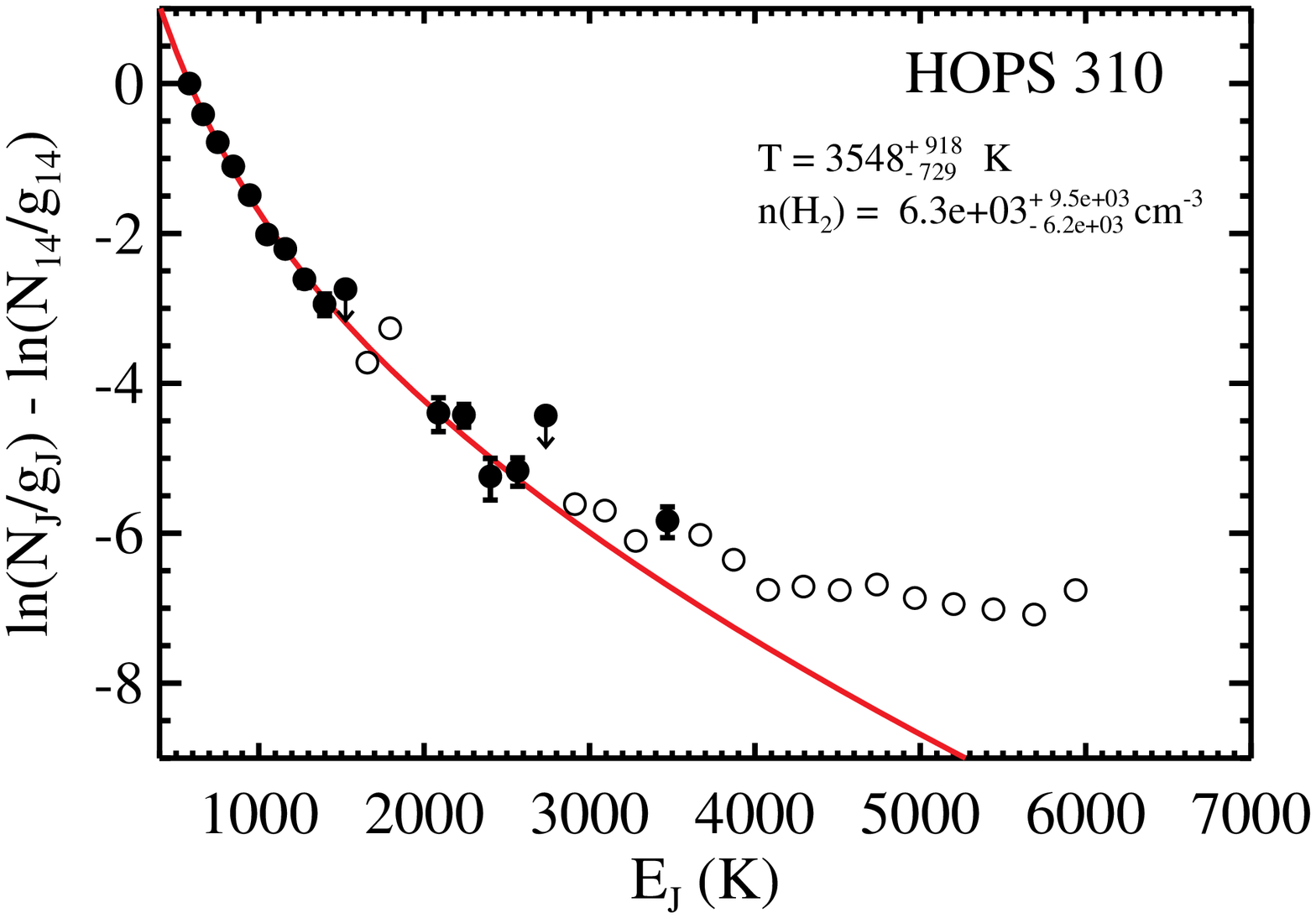}{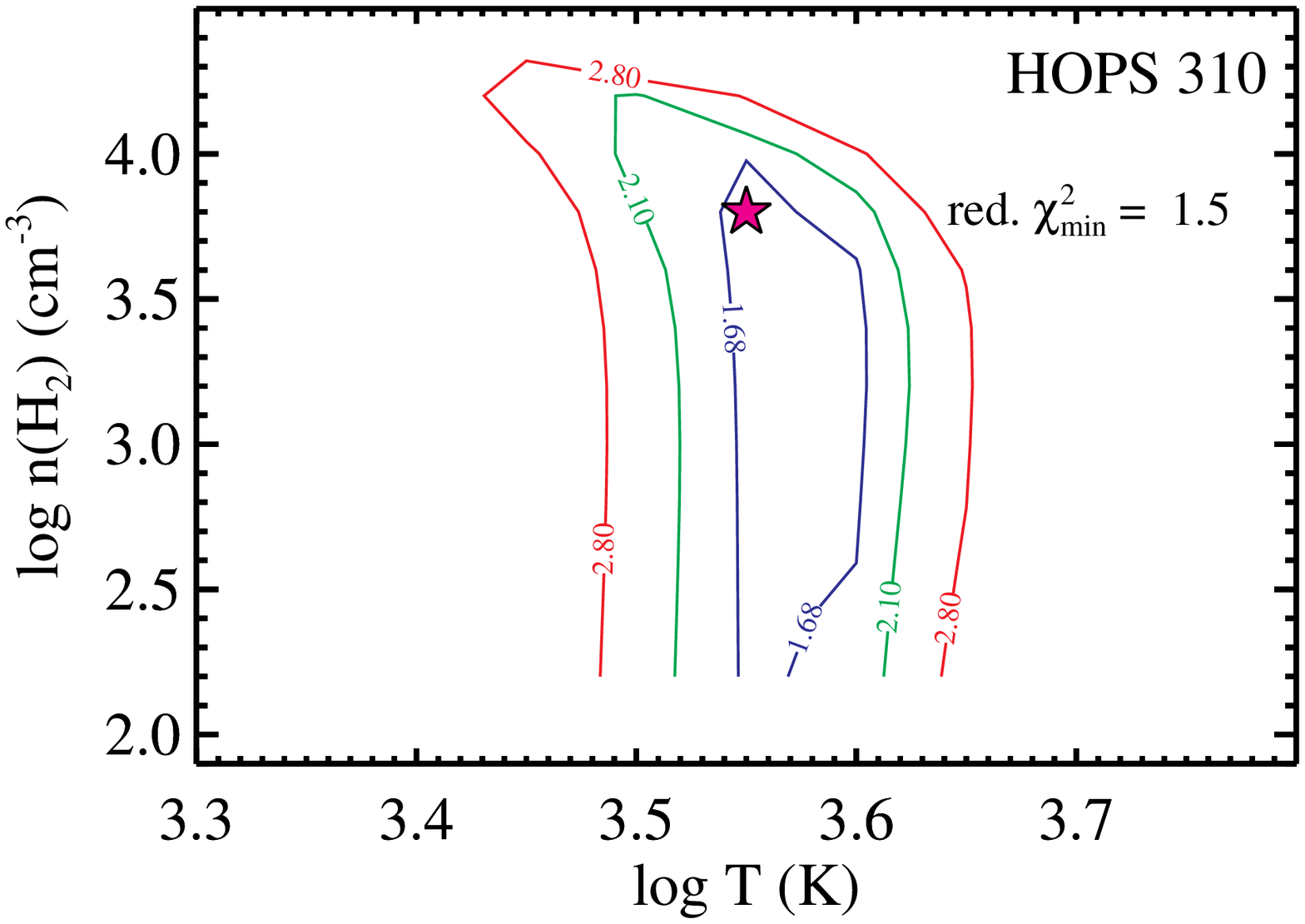}

\caption{ continued ....}
\end{figure*}

\begin{figure*}
\centering
\addtocounter{figure}{-1}
\plottwo{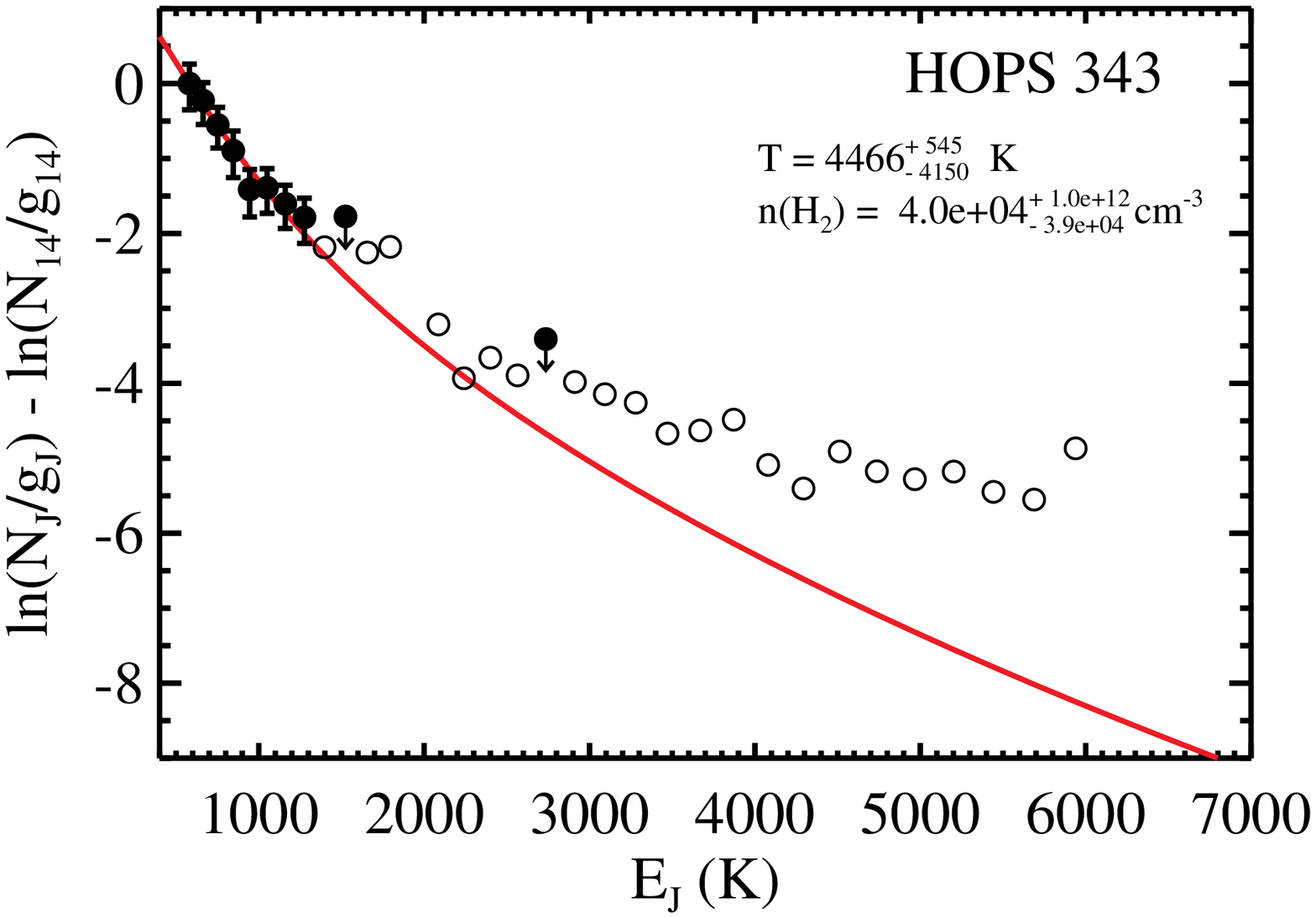}{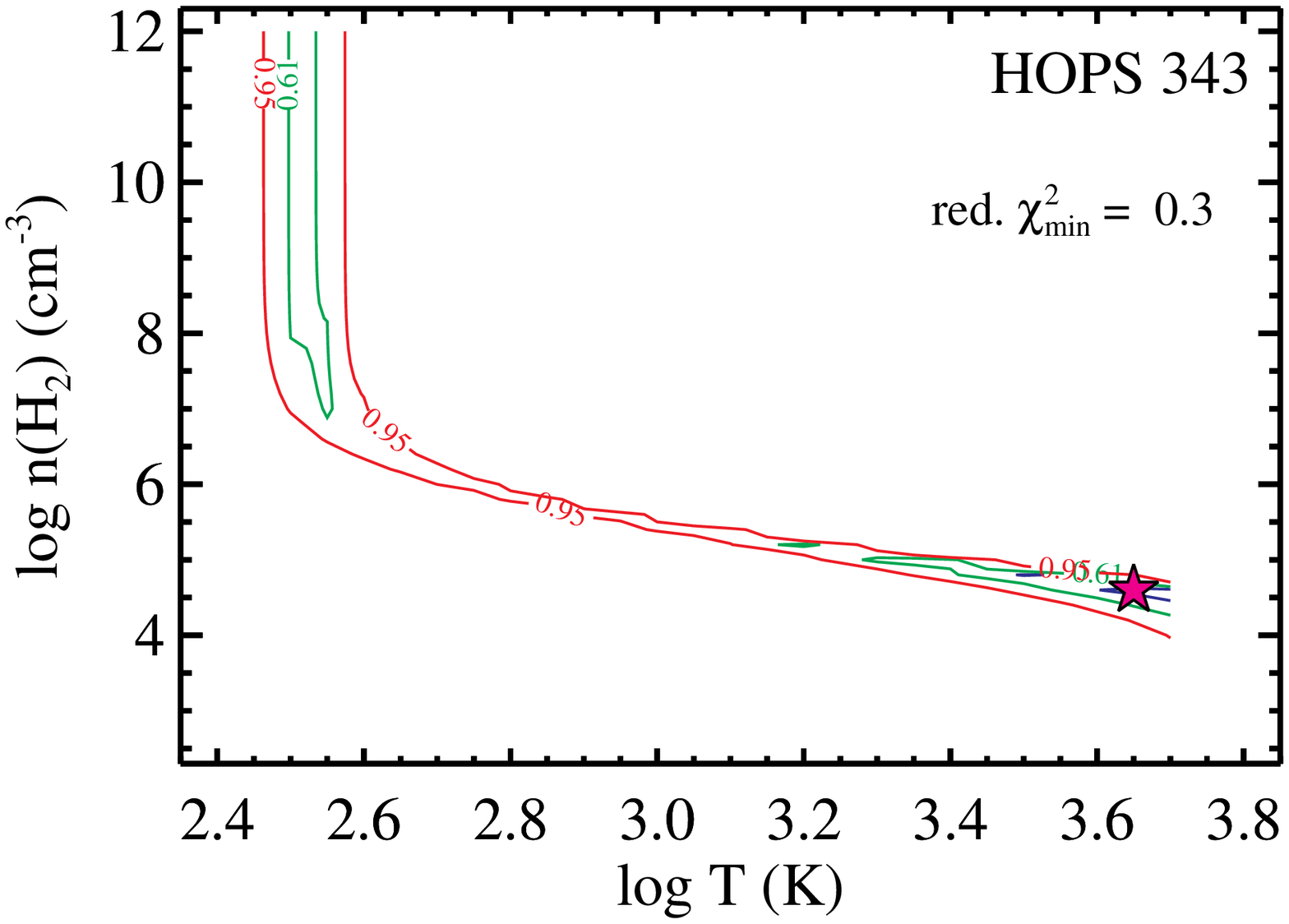}
\plottwo{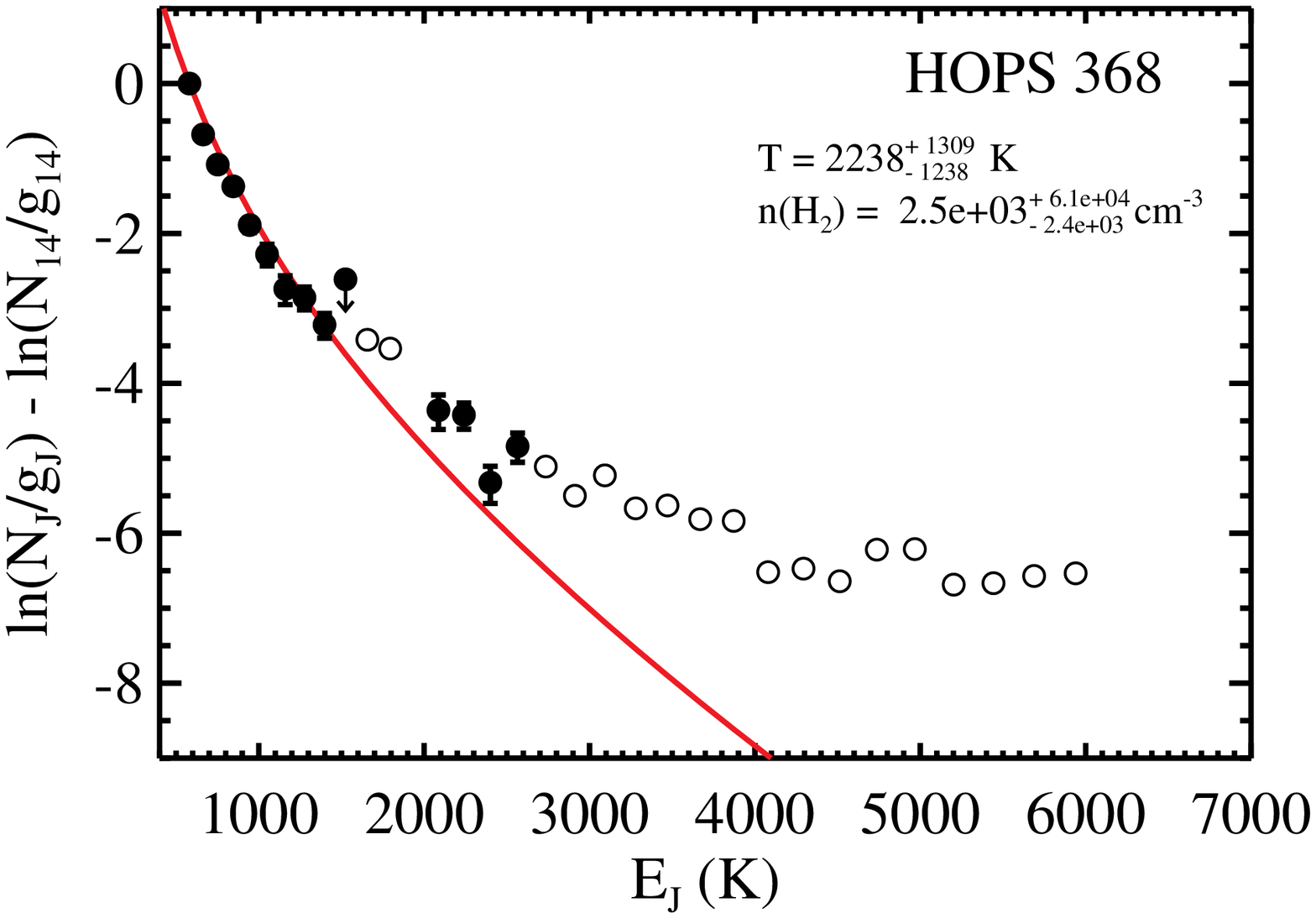}{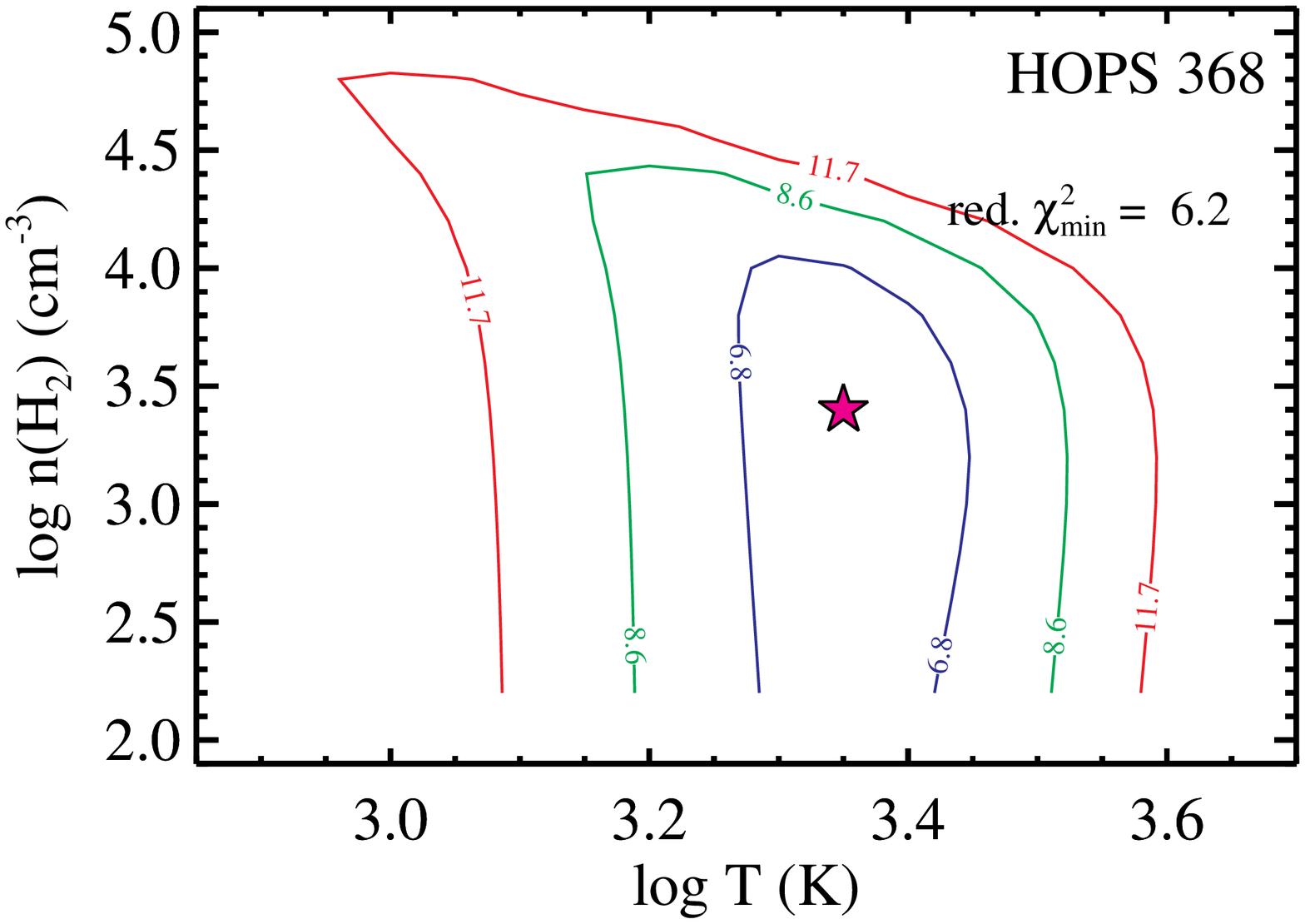}
\plottwo{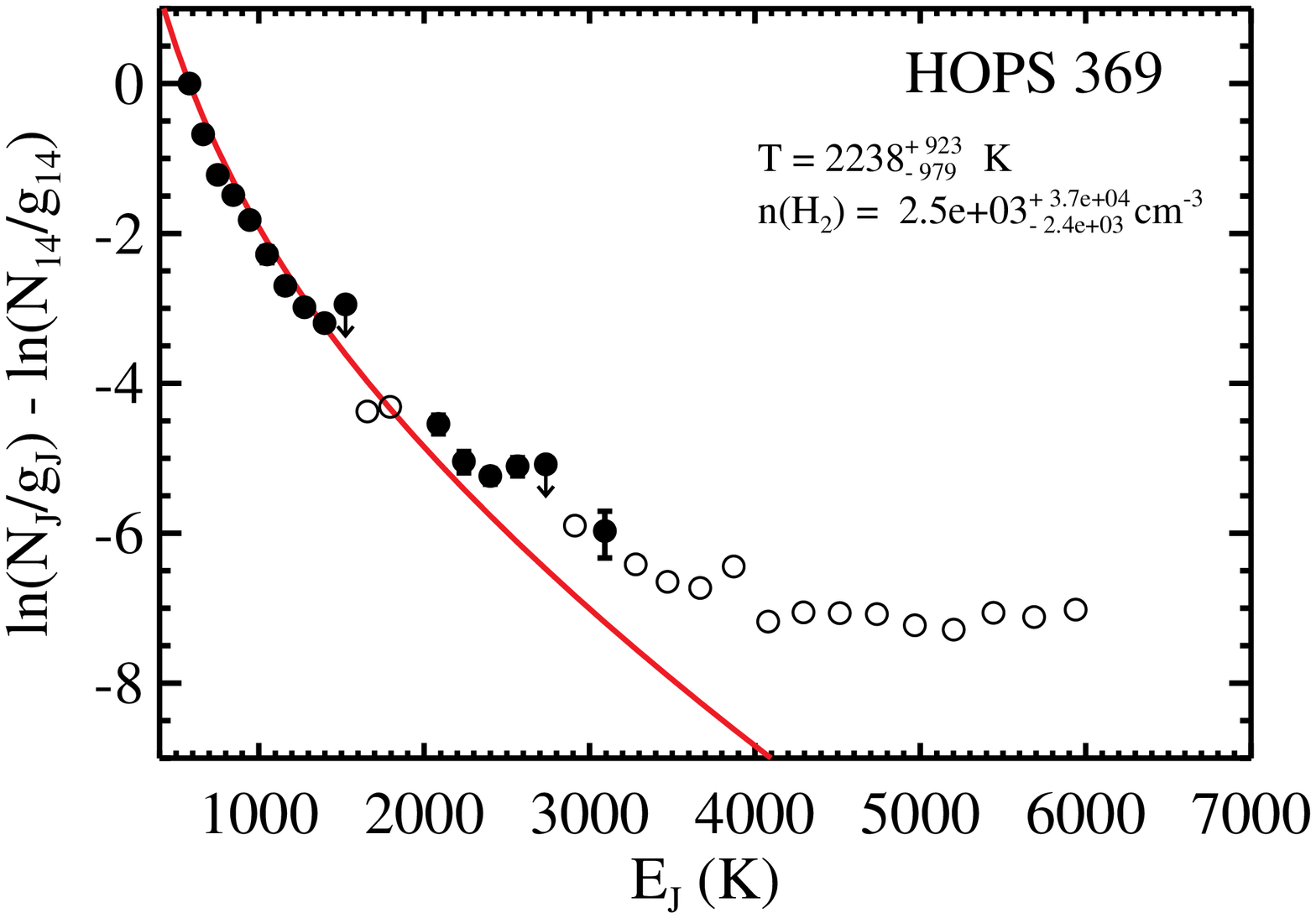}{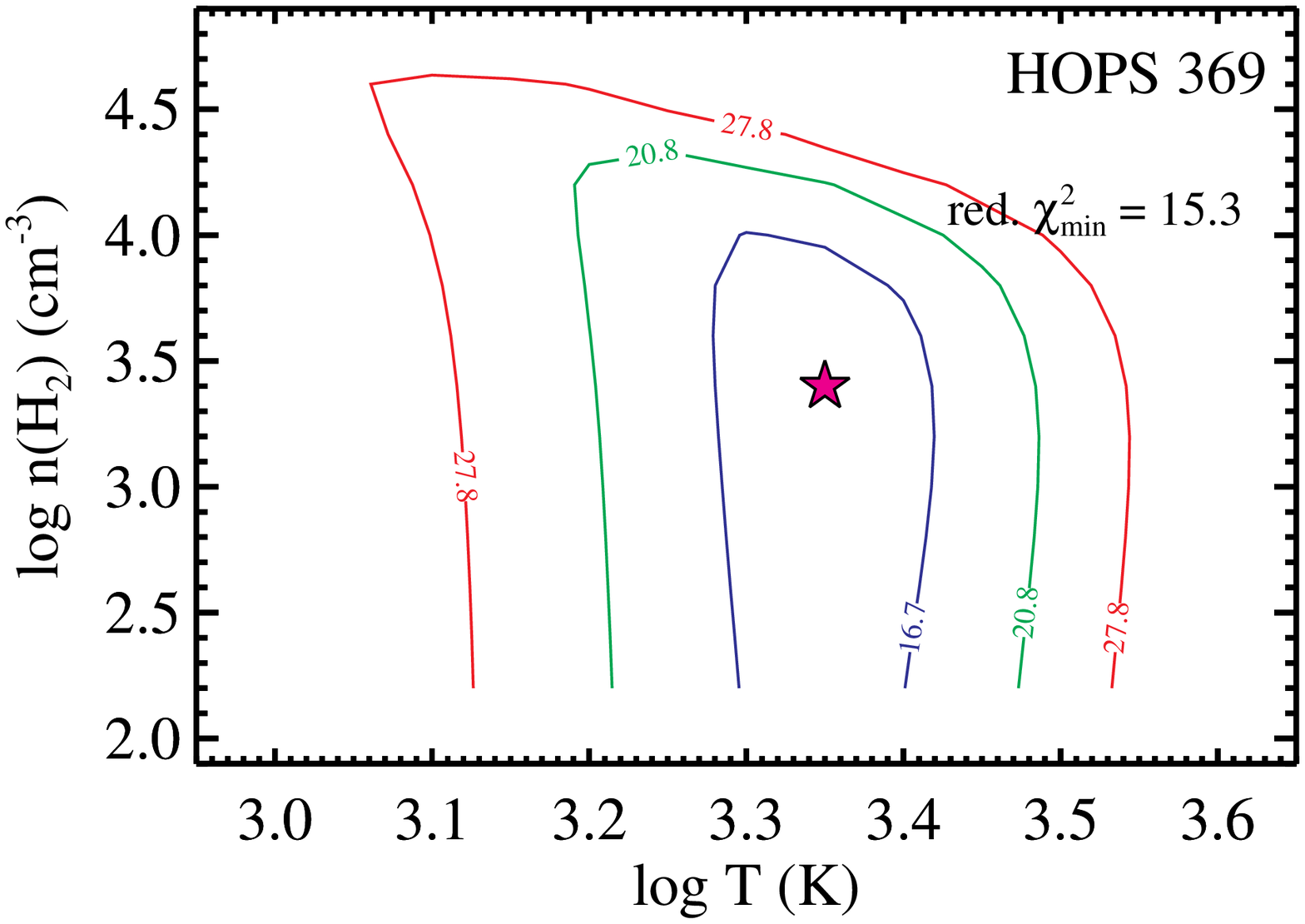}
\plottwo{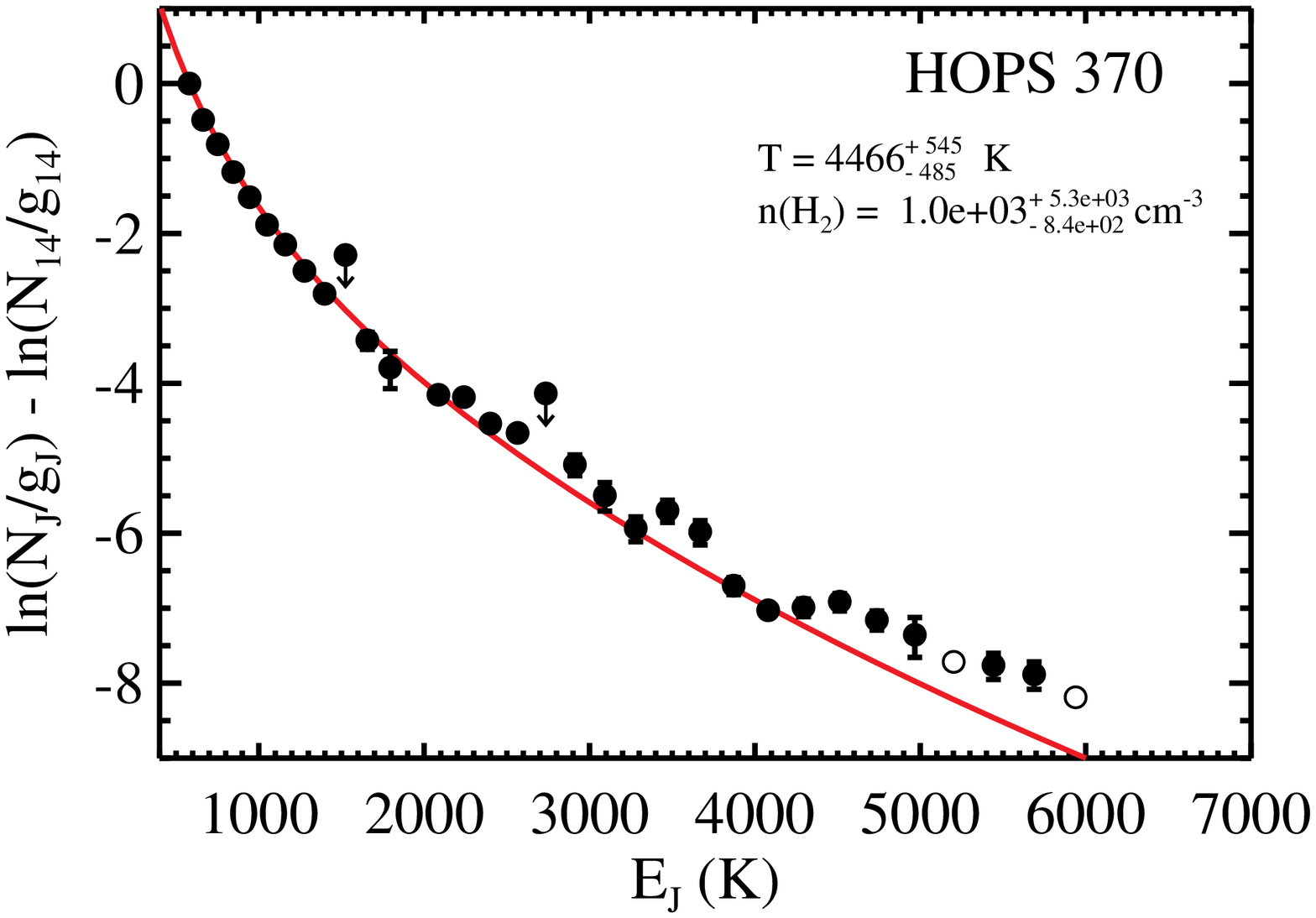}{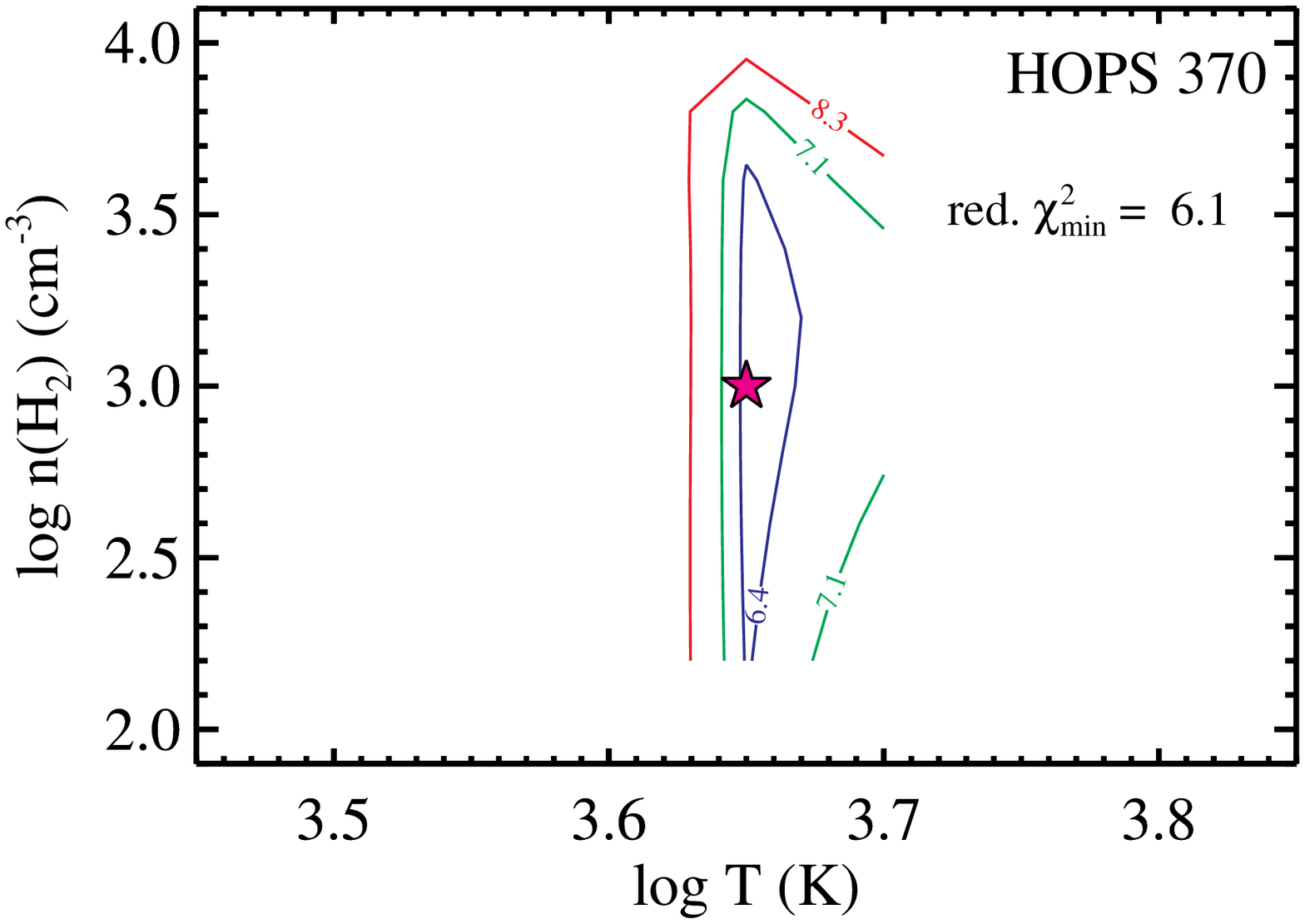}
\caption{ continued ....}
\end{figure*}

\begin{deluxetable*}{lcccccc}
\tablecolumns{7}
\tablewidth{0pt}
\tablecaption{Best fit model parameters \label{model_tbl}}
\tablehead{
\colhead{} & \multicolumn{3}{c}{Isothermal} & \multicolumn{3}{c}{Power law} \\
\cline{2-4} \cline{5-7} \\
\colhead{HOPS~ID} & \colhead{T} & \colhead{$n\mathrm{(H_2)}$} & \colhead{\tablenotemark{a}$L_{\mathrm{CO}}$ (total)} & \colhead{power law} & \colhead{$n\mathrm{(H_2)}$} & \colhead{\tablenotemark{a}$L_{\mathrm{CO}}$ (total)} \\
\colhead{} & \colhead{(K)} & \colhead{($\times$~10$^{3}$~cm$^{-3}$)} & \colhead{($\times$~10$^{-3}$ $L_{\odot}$)} & \colhead{index, $b$} & \colhead{($\times$~10$^{5}$~cm$^{-3}$)} & \colhead{($\times$~10$^{-3}$ $L_{\odot}$)} 
}

\startdata

56&  2511$^{+306}_{-516}$     & 1.6$^{+14.3}_{-1.4}$     &      18  &  \nodata                  &  \nodata                  &   \nodata\\[0.3cm]
60&  2818$^{+1162}_{-1405}$   & 0.2$^{+62.9}$            &      14  &  \nodata                  &  \nodata                  &   \nodata\\[0.3cm]
 87&  5011$_{-1849}$           & 0.4$^{+24.7}_{-0.2}$     &      17  &  \nodata                  &  \nodata                  &   \nodata\\[0.3cm]
108&  5011$_{-545}$            & 15.8$^{+9.3}_{-5.8}$     &     295  &  2.4$^{+0.1}_{-0.1}$      &  6.3$^{+3.7}_{-2.3}$      &       332\\[0.3cm]
182&  3981$^{+485}_{-432}$     & 1.6$^{+4.7}_{-1.4}$      &      76  &  2.7$^{+0.1}_{-0.1}$      &  6.3$^{+3.7}_{-2.3}$      &        93\\[0.3cm]
203&  5011$_{-1030}$           & 15.8$^{+9.3}_{-5.8}$     &      13  &  \nodata                  &  \nodata                  &   \nodata\\[0.3cm]
288&  3162$^{+385}_{-1167}$    & 0.2$^{+39.7}$            &      22  &  \nodata                  &  \nodata                  &   \nodata\\[0.3cm]
310&  3548$^{+918}_{-729}$     & 6.3$^{+9.5}_{-6.2}$      &      16  &  \nodata                  &  \nodata                  &   \nodata\\[0.3cm]
368&  2238$^{+1309}_{-1238}$   & 2.5$^{+60.6}_{-2.4}$     &      13  &  \nodata                  &  \nodata                  &   \nodata\\[0.3cm]
369&  2238$^{+923}_{-979}$     & 2.5$^{+37.3}_{-2.4}$     &      24  &  \nodata                  &  \nodata                  &   \nodata\\[0.3cm]
370&  4466$^{+545}_{-485}$     & 1.0$^{+5.3}_{-0.8}$      &      84  &  2.7$^{+0.1}_{-0.1}$      &  10.0$^{+5.8}_{-3.7}$     &       102\\

\enddata
\tablenotetext{a}{$L_{\mathrm{CO}}$ listed here is the total CO luminosity derived from best fit models for rotational transitions ranging from $J=1-0$ up to $J=80-79$.}

\end{deluxetable*}

\begin{figure}
\centering
\epsscale{1.0}
\plotone{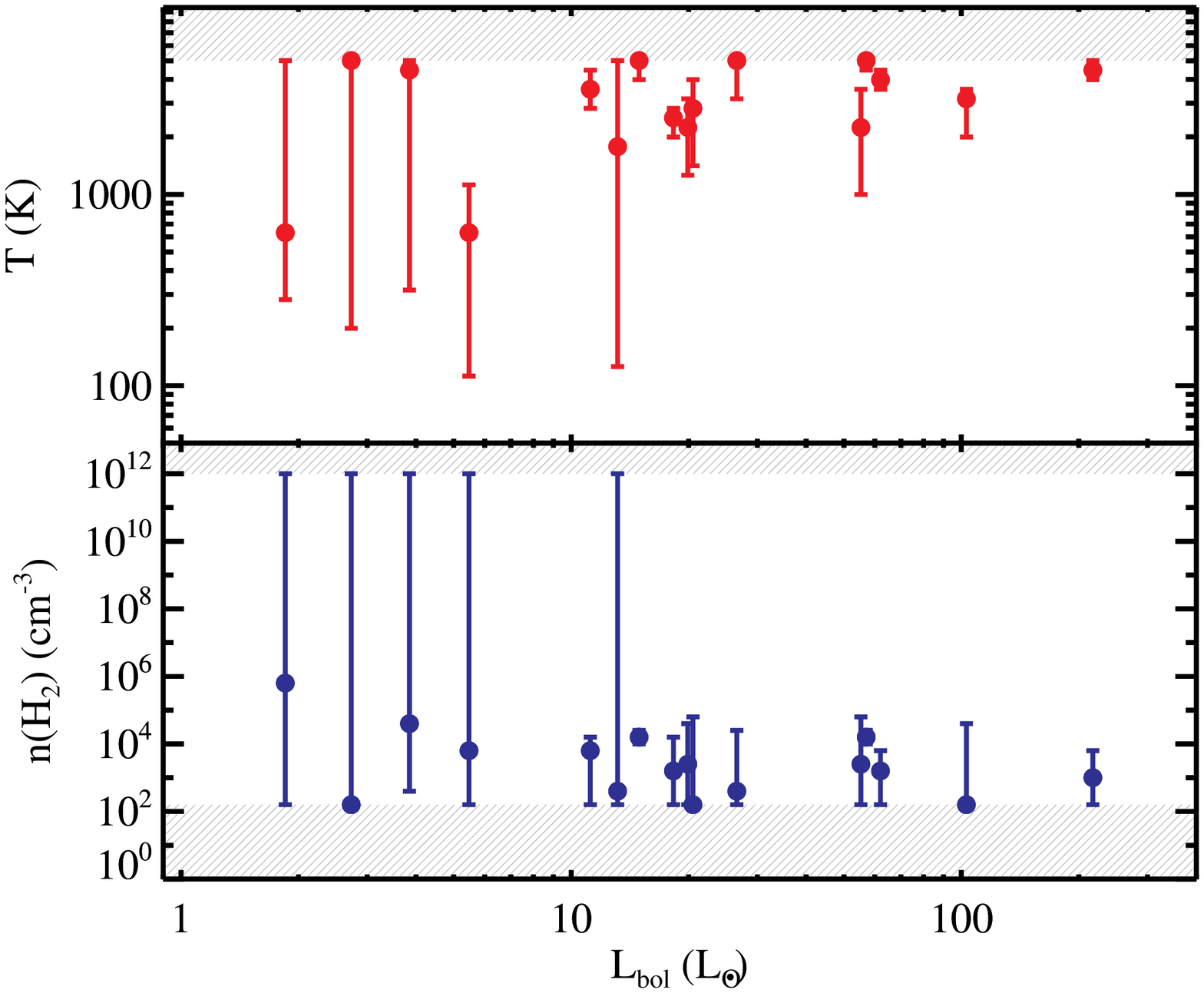}
\caption{ Best fit values of $T$ and $n\mathrm{(H_2)}$ for CO emission
  from an isothermal medium plotted against $L_{bol}$. The
  uncertainties shown correspond to 99.7\% confidence interval. The
  gray shaded regions indicate $T$ \& $n\mathrm{(H_2)}$
  parameter space not explored by our models. \label{iso_fit_lbol}}
\end{figure}

The best fit models and physical parameters (corresponding to the
minimum in reduced-$\chi^2$) obtained for CO emitting gas for all the
protostars are shown in Figure~\ref{model_iso_rot}. We only modeled
sixteen sources for which more than three CO lines are detected. The
best fit values of $T$ and $n\mathrm{(H_2)}$ are shown as a function
of $L_{bol}$ in Figure~\ref{iso_fit_lbol}. For 5 sources for which
less than 8 CO lines were detected, the density and temperature of the
emitting medium are not constrained by the models. For all the other
protostars, these parameters are well constrained; they are listed in
Table~\ref{model_tbl}. The optimal solutions have gas temperatures
$T$~$\ga$~2000~K and densities $n\mathrm{(H_2)}$ $\la$ 10$^{4.5}$
cm$^{-3}$. While gas temperatures are well constrained for these
sources, for many of them only an upper limit could be obtained for
densities. Also, several of them have the best fit densities well
below 10$^4$ cm$^{-3}$.  This is because at high temperatures ($T$
$\ga$ 2000$-$3000~K) and low densities ($n\mathrm{(H_2)}$ $\la$
10$^{4.5}$ cm$^{-3}$), the CO line ratios (or rotational diagrams) are
insensitive to density; at these high gas temperatures the rotational
diagrams are indistinguishable for densities below 10$^{4.5}$
cm$^{-3}$ \citep{neufeld12}. Thus, our modeling shows that if the
observed CO emission originates from a single isothermal component,
then the CO excitation is {\it sub-thermal}.

For sources where the density and temperature of the CO emitting
medium are well constrained, the total CO luminosities (for
transitions $J=1-0$ up to $J=80-79$) obtained from the best fit models
range from 3.0$\times$10$^{-1}$ $L_{\odot}$ for HOPS~108 to
1.3$\times$10$^{-2}$ $L_{\odot}$ for HOPS~368 \& 203 (see
Table~\ref{model_tbl}).  The observed $L_{\mathrm{CO}}$ (see
Table~\ref{rot_tbl} \& Figure~\ref{CO_lum}) computed from the FIR CO lines detected with PACS
($J=14-13$ up to $J=46-45$) is found to be $\sim$ 44\% to 81\% of the
total CO luminosity. Thus, if the CO emission from protostars is dominated
by hot gas ($T$ $\ga$ 2000$-$3000~K) at low densities
($n\mathrm{(H_2)}$ $\la$ 10$^{4.5}$ cm$^{-3}$), most of the CO
luminosity is emitted within the PACS wavelength range. The low
density, high temperature solutions obtained from the fits to single
component isothermal models do not appear to be driven by the high$-J$
CO lines. We modeled rotational diagrams for CO lines observed in the
R1 spectral band ($J_{up}$ = 14$-$25) separately, and found that, for
most sources, a single temperature component can explain the observed
CO emission in the R1 band only for densities $n\mathrm{(H_2)}$ $<$
10$^5$ cm$^{-3}$ and temperatures $T$ $\ga$ 2000~K.

\begin{figure*}
\centering
\epsscale{1.0}

\plottwo{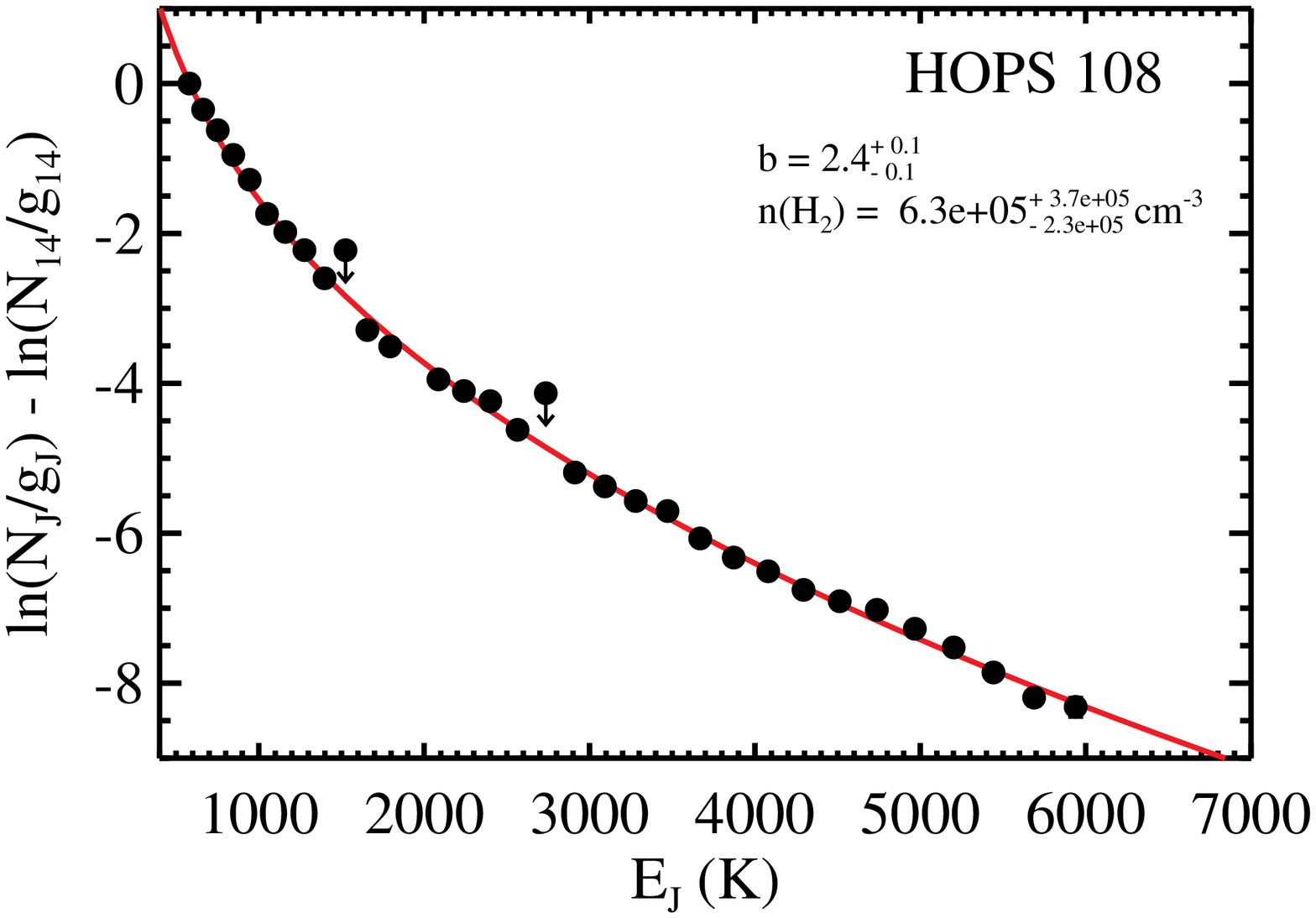}{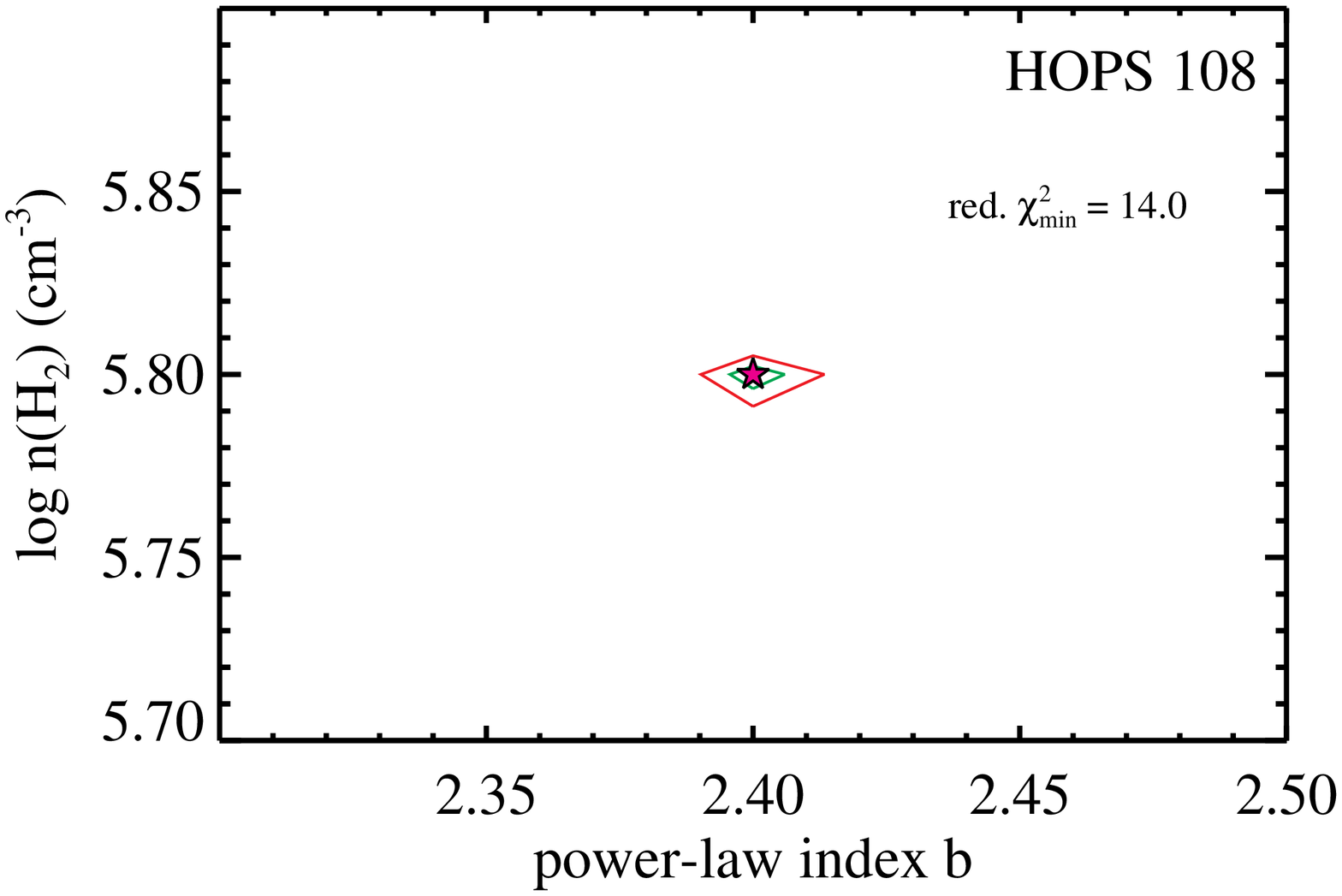}
\plottwo{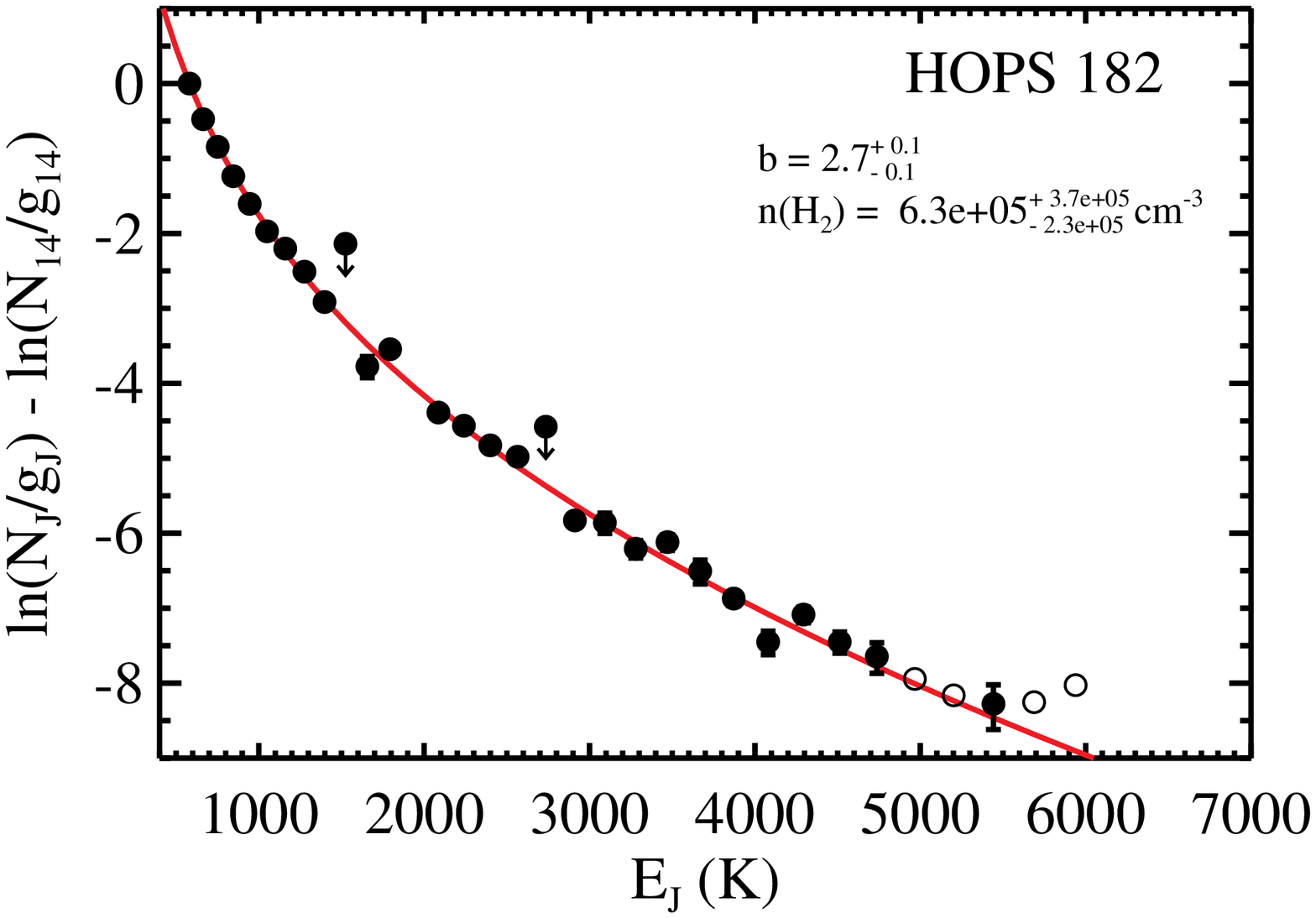}{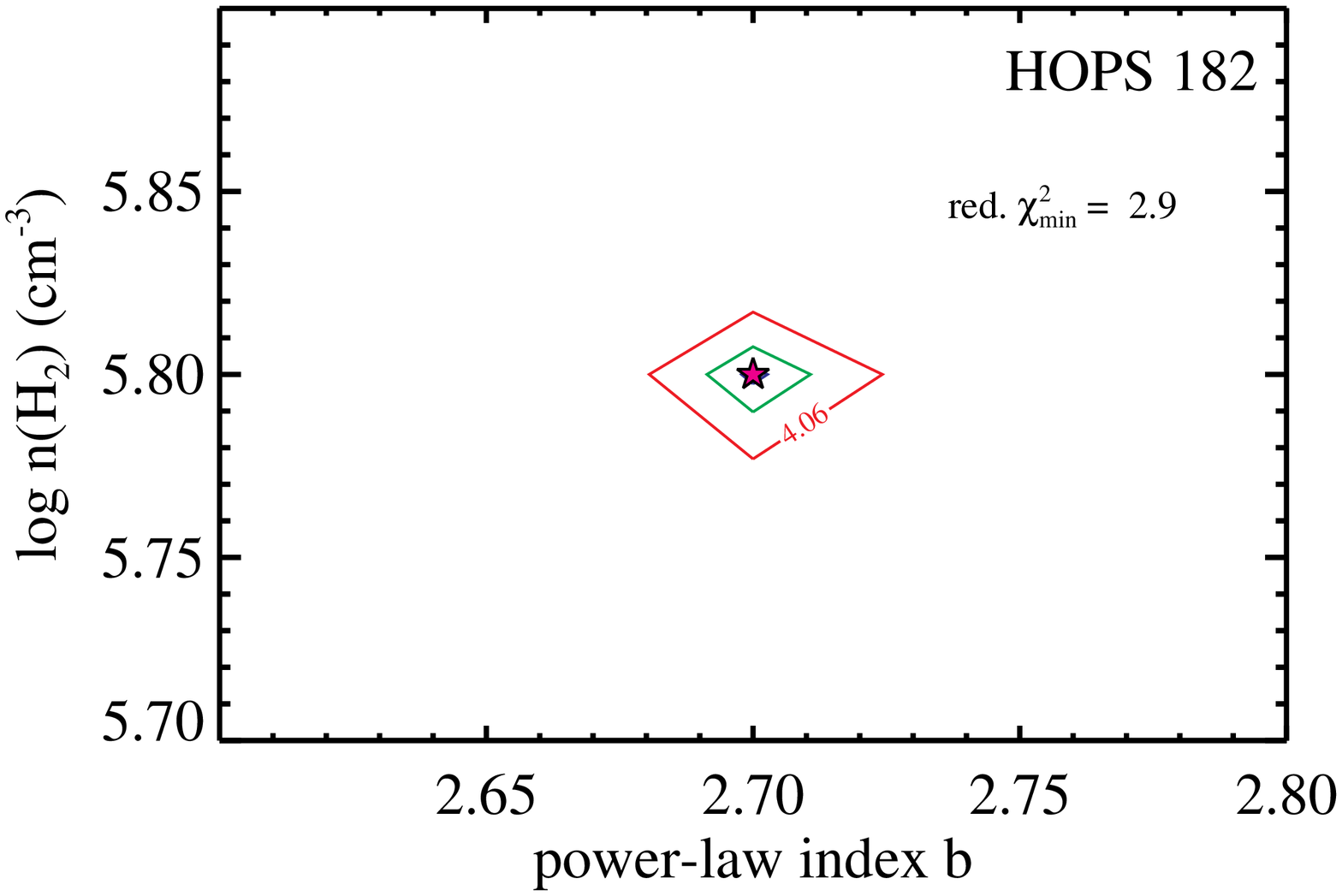}
\plottwo{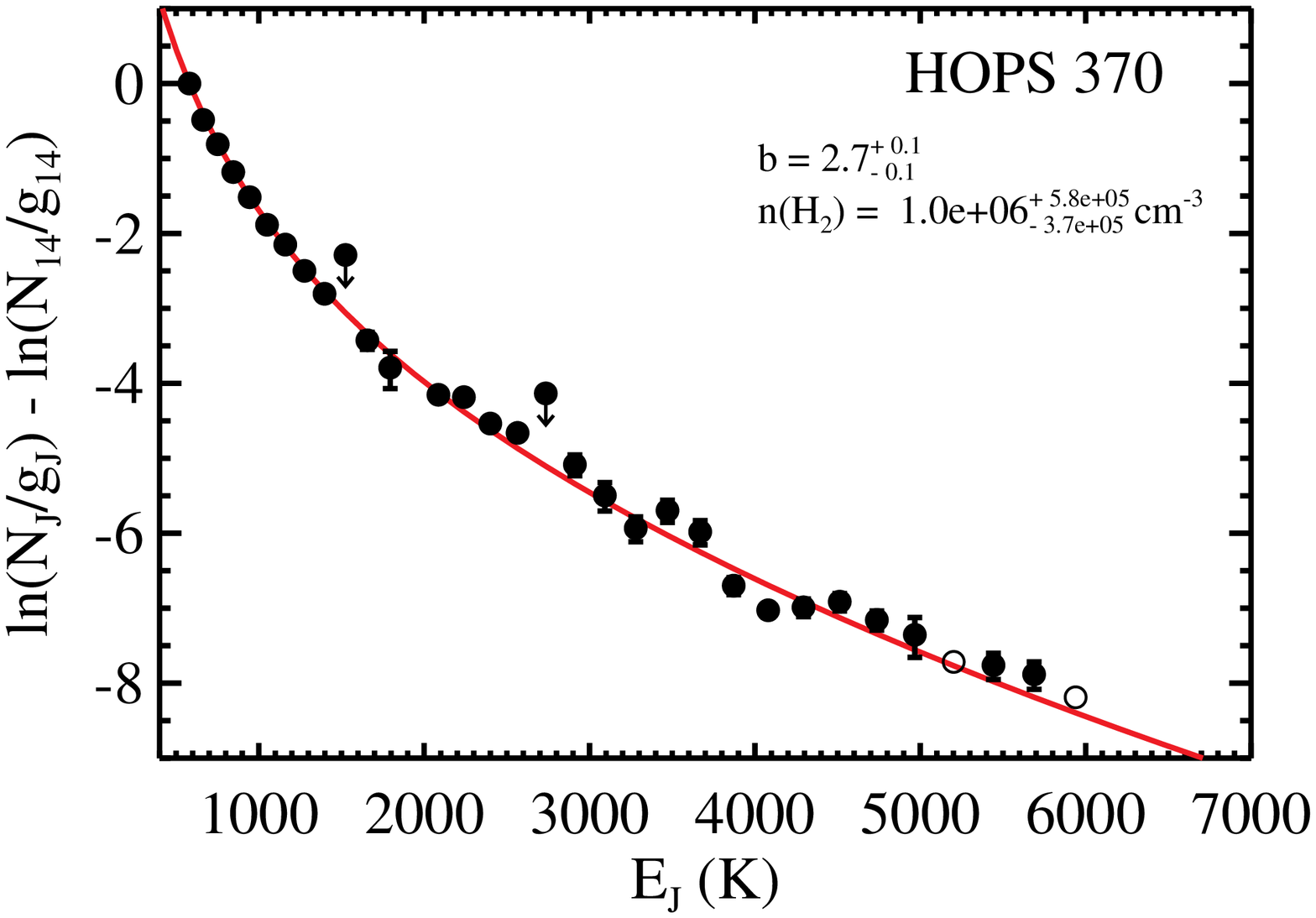}{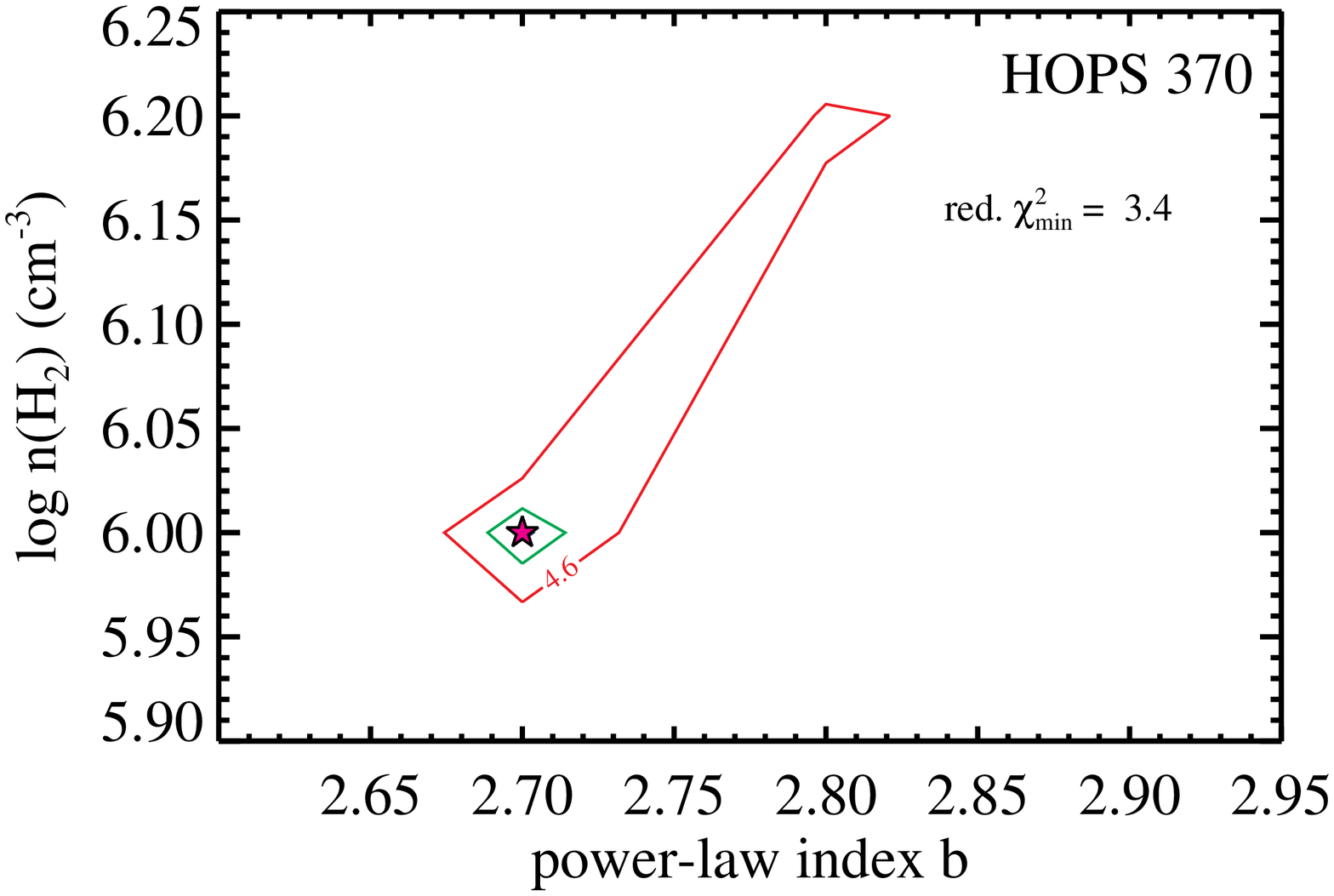}

\caption{ {\it (Left)}~Rotational diagrams of the observed CO emission
  with the best fits for the power-law temperature model overlaid
  (solid red line). Downward arrows indicate upper limits to the
  fluxes of the CO lines which are blended with a nearby line. Open
  circles correspond to 3~$\sigma$ upper limits for the
  non-detections. {\it (Right)} Reduced-$\chi^2$ contours for $T$ and
  $n\mathrm{(H_2)}$. The star symbol marks the minimum value of the
  reduced-$\chi^2$. The contours corresponding to 68.3\% (blue),
  95.4\% (green) and 99.7\% (red) confidence levels are
  shown.  \label{model_power_rot} }
\end{figure*}

For five sources -- HOPS 10, 32, 68, 85 and 343 -- where fewer than 8
CO lines were detected ($J_{up} \le 21$), the densities are not
constrained by the isothermal models.  These objects are among the
lowest $L_{bol}$ sources in our sample (see
Figure~\ref{iso_fit_lbol}). Although high density, low temperature LTE
solutions cannot be ruled out for these sources, they are also
consistent with the low density ($n\mathrm{(H_2)}$ $<$ 10$^5$
cm$^{-3}$), high temperature ($T$ $\ga$ 2000~K) solutions in the
context of single component isothermal models.

Two sources in which several high$-J$ CO lines are detected (HOPS 182
and 370), the highest$-J$ lines ($J_{up} \ge 38$) are not well fit by
the single isothermal component models. This difficulty was noted by
\citet{neufeld12}, who pointed out that CO rotational diagrams with
large positive curvatures cannot be reproduced by a single temperature
component model; instead, these can be modeled using a medium with a
power-law distribution of gas temperatures.

\subsubsection{Medium with a power-law distribution of gas temperatures}

Next, we model the observed CO emission as arising from an optically
thin medium with a continuous distribution of gas temperatures to
account for the presence of an admixture of gas temperatures along the
line of sight. Following the approach of \citet{neufeld12}, the gas
temperature distribution is approximated as a power-law in these
models: the column density of the medium in the temperature interval
between $T$ and $T$ + $dT$ has the form $dN~\propto~T^{-b}\:dT$, over
a temperature range of 10~K to 5000~K \citep[for details
  see][]{neufeld12}. Such power-law temperature distribution models
have been successful in describing the observed rotational diagrams of
molecular hydrogen \citep{neufeld08,yuan11,giannini11}. We computed
the synthetic rotational diagrams for CO emission from such a medium
for a large range in density $n\mathrm{(H_2)}$, the power-law index
$b$, and the column density parameter $\tilde{N}$(CO) \citep[for
  details see][]{neufeld12}. The best fit solutions were obtained by
minimizing the reduced-$\chi^2$.

The power-law temperature models constrain the density of the emitting
gas only for three sources, HOPS~108, 182 \& 370, where more than 25
CO lines are detected. These models are shown in
Figure~\ref{model_power_rot} and the best fit values are listed in
Table~\ref{model_tbl}. For HOPS~108, whose rotational diagram has only
a modest positive curvature, the power-law fit is only marginally
better (in terms of reduced-$\chi^2$) than the isothermal fit. The
rotational diagrams of HOPS~182 and 370 have larger positive curvature
and for these two sources, power-law temperature models provide
significantly better fits (in terms of reduced-$\chi^2$) than those
obtained from a single temperature component model. For these three
sources, the density of the emitting medium is tightly constrained
between $4 \times 10^5 $ cm$^{-3}$ and $10^6 $ cm$^{-3}$, indicating
that the excitation of CO is not {\it thermal}. The total CO
luminosity derived from the best fit power-law solutions (for
transitions with $J_{up}$ = 1 to 80) is slightly higher than that
obtained from the isothermal model fits for HOPS~108, 182 \& 370 (see
Table~\ref{model_tbl}).  Comparison with the observed
$L_{\mathrm{CO}}$ (Table~\ref{rot_tbl}) shows that 52$-$72\% of the
total CO luminosity in these objects is emitted within the PACS
wavelength range.

\begin{figure}
\centering
\epsscale{1.15}
\plottwo{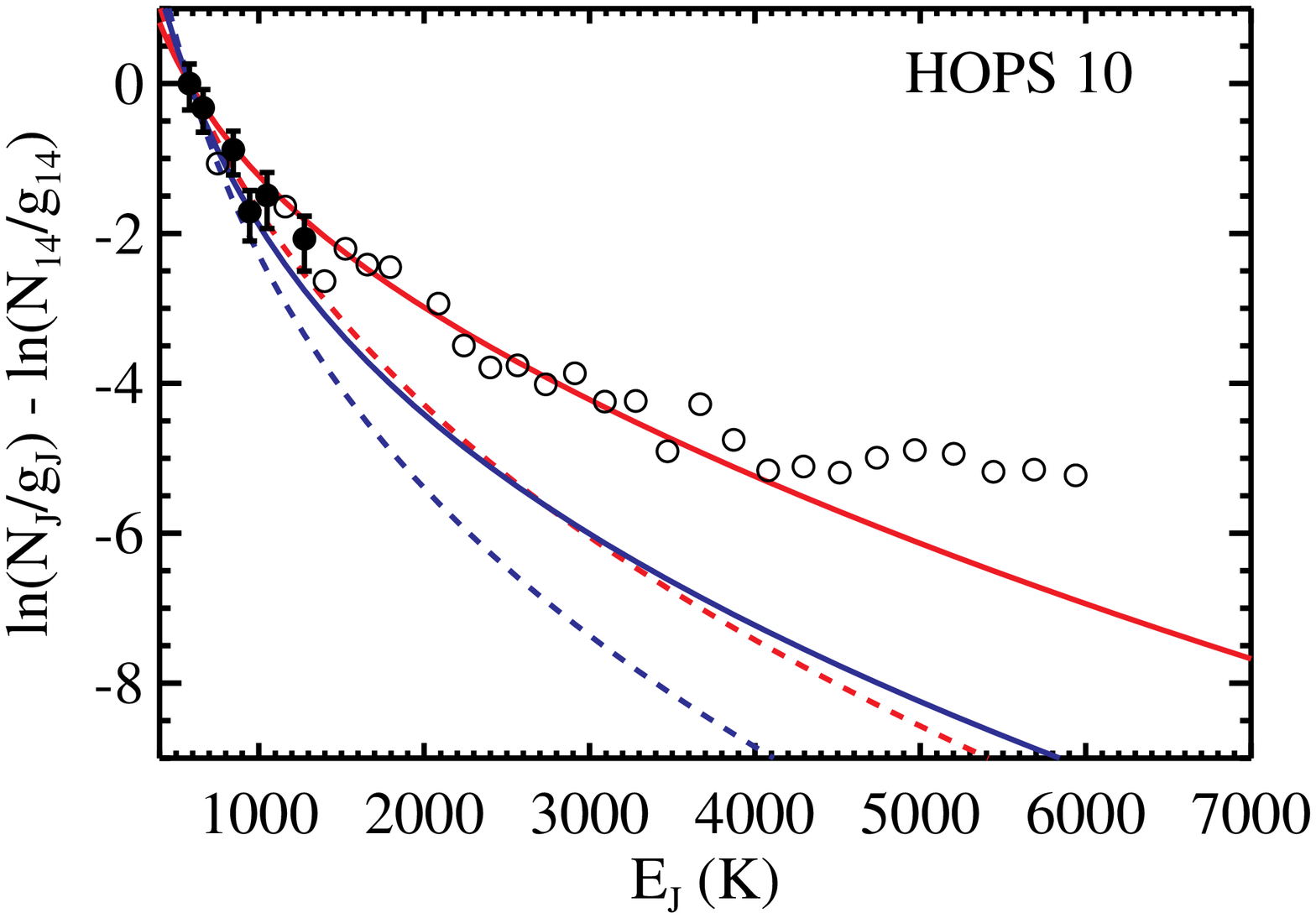}{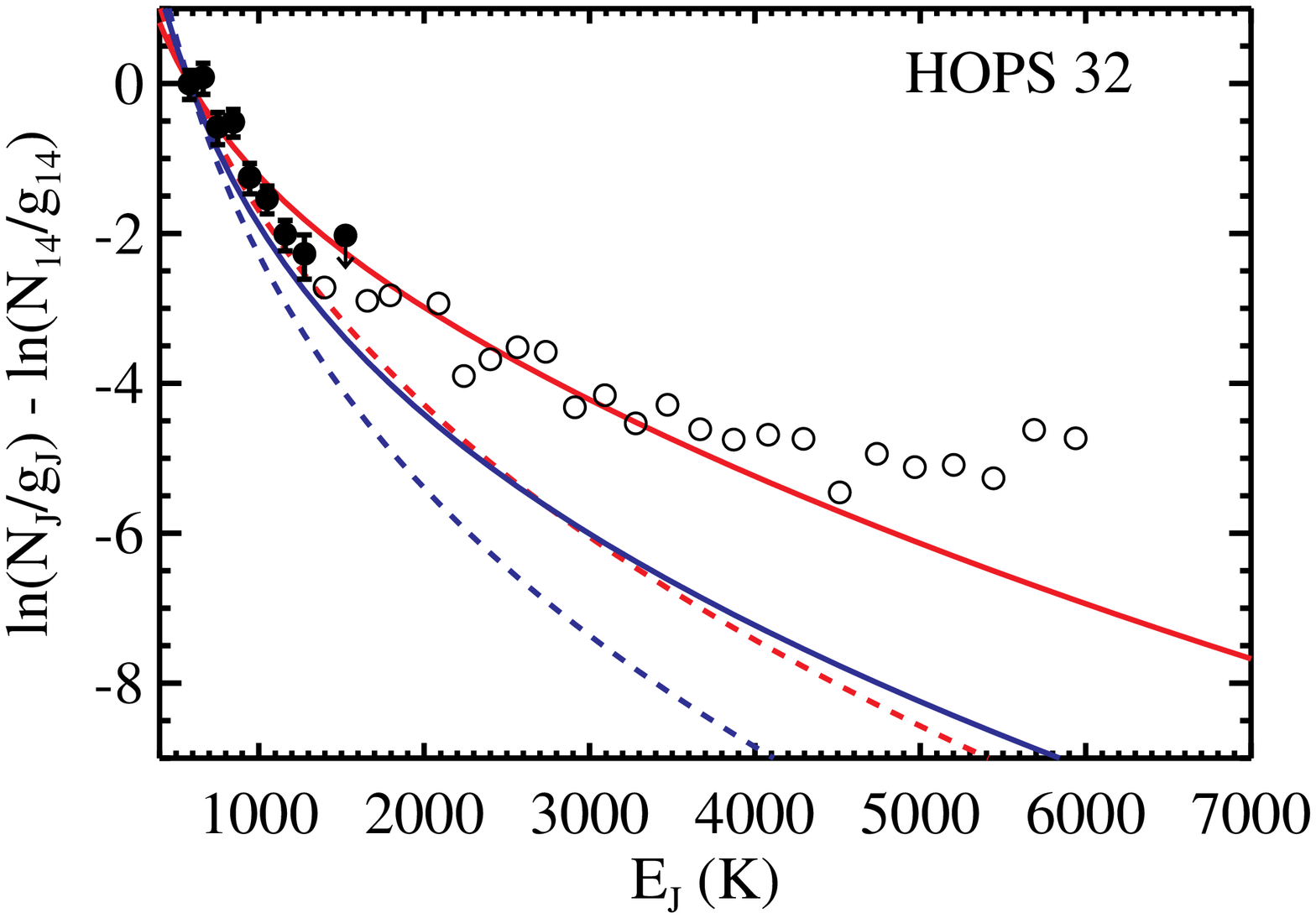}
\plottwo{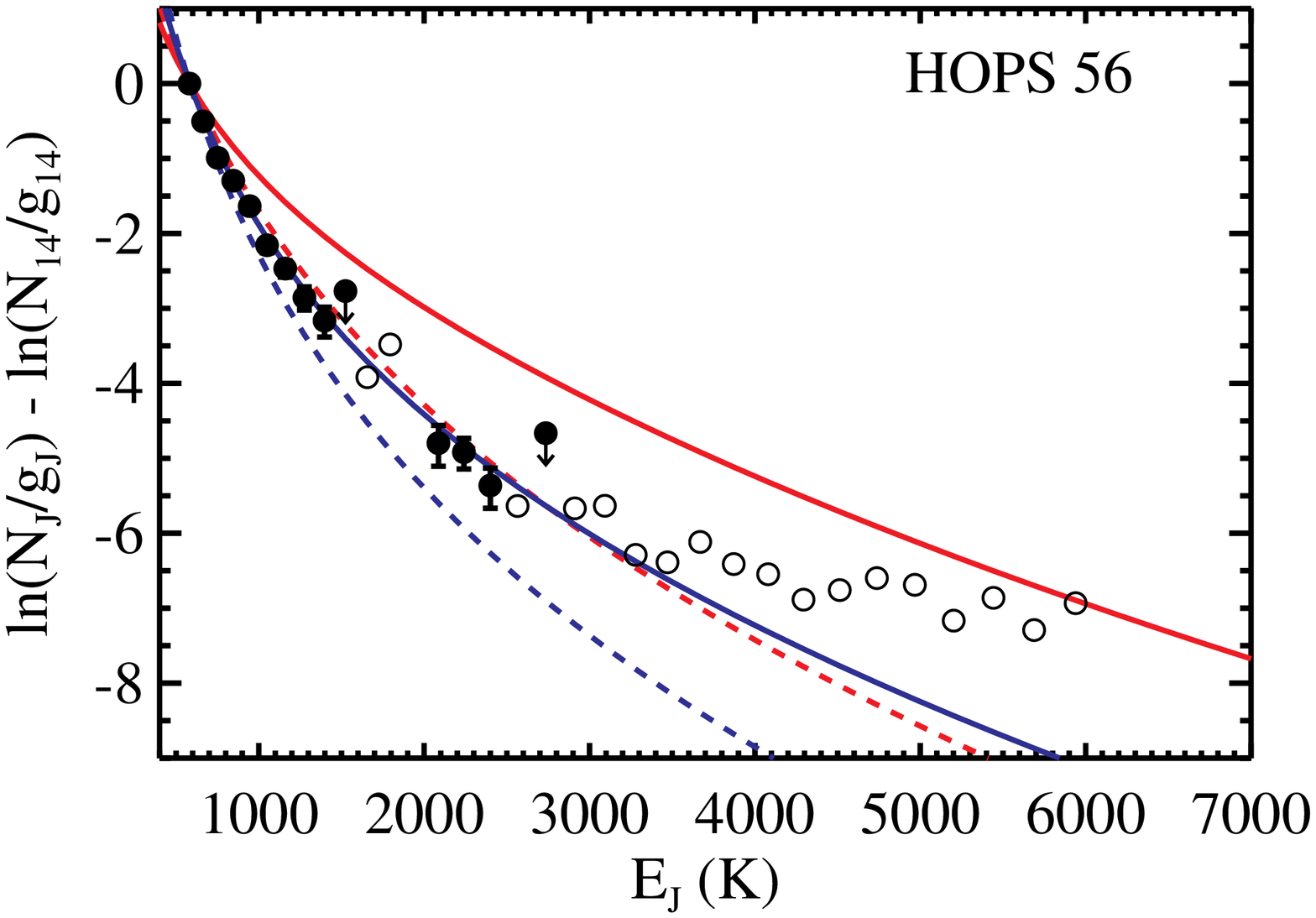}{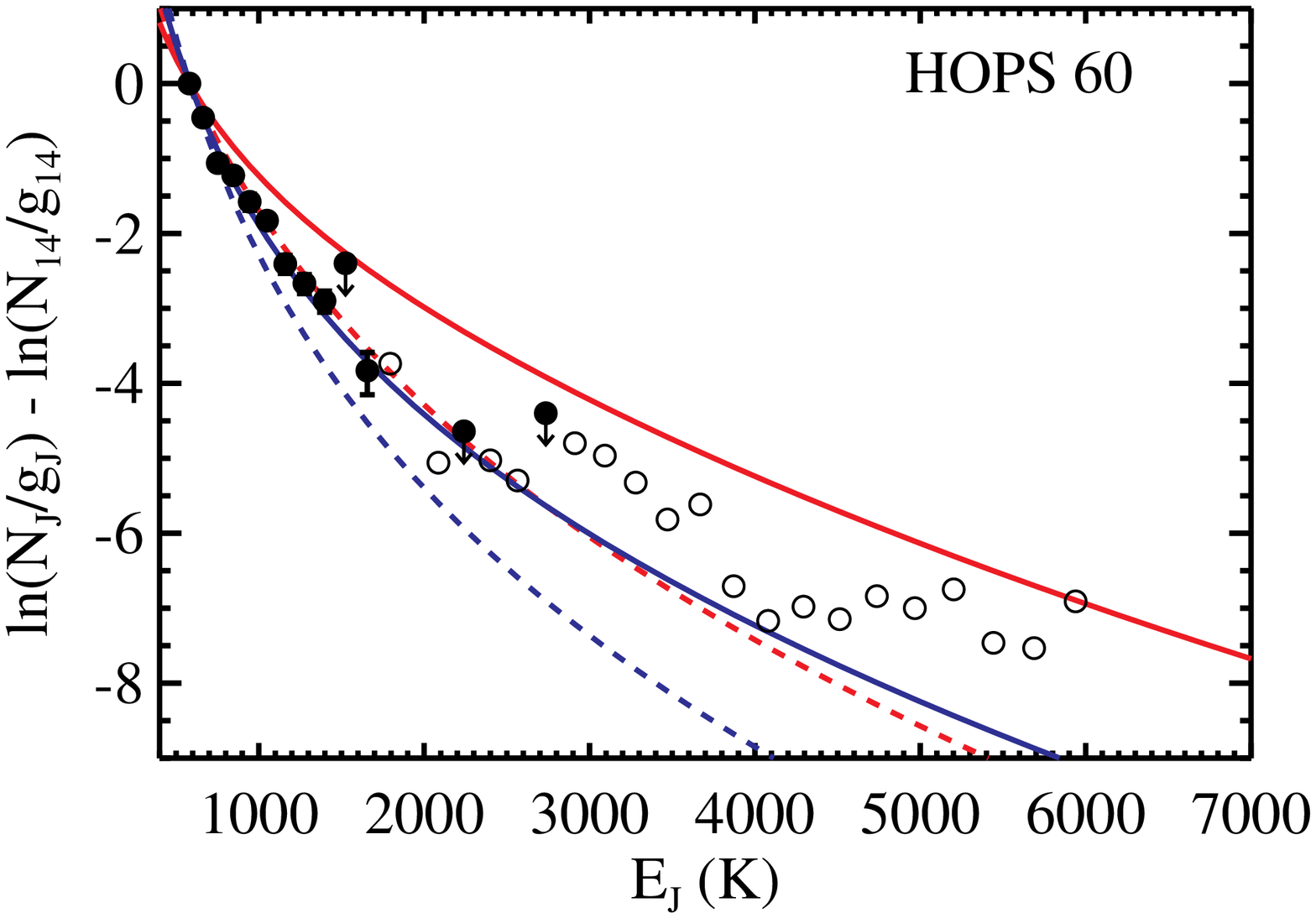}
\plottwo{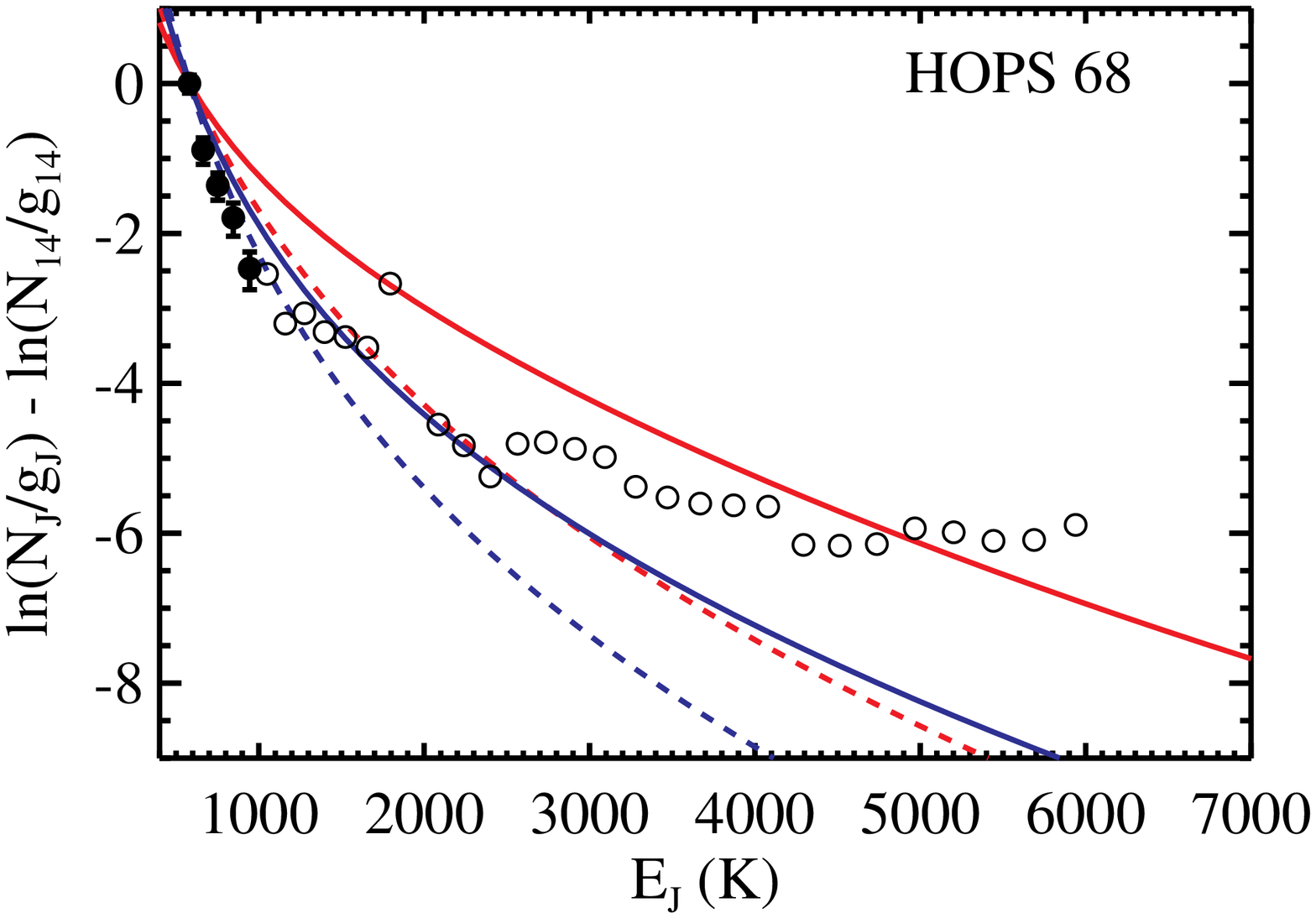}{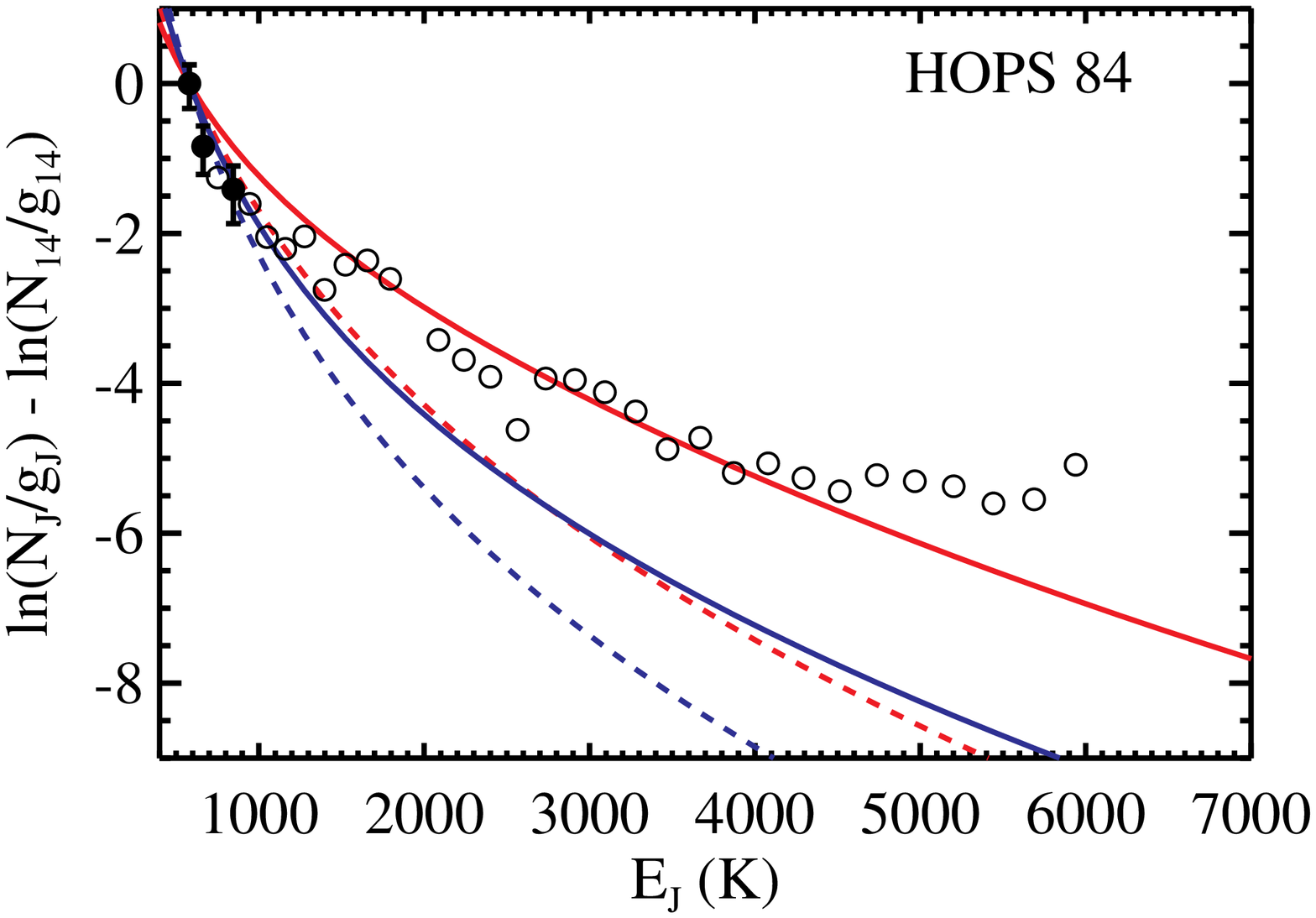}
\plottwo{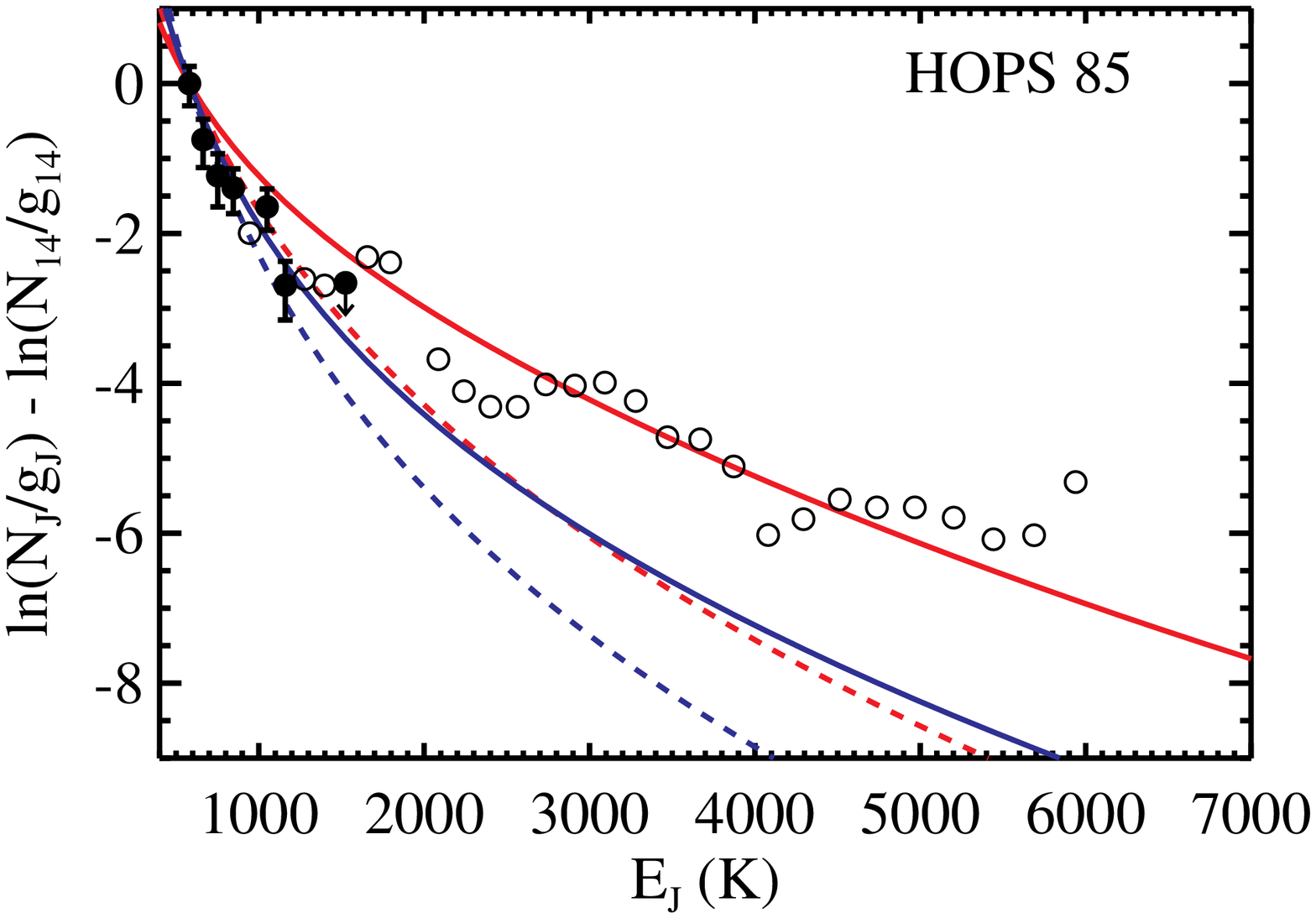}{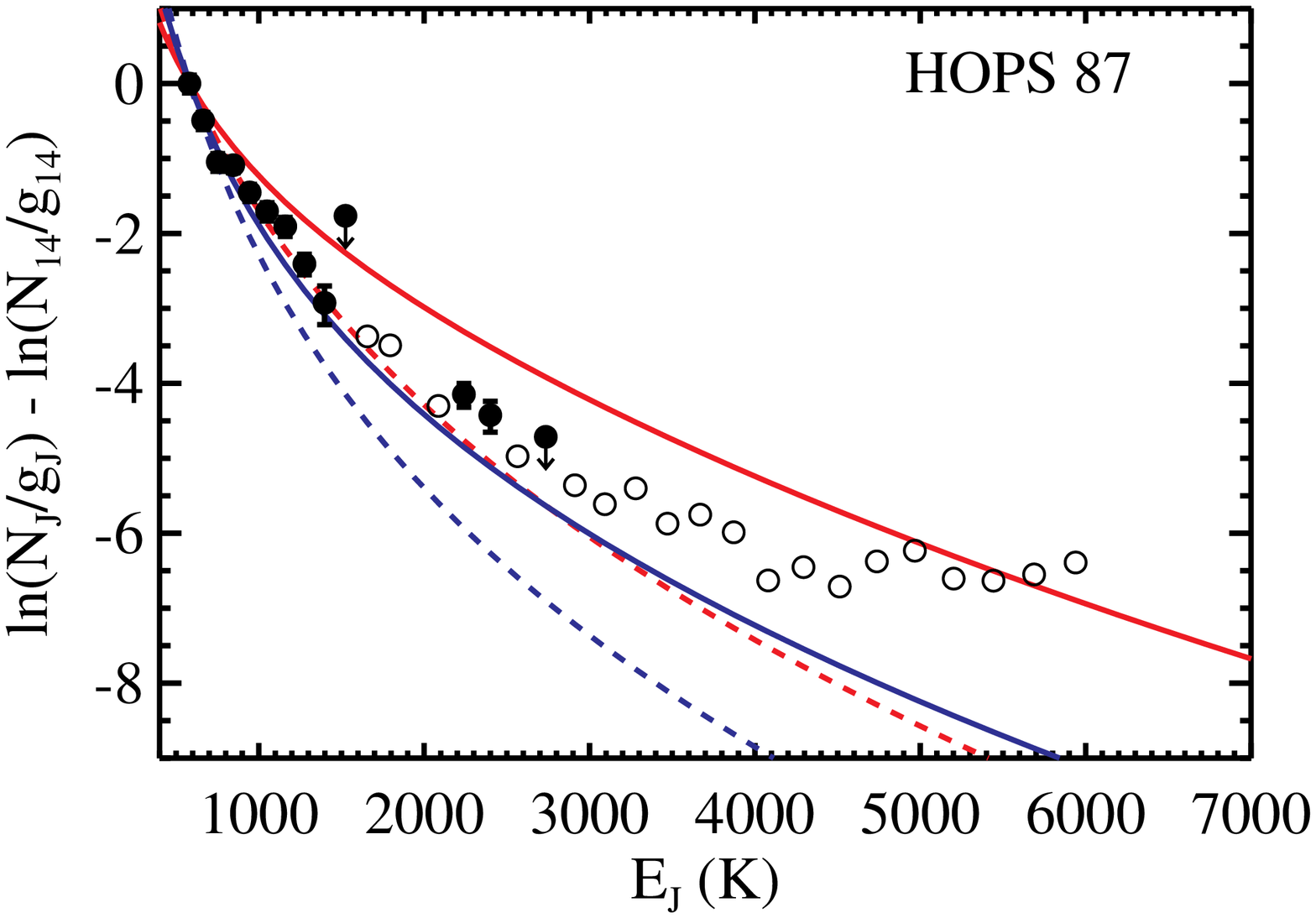}
\plottwo{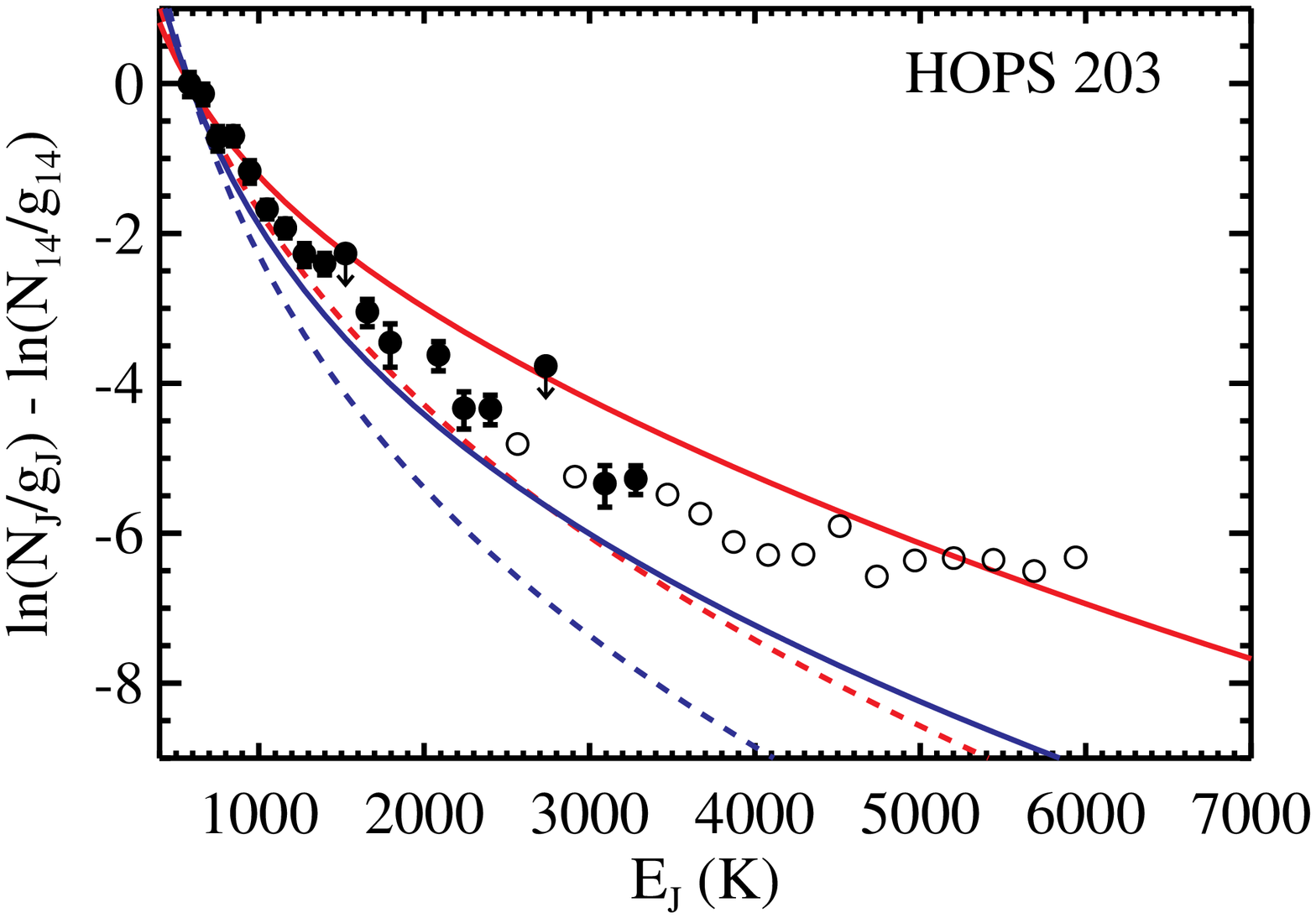}{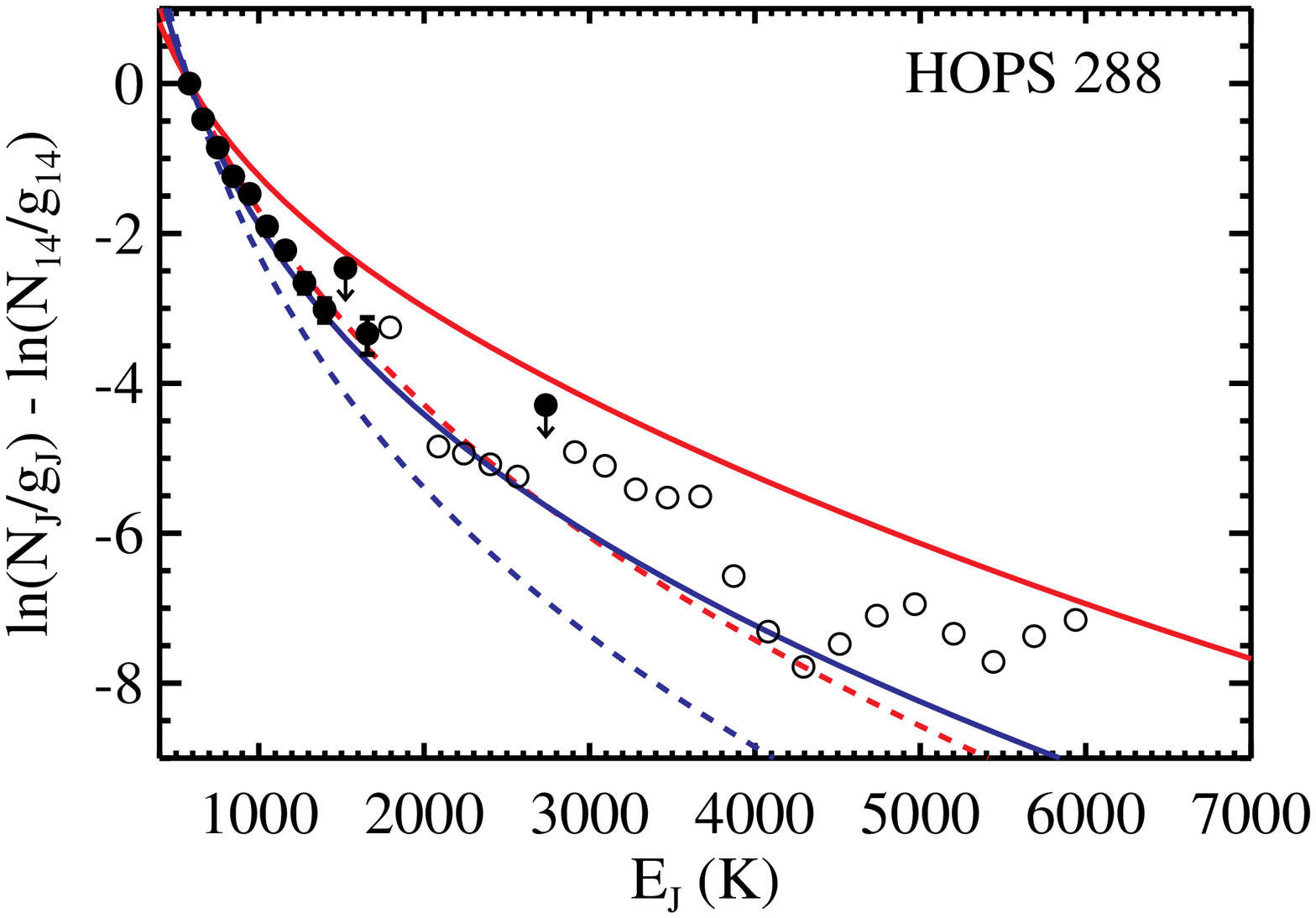}
\plottwo{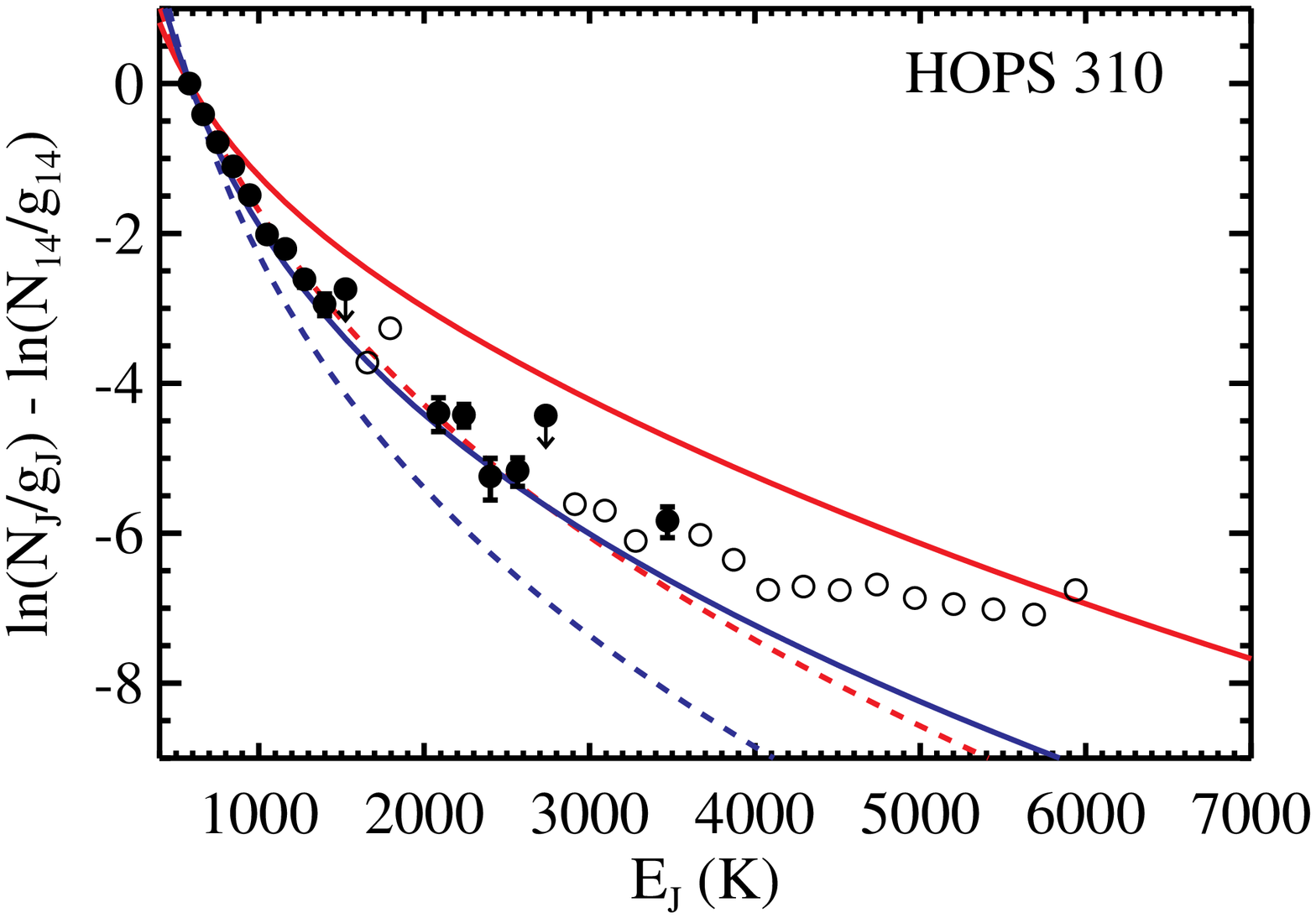}{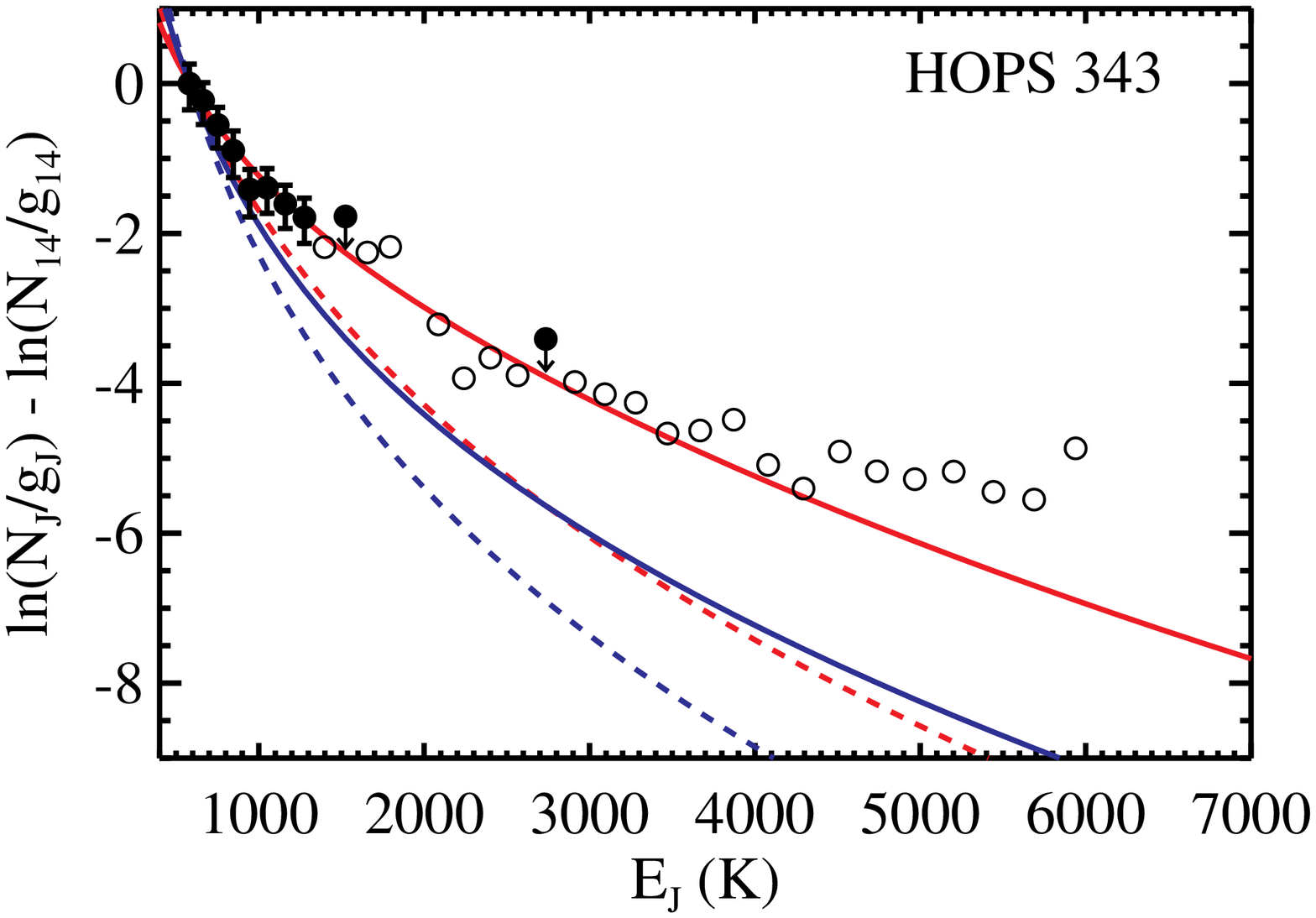}
\plottwo{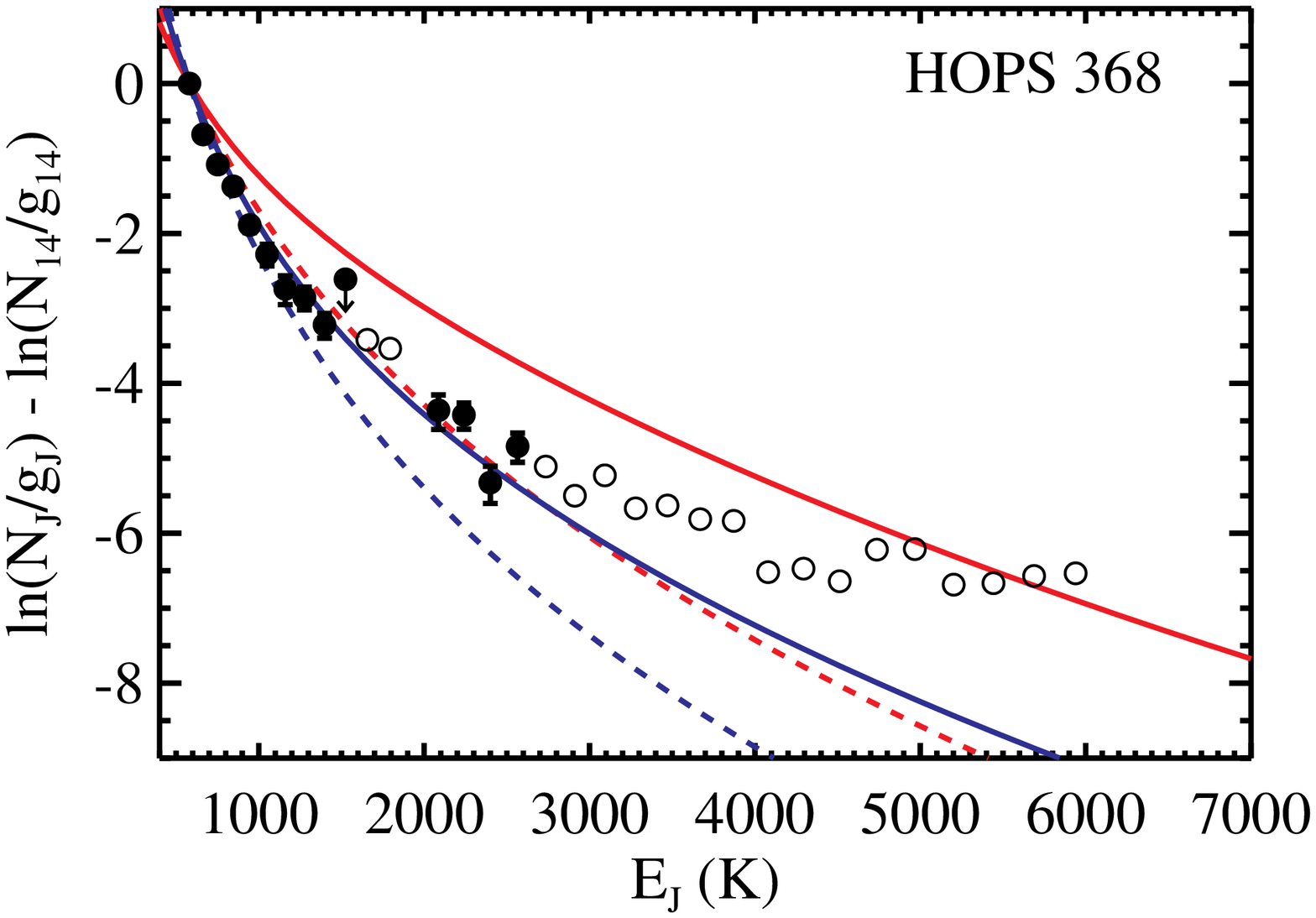}{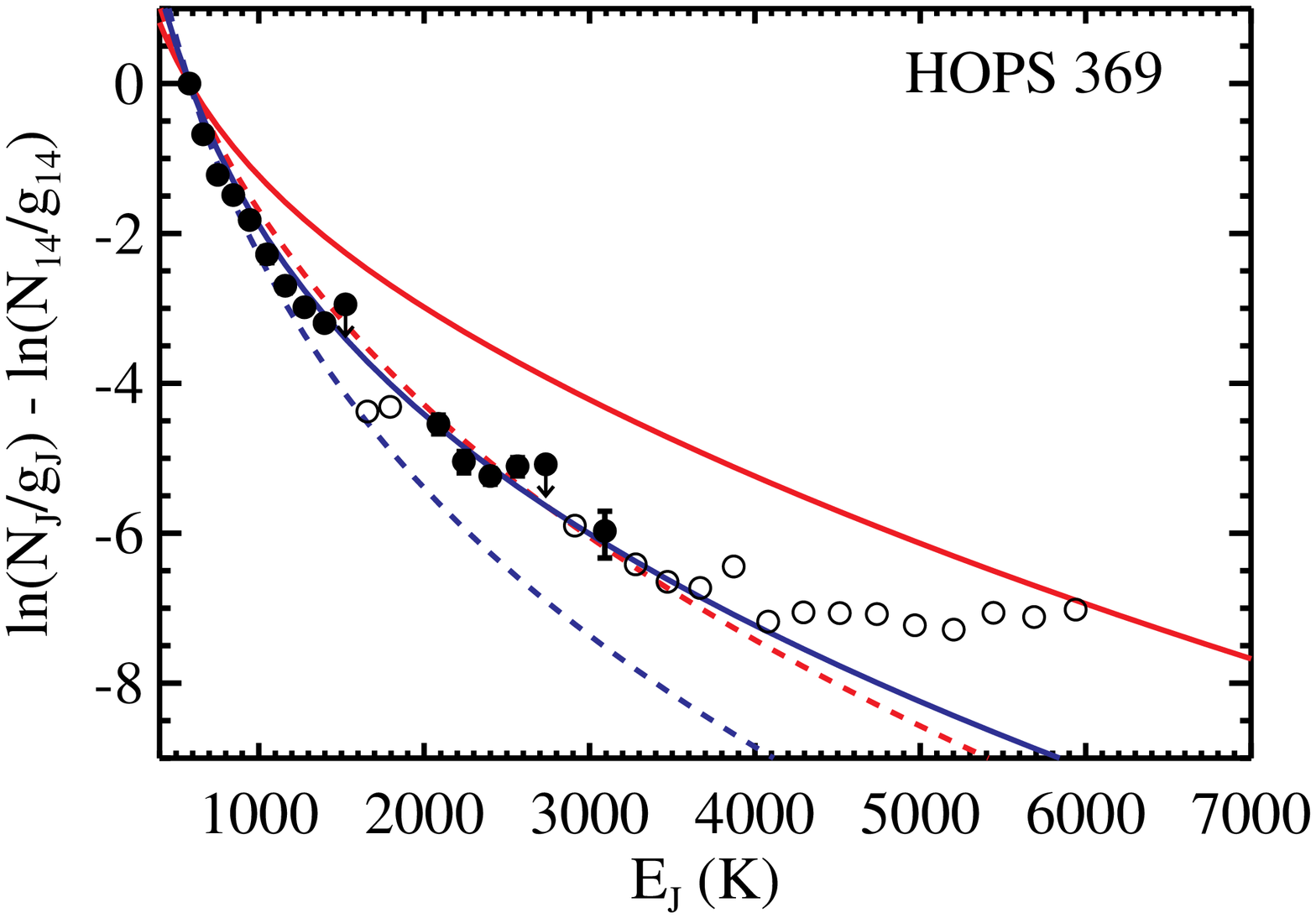}

\caption{Synthetic rotational diagrams for a medium with a power-law
  gas temperature distribution with $b$ = 3.0 (blue)
  and $b$=2.0 (red) for densities $n\mathrm{(H_2)} = 10^5 $
  cm$^{-3}$ (dashed line) and $n\mathrm{(H_2)} = 10^6 $ cm$^{-3}$
  (solid line) overlaid on the observed rotational
  diagrams.  Downward
  arrows indicate upper limits to the fluxes of the CO lines which are
  blended with a nearby line. Open circles correspond to 3$\sigma$
  upper limits for the non-detections. \label{power_law_consistency}}
\end{figure}

For all the other sources with fewer number of detected CO lines, gas
densities are not well constrained. Acceptable solutions could be
found for high densities ($n\mathrm{(H_2)}$ $>$ 10$^6$~cm$^{-3}$) with
steeper power-law indices ($b>$ 3) and for low denisities
($n\mathrm{(H_2)}$ $<$ 10$^5$~cm$^{-3}$) with $b<$ 2.  The low density
solutions with $b\ll$ 2 approach the isothermal solutions described in
the previous section.  The degeneracies in the multi-component
temperature fits are primarily because fewer CO lines are detected in
these sources.  When more lines are detected the models tightly
constrain the physical parameters of the emitting medium, as in the
cases of HOPS~108, 182 \& 370. Nevertheless, the ranges of density and
power-law index found for these three sources also appear to be
consistent with the observed CO emission from most of the other
sources in our sample. This is demonstrated in
Figure~\ref{power_law_consistency}. Power-law temperature models with
$b$ in the range of 2.0$-$3.0 and densities in the range of
10$^5-$10$^6$ cm$^{-3}$ can explain the observed rotational diagrams
of all the other sources for which fewer lines are detected.

\subsubsection{Degeneracies in the model fits}

There are significant degeneracies in the model fits. In the five
sources where fewer than 8 CO lines detected, neither models can rule
out high density ($n\mathrm{(H_2)}$ $\ga$ 10$^8 $ cm$^{-3}$) solutions
where the emitting gas is in LTE. The more flexible power-law
temperature models tightly constrain density to 10$^5 -$10$^6
$~cm$^{-3}$ only for the three sources where more than 25 lines are
detected. For other sources, even though the power-law models with
densities in the range of 10$^5 -$10$^6 $ cm$^{-3}$ are consistent
with the observed emission, we cannot rule out higher density
solutions. Moreover, the power-law temperature models that we use are
at uniform density. In reality, the CO emitting medium is likely to
have a range of temperature and density. Models with multiple
temperature and density components can possibly reproduce the observed
CO emission for LTE conditions.

\subsection{Thermal {\it vs} sub-thermal excitation}

\begin{figure}
\centering
\epsscale{1.1}
\plotone{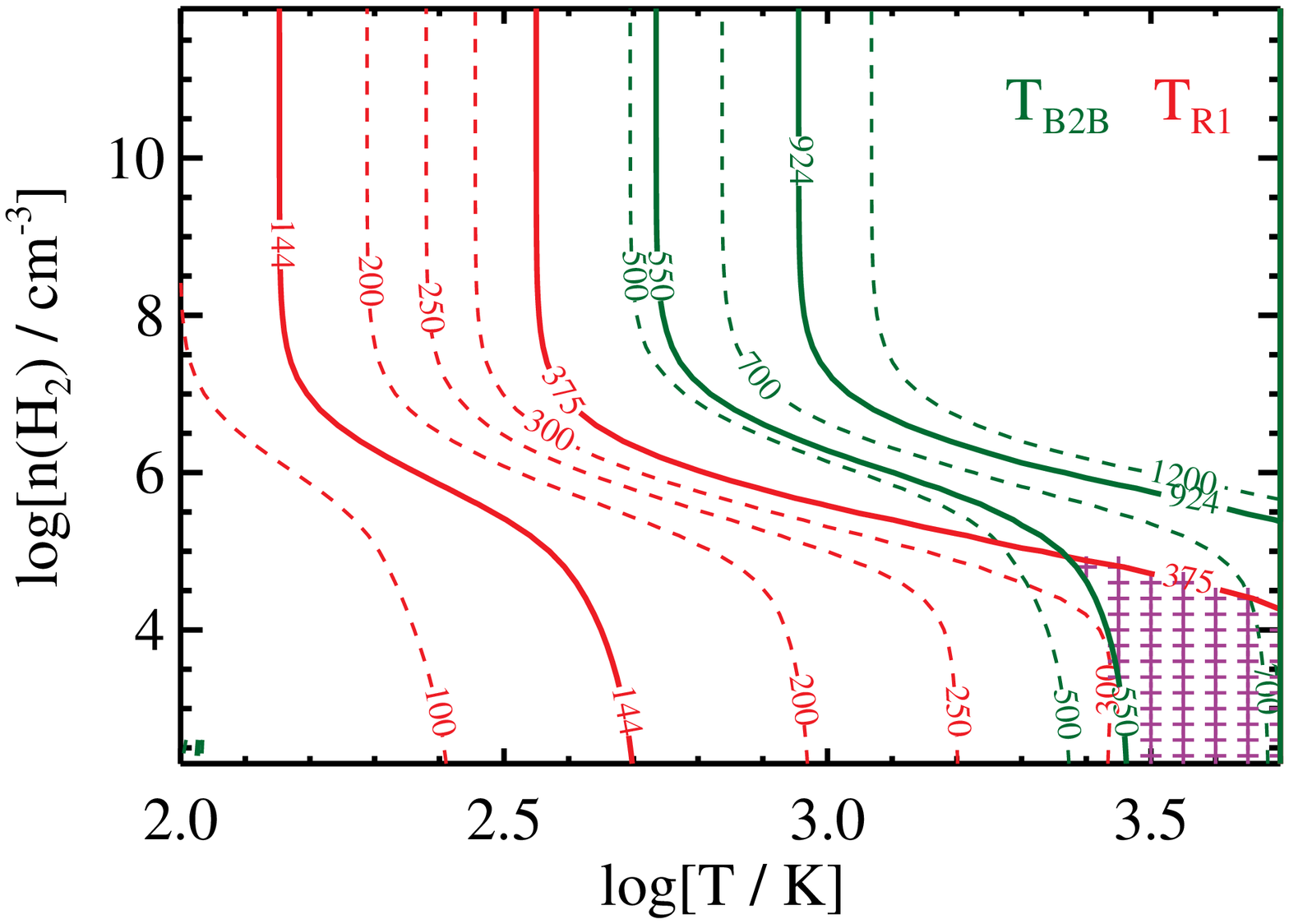}
\caption{Synthetic rotational temperatures, $T_{\mathrm{R1}}$ (red) and
  $T_{\mathrm{B2B}}$ (green), computed for CO emission from an
  optically thin, isothermal, uniform density medium (dashed
  contours). The solid contours represent the upper and lower bounds
  for the observed values of $T_{\mathrm{R1}}$ (red) and
  $T_{\mathrm{B2B}}$ (green). The shaded region (purple crosses)
  indicates  gas temperatures and densities for which a single
  component model can simultaneously reproduce both the observed
  $T_{\mathrm{R1}}$ and $T_{\mathrm{B2B}}$. \label{model_trb} }
\end{figure}

The `universality' of CO rotational temperatures (line ratios),
and the degeneracies of the simple models, suggest strongly that the
CO excitation is largely either {\it thermal} at low temperatures
($\sim$~300~K for the lines observed in the $R1$ band), or {\it
  sub-thermal} at high temperatures ($\ga$~1000$-$2000~K). If the CO
excitation is {\it thermal},
i.e. $n\mathrm{(H_2)}$~$\ga$~10$^8$~cm$^{-3}$, the rotational
temperature should approach the physical temperature of the gas and is
insensitive to density. This can be seen from Figure~\ref{model_trb},
where the average rotational temperatures predicted for the CO
emission from an optically thin, isothermal medium are shown for
various gas temperatures and densities \citep[also see][]{neufeld12}.
At these high densities, the observed distributions of
$T_{\mathrm{R1}}$ and $T_{\mathrm{B2B}}$ indicate a narrow range in
the gas temperature for both the components (see
Figure~\ref{model_trb}). This implies that, if the CO is {\it
  thermally} excited, the physical temperature of the emitting gas
should be very similar for protostars whose luminosities differ by two
orders of magnitude. Additionally, if the CO excitation is {\it
  thermal}, the observed positive curvature of the rotational diagrams
indicates the presence of multiple temperature components.  The average
rotational temperatures, $T_{\mathrm{R1}}$, $T_{\mathrm{B2B}}$ and
$T_{\mathrm{B3A}}$ (see Table~\ref{rot_tbl}), are significantly
different in each source, which suggests the presence of at least 3
temperature components. Moreover, the observed distribution of the
rotational temperatures $T_{\mathrm{LR1}}$ and $T_{\mathrm{SR1}}$ (see
Section~\ref{trot}) indicate that even the CO lines in the R1 spectral
band ($J_{up}$=14$-$25) require at least two temperature components if
the excitation is {\it thermal}. Thus for most sources in our sample 4
or more components would be required to explain the observed emission
if the emitting gas is in LTE. The narrow range in the rotational
temperatures observed for the protostars in our sample then requires
that these multiple temperature components of the CO emitting gas
remain nearly identical for sources with $L_{bol}$ in the range of
$\sim$~2$-$217 $L_{\odot}$. Thus, the primary weakness of LTE
solutions is that they require heating mechanisms that maintain
relatively constant temperatures in multiple gas components over a
range of protostellar luminosity spanning two orders of magnitude.

However, if the CO excitation is {\it sub-thermal}, gas at high
temperatures (T~$\ga$~1000$-$2000~K) and moderate densities
($n\mathrm{(H_2)}$~$\la$~10$^6$~cm$^{-3}$) can reproduce the observed
CO emission from protostars. For low densities,
$n\mathrm{(H_2)}$~$\la$~10$^{4.5}$~cm$^{-3}$, the CO rotational
diagrams are well fit by emission from an isothermal medium with
temperatures in the range of 2000$-$5000~K. One attractive feature of
this set of solutions is that in the low density limit, the rotational
temperatures are insensitive to density and depends only on the
temperature \citep[][also see Figure~\ref{model_trb}]{neufeld12}. At
densities below 10$^{4.5}$~cm$^{-3}$, a large range in the gas
temperatures results only in a narrow range in rotational
temperatures.  For example, emitting gas at temperatures in the range
of 500~K to 5000~K can reproduce the observed range in
$T_{\mathrm{R1}}$ as can be seen from Figure~\ref{model_trb}. In this
case, it is somewhat easier to explain the observed behavior of
rotational temperatures as a function of $L_{bol}$. In protostars with
higher $L_{bol}$ the CO emitting gas is possibly hotter, but the
resulting rotational temperatures are not very different from those
produced by cooler CO gas in low $L_{bol}$ sources. Note, however,
that the isothermal solutions are only illustrative and multiple
temperature components must be present; sources which show large
positive curvature in their rotational diagrams (HOPS 182 and 370),
the high$-J$ ($J_{up}$~$\ga$ 38) lines are not well fit by a single
temperature component.

Alternatively, we can model the temperature of the emitting medium as
a continuous power-law distribution where the parameters are now the
power-law index and the density of the gas, as has been done
successfully for the excitation H$_2$
\citep{neufeld08,yuan11,giannini11}. The observed line ratios of all
the sources are consistent with power-law models with power-law index,
$b$ = 2.0$-$3.0 and with densities in the range of 10$^5 -$10$^6 $
cm$^{-3}$. These densities are also below the critical densities of
the observed CO transitions and the excitation is {\it
  sub-thermal}. Even when multiple gas components with a power-law
distribution of temperatures are present, a range of excitation
conditions ($n\mathrm{(H_2)}$ and $b$) can explain the narrow range in
the observed rotational temperatures as long as the excitation is {\it
  sub-thermal}. Additionally, the CO emission observed with PACS is
dominated by the high temperature ($T$ $\ga$ 1000~K) components, even
when lower temperature components are present. For the best
fit model for HOPS~108 ($n\mathrm{(H_2)}$ =
6.0$\times$10$^5$~cm$^{-3}$ and $b$=2.4), the gas components with
$T$~$\ga$~1000~K dominates the emission even in the R1 spectral band
for CO transitions $J_{up}$~$\ge$~18, and components with
$T$~$\ga$~2000~K contributes more than 50\% of the observed flux for
lines ($J_{up}$~$\ge$~27 ) observed in B2B and B3A spectral bands.

In reality, a solution intermediate to those obtained with the
power-law and isothermal models, with a range of temperatures in which
the lower temperature (T $\la$ 1000~K) components do not contribute
significantly to the observed emission is likely to be more
robust. Thus, if the FIR CO emission observed with PACS is dominated
by hot (T $\ga$ 1000~K), moderate density ($n\mathrm{(H_2)}$ $\la$
10$^6$~cm$^{-3}$) gas present in the vicinity of protostars, the
observed rotational temperatures would allow for a relatively wide
range in the temperature and density of the emitting gas. The ability
to find solutions in the {\it sub-thermal} limit that do not require
multiple temperature components, each of which must maintain relatively
constant temperatures over a range of protostellar luminosity, makes
the high temperature, {\it sub-thermal} excitation the more attractive
solution for the observed FIR CO emission.

\section{Physical origin of FIR CO emission in protostars} \label{origin}

Observations of FIR CO lines ($J_{up}$ $\la$ 25) from protostars prior
to {\it Herschel} have suggested that the CO emission arises from
molecular gas heated by non-dissociative shocks \citep{watson85b,
  ceccarelli98, giannini01, nisini02}.  Studies with {\it Herschel},
however, have invoked two independent mechanisms to explain the CO
emission observed with PACS: {\it (i)} UV-heating of the envelope
cavity walls producing PDRs and {\it (ii)} small scale shocks along
the cavity walls \citep{vankemp10a,fich10, visser12}. The primary
motivation for invoking the PDRs has been the narrow
(FWHM~$\la$~2~km~s$^{-1}$) velocity component seen in the spectrally
resolved line profiles of the low$-J$ ($J_{up}~\le~10$) CO lines
observed towards low-mass protostars \citep{vankemp09a, vankemp09b,
  vd09, yildiz10, yildiz12}. Since PACS lines are spectrally
unresolved, it was assumed that this narrow component due to UV-heated
cavity walls also contributes to the line flux of the higher$-J$
($J_{up}~\ge~14$) CO transitions observed with PACS. Both the
UV-heated and shock-heated gas components were found to be necessary
to explain the observed CO emission in three particular protostars
modeled in detail by \citet{visser12}. In the following, we discuss
how well these two mechanisms can explain the observed properties of
the FIR CO emission from a larger sample of protostars. We argue that
emission from the UV-heated PDRs along the cavity walls is unlikely to
be the dominant component of the observed FIR CO emission. Most of the
observed emission, instead, is likely to originate from low density
($n\mathrm{(H_2)} \la 10^6$ cm$^{-3}$) molecular gas, heated in
outflow shocks, within 2000~AU from the protostars.

\subsection{Origin in  PDRs along envelope cavity walls}
  
In PDR models, energy released from the mass accretion onto the
protostar from the surrounding disk is the source of the UV flux which
heats the cavity walls \citep{visser12}. The accretion luminosity of
the protostars scales with $L_{bol}$ \citep{kenhart95,evans09} and
therefore $L_{\mathrm{UV}}$ is likely to increase with increasing
$L_{bol}$. In the PDR models of \citet{visser12}, most of the
UV-excited emission originates at densities between 10$^6$ and 10$^9$
cm$^{-3}$ and is therefore close to LTE. The total CO luminosity in
this case is proportional to the total number of the CO molecules
present (or equivalently mass of the CO gas) for a constant gas
temperature. Thus, in protostars with higher $L_{bol}$ (and therefore
higher $L_{\mathrm{UV}}$), a larger amount of gas will be heated, in
turn resulting in a higher $L_{\mathrm{CO}}$. The PDR models can
explain the observed correlation between $L_{\mathrm{CO}}$ and
$L_{bol}$, but only if the temperatures of the gas components remain
the same in all the sources.  Indeed, the various rotational
temperatures computed -- $T_{\mathrm{LR1}}$, $T_{\mathrm{SR1}}$,
$T_{\mathrm{R1}}$ and $T_{\mathrm{B2B}}$ -- for the protostars in our
sample are found to be within narrow ranges. If the CO excitation is
{\it thermal}, as is the case in PDRs, the average rotational
temperature computed for a small range in $E_J$ should be
representative of the physical temperature of the gas component
contributing to those CO lines. It is difficult, however, to obtain
similar gas temperatures in PDRs along the envelope cavity walls for
protostars with $L_{bol}$ ranging over two orders of magnitude. In
PDRs, gas-grain collisions and the subsequent thermal emission from
the dust grains is the dominant cooling mechanism for molecular gas
densities $n\mathrm{(H_2)} \ga 10^6$ cm$^{-3}$, and the cooling rate
is $\propto$ $n\mathrm{(H_2)}^{2}$. Since the photo-electric heating
rate is $\propto$ $L_{UV}~n\mathrm{(H_2)}$, the gas temperature
roughly scales as $T \propto L_{UV} / n\mathrm{(H_2)}$. The CO gas
temperature can remain nearly constant in protostars with widely
ranging $L_{bol}$ values only if the envelope density scales linearly
with $L_{bol}$. Figure~\ref{rho1_lbol_tbol}, however, shows that the
envelope density at 1~AU, $\rho_{1}$, is uncorrelated with $L_{bol}$,
so it is unlikely that the gas temperatures along the cavity walls are
similar in all the protostars.

In Section~\ref{lobe} we showed that the luminosity of the observed CO
emission from an outflow lobe $\sim$ 19$\arcsec$ (projected distance
$\sim$ 8000~AU) away from the protostar is a factor of two higher than
that observed for the on-source emission.  If the UV-radiation from
the protostar were responsible for the heating of the gas, then hotter
and more luminous CO emission would be expected from the on-source
position, the opposite of what is observed. Moreover, the rotational
temperature and luminosity of the CO emission from the outflow lobe
far away from the protostar (projected distance $\sim$ 8000~AU) is
indistinguishable from that observed at the on-source positions
(projected distance $\la$ 2000~AU) of all the other sources in our
sample, pointing to a common origin in both cases. The CO excitation
at the outflow lobe is likely due to heating of the low density
($n\mathrm{(H_2)} \la 10^6$ cm$^{-3}$) gas in the outflow shocks and
not from the protostellar accretion-generated PDRs.

The UV flux generated by protostellar accretion will heat up the dense
molecular material close to the protostar and this material is likely
to contribute to the CO emission. Indeed, the narrow velocity
  component (FWHM~$\la$~2~km~s$^{-1}$) which originates in the
  UV-heated cavity walls is seen in the line profiles of the
  CO~$J$~=~6$-$5 and CO~$J$~=~7$-$6 transitions observed towards
  low-mass protostars \citep[e.g.][]{vankemp09a, vankemp09b,
    yildiz12}. For these low$-J$ lines, this narrow component
  dominates the line flux. Similar narrow component is also seen in
  the CO~$J=10-9$ line for the protostars observed with {\it
    Herschel}/HIFI \citep{yildiz10}. However, in most cases the line
  flux is dominated by emission from the broad (FWHM = 25-30
  km~s$^{-1}$) component \citep{yildiz10}. Moreover, for the spectrally
  resolved lines, the broad component originating in shock heated gas
  is seen to increase in strength going from $J_{up}$~=~1 to
  $J_{up}$~=~10 \citep{yildiz12, yildiz10, vankemp09a}. Therefore, for
  the higher excitation CO ($J_{up}$~$\ge$~14) lines observed with
  PACS, emission from PDRs is unlikely to be the dominant contributor
  to the total flux.

\subsection{Origin in outflow shocks}

\subsubsection{Shocks along envelope cavity walls}

Models for CO emission from shocks along envelope cavity walls predict
most of the emission that contributes to the observed CO line fluxes
originates in the high density (preshock density $\ga$ 10$^6 $
cm$^{-3}$) gas \citep{visser12}. Our modeling here, however, suggests
that the CO excitation is {\it sub-thermal} and that the observed
emission likely originates in low density gas ($n\mathrm{(H_2)}$ $\la$
10$^6 $ cm$^{-3}$).  Moreover, we find that the CO emission properties
observed for the outflow position $\ga$~8000~AU away from the
protostar and for the on-source positions of the protostars in our
sample are essentially indistinguishable.  All these results suggest
that the shocked gas producing CO emission is likely not located at
the denser part of the cavity walls close to the protostar.

\subsubsection{CO emission from low density shock-heated gas}

The optimal solutions obtained from both the isothermal and power-law
temperature models indicate that the observed CO emission from
protostars in our sample arises in low density ($n\mathrm{(H_2)}$
$\la$ 10$^6 $ cm$^{-3}$) molecular gas which has high temperature ($T$
$>$ 2000~K) components. Such high temperatures and low densities are
typical of gas heated by outflow shocks and have been reported by
previous studies of FIR CO lines from protostars \citep{ceccarelli98,
  nisini99a, giannini01}. The low densities inferred for the observed
CO emission are more compatible with shock-heated gas within the
envelope cavity along the molecular outflow or at envelope radius
$\ga$ several 100$-$1000~AU. Since the size of the central spaxel,
from which the analyzed spectra are extracted, corresponds to a
projected radius of 2000~AU from the protostars, the shock emission
must come from within this distance. Indeed, compact molecular
outflows on $\la$ 1000~AU scales with high outflow momentum rates have
been observed toward protostars \citep[e.g.,][]{santiago09, leecf09,
  takahashi12}.

Non-dissociative shocks with shock speed $v_s$ $>$~25~km~s$^{-1}$ can
heat the gas to temperatures $>$~2000~K \citep{kn96a}. In general,
outflows in protostars produce multiple shocks with varying velocities
and different shock-front geometries.  If the densities of the
post-shock gas are low, e.g., $n\mathrm{(H_2)}$ $<$ 10$^{4.5} $
cm$^{-3}$, then the isothermal model predicts that the hottest
($T$~$\ga$~2000~K) gas component dominates the CO emission in all the
transitions observed with PACS. Emitting gas at temperatures of
2000$-$5000~K can also produce the narrow range in the observed
rotational temperatures. For slightly higher densities, e.g., 10$^5 $
cm$^{-3}$ $\la$ $n\mathrm{(H_2)}$ $\la$ 10$^6 $ cm$^{-3}$, multiple
gas components whose temperature follows a power-law distribution with
index $b$ ranging from 2$-$3 can explain the observed FIR CO
rotational diagrams.  Even in this case, the hotter ($T$~$\ga$~1000~K)
gas components produce most of the CO emission ($J_{up}$ $\ga$ 18)
observed with PACS. The post-shock gas in bow-shaped C-shocks can have
a power-law temperature distribution
\citep{sbm91,neufeld08,yuan11,giannini11}. Several such bow shocks are
likely to be present along the supersonic molecular outflow from
protostars even within a projected radius of 2000~AU. Classical bow
shocks with a parabolic shape are expected to produce a power-law
index, $b$ $\sim$ 3.8 \citep{sb90,neufeld06,neufeld08}. Our results,
however, indicate that $b$ is in the range of 2$-$3 for the observed
FIR CO emission from protostars; a similar range for the power-law
index have also been found for H$_2$ emission from shocked gas
\citep{neufeld09,yuan11}. As pointed out by \citet{yuan11}, the lower
values of $b$ suggest that either the curvature of the shock front is
smaller than that of a parabola, or a mixture of shocks with different
shock-front geometries ranging from planar to bow is present along the
flow.

The observed outflow momentum rates ($\dot{M}v$) in protostars are
found to scale with $L_{bol}$ \citep{bontemps96,takahashi12}. This
correlation, which extends over 3 orders of magnitude in $L_{bol}$, is
primarily driven by the increase in the mass loss rate in the outflows
\citep{bontemps96}. There is also evidence that the mass loss rate
($\dot{M}$) from young T~Tauri stars and protostars scales with the
accretion rate \citep[][; Remming et al. in
  preparation]{hartigan95}. Since $L_{bol}$ increases with accretion
luminosity in protostars, the mass loss rates from outflows is
expected to scale with $L_{bol}$, and therefore the total mechanical
luminosity in shocks should also scale with $L_{bol}$.  Higher cooling
rates are therefore expected for higher $L_{bol}$ sources, which, as
Figure~\ref{CO_lum} shows, is tracked by $L_{\mathrm{CO}}$. Thus,
shock heating of a lower density ($n\mathrm{(H_2)}$ $\la$ 10$^6 $
cm$^{-3}$) gas can at least qualitatively explain the observed trend
between $L_{\mathrm{CO}}$ and $L_{bol}$ as well as the observed lack
of dependence of $T_{rot}$ on $L_{bol}$ and $T_{bol}$.

\section{Conclusions} \label{conclude}

We have analyzed the emission lines due to the rotational transitions
of CO ($J$=14$-$13 up to $J$ = 46$-$45) in the FIR (57$-$196
$\micron$) spectra of 21 protostars in Orion, obtained with {\it
  Herschel}/PACS. The observed CO lines originate in the warm and hot
gas within a projected distance of $\la$~2000~AU from the
protostars. We searched for correlations between the observed CO
emission properties and protostellar luminosity, evolutionary status,
and envelope density. We modeled the CO lines to constrain the
excitation conditions in the emitting gas and identify possible
heating mechanisms. Our main conclusions are listed below.

\begin{enumerate}

\item 
The total luminosity of the CO lines observed with PACS increases with
protostellar luminosity over a large range (2$-$217~$L_{\odot}$).

\item
The CO rotational diagrams of protostars obtained with PACS show a
positive curvature and the rotational temperature implied by the line
ratios increases with increasing rotational quantum number $J$. A
minimum of 3$-$4 rotational temperature components are required to fit
the observed rotational diagram in the PACS wavelength range. The
rotational temperature computed from the lowest$-J$ lines ($J_{up}$ =
14$-$25) observed, $T_{\mathrm{R1}}$, is found to lie within a narrow
range for protostars while $L_{bol}$ values range over two orders
of magnitude. A similar lack of dependence with $L_{bol}$ is also
observed for the rotational temperatures determined for the higher$-J$
lines.

\item
The observed CO emission properties ($L_{\mathrm{CO}}$, $T_{rot}$) are
uncorrelated with evolutionary indicators and envelope properties of
the protostars such as $T_{bol}$ and envelope density.

\item
If the CO emitting gas is {\it thermally} excited, i.e. in LTE,
multiple gas components at different temperatures are required to
reproduce the observed emission. The temperatures of these components
must somehow remain independent of protostellar luminosity over two
orders of magnitude.

\item
The observed CO emission can also be modeled as arising from {\it
 sub-thermally} excited gas with high temperature (T~$\ga$~2000~K)
components. An isothermal medium at uniform density can reproduce the
observed emission for densities, $n\mathrm{(H_2)}$ $\la$ 10$^{4.5} $
cm$^{-3}$ and temperatures, T $\ga$ 2000~K. Observed emission can also
be reproduced by a uniform density medium with a power-law temperature
distribution (T$_{min}$ = 10~K \& T$_{max}$ = 5000~K) for densities in
the range of 10$^5 - $ 10$^6 $ cm$^{-3}$.

\item
 Emission from PDRs, produced along the UV-heated envelope cavity
 walls, is unlikely to be the dominant component of the CO emission
 observed in the PACS wavelength range. We argue that the simplest
 explanation for both the observed correlation between
 $L_{\mathrm{CO}}$ and $L_{bol}$ and the invariance of CO rotational
 temperatures between protostars is that the CO emission is dominated
 by {\it sub-thermally} excited, shock-heated gas at high temperatures
 (T~$\ga$~2000~K) and low densities ($n\mathrm{(H_2)}$ $\la$ 10$^6 $
 cm$^{-3}$), located within the outflow cavities along the molecular
 outflow or along the cavity walls at radii $\ga$ several
 100$-$1000~AU.

\end{enumerate}

\acknowledgements
This work is based on observations made with Herschel, a European
Space Agency Cornerstone Mission with significant participation by
NASA. Support for the HOPS program was provided by NASA through an
award issued by JPL/Caltech.  

The Herschel spacecraft was designed, built, tested, and launched
under a contract to ESA managed by the Herschel/Planck Project team by
an industrial consortium under the overall responsibility of the prime
contractor Thales Alenia Space (Cannes), and including Astrium
(Friedrichshafen) responsible for the payload module and for system
testing at spacecraft level, Thales Alenia Space (Turin) responsible
for the service module, and Astrium (Toulouse) responsible for the
telescope, with in excess of a hundred subcontractors. 

HIPE is a joint development by the Herschel Science Ground Segment
Consortium, consisting of ESA, the NASA Herschel Science Center, and
the HIFI, PACS and SPIRE consortia.




\appendix

\section{Spectral energy distirbutions of Protostars} \label{appen_a}
The observed SEDs of the protostars in our sample are shown in
Figure~\ref{appen_fig1}. The $J$, $H$, and $K_s$ photometry, when
available, is from the Two Micron All Sky Survey
\citep[2MASS;][]{skrut06}. Broad band fluxes at 3.6, 4.5, 5.8, 8.0 and
24.0~$\micron$ were obtained in a joint survey of the Orion A and B
molecular clouds by the Infrared Array Camera \citep[IRAC;][]{fazio04}
and Multiband Imaging Photometer \citep[MIPS;][]{rieke04} aboard
Spitzer \citep{kry12,tm12}. {\it Spitzer} Infrared Spectrograph
\citep[IRS;][]{houck04} spectra also shown in Figure~\ref{appen_fig1}
were obtained as part of the GO programs 30859 and 50374 \citep[PI:
  Megeath, S. T.; Poteet et al. in preparation;][]{poteet11}. For five
sources toward the OMC-2 region (HOPS~60, 108, 368, 369 \& 370),
photometry at 19.7, 31.4 and 37.1 $\micron$ was obtained with
SOFIA/FORCAST \citep{adams12}. Images at 70 and 160~$\micron$ of all
the protostars in our sample were obtained with {\it Herschel}/PACS as
part of the HOPS program \citep{fischer10,stanke10,fischer12}. Fluxes
at these wavelenths were measured within an aperture of 9.6$\arcsec$
radius at 70~$\micron$ and a 12.8$\arcsec$ radius at 160~$\micron$
after subtracting the median sky background estimated from an annulus
extending from the aperture limit to twice that value
\citetext{Fischer et al. in preparation}. In addition, we also
obtained submillimeter observations of our sample sources at
350~$\micron$ with the Submillimeter APEX Bolometer Camera (SABOCA)
and at 870~$\micron$ with the Large APEX Bolometer Camera (LABOCA) on
the Atacama Pathfinder Experiment \citetext{Stanke et al. in
  preparation}. For the submillimeter data, we used the peak flux
density per beam observed toward the source to construct the SEDs; the
FWHMs of the beams were 7.3$\arcsec$ at 350~$\micron$ and 19$\arcsec$
at 870~$\micron$.

\begin{figure}[!h]
\centering
\resizebox{0.45\textwidth}{!}{\includegraphics{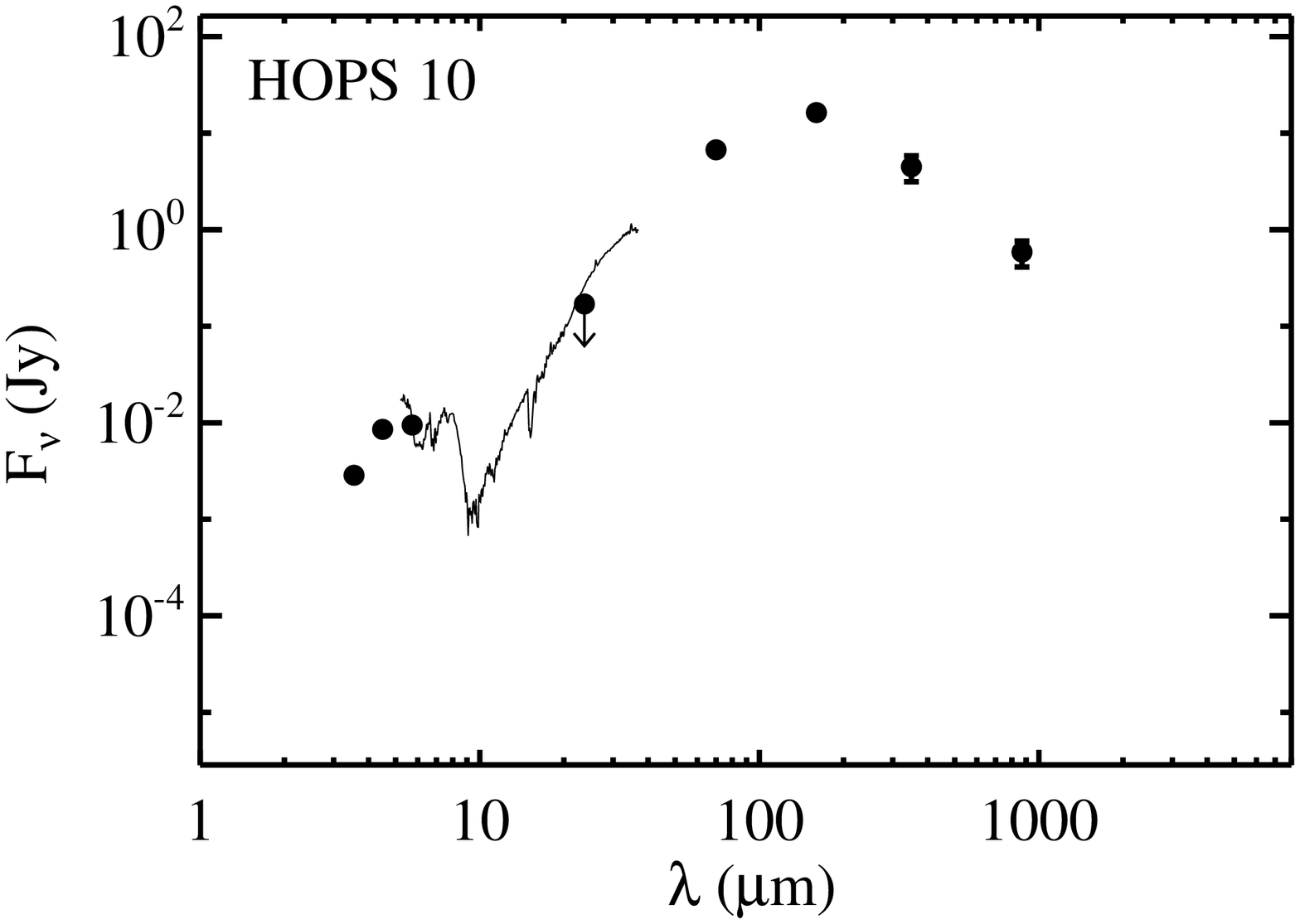}}
\resizebox{0.45\textwidth}{!}{\includegraphics{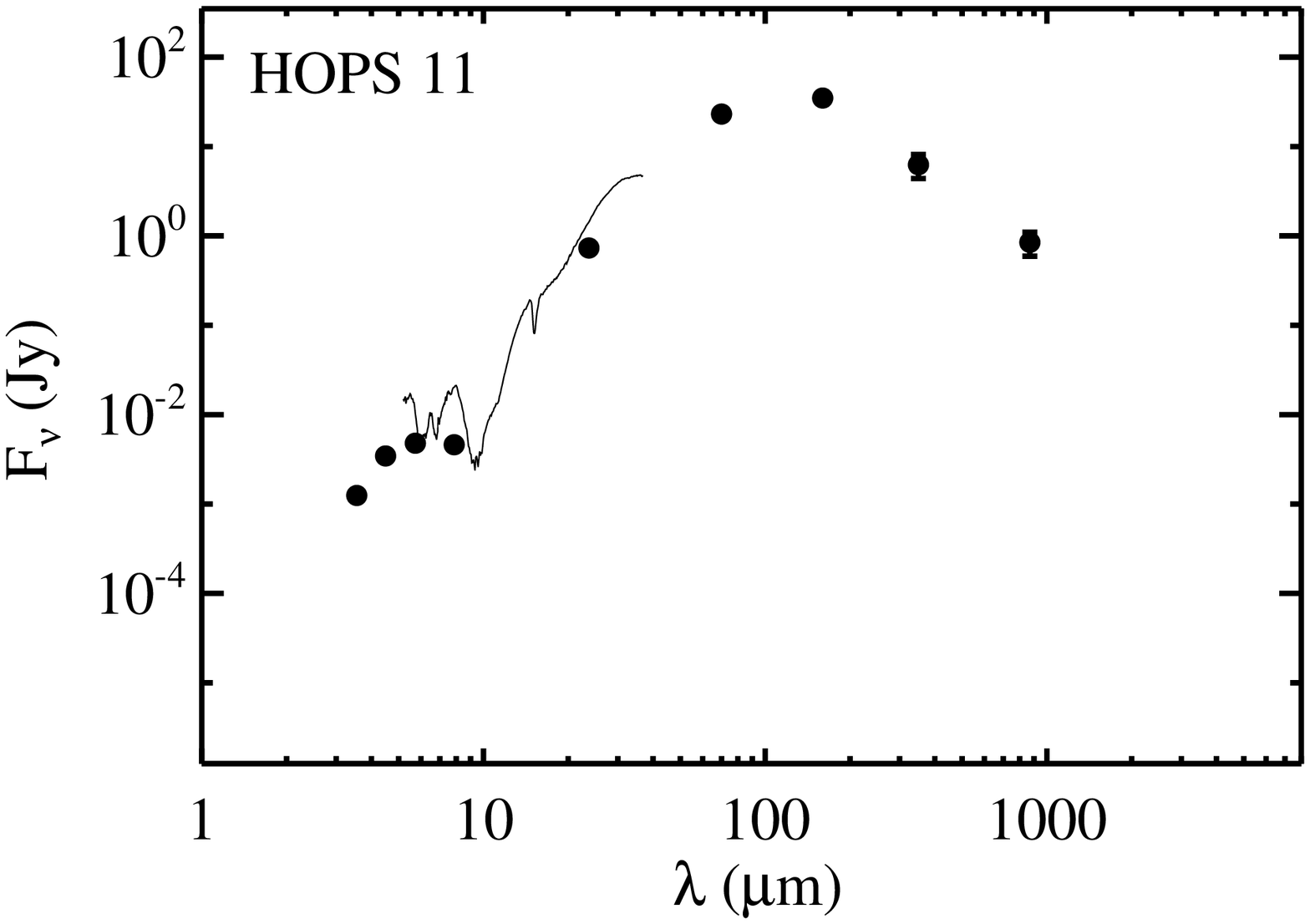}}
\resizebox{0.45\textwidth}{!}{\includegraphics{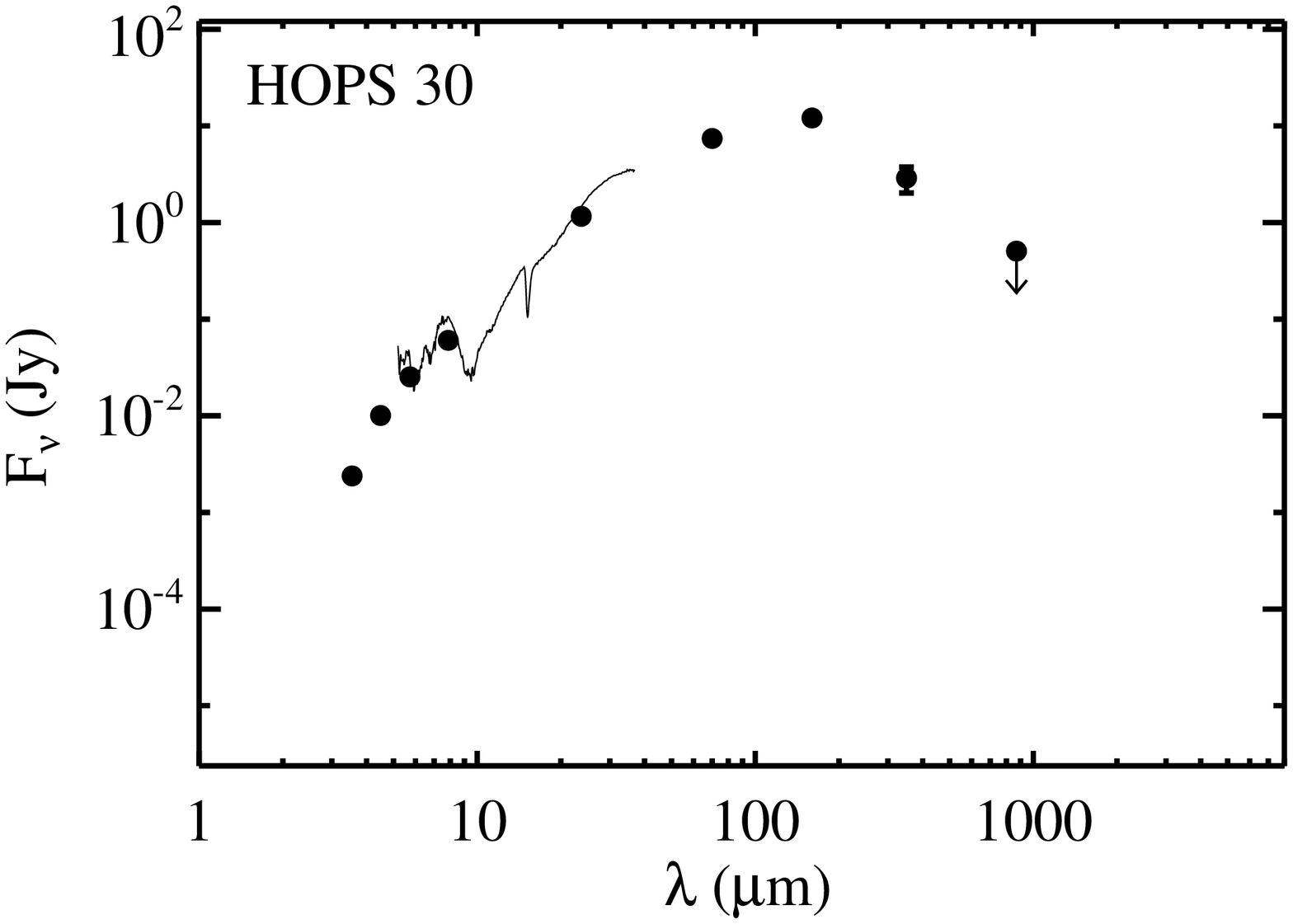}}
\resizebox{0.45\textwidth}{!}{\includegraphics{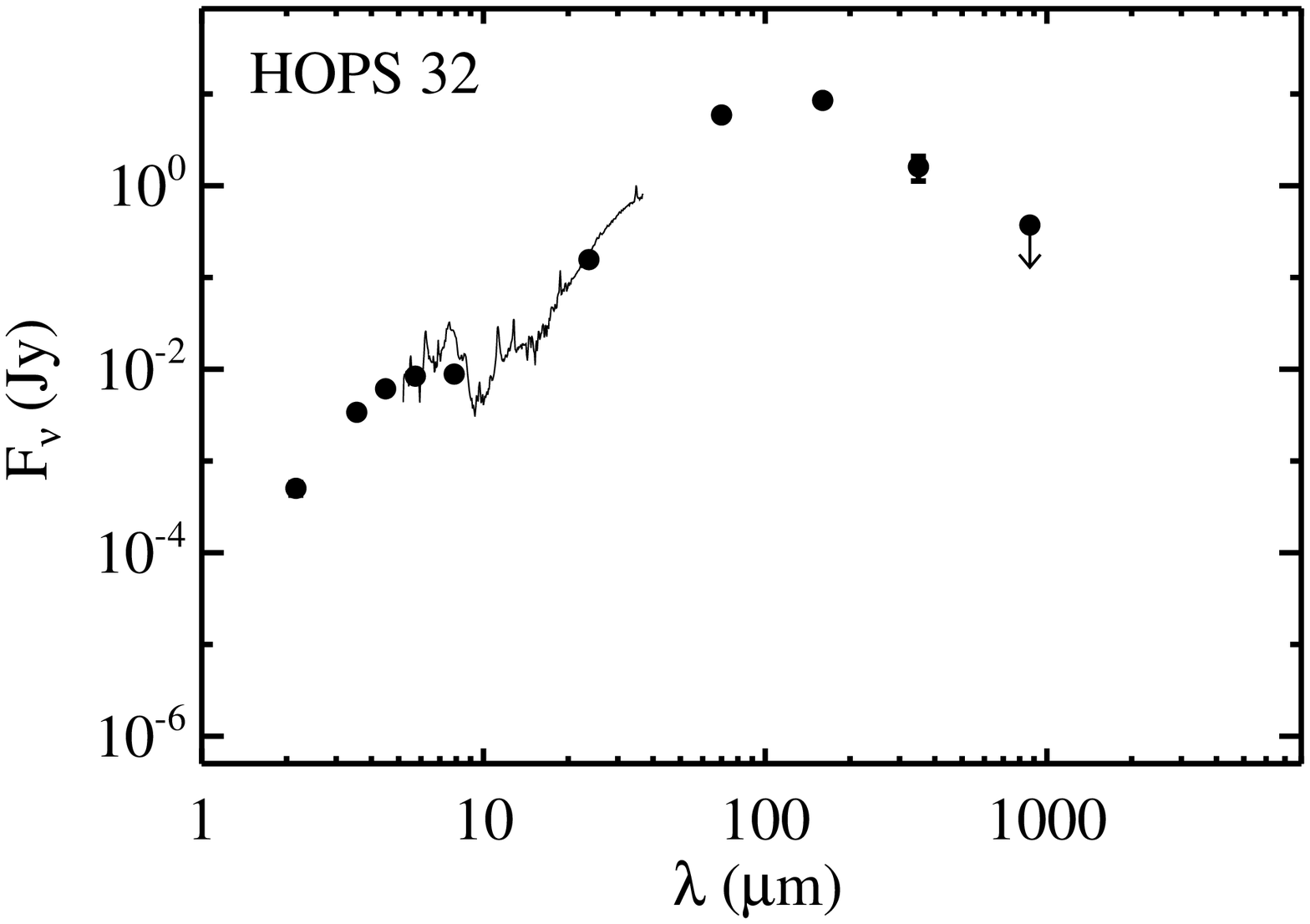}}
\caption{Observed SEDs of the protostars in our sample. The solid
  circles represent broad band photometry from 2MASS, {\it
    Spitzer}/IRAC \& MIPS, SOFIA/FORCAST, {\it Herschel}/PACS and
  APEX/SABOCA \& LABOCA. The downward arrow indicates upper
  limits. The solid line represent {\it Spitzer}/IRS low resolution
  spectra.  \label{appen_fig1}}
\end{figure}

\clearpage

\begin{figure}
\centering
\addtocounter{figure}{-1}
\resizebox{0.45\textwidth}{!}{\includegraphics{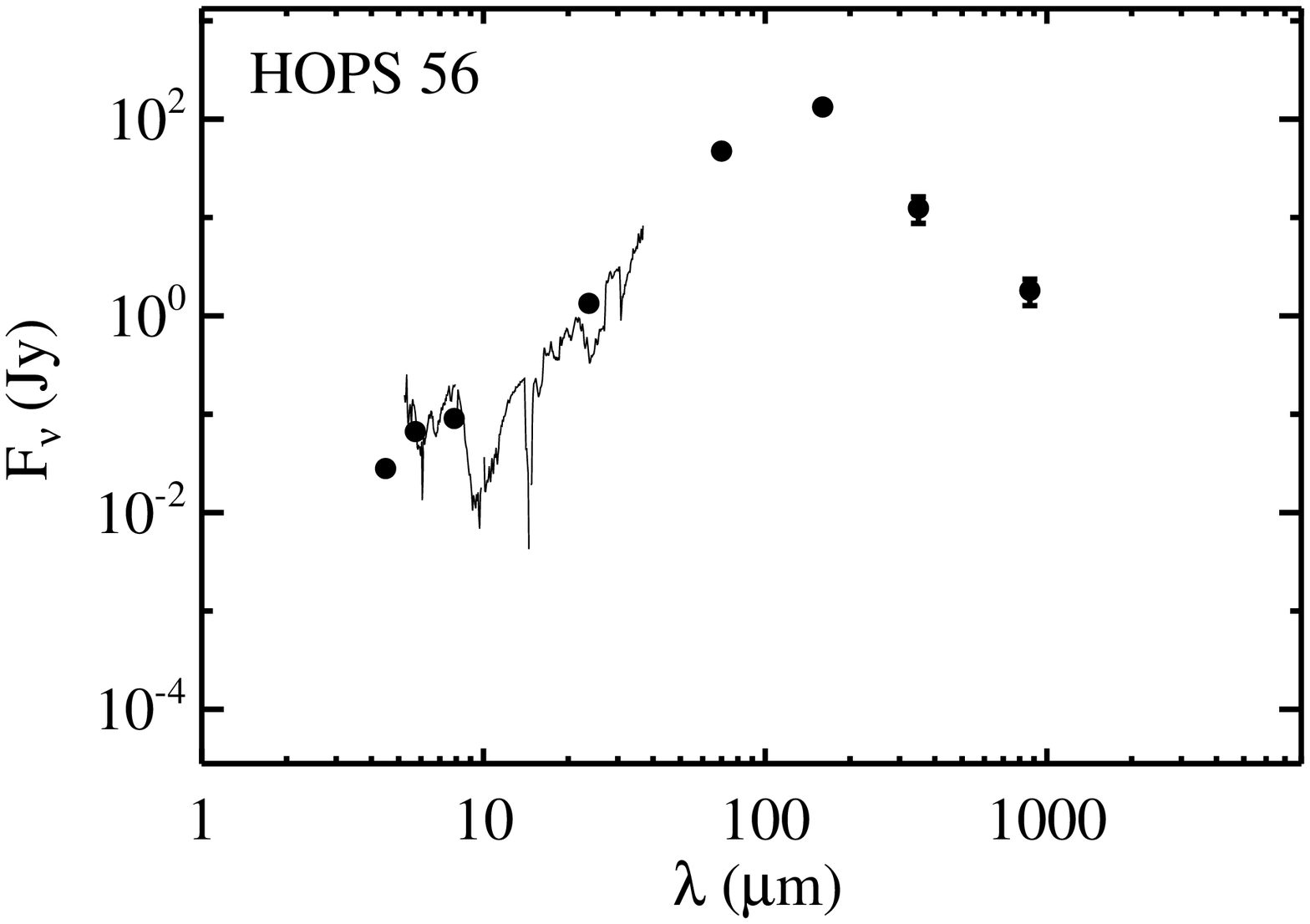}}
\resizebox{0.45\textwidth}{!}{\includegraphics{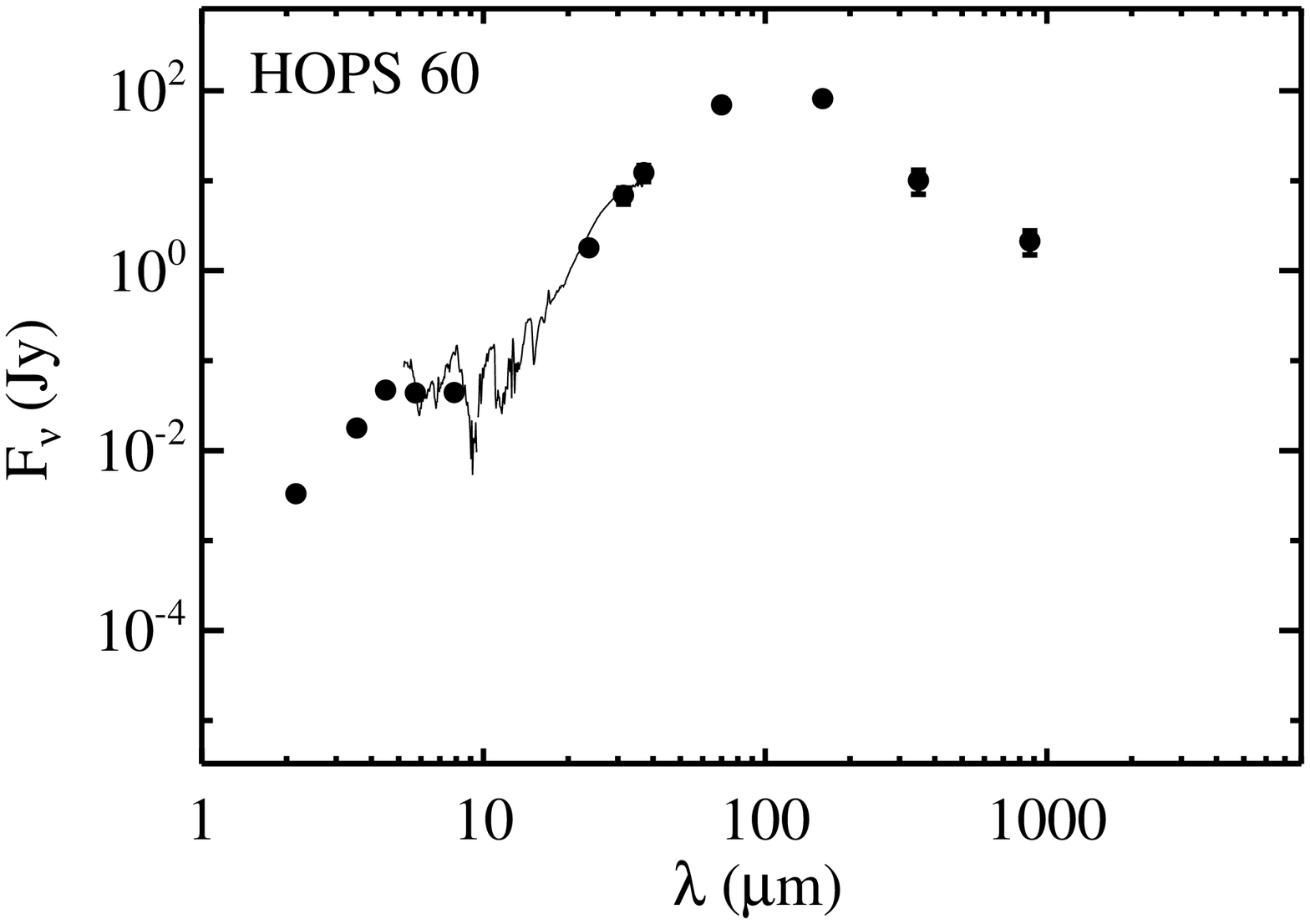}}
\resizebox{0.45\textwidth}{!}{\includegraphics{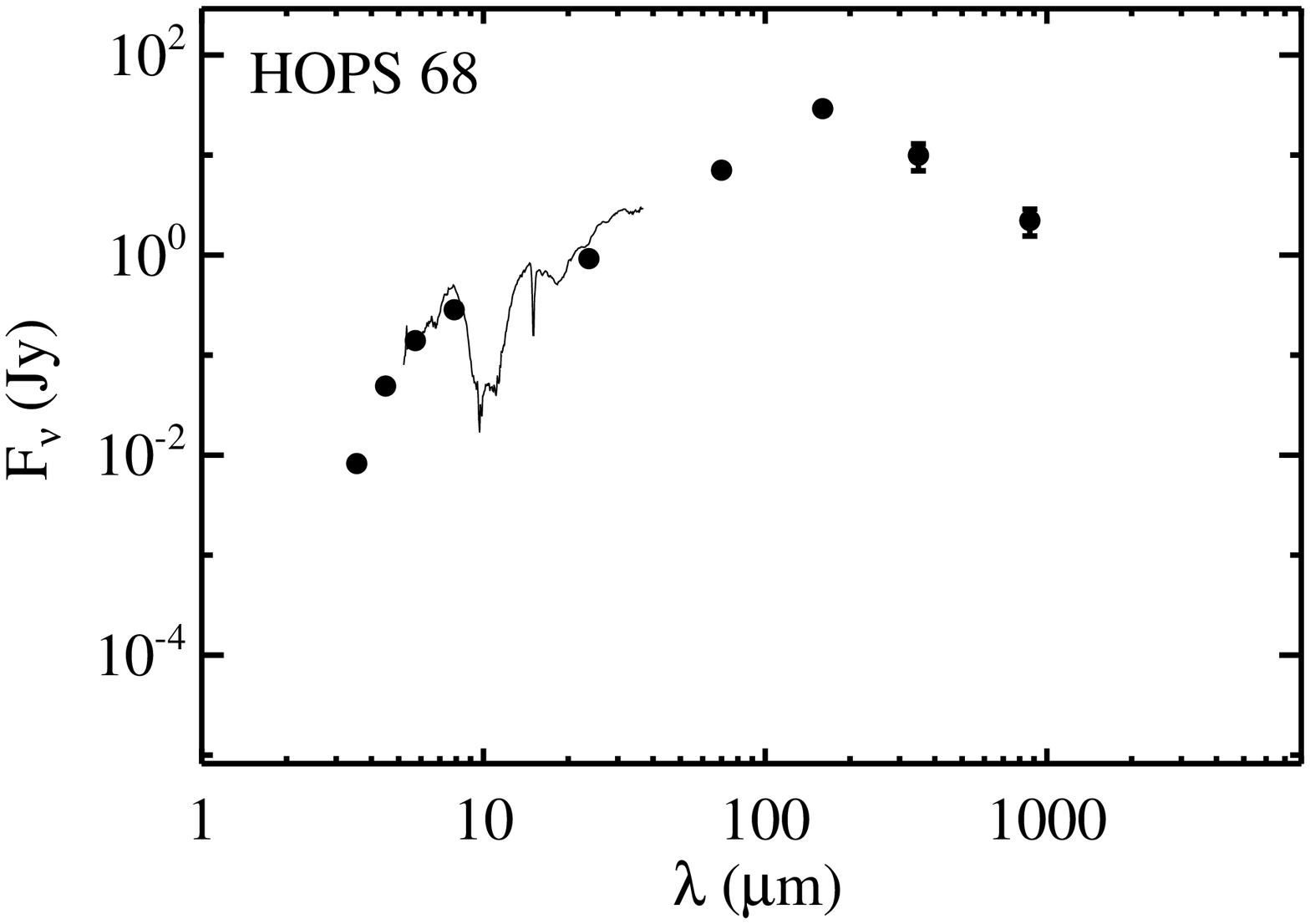}}
\resizebox{0.45\textwidth}{!}{\includegraphics{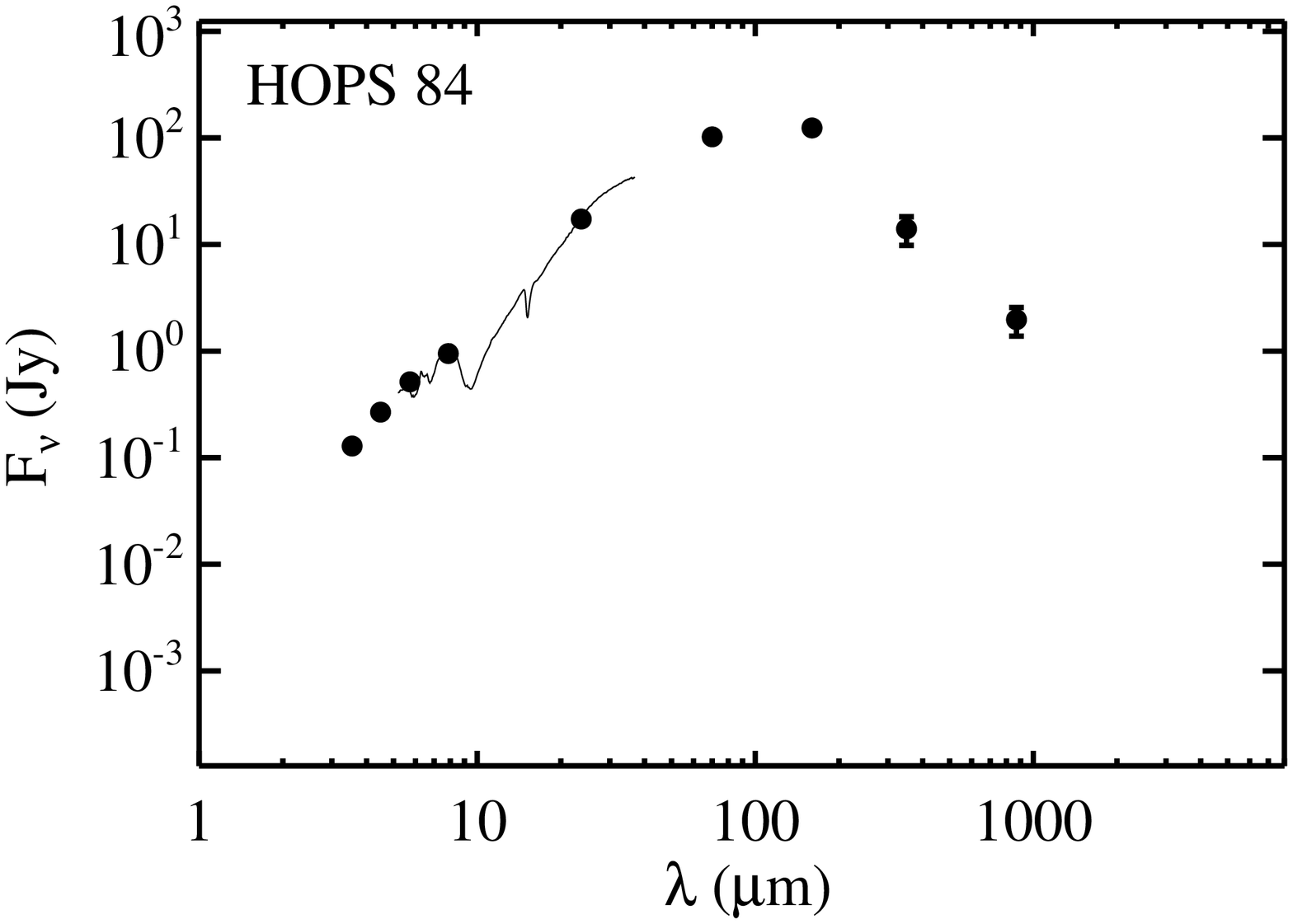}}
\resizebox{0.45\textwidth}{!}{\includegraphics{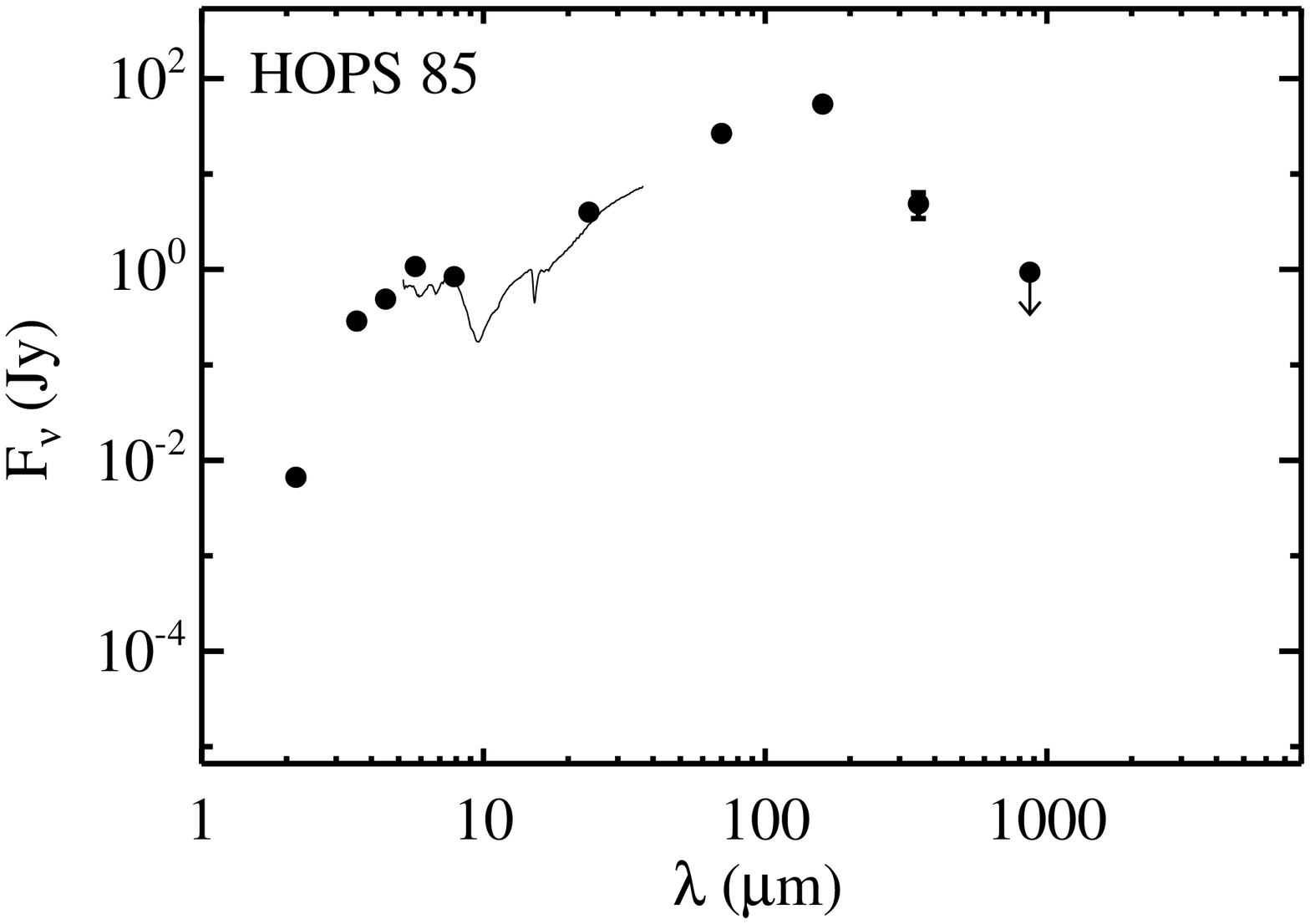}}
\resizebox{0.45\textwidth}{!}{\includegraphics{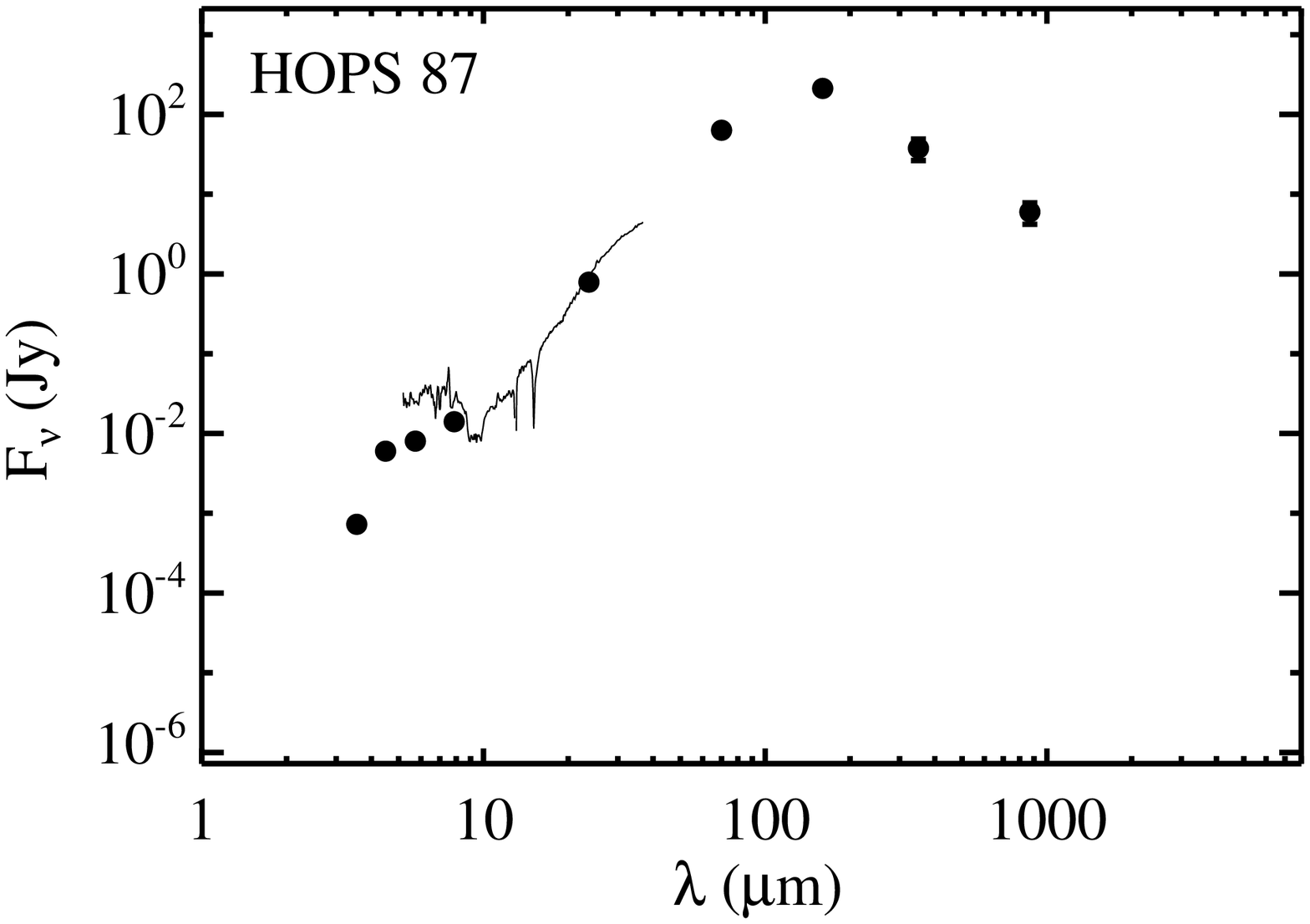}}
\caption{continued ... }
\end{figure}

\clearpage

\begin{figure}
\centering
\addtocounter{figure}{-1}
\resizebox{0.45\textwidth}{!}{\includegraphics{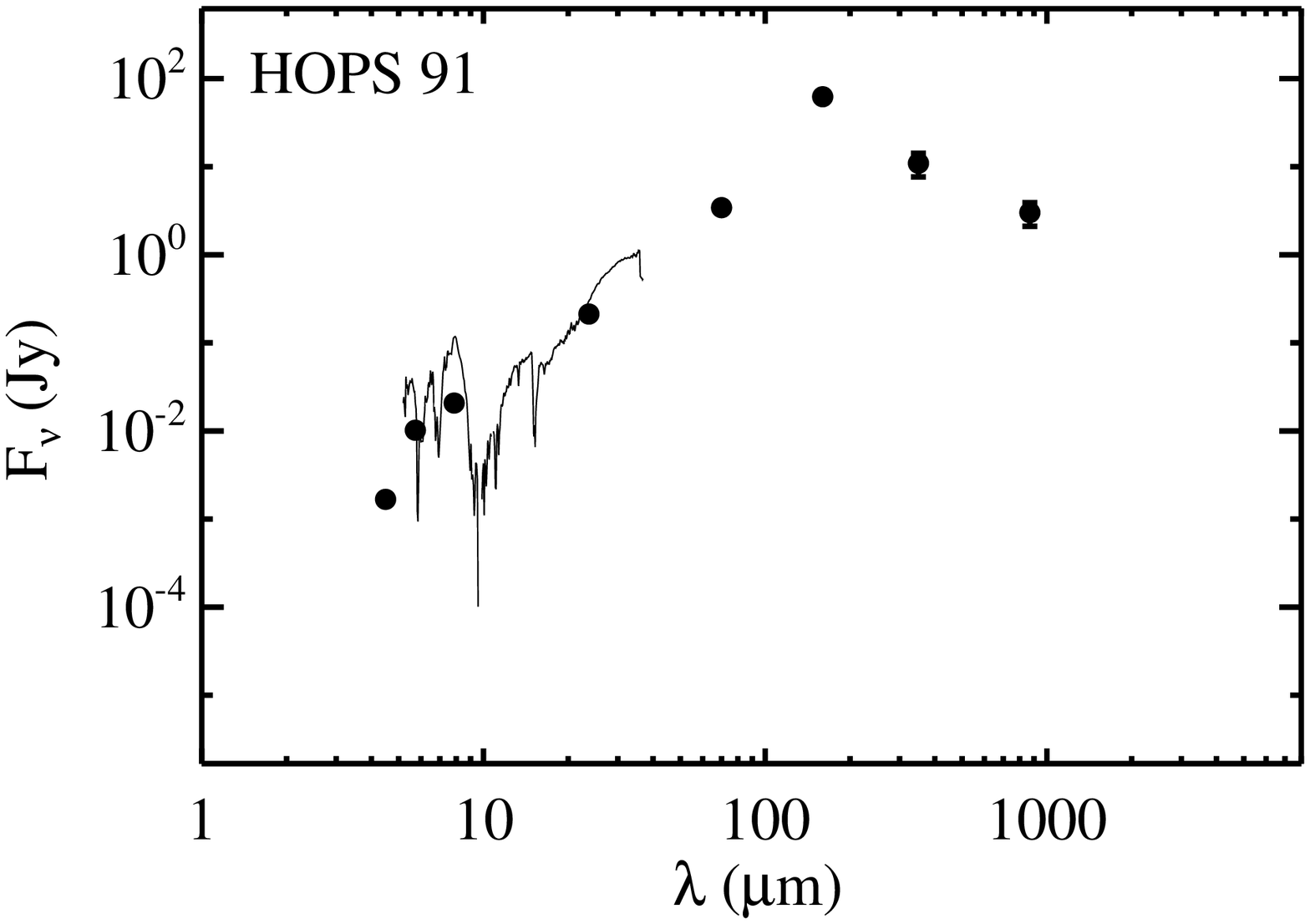}}
\resizebox{0.45\textwidth}{!}{\includegraphics{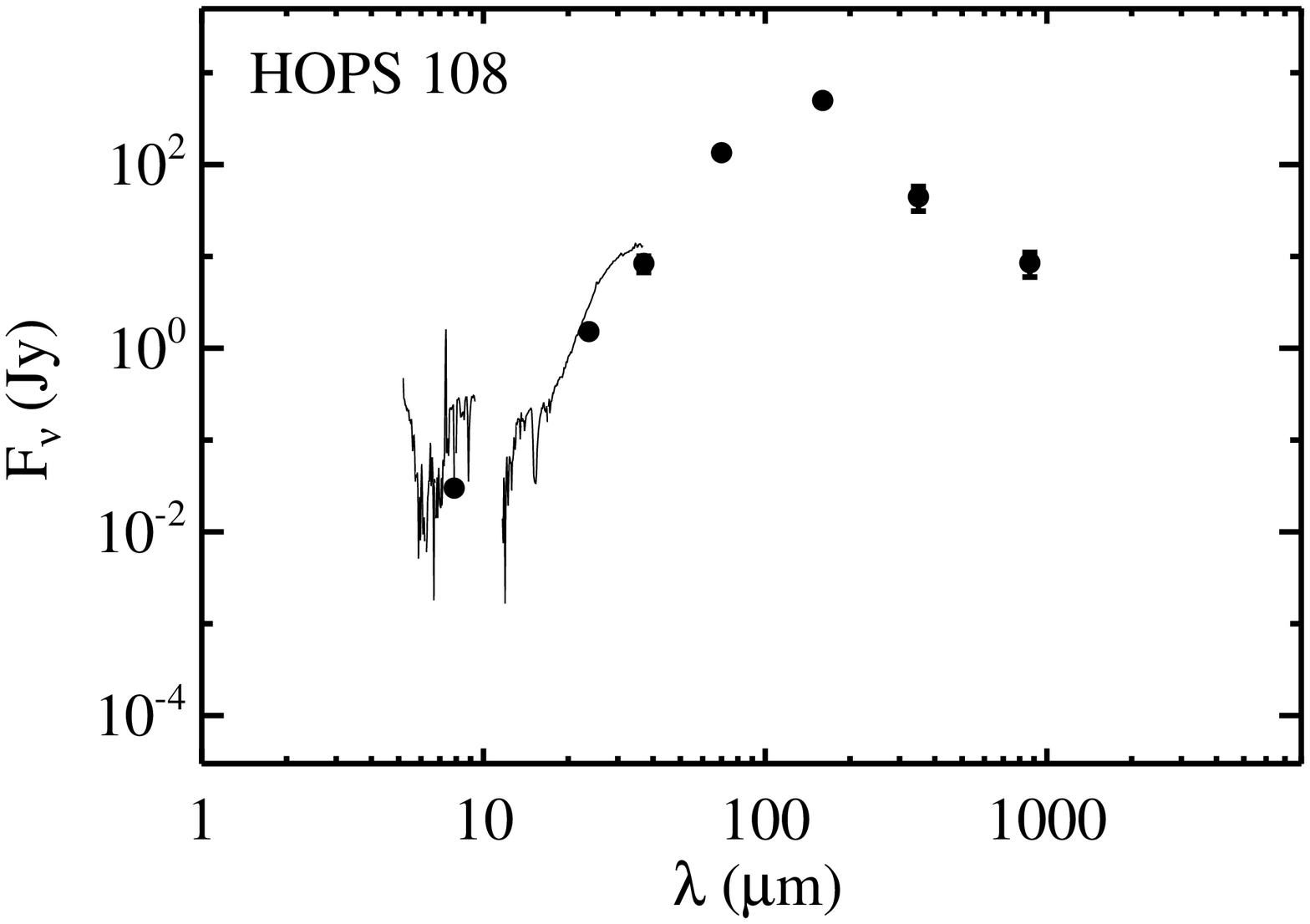}}
\resizebox{0.45\textwidth}{!}{\includegraphics{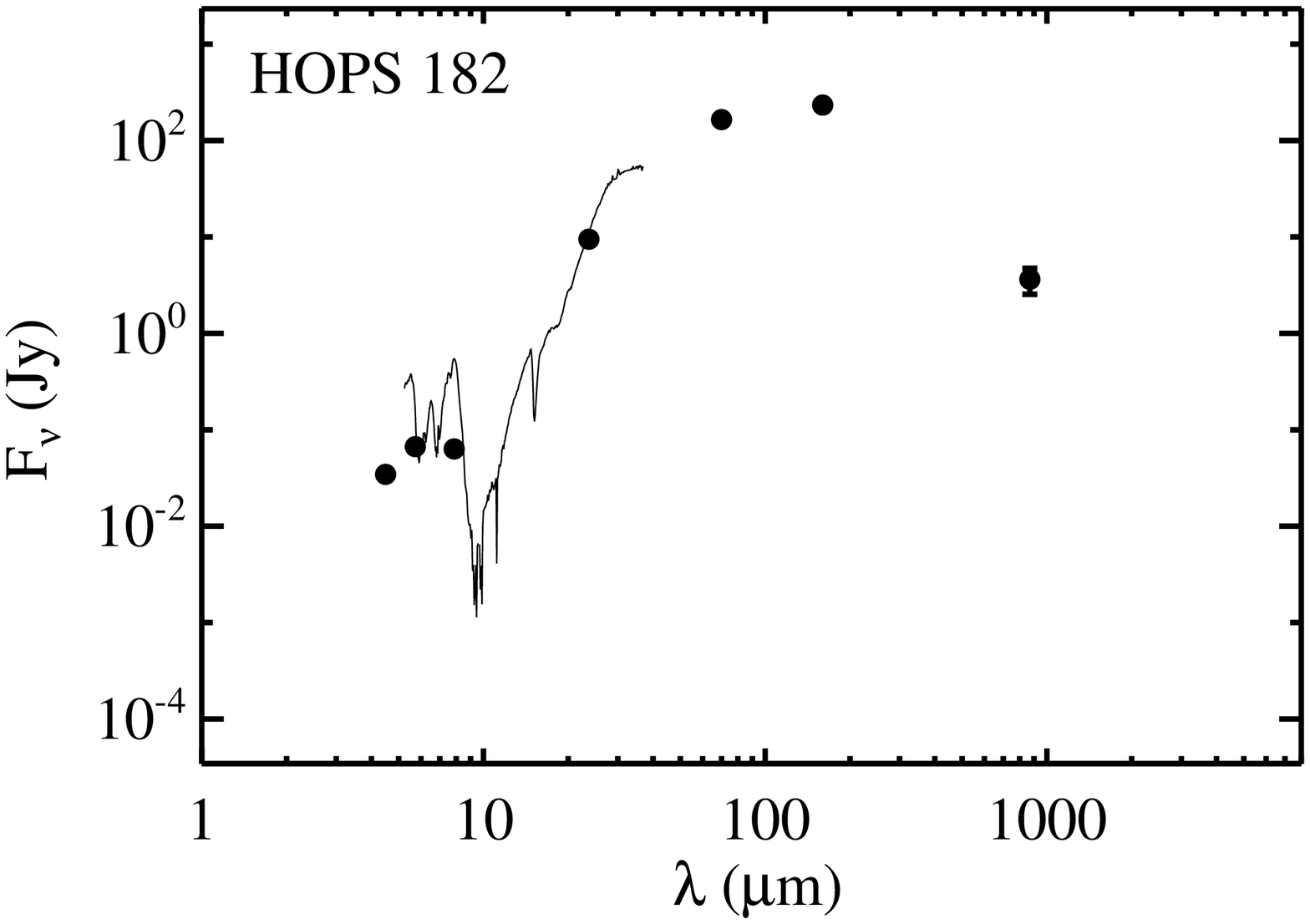}}
\resizebox{0.45\textwidth}{!}{\includegraphics{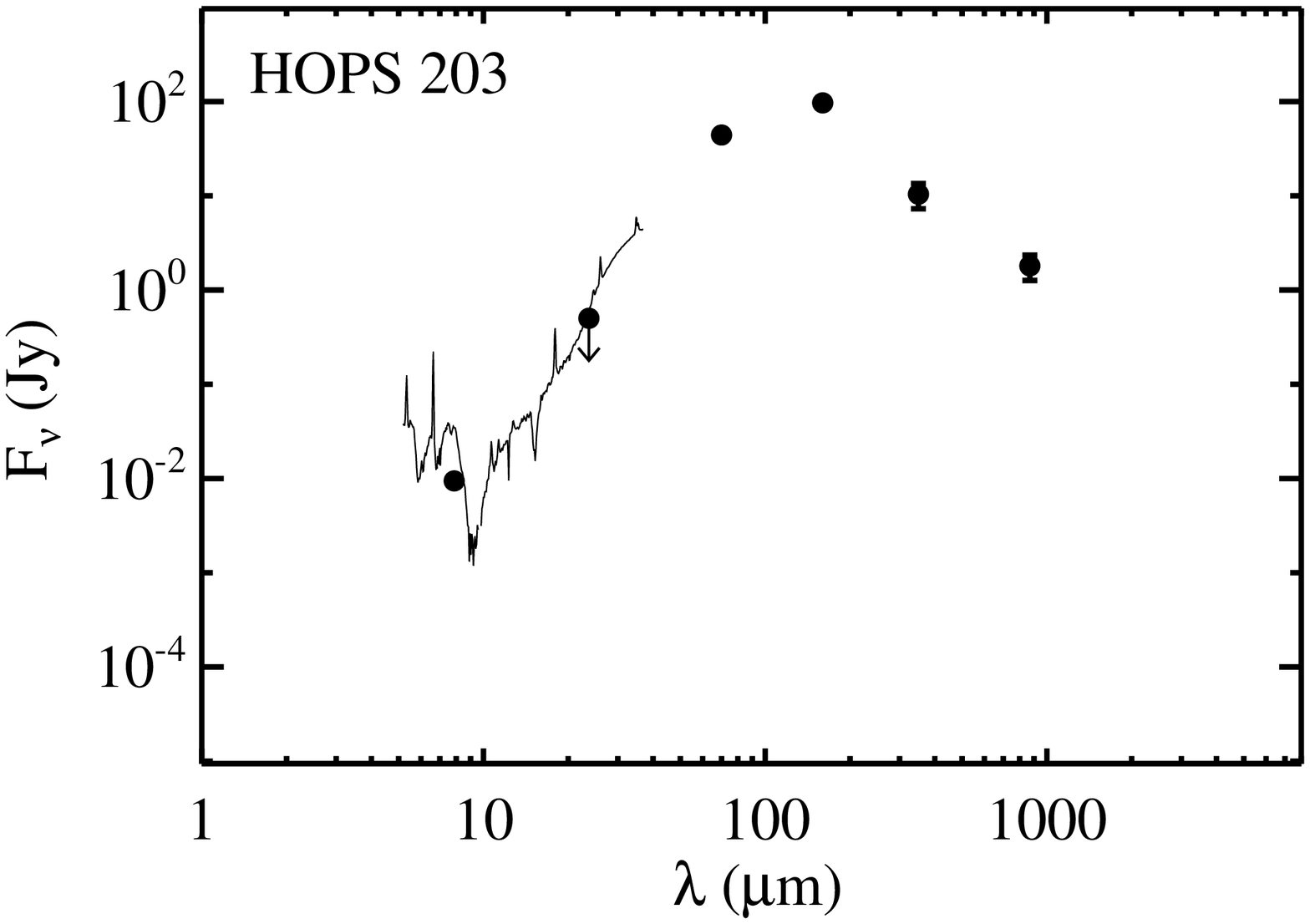}}
\resizebox{0.45\textwidth}{!}{\includegraphics{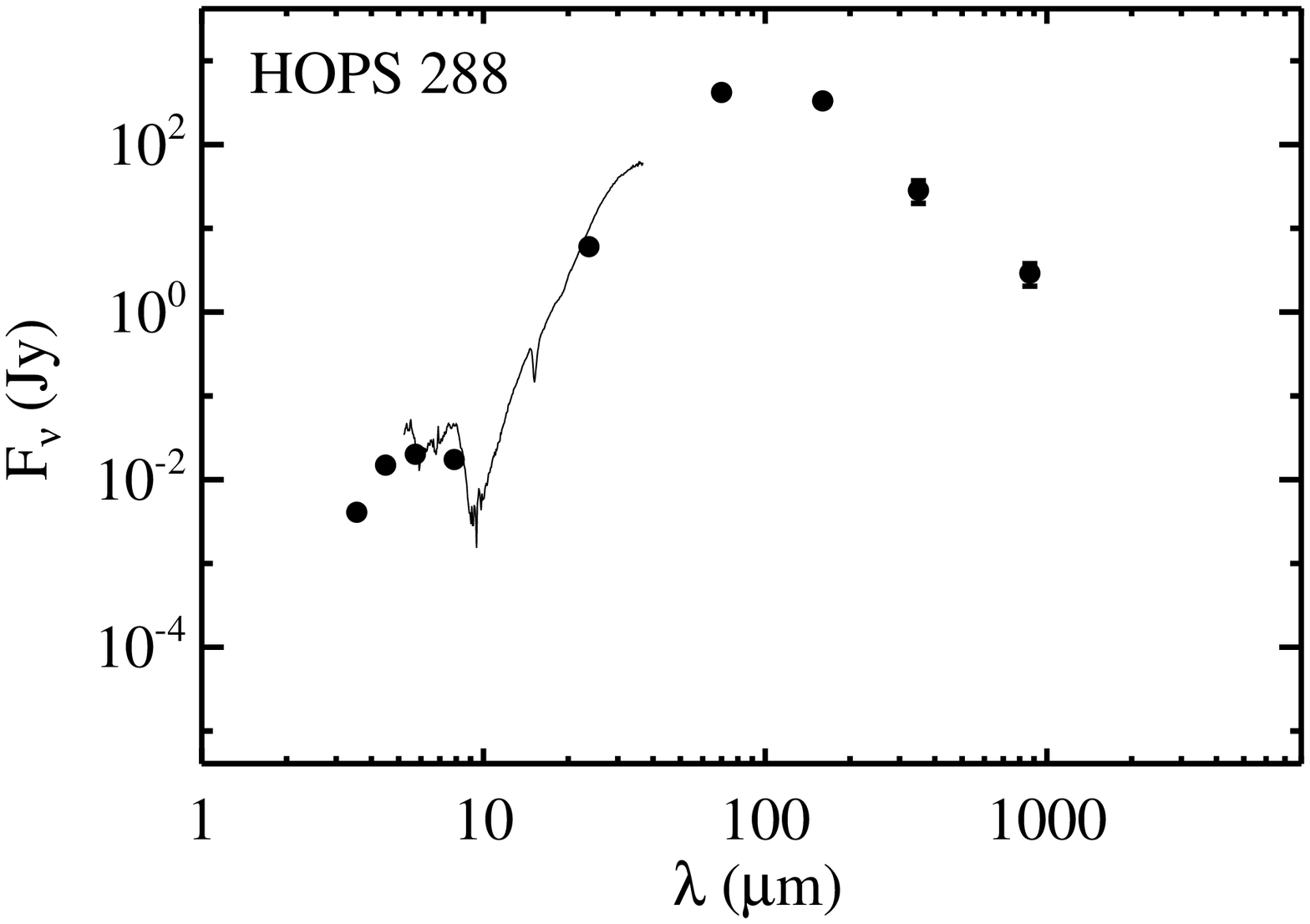}}
\resizebox{0.45\textwidth}{!}{\includegraphics{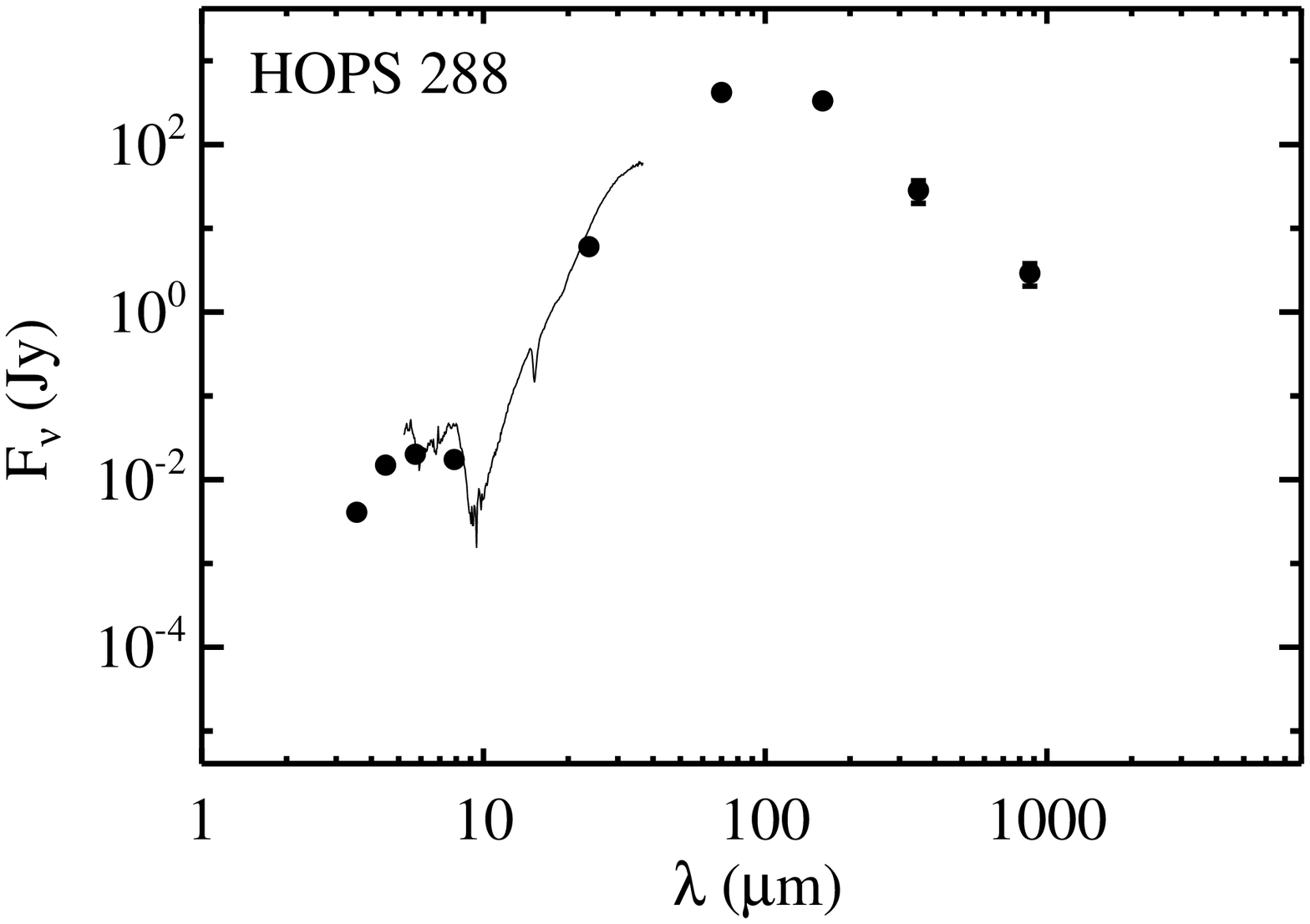}}
\caption{continued ... }
\end{figure}

\clearpage

\begin{figure}
\centering
\addtocounter{figure}{-1}
\resizebox{0.45\textwidth}{!}{\includegraphics{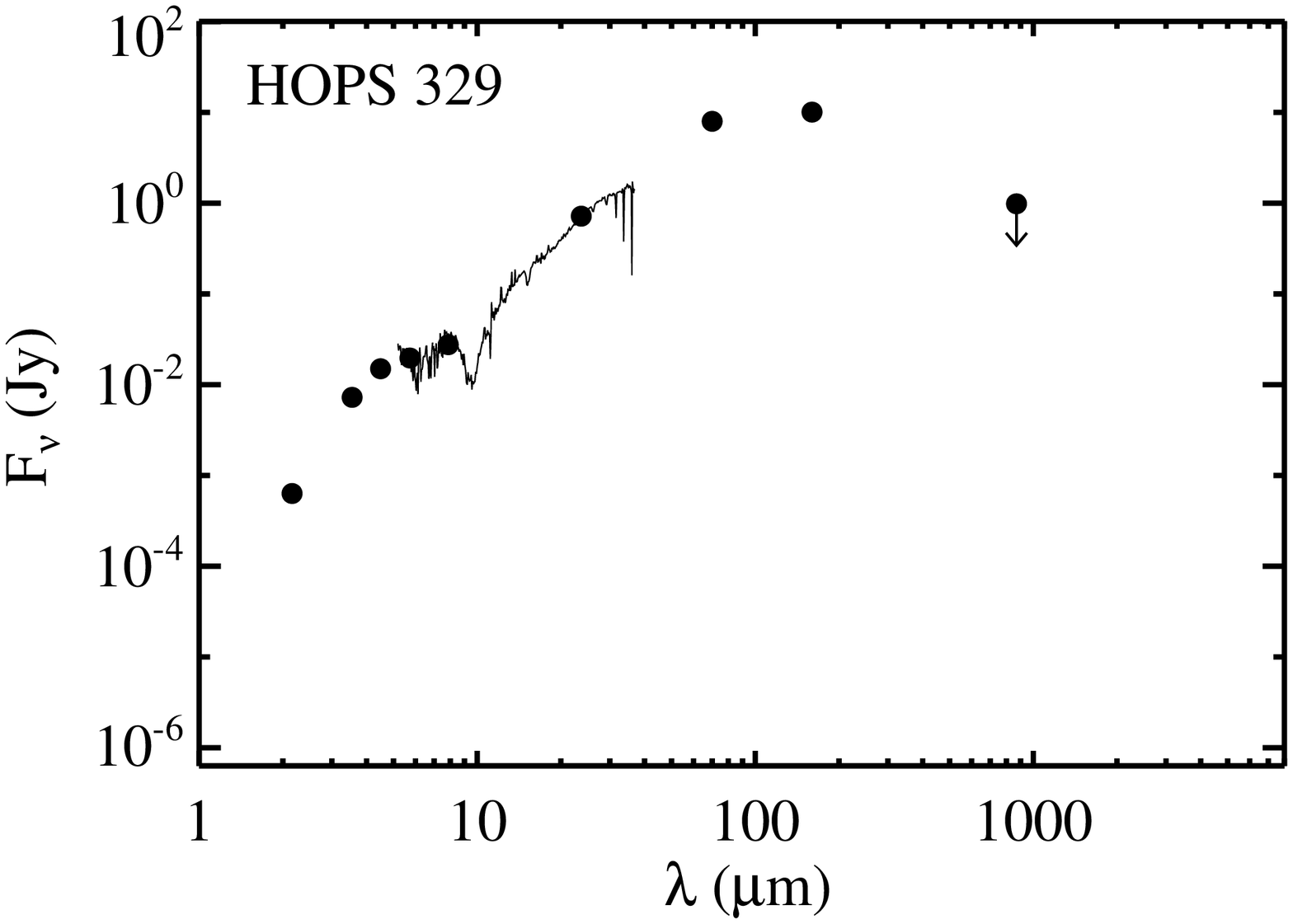}}
\resizebox{0.45\textwidth}{!}{\includegraphics{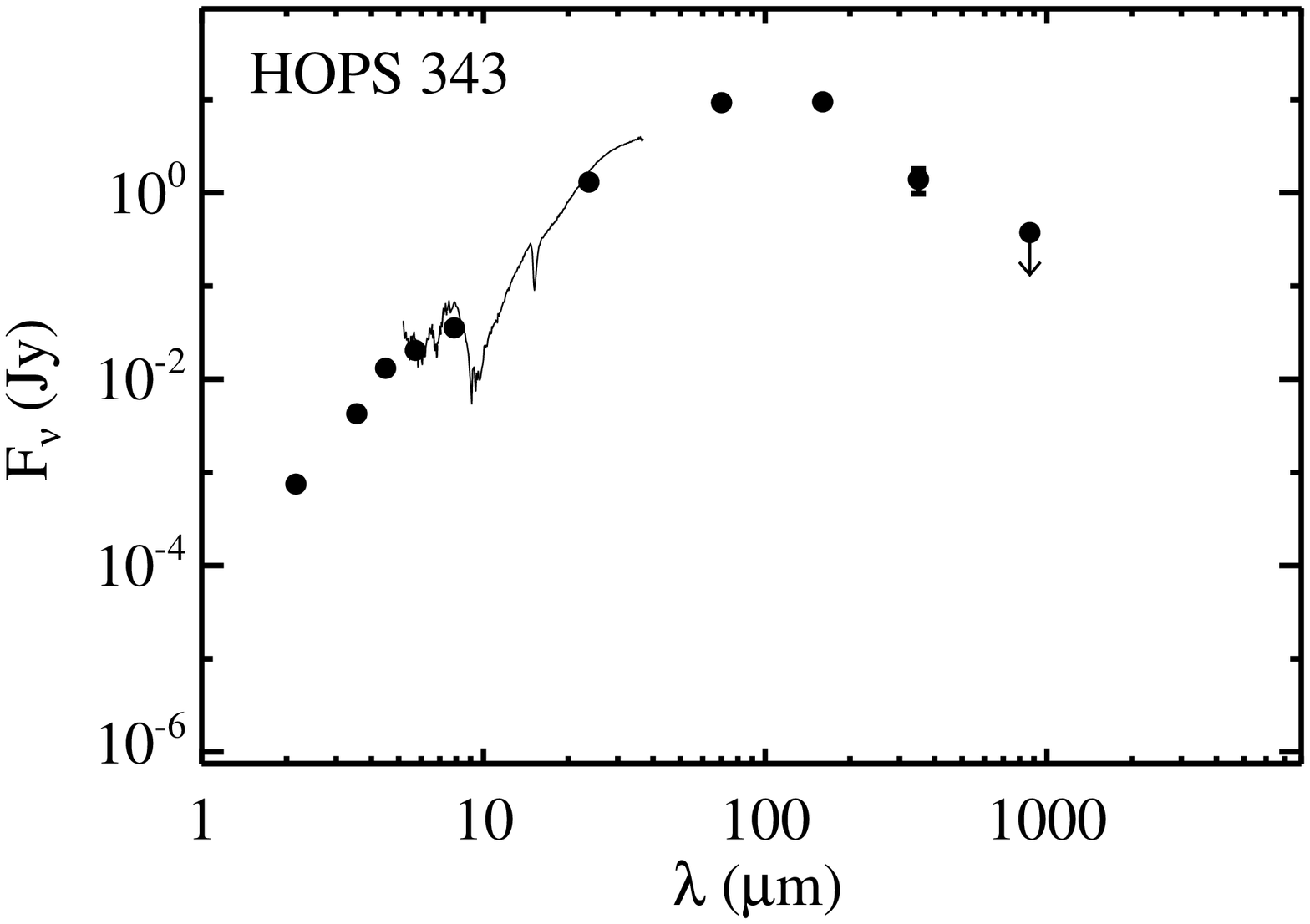}}
\resizebox{0.45\textwidth}{!}{\includegraphics{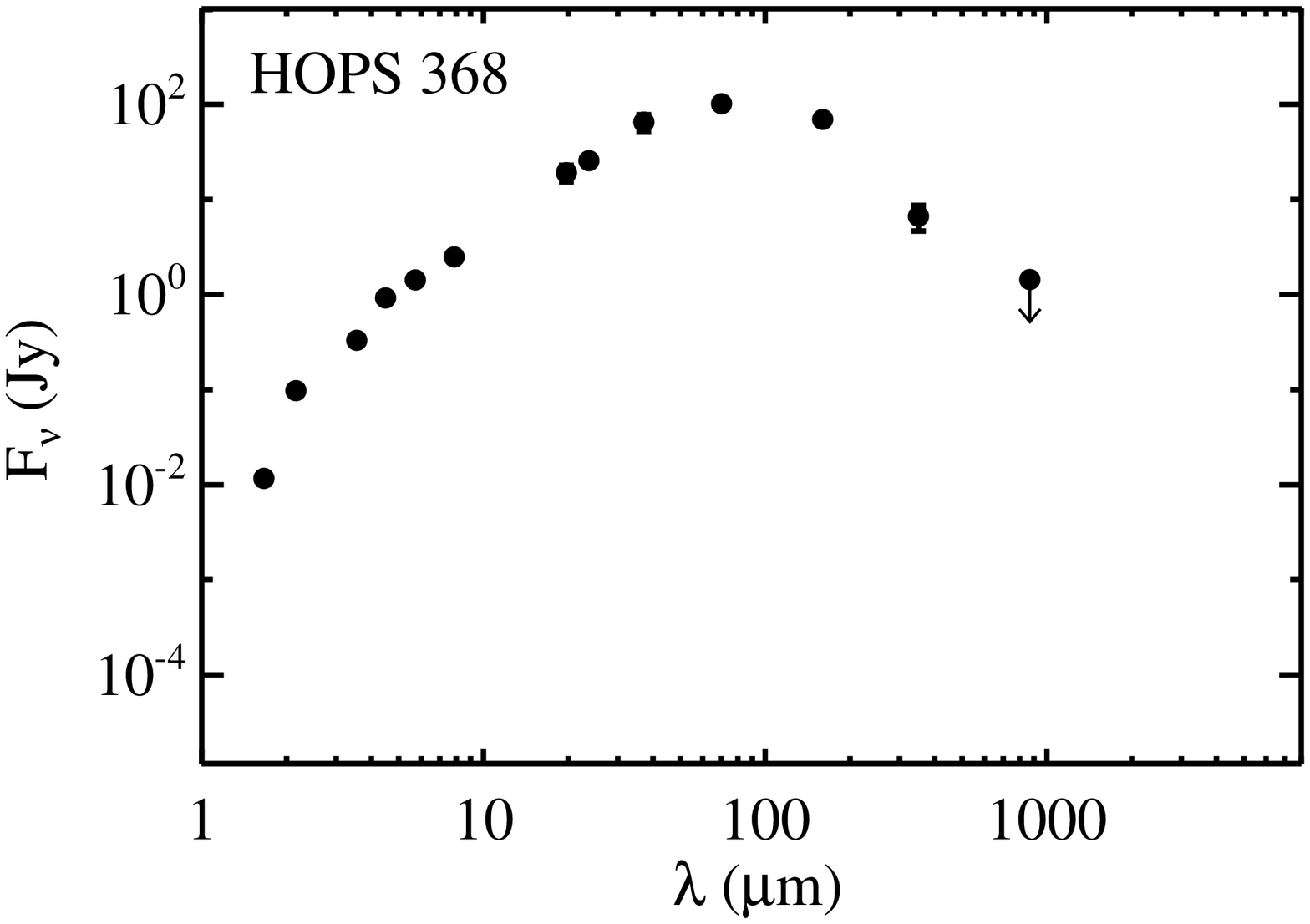}}
\resizebox{0.45\textwidth}{!}{\includegraphics{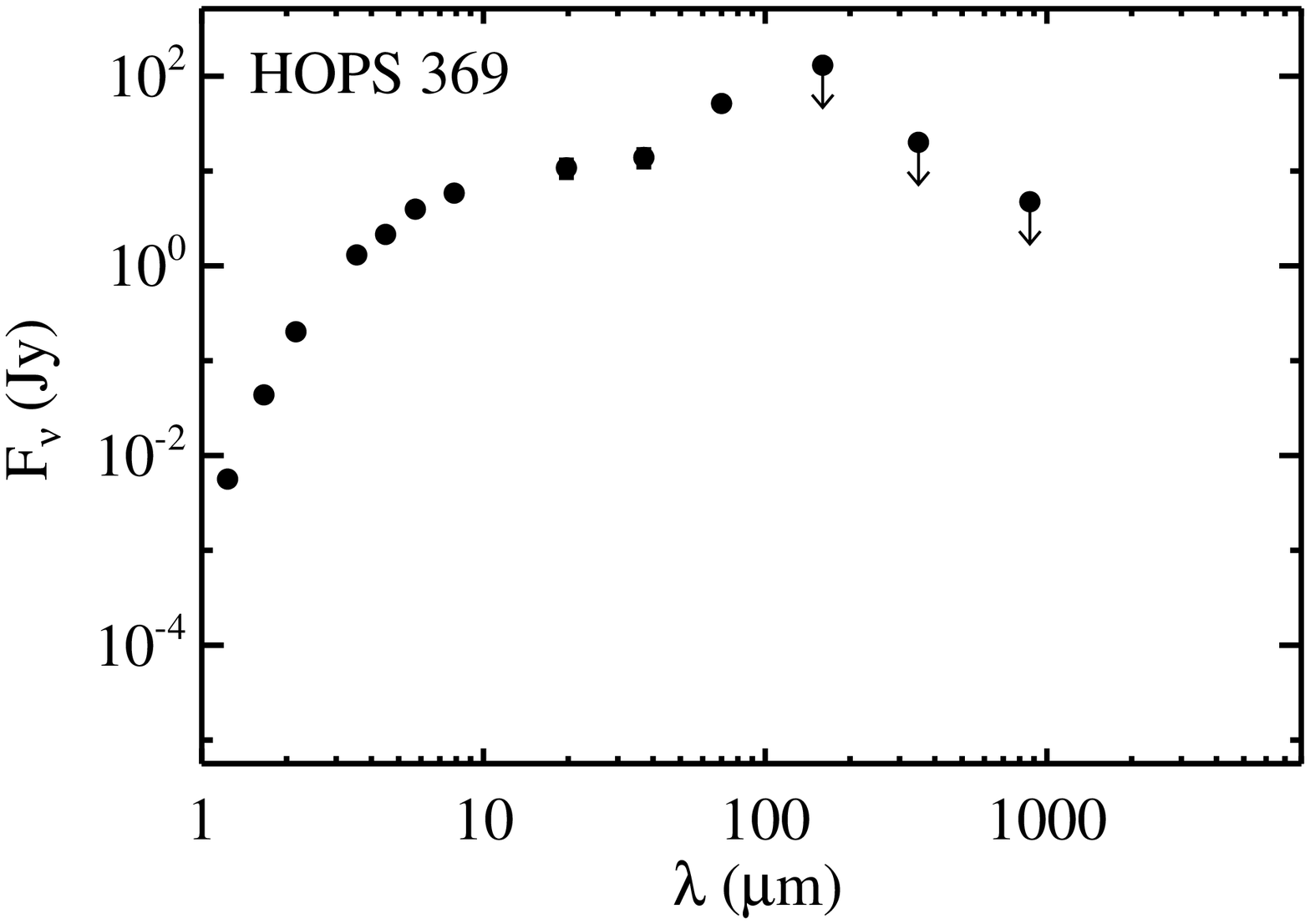}}
\resizebox{0.45\textwidth}{!}{\includegraphics{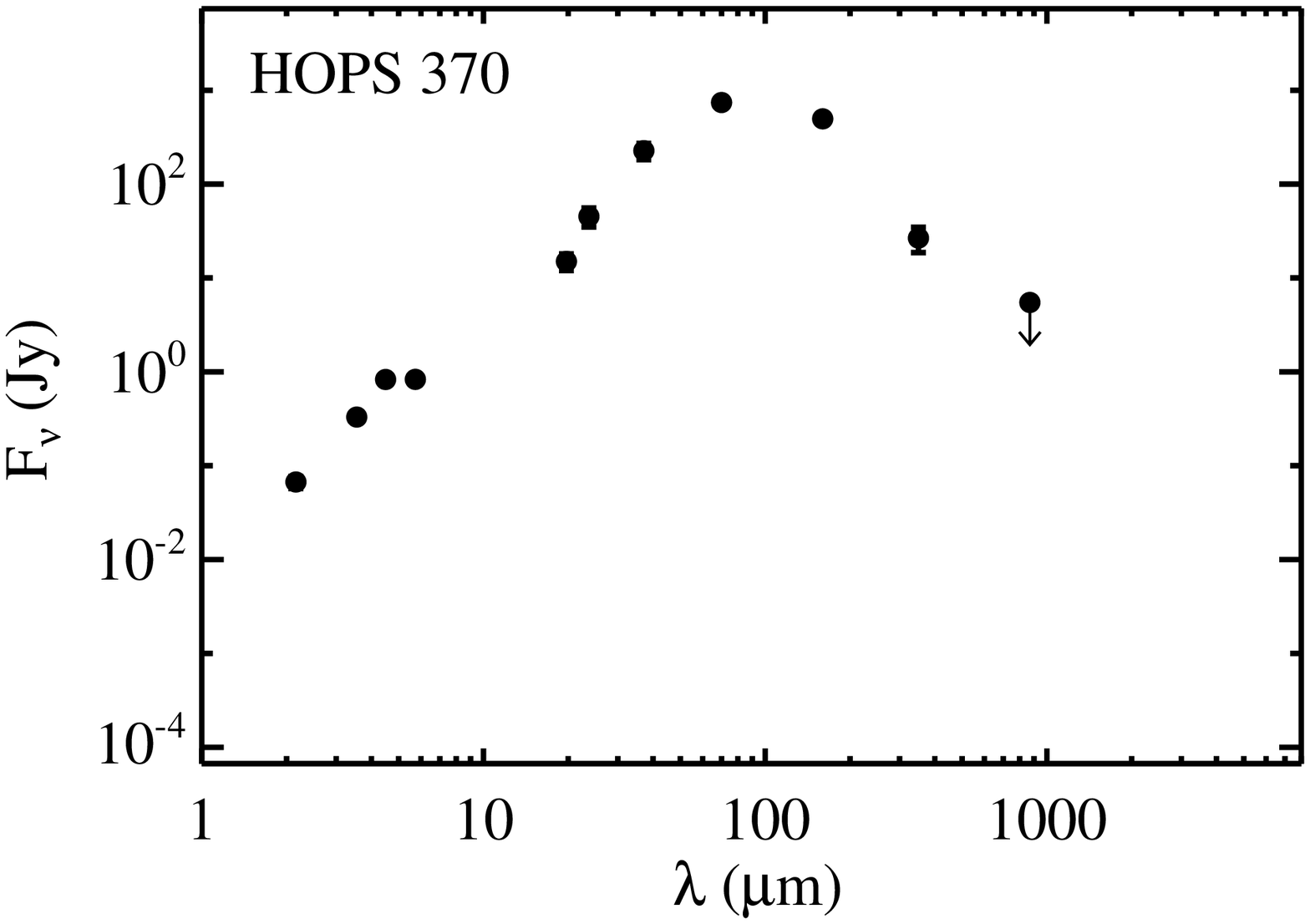}}
\caption{continued ... }
\end{figure}


\begin{thebibliography}{89}
\expandafter\ifx\csname natexlab\endcsname\relax\def\natexlab#1{#1}\fi

\bibitem[{{Adams} {et~al.}(2012){Adams}, {Herter}, {Osorio}, {Macias},
  {Megeath}, {Fischer}, {Ali}, {Calvet}, {D'Alessio}, {De Buizer}, {Gull},
  {Henderson}, {Keller}, {Morris}, {Remming}, {Schoenwald}, {Shuping},
  {Stacey}, {Stanke}, {Stutz}, \& {Vacca}}]{adams12}
{Adams}, J.~D., {Herter}, T.~L., {Osorio}, M., {Macias}, E., {Megeath}, S.~T.,
  {Fischer}, W.~J., {Ali}, B., {et~al.} 2012, \apjl, 749, L24

\bibitem[{{Ali} {et~al.}(2010){Ali}, {Tobin}, {Fischer}, {Poteet}, {Megeath},
  {Allen}, {Hartmann}, {Calvet}, {Furlan}, \& {Osorio}}]{ali10}
{Ali}, B., {Tobin}, J.~J., {Fischer}, W.~J., {Poteet}, C.~A., {Megeath}, S.~T.,
  {Allen}, L., {Hartmann}, L., {et~al.} 2010, \aap, 518, L119

\bibitem[{{Andr\'{e}} \& {Montmerle}(1994)}]{andremont94}
{Andr\'{e}}, P. \& {Montmerle}, T. 1994, \apj, 420, 837

\bibitem[{{Andr\'{e}} {et~al.}(1993){Andr\'{e}}, {Ward-Thompson}, \&
  {Barsony}}]{andre93}
{Andr\'{e}}, P., {Ward-Thompson}, D., \& {Barsony}, M. 1993, \apj, 406, 122

\bibitem[{{Bachiller}(1996)}]{bach96}
{Bachiller}, R. 1996, \araa, 34, 111

\bibitem[{{Bachiller} \& {Tafalla}(1999)}]{bt99}
{Bachiller}, R. \& {Tafalla}, M. 1999, in NATO ASIC Proc. 540: The Origin of
  Stars and Planetary Systems, 227

\bibitem[{{Benedettini} {et~al.}(2000){Benedettini}, {Giannini}, {Nisini},
  {Tommasi}, {Lorenzetti}, {Di Giorgio}, {Saraceno}, {Smith}, \&
  {White}}]{benedettini00}
{Benedettini}, M., {Giannini}, T., {Nisini}, B., {Tommasi}, E., {Lorenzetti},
  D., {Di Giorgio}, A.~M., {Saraceno}, P., {Smith}, H.~A., \& {White}, G.~J.
  2000, \aap, 359, 148

\bibitem[{{Bontemps} {et~al.}(1996){Bontemps}, {Andr\'{e}}, {Terebey}, \&
  {Cabrit}}]{bontemps96}
{Bontemps}, S., {Andr\'{e}}, P., {Terebey}, S., \& {Cabrit}, S. 1996, \aap,
  311, 858

\bibitem[{{Cassen} \& {Moosman}(1981)}]{cm81}
{Cassen}, P. \& {Moosman}, A. 1981, Icarus, 48, 353

\bibitem[{{Ceccarelli}(2000)}]{ceccarelli00}
{Ceccarelli}, C. 2000, in ESA Special Publication, Vol. 456, ISO Beyond the
  Peaks: The 2nd ISO Workshop on Analytical Spectroscopy, ed. {A.~Salama,
  M.~F.~Kessler, K.~Leech, \& B.~Schulz}, 141

\bibitem[{{Ceccarelli} {et~al.}(1998){Ceccarelli}, {Caux}, {White}, {Molinari},
  {Furniss}, {Liseau}, {Nisini}, {Saraceno}, {Spinoglio}, \&
  {Wolfire}}]{ceccarelli98}
{Ceccarelli}, C., {Caux}, E., {White}, G.~J., {Molinari}, S., {Furniss}, I.,
  {Liseau}, R., {Nisini}, B., {et~al.} 1998, \aap, 331, 372

\bibitem[{{Chen} {et~al.}(1995){Chen}, {Myers}, {Ladd}, \& {Wood}}]{chen95}
{Chen}, H., {Myers}, P.~C., {Ladd}, E.~F., \& {Wood}, D.~O.~S. 1995, \apj, 445,
  377

\bibitem[{{Chini} {et~al.}(1997){Chini}, {Reipurth}, {Ward-Thompson}, {Bally},
  {Nyman}, {Sievers}, \& {Billawala}}]{chini97}
{Chini}, R., {Reipurth}, B., {Ward-Thompson}, D., {Bally}, J., {Nyman}, L.-A.,
  {Sievers}, A., \& {Billawala}, Y. 1997, \apjl, 474, L135+

\bibitem[{{Draine} {et~al.}(1983){Draine}, {Roberge}, \& {Dalgarno}}]{drd83}
{Draine}, B.~T., {Roberge}, W.~G., \& {Dalgarno}, A. 1983, \apj, 264, 485

\bibitem[{{Enoch} {et~al.}(2009){Enoch}, {Evans}, {Sargent}, \&
  {Glenn}}]{enoch09}
{Enoch}, M.~L., {Evans}, N.~J., {Sargent}, A.~I., \& {Glenn}, J. 2009, \apj,
  692, 973

\bibitem[{{Evans} {et~al.}(2009){Evans}, {Dunham}, {J{\o}rgensen}, {Enoch},
  {Mer{\'{\i}}n}, {van Dishoeck}, {Alcal{\'a}}, {Myers}, {Stapelfeldt},
  {Huard}, {Allen}, {Harvey}, {van Kempen}, {Blake}, {Koerner}, {Mundy},
  {Padgett}, \& {Sargent}}]{evans09}
{Evans}, N.~J., {Dunham}, M.~M., {J{\o}rgensen}, J.~K., {Enoch}, M.~L.,
  {Mer{\'{\i}}n}, B., {van Dishoeck}, E.~F., {Alcal{\'a}}, J.~M., {et~al.}
  2009, \apjs, 181, 321

\bibitem[{{Fazio} {et~al.}(2004){Fazio}, {Hora}, {Allen}, {Ashby}, {Barmby},
  {Deutsch}, {Huang}, {Kleiner}, {Marengo}, {Megeath}, {Melnick}, {Pahre},
  {Patten}, {Polizotti}, {Smith}, {Taylor}, {Wang}, {Willner}, {Hoffmann},
  {Pipher}, {Forrest}, {McMurty}, {McCreight}, {McKelvey}, {McMurray}, {Koch},
  {Moseley}, {Arendt}, {Mentzell}, {Marx}, {Losch}, {Mayman}, {Eichhorn},
  {Krebs}, {Jhabvala}, {Gezari}, {Fixsen}, {Flores}, {Shakoorzadeh}, {Jungo},
  {Hakun}, {Workman}, {Karpati}, {Kichak}, {Whitley}, {Mann}, {Tollestrup},
  {Eisenhardt}, {Stern}, {Gorjian}, {Bhattacharya}, {Carey}, {Nelson},
  {Glaccum}, {Lacy}, {Lowrance}, {Laine}, {Reach}, {Stauffer}, {Surace},
  {Wilson}, {Wright}, {Hoffman}, {Domingo}, \& {Cohen}}]{fazio04}
{Fazio}, G.~G., {Hora}, J.~L., {Allen}, L.~E., {Ashby}, M.~L.~N., {Barmby}, P.,
  {Deutsch}, L.~K., {Huang}, J.-S., {et~al.} 2004, \apjs, 154, 10

\bibitem[{{Feigelson} \& {Nelson}(1985)}]{feig85}
{Feigelson}, E.~D. \& {Nelson}, P.~I. 1985, \apj, 293, 192

\bibitem[{{Fich} {et~al.}(2010){Fich}, {Johnstone}, {van Kempen}, {McCoey},
  {Fuente}, {Caselli}, {Kristensen}, {Plume}, {Cernicharo}, {Herczeg}, {van
  Dishoeck}, {Wampfler}, {Gaufre}, {Gill}, {Javadi}, {Justen}, {Laauwen},
  {Luinge}, {Ossenkopf}, {Pearson}, {Bachiller}, {Baudry}, {Benedettini},
  {Bergin}, {Benz}, {Bjerkeli}, {Blake}, {Bontemps}, {Braine}, {Bruderer},
  {Codella}, {Daniel}, {di Giorgio}, {Dominik}, {Doty}, {Encrenaz}, {Giannini},
  {Goicoechea}, {de Graauw}, {Helmich}, {Herpin}, {Hogerheijde}, {Jacq},
  {J{\o}rgensen}, {Larsson}, {Lis}, {Liseau}, {Marseille}, {Melnick}, {Nisini},
  {Olberg}, {Parise}, {Risacher}, {Santiago}, {Saraceno}, {Shipman}, {Tafalla},
  {van der Tak}, {Visser}, {Wyrowski}, \& {Y{\i}ld{\i}z}}]{fich10}
{Fich}, M., {Johnstone}, D., {van Kempen}, T.~A., {McCoey}, C., {Fuente}, A.,
  {Caselli}, P., {Kristensen}, L.~E., {et~al.} 2010, \aap, 518, L86

\bibitem[{{Fischer} {et~al.}(2010){Fischer}, {Megeath}, {Ali}, {Tobin},
  {Osorio}, {Allen}, {Kryukova}, {Stanke}, {Stutz}, {Bergin}, {Calvet}, {di
  Francesco}, {Furlan}, {Hartmann}, {Henning}, {Krause}, {Manoj}, {Maret},
  {Muzerolle}, {Myers}, {Neufeld}, {Pontoppidan}, {Poteet}, {Watson}, \&
  {Wilson}}]{fischer10}
{Fischer}, W.~J., {Megeath}, S.~T., {Ali}, B., {Tobin}, J.~J., {Osorio}, M.,
  {Allen}, L.~E., {Kryukova}, E., {et~al.} 2010, \aap, 518, L122

\bibitem[{{Fischer} {et~al.}(2012){Fischer}, {Megeath}, {Tobin}, {Stutz},
  {Ali}, {Remming}, {Kounkel}, {Stanke}, {Henning}, {Manoj}, \&
  {Wilson}}]{fischer12}
{Fischer}, W.~J., {Megeath}, S.~T., {Tobin}, J.~J., {Stutz}, A.~M., {Ali}, B.,
  {Remming}, I., {Kounkel}, M., {et~al.} 2012, \apj

\bibitem[{{Giannini} {et~al.}(2001){Giannini}, {Nisini}, \&
  {Lorenzetti}}]{giannini01}
{Giannini}, T., {Nisini}, B., \& {Lorenzetti}, D. 2001, \apj, 555, 40

\bibitem[{{Giannini} {et~al.}(2011){Giannini}, {Nisini}, {Neufeld}, {Yuan},
  {Antoniucci}, \& {Gusdorf}}]{giannini11}
{Giannini}, T., {Nisini}, B., {Neufeld}, D., {Yuan}, Y., {Antoniucci}, S., \&
  {Gusdorf}, A. 2011, \apj, 738, 80

\bibitem[{{Goicoechea} {et~al.}(2012){Goicoechea}, {Cernicharo}, {Karska},
  {Herczeg}, {Polehampton}, {Wampfler}, {Kristensen}, {van Dishoeck},
  {Etxaluze}, {Berne}, \& {Visser}}]{goico12}
{Goicoechea}, J.~R., {Cernicharo}, J., {Karska}, A., {Herczeg}, G.~J.,
  {Polehampton}, E.~T., {Wampfler}, S.~F., {Kristensen}, L.~E., {et~al.} 2012,
  ArXiv e-prints

\bibitem[{{Goldsmith} \& {Langer}(1999)}]{gl99}
{Goldsmith}, P.~F. \& {Langer}, W.~D. 1999, \apj, 517, 209

\bibitem[{{Hartigan} {et~al.}(1995){Hartigan}, {Edwards}, \&
  {Ghandour}}]{hartigan95}
{Hartigan}, P., {Edwards}, S., \& {Ghandour}, L. 1995, \apj, 452, 736

\bibitem[{{Hartmann}(2009)}]{hart09}
{Hartmann}, L. 2009, {Accretion Processes in Star Formation: Second Edition},
  ed. {Hartmann, L.} (Cambridge University Press)

\bibitem[{{Herczeg} {et~al.}(2012){Herczeg}, {Karska}, {Bruderer},
  {Kristensen}, {van Dishoeck}, {J{\o}rgensen}, {Visser}, {Wampfler}, {Bergin},
  {Y{\i}ld{\i}z}, {Pontoppidan}, \& {Gracia-Carpio}}]{herczeg12}
{Herczeg}, G.~J., {Karska}, A., {Bruderer}, S., {Kristensen}, L.~E., {van
  Dishoeck}, E.~F., {J{\o}rgensen}, J.~K., {Visser}, R., {et~al.} 2012, \aap,
  540, A84

\bibitem[{{Hollenbach} \& {McKee}(1989)}]{hm89}
{Hollenbach}, D. \& {McKee}, C.~F. 1989, \apj, 342, 306

\bibitem[{{Houck} {et~al.}(2004){Houck}, {Roellig}, {van Cleve}, {Forrest},
  {Herter}, {Lawrence}, {Matthews}, {Reitsema}, {Soifer}, {Watson}, {Weedman},
  {Huisjen}, {Troeltzsch}, {Barry}, {Bernard-Salas}, {Blacken}, {Brandl},
  {Charmandaris}, {Devost}, {Gull}, {Hall}, {Henderson}, {Higdon}, {Pirger},
  {Schoenwald}, {Sloan}, {Uchida}, {Appleton}, {Armus}, {Burgdorf},
  {Fajardo-Acosta}, {Grillmair}, {Ingalls}, {Morris}, \& {Teplitz}}]{houck04}
{Houck}, J.~R., {Roellig}, T.~L., {van Cleve}, J., {Forrest}, W.~J., {Herter},
  T., {Lawrence}, C.~R., {Matthews}, K., {et~al.} 2004, \apjs, 154, 18

\bibitem[{{Isobe} {et~al.}(1986){Isobe}, {Feigelson}, \& {Nelson}}]{isobe86}
{Isobe}, T., {Feigelson}, E.~D., \& {Nelson}, P.~I. 1986, \apj, 306, 490

\bibitem[{{Jaffe} {et~al.}(1987){Jaffe}, {Harris}, \& {Genzel}}]{jaffe87}
{Jaffe}, D.~T., {Harris}, A.~I., \& {Genzel}, R. 1987, \apj, 316, 231

\bibitem[{{Kaufman} \& {Neufeld}(1996)}]{kn96a}
{Kaufman}, M.~J. \& {Neufeld}, D.~A. 1996, \apj, 456, 611

\bibitem[{{Kenyon} {et~al.}(1993){Kenyon}, {Calvet}, \& {Hartmann}}]{kch93}
{Kenyon}, S.~J., {Calvet}, N., \& {Hartmann}, L. 1993, \apj, 414, 676

\bibitem[{{Kenyon} \& {Hartmann}(1995)}]{kenhart95}
{Kenyon}, S.~J. \& {Hartmann}, L. 1995, \apjs, 101, 117

\bibitem[{{Kryukova} {et~al.}(2012){Kryukova}, {Megeath}, {Gutermuth},
  {Pipher}, {Allen}, {Allen}, {Myers}, \& {Muzerolle}}]{kry12}
{Kryukova}, E., {Megeath}, S.~T., {Gutermuth}, R.~A., {Pipher}, J., {Allen},
  T.~S., {Allen}, L.~E., {Myers}, P.~C., \& {Muzerolle}, J. 2012, \aj, 144, 31

\bibitem[{{Lada}(1987)}]{lada87}
{Lada}, C.~J. 1987, in IAU Symposium, Vol. 115, Star Forming Regions, ed.
  M.~{Peimbert} \& J.~{Jugaku}, 1--17

\bibitem[{{Lavalley} {et~al.}(1992){Lavalley}, {Isobe}, \& {Feigelson}}]{lav92}
{Lavalley}, M., {Isobe}, T., \& {Feigelson}, E. 1992, in Astronomical Society
  of the Pacific Conference Series, Vol.~25, Astronomical Data Analysis
  Software and Systems I, ed. D.~M. {Worrall}, C.~{Biemesderfer}, \&
  J.~{Barnes}, 245

\bibitem[{{Lee} {et~al.}(2009){Lee}, {Hirano}, {Palau}, {Ho}, {Bourke},
  {Zhang}, \& {Shang}}]{leecf09}
{Lee}, C.-F., {Hirano}, N., {Palau}, A., {Ho}, P.~T.~P., {Bourke}, T.~L.,
  {Zhang}, Q., \& {Shang}, H. 2009, \apj, 699, 1584

\bibitem[{{McKee} \& {Ostriker}(2007)}]{mo07}
{McKee}, C.~F. \& {Ostriker}, E.~C. 2007, \araa, 45, 565

\bibitem[{{Megeath} {et~al.}(2012){Megeath}, {Gutermuth}, {Muzerolle},
  {Kryukova}, {Flaherty}, {Hora}, {Allen}, {Hartmann}, {Myers}, {Pipher},
  {Stauffer}, {Young}, \& {Fazio}}]{tm12}
{Megeath}, S.~T., {Gutermuth}, R., {Muzerolle}, J., {Kryukova}, E., {Flaherty},
  K., {Hora}, J., {Allen}, L., {et~al.} 2012, ArXiv e-prints

\bibitem[{{Myers} \& {Ladd}(1993)}]{ml93}
{Myers}, P.~C. \& {Ladd}, E.~F. 1993, \apjl, 413, L47

\bibitem[{{Neufeld}(2012)}]{neufeld12}
{Neufeld}, D.~A. 2012, \apj, 749, 125

\bibitem[{{Neufeld} \& {Hollenbach}(1994)}]{nh94}
{Neufeld}, D.~A. \& {Hollenbach}, D.~J. 1994, \apj, 428, 170

\bibitem[{{Neufeld} {et~al.}(2006){Neufeld}, {Melnick}, {Sonnentrucker},
  {Bergin}, {Green}, {Kim}, {Watson}, {Forrest}, \& {Pipher}}]{neufeld06}
{Neufeld}, D.~A., {Melnick}, G.~J., {Sonnentrucker}, P., {Bergin}, E.~A.,
  {Green}, J.~D., {Kim}, K.~H., {Watson}, D.~M., {Forrest}, W.~J., \& {Pipher},
  J.~L. 2006, \apj, 649, 816

\bibitem[{{Neufeld} {et~al.}(2009){Neufeld}, {Nisini}, {Giannini}, {Melnick},
  {Bergin}, {Yuan}, {Maret}, {Tolls}, {G{\"u}sten}, \& {Kaufman}}]{neufeld09}
{Neufeld}, D.~A., {Nisini}, B., {Giannini}, T., {Melnick}, G.~J., {Bergin},
  E.~A., {Yuan}, Y., {Maret}, S., {et~al.} 2009, \apj, 706, 170

\bibitem[{{Neufeld} \& {Yuan}(2008)}]{neufeld08}
{Neufeld}, D.~A. \& {Yuan}, Y. 2008, \apj, 678, 974

\bibitem[{{Nisini} {et~al.}(1999){Nisini}, {Benedettini}, {Giannini}, {Caux},
  {di Giorgio}, {Liseau}, {Lorenzetti}, {Molinari}, {Saraceno}, {Smith},
  {Spinoglio}, \& {White}}]{nisini99a}
{Nisini}, B., {Benedettini}, M., {Giannini}, T., {Caux}, E., {di Giorgio},
  A.~M., {Liseau}, R., {Lorenzetti}, D., {et~al.} 1999, \aap, 350, 529

\bibitem[{{Nisini} {et~al.}(2002){Nisini}, {Giannini}, \&
  {Lorenzetti}}]{nisini02}
{Nisini}, B., {Giannini}, T., \& {Lorenzetti}, D. 2002, \apj, 574, 246

\bibitem[{{Nisini} {et~al.}(1997){Nisini}, {Saraceno}, {Ceccarelli},
  {Giannini}, {Molinari}, {Tommasi}, {Spinoglio}, \& {White}}]{nisini97}
{Nisini}, B., {Saraceno}, P., {Ceccarelli}, C., {Giannini}, T., {Molinari}, S.,
  {Tommasi}, E., {Spinoglio}, L., \& {White}, G.~J. 1997, in ESA Special
  Publication, Vol. 401, The Far Infrared and Submillimetre Universe., ed.
  {A.~Wilson}, 321

\bibitem[{{Ott}(2010)}]{ott10}
{Ott}, S. 2010, in Astronomical Society of the Pacific Conference Series, Vol.
  434, Astronomical Data Analysis Software and Systems XIX, ed. {Y.~Mizumoto,
  K.-I.~Morita, \& M.~Ohishi}, 139

\bibitem[{{Pilbratt} {et~al.}(2010){Pilbratt}, {Riedinger}, {Passvogel},
  {Crone}, {Doyle}, {Gageur}, {Heras}, {Jewell}, {Metcalfe}, {Ott}, \&
  {Schmidt}}]{pilbratt10}
{Pilbratt}, G.~L., {Riedinger}, J.~R., {Passvogel}, T., {Crone}, G., {Doyle},
  D., {Gageur}, U., {Heras}, A.~M., {et~al.} 2010, \aap, 518, L1

\bibitem[{{Poglitsch} {et~al.}(2010){Poglitsch}, {Waelkens}, {Geis},
  {Feuchtgruber}, {Vandenbussche}, {Rodriguez}, {Krause}, {Renotte}, {van
  Hoof}, {Saraceno}, {Cepa}, {Kerschbaum}, {Agn{\`e}se}, {Ali}, {Altieri},
  {Andreani}, {Augueres}, {Balog}, {Barl}, {Bauer}, {Belbachir}, {Benedettini},
  {Billot}, {Boulade}, {Bischof}, {Blommaert}, {Callut}, {Cara}, {Cerulli},
  {Cesarsky}, {Contursi}, {Creten}, {De Meester}, {Doublier}, {Doumayrou},
  {Duband}, {Exter}, {Genzel}, {Gillis}, {Gr{\"o}zinger}, {Henning},
  {Herreros}, {Huygen}, {Inguscio}, {Jakob}, {Jamar}, {Jean}, {de Jong},
  {Katterloher}, {Kiss}, {Klaas}, {Lemke}, {Lutz}, {Madden}, {Marquet},
  {Martignac}, {Mazy}, {Merken}, {Montfort}, {Morbidelli}, {M{\"u}ller},
  {Nielbock}, {Okumura}, {Orfei}, {Ottensamer}, {Pezzuto}, {Popesso},
  {Putzeys}, {Regibo}, {Reveret}, {Royer}, {Sauvage}, {Schreiber}, {Stegmaier},
  {Schmitt}, {Schubert}, {Sturm}, {Thiel}, {Tofani}, {Vavrek}, {Wetzstein},
  {Wieprecht}, \& {Wiezorrek}}]{pog10}
{Poglitsch}, A., {Waelkens}, C., {Geis}, N., {Feuchtgruber}, H.,
  {Vandenbussche}, B., {Rodriguez}, L., {Krause}, O., {et~al.} 2010, \aap, 518,
  L2+

\bibitem[{{Poteet} {et~al.}(2011){Poteet}, {Megeath}, {Watson}, {Calvet},
  {Remming}, {McClure}, {Sargent}, {Fischer}, {Furlan}, {Allen}, {Bjorkman},
  {Hartmann}, {Muzerolle}, {Tobin}, \& {Ali}}]{poteet11}
{Poteet}, C.~A., {Megeath}, S.~T., {Watson}, D.~M., {Calvet}, N., {Remming},
  I.~S., {McClure}, M.~K., {Sargent}, B.~A., {et~al.} 2011, \apjl, 733, L32

\bibitem[{{Reipurth} \& {Bally}(2001)}]{rb01}
{Reipurth}, B. \& {Bally}, J. 2001, \araa, 39, 403

\bibitem[{{Rieke} {et~al.}(2004){Rieke}, {Young}, {Engelbracht}, {Kelly},
  {Low}, {Haller}, {Beeman}, {Gordon}, {Stansberry}, {Misselt}, {Cadien},
  {Morrison}, {Rivlis}, {Latter}, {Noriega-Crespo}, {Padgett}, {Stapelfeldt},
  {Hines}, {Egami}, {Muzerolle}, {Alonso-Herrero}, {Blaylock}, {Dole}, {Hinz},
  {Le Floc'h}, {Papovich}, {P{\'e}rez-Gonz{\'a}lez}, {Smith}, {Su}, {Bennett},
  {Frayer}, {Henderson}, {Lu}, {Masci}, {Pesenson}, {Rebull}, {Rho}, {Keene},
  {Stolovy}, {Wachter}, {Wheaton}, {Werner}, \& {Richards}}]{rieke04}
{Rieke}, G.~H., {Young}, E.~T., {Engelbracht}, C.~W., {Kelly}, D.~M., {Low},
  F.~J., {Haller}, E.~E., {Beeman}, J.~W., {et~al.} 2004, \apjs, 154, 25

\bibitem[{{Santiago-Garc{\'{\i}}a} {et~al.}(2009){Santiago-Garc{\'{\i}}a},
  {Tafalla}, {Johnstone}, \& {Bachiller}}]{santiago09}
{Santiago-Garc{\'{\i}}a}, J., {Tafalla}, M., {Johnstone}, D., \& {Bachiller},
  R. 2009, \aap, 495, 169

\bibitem[{{Saraceno} {et~al.}(1999{\natexlab{a}}){Saraceno}, {Benedettini}, {di
  Giorgio}, {Giannini}, {Nisini}, {Lorenzetti}, {Molinari}, {Pezzuto},
  {Spinoglio}, {Tommasi}, {Clegg}, {Correia}, {Griffin}, {Kaufman}, {Leeks},
  {White}, {Caux}, {Liseau}, \& {Smith}}]{saraceno99a}
{Saraceno}, P., {Benedettini}, M., {di Giorgio}, A.~M., {Giannini}, T.,
  {Nisini}, B., {Lorenzetti}, D., {Molinari}, S., {et~al.} 1999{\natexlab{a}},
  in The Physics and Chemistry of the Interstellar Medium, ed. {V.~Ossenkopf,
  J.~Stutzki, \& G.~Winnewisser}, 279

\bibitem[{{Saraceno} {et~al.}(1999{\natexlab{b}}){Saraceno}, {Nisini},
  {Benedettini}, {di Giorgio}, {Giannini}, {Kaufman}, {Lorenzetti}, {Molinari},
  {Pezzuto}, {Spinoglio}, {Tommasi}, {Caux}, {Ceccarelli}, {Clegg}, {Correia},
  {Griffin}, {Leeks}, {Liseau}, {Smith}, \& {White}}]{saraceno99b}
{Saraceno}, P., {Nisini}, B., {Benedettini}, M., {di Giorgio}, A.~M.,
  {Giannini}, T., {Kaufman}, M.~J., {Lorenzetti}, D., {et~al.}
  1999{\natexlab{b}}, in ESA Special Publication, Vol. 427, The Universe as
  Seen by ISO, ed. {P.~Cox \& M.~Kessler}, 575

\bibitem[{{Shu} {et~al.}(1987){Shu}, {Adams}, \& {Lizano}}]{shu87}
{Shu}, F.~H., {Adams}, F.~C., \& {Lizano}, S. 1987, \araa, 25, 23

\bibitem[{{Skrutskie} {et~al.}(2006){Skrutskie}, {Cutri}, {Stiening},
  {Weinberg}, {Schneider}, {Carpenter}, {Beichman}, {Capps}, {Chester},
  {Elias}, {Huchra}, {Liebert}, {Lonsdale}, {Monet}, {Price}, {Seitzer},
  {Jarrett}, {Kirkpatrick}, {Gizis}, {Howard}, {Evans}, {Fowler}, {Fullmer},
  {Hurt}, {Light}, {Kopan}, {Marsh}, {McCallon}, {Tam}, {Van Dyk}, \&
  {Wheelock}}]{skrut06}
{Skrutskie}, M.~F., {Cutri}, R.~M., {Stiening}, R., {Weinberg}, M.~D.,
  {Schneider}, S., {Carpenter}, J.~M., {Beichman}, C., {et~al.} 2006, \aj, 131,
  1163

\bibitem[{{Smith} \& {Brand}(1990)}]{sb90}
{Smith}, M.~D. \& {Brand}, P.~W.~J.~L. 1990, \mnras, 245, 108

\bibitem[{{Smith} {et~al.}(1991){Smith}, {Brand}, \& {Moorhouse}}]{sbm91}
{Smith}, M.~D., {Brand}, P.~W.~J.~L., \& {Moorhouse}, A. 1991, \mnras, 248, 451

\bibitem[{{Spaans} {et~al.}(1995){Spaans}, {Hogerheijde}, {Mundy}, \& {van
  Dishoeck}}]{spaans95}
{Spaans}, M., {Hogerheijde}, M.~R., {Mundy}, L.~G., \& {van Dishoeck}, E.~F.
  1995, \apjl, 455, L167

\bibitem[{{Stacey} {et~al.}(1983){Stacey}, {Kurtz}, {Smyers}, \&
  {Harwit}}]{stacey83}
{Stacey}, G.~J., {Kurtz}, N.~T., {Smyers}, S.~D., \& {Harwit}, M. 1983, \mnras,
  202, 25P

\bibitem[{{Stacey} {et~al.}(1982){Stacey}, {Kurtz}, {Smyers}, {Harwit},
  {Russell}, \& {Melnick}}]{stacey82}
{Stacey}, G.~J., {Kurtz}, N.~T., {Smyers}, S.~D., {Harwit}, M., {Russell},
  R.~W., \& {Melnick}, G. 1982, \apjl, 257, L37

\bibitem[{{Stahler} \& {Palla}(2005)}]{sp05}
{Stahler}, S.~W. \& {Palla}, F. 2005, {The Formation of Stars}, ed. {Stahler,
  S.~W.~\& Palla, F.}

\bibitem[{{Stanke} {et~al.}(2010){Stanke}, {Stutz}, {Tobin}, {Ali}, {Megeath},
  {Krause}, {Linz}, {Allen}, {Bergin}, {Calvet}, {di Francesco}, {Fischer},
  {Furlan}, {Hartmann}, {Henning}, {Manoj}, {Maret}, {Muzerolle}, {Myers},
  {Neufeld}, {Osorio}, {Pontoppidan}, {Poteet}, {Watson}, \&
  {Wilson}}]{stanke10}
{Stanke}, T., {Stutz}, A.~M., {Tobin}, J.~J., {Ali}, B., {Megeath}, S.~T.,
  {Krause}, O., {Linz}, H., {et~al.} 2010, \aap, 518, L94

\bibitem[{{Storey} {et~al.}(1981){Storey}, {Watson}, {Townes}, {Haller}, \&
  {Hansen}}]{storey81}
{Storey}, J.~W.~V., {Watson}, D.~M., {Townes}, C.~H., {Haller}, E.~E., \&
  {Hansen}, W.~L. 1981, \apj, 247, 136

\bibitem[{{Takahashi} \& {Ho}(2012)}]{takahashi12}
{Takahashi}, S. \& {Ho}, P.~T.~P. 2012, \apjl, 745, L10

\bibitem[{{Takahashi} {et~al.}(2008){Takahashi}, {Saito}, {Ohashi}, {Kusakabe},
  {Takakuwa}, {Shimajiri}, {Tamura}, \& {Kawabe}}]{takahashi08}
{Takahashi}, S., {Saito}, M., {Ohashi}, N., {Kusakabe}, N., {Takakuwa}, S.,
  {Shimajiri}, Y., {Tamura}, M., \& {Kawabe}, R. 2008, \apj, 688, 344

\bibitem[{{Terebey} {et~al.}(1984){Terebey}, {Shu}, \& {Cassen}}]{tsc84}
{Terebey}, S., {Shu}, F.~H., \& {Cassen}, P. 1984, \apj, 286, 529

\bibitem[{{Tobin} {et~al.}(2008){Tobin}, {Hartmann}, {Calvet}, \&
  {D'Alessio}}]{tobin08}
{Tobin}, J.~J., {Hartmann}, L., {Calvet}, N., \& {D'Alessio}, P. 2008, \apj,
  679, 1364

\bibitem[{{van Dishoeck}(2004)}]{vd04}
{van Dishoeck}, E.~F. 2004, \araa, 42, 119

\bibitem[{{van Dishoeck} {et~al.}(2009){van Dishoeck}, {van Kempen}, \&
  {G{\"u}sten}}]{vd09}
{van Dishoeck}, E.~F., {van Kempen}, T.~A., \& {G{\"u}sten}, R. 2009, in
  Astronomical Society of the Pacific Conference Series, Vol. 417,
  Submillimeter Astrophysics and Technology: a Symposium Honoring Thomas G.
  Phillips, ed. {D.~C.~Lis, J.~E.~Vaillancourt, P.~F.~Goldsmith, T.~A.~Bell,
  N.~Z.~Scoville, \& J.~Zmuidzinas}, 203

\bibitem[{{van Kempen} {et~al.}(2010{\natexlab{a}}){van Kempen}, {Green},
  {Evans}, {van Dishoeck}, {Kristensen}, {Herczeg}, {Mer{\'{\i}}n}, {Lee},
  {J{\o}rgensen}, {Bouwman}, {Acke}, {Adamkovics}, {Augereau}, {Bergin},
  {Blake}, {Brown}, {Carr}, {Chen}, {Cieza}, {Dominik}, {Dullemond}, {Dunham},
  {Glassgold}, {G{\"u}del}, {Harvey}, {Henning}, {Hogerheijde}, {Jaffe}, {Kim},
  {Knez}, {Lacy}, {Maret}, {Meeus}, {Meijerink}, {Mulders}, {Mundy}, {Najita},
  {Olofsson}, {Pontoppidan}, {Salyk}, {Sturm}, {Visser}, {Waters}, {Waelkens},
  \& {Y{\i}ld{\i}z}}]{vankemp10b}
{van Kempen}, T.~A., {Green}, J.~D., {Evans}, N.~J., {van Dishoeck}, E.~F.,
  {Kristensen}, L.~E., {Herczeg}, G.~J., {Mer{\'{\i}}n}, B., {et~al.}
  2010{\natexlab{a}}, \aap, 518, L128

\bibitem[{{van Kempen} {et~al.}(2010{\natexlab{b}}){van Kempen}, {Kristensen},
  {Herczeg}, {Visser}, {van Dishoeck}, {Wampfler}, {Bruderer}, {Benz}, {Doty},
  {Brinch}, {Hogerheijde}, {J{\o}rgensen}, {Tafalla}, {Neufeld}, {Bachiller},
  {Baudry}, {Benedettini}, {Bergin}, {Bjerkeli}, {Blake}, {Bontemps}, {Braine},
  {Caselli}, {Cernicharo}, {Codella}, {Daniel}, {di Giorgio}, {Dominik},
  {Encrenaz}, {Fich}, {Fuente}, {Giannini}, {Goicoechea}, {de Graauw},
  {Helmich}, {Herpin}, {Jacq}, {Johnstone}, {Kaufman}, {Larsson}, {Lis},
  {Liseau}, {Marseille}, {McCoey}, {Melnick}, {Nisini}, {Olberg}, {Parise},
  {Pearson}, {Plume}, {Risacher}, {Santiago-Garc{\'{\i}}a}, {Saraceno},
  {Shipman}, {van der Tak}, {Wyrowski}, {Y{\i}ld{\i}z}, {Ciechanowicz},
  {Dubbeldam}, {Glenz}, {Huisman}, {Lin}, {Morris}, {Murphy}, \&
  {Trappe}}]{vankemp10a}
{van Kempen}, T.~A., {Kristensen}, L.~E., {Herczeg}, G.~J., {Visser}, R., {van
  Dishoeck}, E.~F., {Wampfler}, S.~F., {Bruderer}, S., {et~al.}
  2010{\natexlab{b}}, \aap, 518, L121

\bibitem[{{van Kempen} {et~al.}(2009{\natexlab{a}}){van Kempen}, {van
  Dishoeck}, {G{\"u}sten}, {Kristensen}, {Schilke}, {Hogerheijde}, {Boland},
  {Menten}, \& {Wyrowski}}]{vankemp09b}
{van Kempen}, T.~A., {van Dishoeck}, E.~F., {G{\"u}sten}, R., {Kristensen},
  L.~E., {Schilke}, P., {Hogerheijde}, M.~R., {Boland}, W., {Menten}, K.~M., \&
  {Wyrowski}, F. 2009{\natexlab{a}}, \aap, 507, 1425

\bibitem[{{van Kempen} {et~al.}(2009{\natexlab{b}}){van Kempen}, {van
  Dishoeck}, {G{\"u}sten}, {Kristensen}, {Schilke}, {Hogerheijde}, {Boland},
  {Nefs}, {Menten}, {Baryshev}, \& {Wyrowski}}]{vankemp09a}
{van Kempen}, T.~A., {van Dishoeck}, E.~F., {G{\"u}sten}, R., {Kristensen},
  L.~E., {Schilke}, P., {Hogerheijde}, M.~R., {Boland}, W., {et~al.}
  2009{\natexlab{b}}, \aap, 501, 633

\bibitem[{{Visser} {et~al.}(2012){Visser}, {Kristensen}, {Bruderer}, {van
  Dishoeck}, {Herczeg}, {Brinch}, {Doty}, {Harsono}, \& {Wolfire}}]{visser12}
{Visser}, R., {Kristensen}, L.~E., {Bruderer}, S., {van Dishoeck}, E.~F.,
  {Herczeg}, G.~J., {Brinch}, C., {Doty}, S.~D., {Harsono}, D., \& {Wolfire},
  M.~G. 2012, \aap, 537, A55

\bibitem[{{Watson} {et~al.}(1985){Watson}, {Genzel}, {Townes}, \&
  {Storey}}]{watson85b}
{Watson}, D.~M., {Genzel}, R., {Townes}, C.~H., \& {Storey}, J.~W.~V. 1985,
  \apj, 298, 316

\bibitem[{{Watson} {et~al.}(1980){Watson}, {Storey}, {Townes}, {Haller}, \&
  {Hansen}}]{watson80}
{Watson}, D.~M., {Storey}, J.~W.~V., {Townes}, C.~H., {Haller}, E.~E., \&
  {Hansen}, W.~L. 1980, \apjl, 239, L129

\bibitem[{{Whitney} {et~al.}(2003){Whitney}, {Wood}, {Bjorkman}, \&
  {Wolff}}]{whitney03}
{Whitney}, B.~A., {Wood}, K., {Bjorkman}, J.~E., \& {Wolff}, M.~J. 2003, \apj,
  591, 1049

\bibitem[{{Wilking}(1989)}]{wilk89}
{Wilking}, B.~A. 1989, \pasp, 101, 229

\bibitem[{{Williams} {et~al.}(2003){Williams}, {Plambeck}, \&
  {Heyer}}]{williams03}
{Williams}, J.~P., {Plambeck}, R.~L., \& {Heyer}, M.~H. 2003, \apj, 591, 1025

\bibitem[{{Yang} {et~al.}(2010){Yang}, {Stancil}, {Balakrishnan}, \&
  {Forrey}}]{yang10}
{Yang}, B., {Stancil}, P.~C., {Balakrishnan}, N., \& {Forrey}, R.~C. 2010,
  \apj, 718, 1062

\bibitem[{{Y{\i}ld{\i}z} {et~al.}(2012){Y{\i}ld{\i}z}, {Kristensen}, {van
  Dishoeck}, {Belloche}, {van Kempen}, {Hogerheijde}, {G{\"u}sten}, \& {van der
  Marel}}]{yildiz12}
{Y{\i}ld{\i}z}, U.~A., {Kristensen}, L.~E., {van Dishoeck}, E.~F., {Belloche},
  A., {van Kempen}, T.~A., {Hogerheijde}, M.~R., {G{\"u}sten}, R., \& {van der
  Marel}, N. 2012, \aap, 542, A86

\bibitem[{{Y{\i}ld{\i}z} {et~al.}(2010){Y{\i}ld{\i}z}, {van Dishoeck},
  {Kristensen}, {Visser}, {J{\o}rgensen}, {Herczeg}, {van Kempen},
  {Hogerheijde}, {Doty}, {Benz}, {Bruderer}, {Wampfler}, {Deul}, {Bachiller},
  {Baudry}, {Benedettini}, {Bergin}, {Bjerkeli}, {Blake}, {Bontemps}, {Braine},
  {Caselli}, {Cernicharo}, {Codella}, {Daniel}, {di Giorgio}, {Dominik},
  {Encrenaz}, {Fich}, {Fuente}, {Giannini}, {Goicoechea}, {de Graauw},
  {Helmich}, {Herpin}, {Jacq}, {Johnstone}, {Larsson}, {Lis}, {Liseau}, {Liu},
  {Marseille}, {McCoey}, {Melnick}, {Neufeld}, {Nisini}, {Olberg}, {Parise},
  {Pearson}, {Plume}, {Risacher}, {Santiago-Garc{\'{\i}}a}, {Saraceno},
  {Shipman}, {Tafalla}, {Tielens}, {van der Tak}, {Wyrowski}, {Dieleman},
  {Jellema}, {Ossenkopf}, {Schieder}, \& {Stutzki}}]{yildiz10}
{Y{\i}ld{\i}z}, U.~A., {van Dishoeck}, E.~F., {Kristensen}, L.~E., {Visser},
  R., {J{\o}rgensen}, J.~K., {Herczeg}, G.~J., {van Kempen}, T.~A., {et~al.}
  2010, \aap, 521, L40

\bibitem[{{Yuan} \& {Neufeld}(2011)}]{yuan11}
{Yuan}, Y. \& {Neufeld}, D.~A. 2011, \apj, 726, 76

\end{thebibliography}
\end{document}